\documentclass[useAMS,usenatbib]{mn2e}
\pdfoutput=1
\usepackage{graphicx} 
\usepackage{subfigure}
\usepackage{color}
\usepackage{amssymb}
\usepackage{float}

\title[AGN candidates selected by MIPS 24~$\mu$m variability]{Selection of AGN
  candidates in the GOODS-South Field through SPITZER/MIPS 24~$\umu$m
  variability} 

\author[Garc\'ia-Gonz\'alez et al.]{Judit Garc\'ia-Gonz\'alez,$^1$
  Almudena Alonso-Herrero,$^{1,2}$ Pablo G.~P\'erez-Gonz\'alez,$^3$ 
\newauthor 
Antonio Hern\'an-Caballero,$^1$ Vicki~L.~Sarajedini,$^4$ and V\'ictor~Villar$^3$\\
$^1$ Instituto de F\'isica de Cantabria, CSIC-UC, Avenida de los Castros s/n, 39005 Santander, Spain\\
$^2$ Augusto Gonz\'alez Linares Senior Research Fellow\\
$^3$ Departamento de Astrof\'isica, Facultad de CC. F\'isicas, Universidad Complutense de Madrid, 28040 Madrid, Spain\\
$^4$ Departament of Astronomy, University of Florida, Gainesville, FL 32611, USA}

\begin{document}

\date{}

\pagerange{\pageref{firstpage}--\pageref{lastpage}} \pubyear{2014}

\label{firstpage}

\maketitle

\begin{abstract}

  We present a study of galaxies showing mid-infrared variability in data taken in the deepest {\em Spitzer/MIPS} 24~$\umu$m surveys in the GOODS-South field. We divide the dataset in epochs and subepochs to study the long-term (months-years) and the short-term (days) variability. We use a $\chi^2$-statistics method to select AGN candidates with a probability $\leq$ 1\% that the observed variability is due to statistical errors alone. We find 39 (1.7\% of the parent sample) sources that show long-term variability and 55 (2.2\% of the parent sample) showing short-term variability. That is, 0.03 sources $\times$ arcmin$^{-2}$ for both, long-term and short-term variable sources. After removing the expected number of false positives inherent to the method, the estimated percentages are 1.0\% and 1.4\% of the parent sample for the long-term and short-term respectively. We compare our candidates with AGN selected in the X-ray and radio  bands, and AGN candidates selected by their IR emission. Approximately, 50\% of the MIPS 24 $\umu$m variable sources would be identified as AGN with these other methods. Therefore, MIPS 24 $\umu$m variability is a new method to identify AGN candidates, possibly dust obscured and low luminosity AGN, that might be missed by other methods. However, the contribution of the MIPS 24 $\umu$m variable identified AGN to the general AGN population is small ($\leq$ 13\%) in GOODS-South.

\end{abstract}


\section{Introduction}

Supermassive black holes (SMBH) ($M_{SMBH} \textgreater
10^6~M_{\sun}$) are believed to exist in the center of all
galaxies containing a 
significant bulge component \citep{KR}. Furthermore, the bulge
properties, such as stellar mass and velocity dispersion, are
correlated with the black hole mass (\citealt{Ma}; \citealt{MH};
\citealt{HR}). From these relations one may deduce that the bulges of
the galaxies and the supermassive black holes evolve together,
probably mutually regulating each other, or at least, sharing
formation and growing mechanisms (see \citealt{AlexanderHickox} for a review). 

In order to study the connection between galaxy and SMBH formation, we
must study those galaxies where the SMBH is growing. This phenomenon
is revealed in a variety of ways and it is generally referred to
  as active galactic nucleus or AGN. There are different methods to identify
AGN. Conventionally, the most direct and used one has been optical
spectroscopy (\citealt*{BPT}; \citealt{VO}). AGN can be identified
through the presence of broad lines (FWHM  $\gtrsim$
1000~km~s$^{-1}$, Type 1 AGN) or by narrow lines (FWHM $\lesssim$
1000~km~s$^{-1}$) and line ratios in
diagnostic diagrams such as [OIII]/H$\beta$ vs [NII]/H$\alpha$ (Type 2
AGN). Unfortunately, spectroscopy of a large number of objects is very
expensive in terms of observing time, especially for faint sources in
cosmological fields. Furthermore, AGN lines can be obscured and/or
swamped by emission from the host galaxy (\citealt*{MFC}).  

AGN in cosmological fields are routinely identified 
by their X-ray emission (\citealt{Al}; \citealt{Brandt05}),
mid-infrared (IR) emission (\citealt{La}; \citealt{Ste};
\citealt{Almu}; \citealt{Do}; \citealt{Mateos2012,Stern2012,Assef2013}; \citealt{La13}), 
excess radio emission 
(\citealt{Do05,DelMoro}), and combinations of different emissions
\citep[e.g.,][]{MartinezSansigre05}. All these methods 
present their own biases. For example, only about 10\% of the
optically selected QSO are detected as radio sources
(\citealt{SW}), and in general there is little overlap between
  radio and IR/X-ray selected AGN \citep{Hickox09}. X-ray surveys may
miss the most obscured AGN needed to fit the  cosmic X-ray
  background \citep{Gilli07}. The 
IR methods, on the other hand, are only complete for the most luminous AGN (e.g.,
\citealt{Do}), but  they are likely to select obscured AGN not detected in
X-rays (\citealt{Mat}). 

Variability can also be used to select AGN. Practically all AGN vary
on time-scales from hours to millions of years (\citealt*{UMU}; \citealt{Hickox2014}). Any variability
detected in galaxies on human time-scales must originate in the
nuclear region, because the typical timescale for star formation variability is $\geq$ 100 Myr (\citealt{Hickox2014}). In particular low-luminosity AGN are expected to
  show stronger 
variability than the luminous ones (\citealt{Tre}). Therefore, 
  variability is likely to be an effective method to select
low-luminosity  AGN. Although the 
mechanisms that produce variability are not well understood, the main
explanations involve disk instabilities (\citealt{Pe}) or changes in
the amount of accreting material (\citealt{HB}).  

All these methods of AGN selection are complementary and each one can detect sources other
methods miss. It is therefore important to study the same region of
the sky with different methods of AGN selection. 

In the last decade a number of studies have identified AGN in the
Great Observatory Origins Deep Survey (GOODS; \citealt{Giavalisco2004}) fields. The GOODS fields
are two fields of 150~arcmin$^2$ centered 
around of the Hubble Deep Field North (HDFN; \citealt{Williams1996}) and the Chandra Deep
Field South (CDFS; \citealt{Giacconi2001}). The observations in the  GOODS fields
are amongst the deepest at all wavelengths, from
X-rays to radio. In particular, these fields have observations
from {\em Spitzer}, {\em Hubble Space
  Telescope} ({\em HST}), {\em Chandra}, {\em Herschel}, {\em XMM-Newton} and many
ground-based facilities. 

There are a number of variability studies in the GOODS fields, most of
them using optical data. The first one was made by \citet*{SGK} using
{\em V}-band data ($\lambda_c=550$~nm) from {\em HST} in two epochs separated by five years. They
found nuclear variability evidence in 16 of 217 galaxies (7\% of the
sample) with magnitudes down to 27.5. \citet{Co} conducted a similar
study using the {\em HST} {\em i}-band ($\lambda_c=775$~nm) data
from the Hubble Ultra Deep Field (HUDF; \citealt{Beckwith2006}). They determined that 1\% of the sources
(45 sources) presented significant variability. \citet{KS} conducted a study of five epoch {\em V}-band data in the GOODS
South field. They 
selected a sample of 22 mid-IR power-law sources \citep[using
the criteria of][]{Almu} and 102 X-ray
sources and found that 26\% of the sample were variable in the
optical. \citet{Tre8} used ground-based data, also in the {\em
  V}-band and obtained 132
variable AGN candidates (2.6\% of the sample). \citet*{VKG} selected
all the objects in the {\em z}-band ($\lambda_c=850$~nm) catalog in
the GOODS fields in five epochs. They found 139 variable AGN candidates
($\sim$1.3\% of the sample) in the North and South fields. \citet{Sa}
identified 85 variable galaxies ($\sim$2\% of the sample) in the North
and South fields using five epochs {\em V}-band images from the {\em
  Hubble Space Telescope} Advanced Camera for Surveys. 

X-ray variability of low luminosity X-ray sources has also been used
to identify additional AGN in the CDFS. \citet{Pa} studied 346 sources
and found that 45\% of the sources with more than 100 counts presented
X-ray variability. \citet{Yo} found that 185 of
369 AGN and 20 of 92 galaxies (i.e., low-luminosity AGN with
  $L_{\rm 0.5-8keV}<10^{42}\,{\rm erg \,s}^{-1}$)
  presented X-ray variability. 

\citet{Mo} studied radio
variability in the Extended-CDFS. They found that 1.2\% of the point sources
presented radio variability  associated with the central regions of AGN or
star-forming galaxies. 

The aim of this paper is to identify AGN through mid-IR
variability in the GOODS-South field  using 24~$\umu$m 
observations taken with  the Multiband Imaging Photometer for {\it
  Spitzer} \citep[MIPS,][]{Rie} 
on board the {\em Spitzer} Space Telescope (\citealt{We}).   
The near and mid-IR nuclear emission of AGN,  once the
  stellar component is subtracted, is believed to be due to 
hot and warm dust ($200-2000\,$K) in the dusty torus of the AGN, according to the 
  Unified Model \citep{Antonucci1993}. In this context, variability in the accretion
disk emission would cause delayed variability in the near and mid-IR as the hot and
warm dust, respectively, in the torus react to this change (see \citealt{HK} and references therein).

Our choice of using mid-IR variability allows a novel way to select low luminosity and possibly
  obscured AGN that might be otherwise missed by other techniques. Apart
  from this work, there is only other IR variability study in the
  Bo\"otes cosmological field using IRAC data (\citealt{Ko}). They used the most sensitive IRAC bands at 3.6 and 4.5 
$\umu$m and found that 1.1\% of the sources
satisfied their variability criteria. 

The paper is organized as follows: in Section \ref{data} we present
the MIR data used to detect variable sources. In Section
\ref{data-photometry} we explain the procedures followed to get
photometry of the data. In Section \ref{selection-variable-sources} we
present the statistical method used to select the variable candidates. In Section \ref{properties-variable-candidates} we present the general properties of these candidates, as well as their IRAC
properties. In Section \ref{combo-section} we present the candidates
in the Extended Chandra Deep Field South (E-CDFS), their properties, and a cross-correlation with
other AGN catalogs in the same field. The discussion 
and conclusions are given in  Section~\ref{summary}. Throughout this
work we use a cosmology
with 
H$_0=70$~km~s$^{-1}$~Mpc$^{-1}$, $\Omega_m=0.3$ and
$\Omega_{\Lambda}=0.7$.

\begin{figure}
  \includegraphics[width=0.52\textwidth]{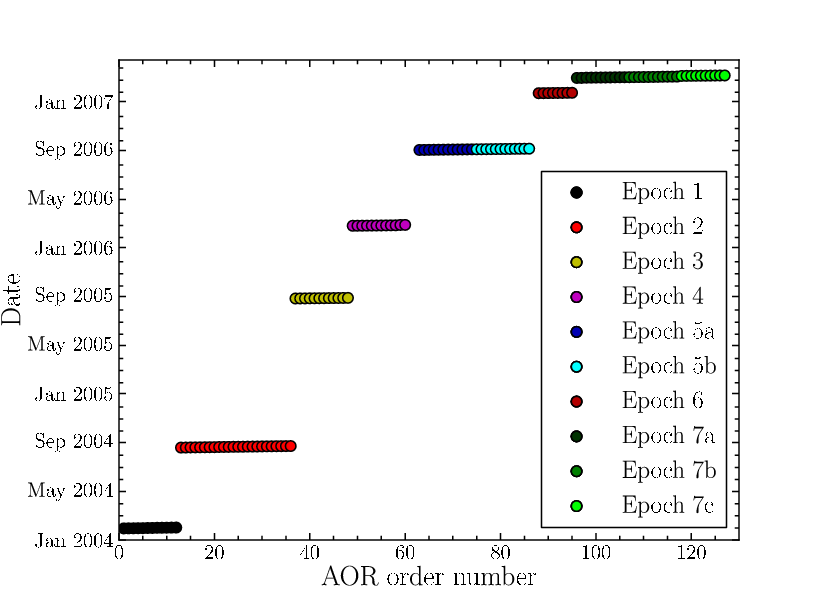}
  \caption{Summary of the different epochs available in
      the GOODS-South field with deep {\em Spitzer}/MIPS
24~$\umu$m observations. The epochs with the longest durations 
can also be divided in subepochs  to 
study short-term variability (time scales of days). The number of AORs
per epoch and the exposure time per BCD are given in Table
\ref{table-epochs}.} 
  \label{epochs}
\end{figure}

\begin{table*}
  \begin{minipage}{180mm}
    \caption{MIPS 24~$\umu$m observing programs in GOODS-South.}
    \label{table-epochs}
    \begin{tabular}{p{0.8cm}@{\hspace{0.4cm}}p{2cm}@{\hspace{0.4cm}}p{1cm}@{\hspace{0.4cm}}p{4cm}@{\hspace{0.4cm}}p{1.5cm}@{\hspace{0.4cm}}p{1.5cm}@{\hspace{0.4cm}}p{1.cm}@{\hspace{0.4cm}}p{1.cm}@{\hspace{0.4cm}}p{1.3cm}@{\hspace{0.2cm}}p{1.3cm}}
      \hline
      Epoch & PI & Program  & Program name & Initial date & Final
      date & AORs$^1$ & BCDs$^2$ & $t_{\rm exp}$$^3$ & Area \\
      &&ID&&&&&&(s)&(arcmin$^2$)\\
      \hline \hline
      1 & Rieke, G. & 81 & The Deep Infrared Sky & 29-01-2004 & 01-02-2004 & 12 & 7660 & 10 &2653\\
      2 & Dickinson, M. & 194 & Great Observatories Origins Deep Survey (GOODS), continued & 19-08-2004 & 23-08-2004 & 24 & 14974 & 30&255\\
      3 & Rieke, G. & 81 & The Deep Infrared Sky & 26-08-2005 & 28-08-2005 & 12 & 6660 & 10&1706\\
      4 & Frayer, D. T. & 20147 & Ultra-Deep MIPS-70 Imaging of GOODS CDF-S  & 24-02-2006 & 26-02-2006 & 12 & 19968 & 10&226\\
      5 & Dickinson, M. & 30948 & A deep-Wide Far-Infrared Survey of Cosmological Star Formation and AGN Activity & 01-09-2006 & 05-09-2006 & 24 & 39633 & 10&756\\
      6 & Dickinson, M. & 30948 & A deep-Wide Far-Infrared Survey of Cosmological Star Formation and AGN Activity & 22-01-2007 & 23-01-2007 & 8 & 6216 & 10&2293\\
      7 & Dickinson, M. & 30948 & A deep-Wide Far-Infrared Survey of Cosmological Star Formation and AGN Activity& 01-03-2007 & 07-03-2007 & 32 & 44274 & 10&2265\\
      \hline
    \end{tabular}
    $^1$ AOR: Astronomical Observation Request.

       $^2$ BCD: Basic Calibrated Data.

    $^3$ $t_{\rm exp}$: Exposure time per BCD.
  \end{minipage}
\end{table*}

\section{The data}
\label{data}

We compiled all the data taken around the GOODS South field with the
MIPS instrument at 24~$\umu$m by querying the {\em Spitzer} Heritage
Archive\footnote
{http://sha.ipac.caltech.edu/applicatio ns/Spitzer/SHA}. This field was
observed by {\em Spitzer} during several campaigns from January 2004
to March 2007. We focused our study on a region around
RA=$3^h32^m36^s$ (J2000) and DEC=$-27\degr48\arcmin39\arcsec$
(J2000). These data correspond to different observing proposals from
different PI, including the Guaranteed Time Observations program (GTO, PI: G. Rieke) and the GOODS program (PI: M. Dickinson). We refer the reader to  Table~\ref{table-epochs} for a detailed description of all the MIPS 24 $\umu$m observing programs in GOODS-South. We have obtained 151
AORs (Astronomical Observation Request) but only
downloaded 127 because the others were from the SWIRE (Spitzer
Wide-area InfraRed Extragalactic) survey and were not sufficiently
deep. For each epoch and subepoch, we built a mosaic with the AORs
using the software {\sevensize
  MOPEX}\footnote{http://ssc.spitzer.caltech.edu} provided by the {\em Spitzer}
Science Center (SSC).

We divided these data sets  into 7 different epochs in order to detect
variable sources (see Figure~\ref{epochs}). 
Epochs 5 and 7, which have the longest durations, can also be
divided into subepochs  to detect short term variability in time scales of
days and even of hours. In Table~\ref{table-epochs} we list the 
main information about the epochs. Each epoch has both a
  different field of view (FoV) and a different depth. As can be seen
from Table \ref{table-epochs}, Epoch 2 
has the longest exposure time per Basic Calibrated Data (BCD), resulting in the deepest
MIPS 24~$\umu$m exposure in our data sets (see below). 

\begin{figure}
  \includegraphics[width=0.52\textwidth]{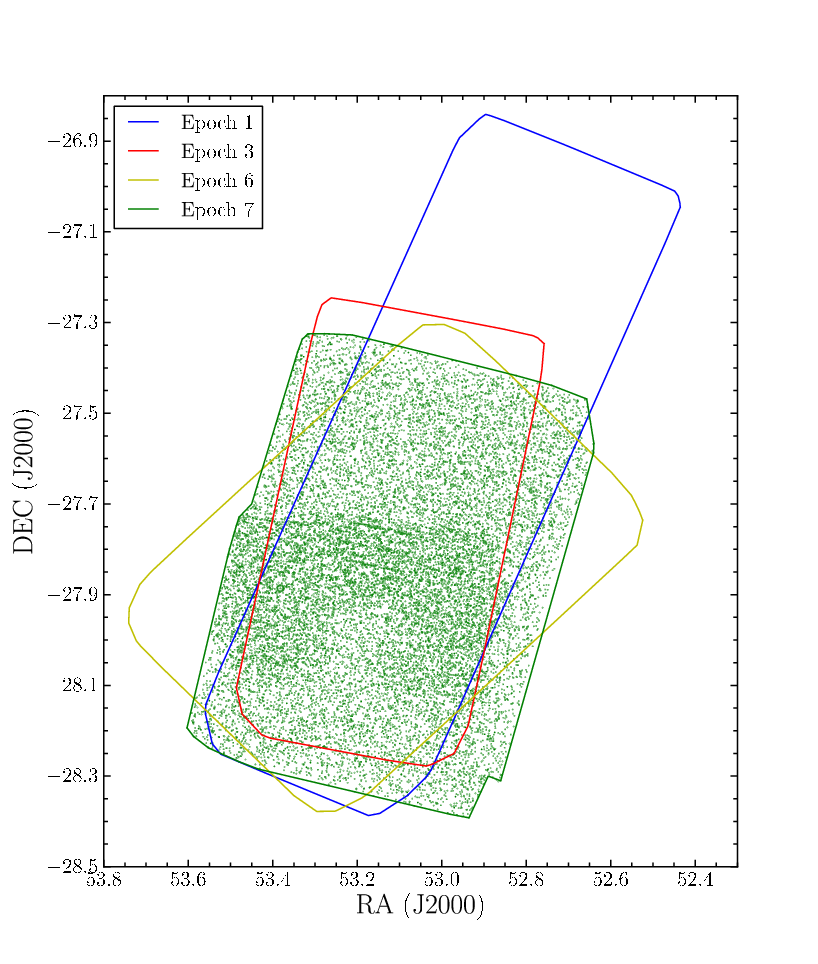}
  \caption{FoV of Epochs 1, 3, 6, and 7, and their overlap
      region. These four epochs are 
    used to study the long-term variability. Epoch 7 is subdivided in
    three epochs to study the short-term variability. The filled
    dots indicate the MIPS $24\,\umu$m sources detected in epoch 7. This epoch is deeper in the center region.} 
  \label{superposicion}
\end{figure}

For this study we decided to exclude Epochs 2, 4, and 5 because their
  FoV is small when compared to the other epochs (see
  Table~\ref{table-epochs}). Figure~\ref{superposicion} shows the FoV of Epochs 1, 3, 6, and 7 and 
how they overlap. The common area for the four epochs is
$\sim$1360~arcmin$^2$. They probe time scales of months up to
  three years,
  and henceforth are used to study the long-term variability covering
  a period of over three years. We also
subdivided Epoch 7 in three epochs, namely Epochs 7a, 
7b, and 7c to study the short-term 
variability. The short-term variability epochs have a common area of
$\sim$1960~arcmin$^2$ and probe time scales of days, covering a
  period of 7 days.  

\section{MIPS $24\,\umu$m photometry}
 \label{data-photometry}

To study the temporal variability of MIPS $24\,\umu$m sources detected
in the common  regions we built a source catalog for each epoch
  and subepoch. We used SExtractor 
(Source-Extractor, \citealt{BA}) to detect sources and the Image
Reduction and Analysis Facility ({\sevensize IRAF})\footnote{IRAF is
  distributed by the National Optical Astronomy Observatory, which is
  operated by the Association of Universities for Research in
  Astronomy (AURA), Ibc., under cooperative agreement with the
  National Science Foundation.} to perform the photometry following
the procedure explained in P\'erez Gonz\'alez et al. (2005,
2008). Sources were detected in five passes to recover the faintest
ones, possibly hidden (i.e., more difficult to detect) by brighter
sources. All the measurements were made by PSF fitting. To obtain the
photometry, we used a circular aperture of radius $\sim$12$\arcsec$
and then applied an aperture correction of 17\% as in \citet{PG} to
obtain the total flux. We calculated the uncertainties in the flux
taking into account the correlation of the pixel-to-pixel noise
introduced by the reduction method and mosaic construction, as
described in  appendix A.3 of \citet{PG2}. We obtained a 24~$\umu$m
source catalog for each epoch. In this work we restrict the
analysis to sources above the 5$\sigma$ detection limit in the shallowest data in the mosaics. This corresponds to MIPS
  $24\,\umu$m fluxes of $80\,\umu$Jy and $100\,\umu$Jy for the long-term
  and the short-term epochs, respectively.  We also discarded sources with neighbours at distances of less than 10$\arcsec$ to minimize crowding effects in the photometry that could affect the flux measurements and produce false variability positives. 
In Table \ref{source-counts} we list for each epoch
the total number of detections,  the number of $>5\sigma$
  detections, the number of $>5\sigma$ detections without neighbours, the area covered, and the density of objects. The positional accuracy of our catalogs is better than 0.7$\arcsec$.

\begin{table*}
  \begin{minipage}{180mm}
    \begin{center}
    \caption{MIPS 24~$\umu$m source counts for the different epochs. }
    \label{source-counts}
    \begin{tabular}{cccccc}
      \hline
      Epoch & $N_{\rm detections}$ & $N_{\rm detections}$ & $N_{\rm detections}$ & Area & Density \\
      && F\textgreater5$\sigma$ & F\textgreater5$\sigma$ without neighbours& (arcmin$^2$)& (objects/${\rm arcmin^2}$)\\
      \hline
      \multicolumn{6}{c}{Long-term variability (5$\sigma$=80~$\umu$Jy)}\\
      \hline \hline
       1 & 19742 & 11467 & 8017 & 2653 & 3.02\\
       3 & 11843 & 7116 & 5041 & 1706 & 2.95 \\
       6 & 15707 & 9697 & 6870 & 2293 & 3.00 \\
       7 & 18406 & 9320 & 6605 & 2265 & 2.92 \\
      \hline 
       \multicolumn{6}{c}{Short-term variability (5$\sigma$=100~$\umu$Jy)}\\
      \hline \hline
       7a & 14453 & 7607 & 5715 & 2027 & 2.82 \\
       7b & 15629 & 7348 & 5551 & 2107 & 2.63\\
       7c & 14659 & 7673 & 5789 & 2046 & 2.83\\
       \hline
    \end{tabular}
    \end{center}
  \end{minipage}
\end{table*}

To identify the common sources in all the epochs we cross-matched the
catalogs using a 2$\arcsec$ radius, imposing additionally that
  the $2\arcsec$ criterion was fulfilled in each pair of epochs. Due to this criterion, we missed 316 sources cross-matching Epochs 1, 3, 6, and 7 and 282 
sources cross-matching Epochs 7a, 7b, and 7c.  For the
long-term variability (Epochs 1, 3, 6, and 7) there are 2277 sources (1.67 objects $\times$ arcmin$^{-2}$) in
common with 24~$\umu$m flux \textgreater 80~$\umu$Jy (5$\sigma$
detection) without neighbours within 10$\arcsec$ and satisfying the $2\arcsec$ criterion. For the short-term variability (Epochs 7a, 7b, and 7c)
there are 2452 sources (1.25 objects $\times$ arcmin$^{-2}$) in common with 24~$\umu$m flux \textgreater
100~$\umu$Jy (5$\sigma$ detection) without neighbours within 10$\arcsec$ and satisfying the $2\arcsec$ criterion. Our final catalogs contain 2277 MIPS 24 $\umu$m sources detected in  Epochs 1, 3, 6, and 7 and 2452 MIPS 24 $\umu$ sources in Epochs 7a, 7b, and 7c, covering an area of 1360 and 1960 arcmin$^2$, respectively. 

\section{Selection of MIPS $24\,\umu$m variable sources}
\label{selection-variable-sources}

In this section we describe the method used to select the 24 $\umu$m variable sources. To do so
we used a $\chi^2$-statistics method to account for the variations of intrinsic flux uncertainties of each epoch (related to differences in depth). This is the case for our study as different epochs have different depths and within a given mosaic there are some variations in depth. The latter effect is most prominent in epoch 7, which is deeper in the center.

This method associates each flux with its error. The $\chi^2$-statistics is defined as follows:
\begin{equation}
  \chi^2=\sum_{i=1}^n{(F_i-\bar{F})^2\over{\sigma_i^2}}
  \label{ec-Xcuadrado}
\end{equation}
where n is the number of epochs, $F_i$ is the flux in a given epoch, $\sigma_i$ is the associated error in the $i^{th}$ epoch, and $\bar{F}$ is the mean flux.

As errors are essential in this method, we checked them for each
epoch. Errors in the parent photometric catalog could be affected by
correlation of the noise due to the reduction method. We compared our
estimated errors with the uncertainties resulting from the
scatter of points with the fluxes estimated
in images from different epochs  (see Figure A1 and Appendix A). As fluxes in different epochs are
measured independently, the scatter must account for the real
uncertainties of the  measured fluxes. The scatter in this plot 
is entirely consistent with the photometric errors estimated in
Section \ref{data-photometry}.  We refer the reader to
  Appendix~\ref{photometry-errors} 
for more details.  

\begin{figure}
  \subfigure {\includegraphics[width=0.52\textwidth]{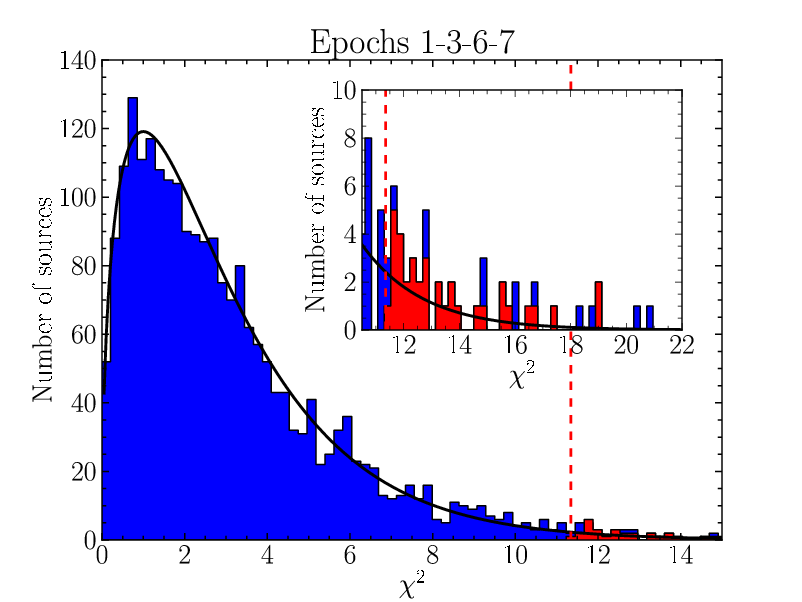}}
  \subfigure {\includegraphics[width=0.52\textwidth]{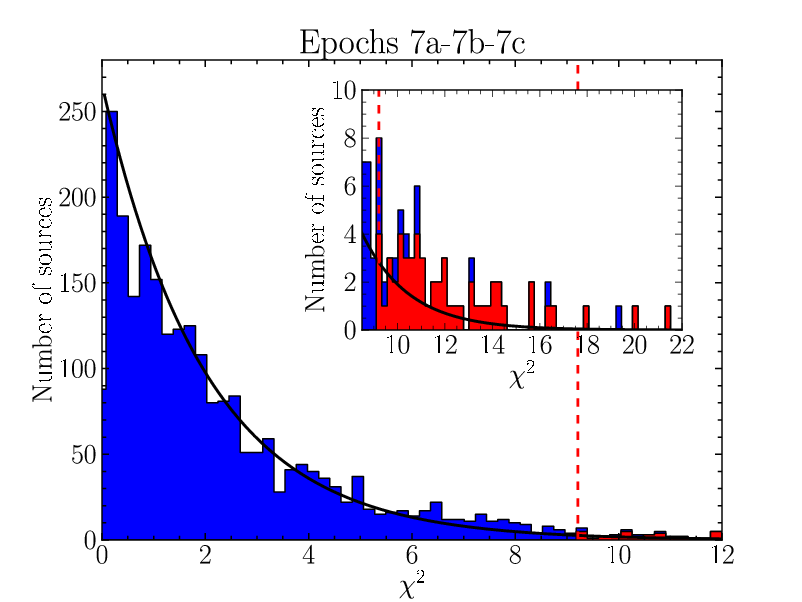}}
  \caption{ Observed $\chi^2$ distributions (filled histograms) for all sources without 
    neighbours within 10$\arcsec$. The top panel is the distribution
    for the four epochs used for the long-term variability, whereas
    the bottom is for the three epochs used for the short-term
    variability. The black line is the theoretical $\chi^2$
    distribution for 3 degrees of freedom and 2 degrees of freedom,
    respectively. The dashed red line marks the 99$^{th}$ percentile for
    $\chi^2$ due to random photometric errors alone. In the insets we
    zoom on the high $\chi^2$ region. The red histograms show
    the $\chi^2$ distribution for the final candidates after
    discarding problematic sources visually.} 
  \label{Xcuadrado}
\end{figure}

We calculated the $\chi^2$ value for each source without neighbours.  We selected as variable candidates those sources above the 99$^{th}$ percentile of the $\chi^2$ distribution expected from photometric errors alone. That is, only 1\% of non-variable sources satisfy the selection criteria. This
value corresponds to $\chi^2\geq11.34$ for the 4 epochs sample (3
degrees of freedom) and $\chi^2\geq9.21$ for the 3 epochs one (2
degrees of freedom). In Figure~\ref{Xcuadrado} we show the observed
$\chi^2$ distribution (filled histograms), the theoretical
distribution (black line), and 
the threshold (red dashed line) for the four epochs (top panel)
and three epochs (bottom panel). As can be seen from these figures,
the calculated values of $\chi^2$ follow well the expected theoretical
distribution for gaussian photometric errors, indicating that our estimates of the flux uncertainties are
accurate (see Appendix \ref{KS-test-appendix} for more details).

\begin{figure*}
 \includegraphics[width=180mm]{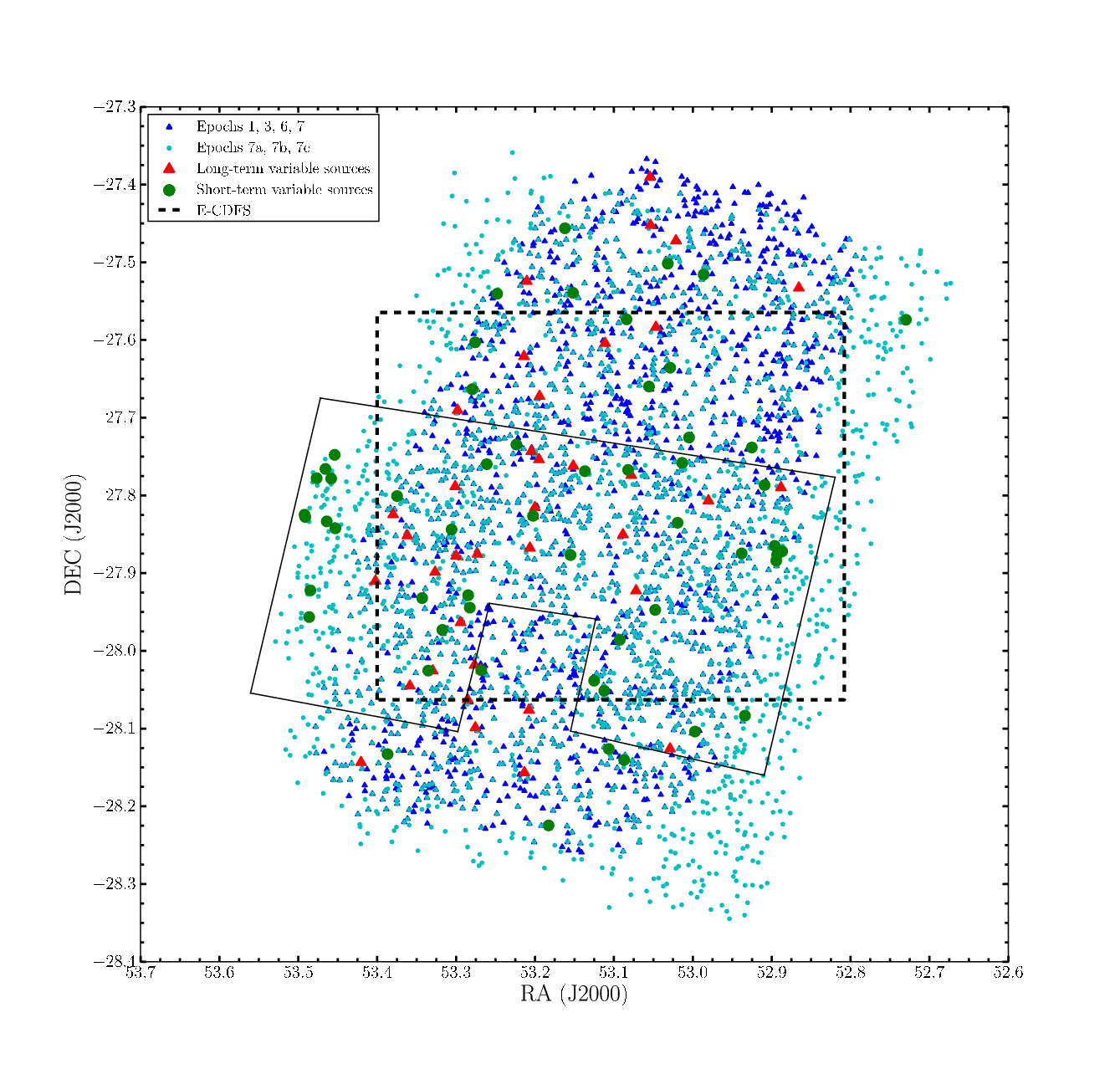}
  \caption{Location of the MIPS $24\,\umu$m long-term variable
    sources (red large triangles) and the short-term variable sources
    (green large circles) in GOODS-South. The black dashed line encloses the
    E-CDFS (see Section~5 for more details). The small
    blue triangles and the small cyan circles 
    are all the MIPS 24~$\umu$m common sources  to Epochs 1, 3,
    6, and 7 and to Epochs 7a, 7b, and 7c, respectively. The solid
    line encloses the 
    deepest region in Epochs 7a, 7b, and 7c.} 
  \label{candidatos}
\end{figure*}

Every object with a $\chi^2$ value higher than the threshold was
visually inspected to remove artefacts. We also discarded objects that
fell close to the edge of the mosaic. We also compared the
candidates with the supernova (SN) catalog of \citet{St} and found that none of
our candidates was in the SN catalog. The original number of selected variable sources before the removal of artefacts/objects close to the edge of mosaics was 52 and 64 for long-term and short-term variable sources.  
In the insets of Figure
\ref{Xcuadrado} the red histogram shows the distribution for the final
candidates. After discarding problematic objects,  our final
  sample contains 39
MIPS $24\,\umu$m long-term variable sources (0.03 sources $\times$ arcmin$^{-2}$) and 55 MIPS
$24\,\umu$m short-term 
variable sources (0.03 sources $\times$ arcmin$^{-2}$). Only two sources are identified as having both, long and short-term 
variability. The spatial
distribution of the  MIPS $24\,\umu$m variable sources in the
GOODS-South field is shown in Figure~\ref{candidatos}.

 The $\chi^2$ cut means that we would
  expect that 1\% of the parent samples of MIPS
  $24\,\umu$m sources would be incorrectly identified as variable (i.e., false positives). The
  expected numbers of false positives are then 23 and 25 sources for 
  long and short-term variable sources. Taking the original number of selected variable sources before the removal into account we expect that the
  fraction of false positives in our final 
  sample of variable sources would be $\sim$44\% for long-term and $\sim$39\% for short-term. We detect many more variable source candidates than expected by random errors, so our selection is statistically meaningful.

 The selected MIPS $24\,\umu$m long-term and short-term variable
  sources represent 1.7 and 2.2\% of the original parent samples, 
respectively. After removing the expected number of false positives, the estimated
percentages are 1.0\% and 1.4\%. These fractions of variable sources at 24~$\umu$m are
similar to those found in the same cosmological field at other
wavelengths, mostly optical and near IR (e.g. \citealt{Co};
\citealt{VKG}; \citealt{Sa}; \citealt{Ko}).  The higher fraction of 
short-term variable sources is due to the presence of a deeper region
in 
Subepochs 7a, 7b, and 7c (shown as the area enclosed by the
solid line in Figure~\ref{candidatos}). This means that the
photometric errors of sources in this region are smaller and then if
variable, they present higher values of $\chi^2$ than sources in
shallower areas. 

We note that the presence of intense (obscured)  star-formation in the host
galaxy would impair the detection of AGN variability at 24 $\umu$m, so only sources with 
the highest variability might be detected. We refer the reader to Sections 6.1 and 6.3 for
further discussion on this issue.

  In Appendix C we show an example of the MIPS 24 $\umu$m images of four variable sources, two long-term and two short-term, in each of the epochs of our study. In Tables D1 and E1 in the Appendices  
 we list the flux  and corresponding error 
at each epoch, median flux, and $\chi^2$ value, for the long-term and
short-term MIPS $24\,\umu$m variable sources, respectively. 

\section{Properties of the MIPS $24\,\umu$m variable sources}
\label{properties-variable-candidates} 

In this section we analyze the different properties of the 24 $\umu$m variable sources, such as their median 24 $\umu$m fluxes, variability properties, and their IRAC colours. 

\subsection{MIPS 24 $\umu$m properties}

\begin{figure}
  \subfigure {\includegraphics[width=0.5\textwidth]{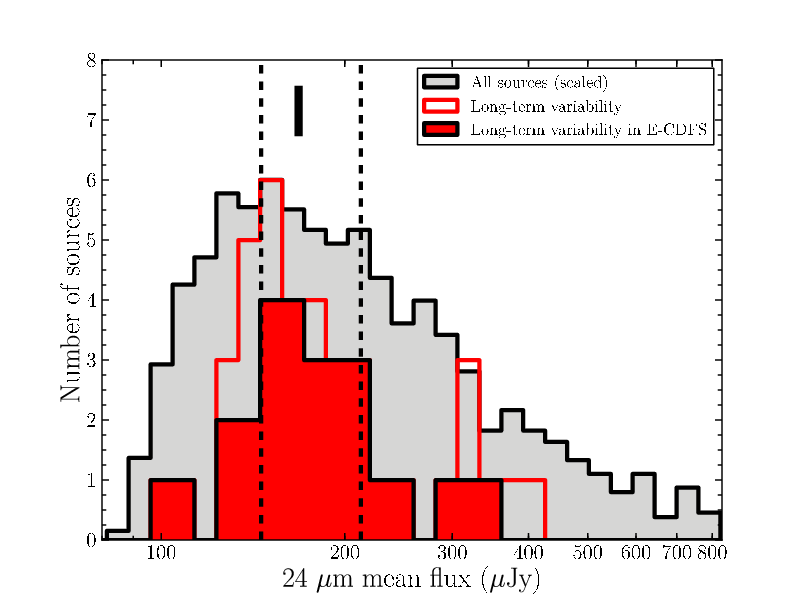}}
  \subfigure {\includegraphics[width=0.5\textwidth]{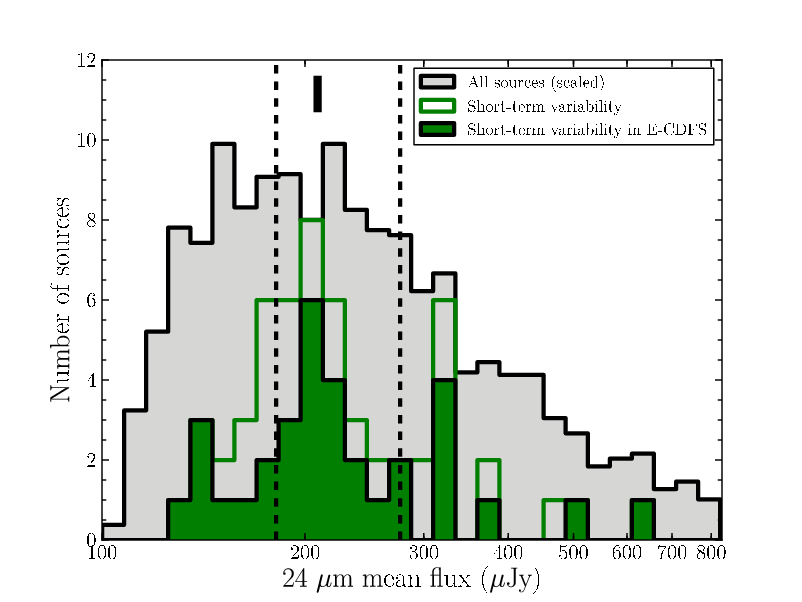}}
  \caption{Distribution of the 24~$\umu$m mean flux. The empty histograms are the distribution of the mean flux
    at 24~$\umu$m for long-term 
    variable sources (top panel) and short-term variable sources
    (bottom panel) for the full sample, whereas the filled histogram
    are for sources in the E-CDFS (see Section
    \ref{combo-section}).  The black lines
are  the median and the dashed lines the first and third quartiles for the variable sources. The grey
histograms shows the scaled distributions of all the sources detected
in the four epochs 
(top) and three epochs (bottom)
    without neighbours within 10$\arcsec$.} 
  \label{distribucion-flujos}
\end{figure}

In Figure \ref{distribucion-flujos} we show the distribution of the
mean flux (over the 3 or 4 different epochs) at 24~$\umu$m for the long-term variable sources (top
panel) and short-term variable sources (bottom panel) compared with the corresponding flux distribution of the parent
  sample for $>5\sigma$ detections. In both
cases, the $24\,\umu$m fluxes of the variable sources are
  dominated by sources with mean fluxes below 300~$\umu$Jy. The
median 24 $\umu$m flux is 168~$\umu$Jy for the long-term variable
sources and 209~$\umu$Jy 
for the short-term variable sources (see Table
\ref{properties-candidates}). This slight difference in the
median values of the 24~$\umu$m fluxes for long and short-term
variability is likely reflecting the different depths (i.e., 5$\sigma$
  detection limits) of the epochs
rather than different intrinsic properties of the sources (see Section~6).

 We cross-correlated our parent MIPS 24 $\umu$m catalogs with the \cite{Xue} deep  
X-ray catalog of AGN and galaxies using a search radius of 2.5$\arcsec$(see Section 6.2). There are 211 X-ray sources that are not stars 
in \cite{Xue} catalog detected in 24 $\umu$m satisfying our criteria, that is, they
 have 24 $\umu$m fluxes over our  5$\sigma$ limit and have no neighbours 
within 10$\arcsec$. Of the 211 sources, 149 are classified as AGN in the \cite{Xue} 
catalog. These X-ray selected AGN in our parent catalogs have a 
24 $\umu$m median flux of $\sim$240 $\umu$Jy. This implies that our selected
 24 $\umu$m variable sources are typically fainter at 24 $\umu$m than X-ray selected AGN.
Since the redshift distributions are similar (see Section 6.1), this may indicate
that the 24 $\umu$m variable sources, if they were AGN, are less luminous, as predicted by \cite{Tre}.

\subsection{Variability Properties}
\label{subsection-properties}

\begin{figure*}
  \begin{minipage}{180mm}
    \begin{center}
    \subfigure {\includegraphics[width=87mm]{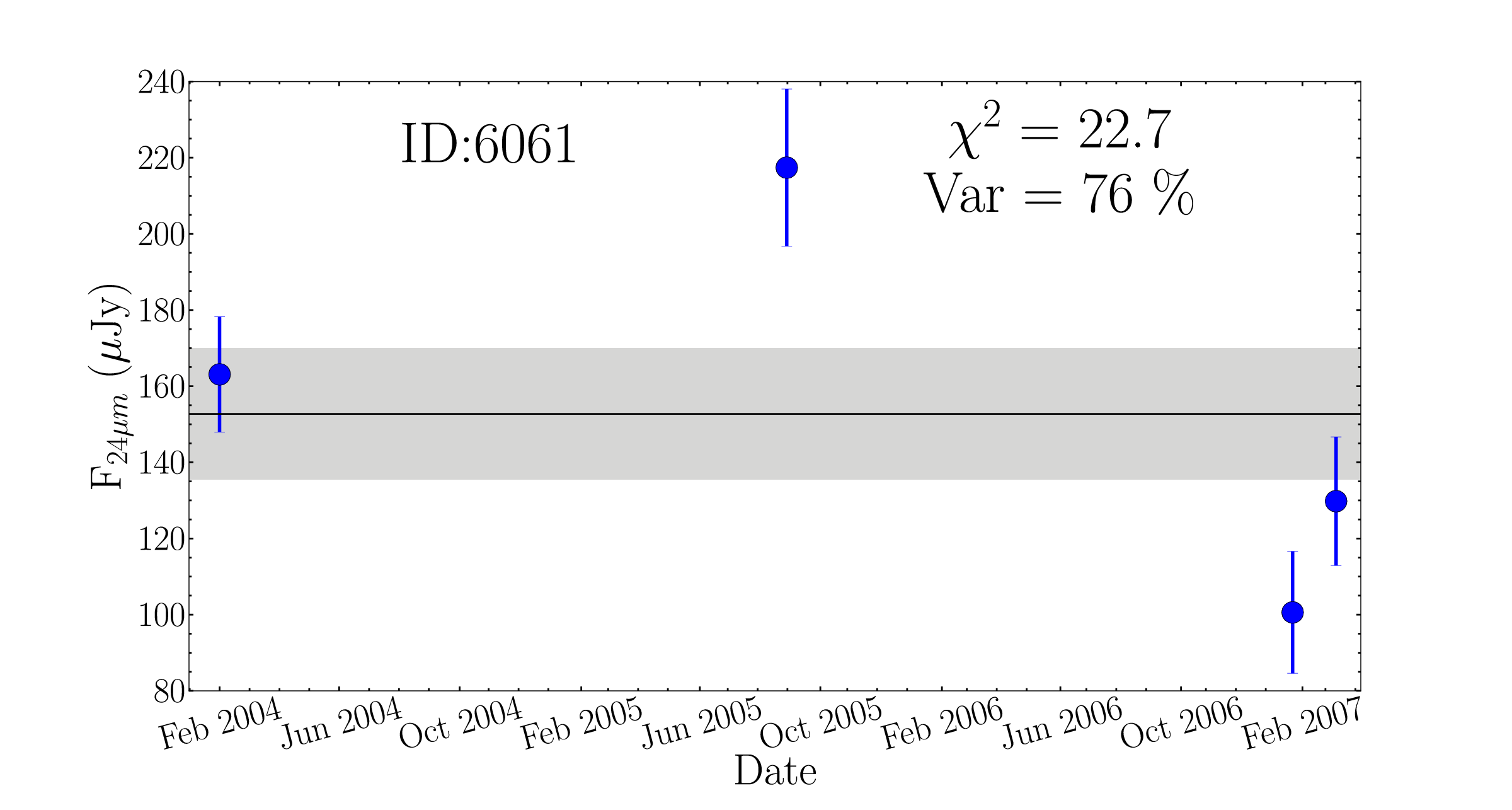}}{\hspace{0cm}}
    \subfigure {\includegraphics[width=87mm]{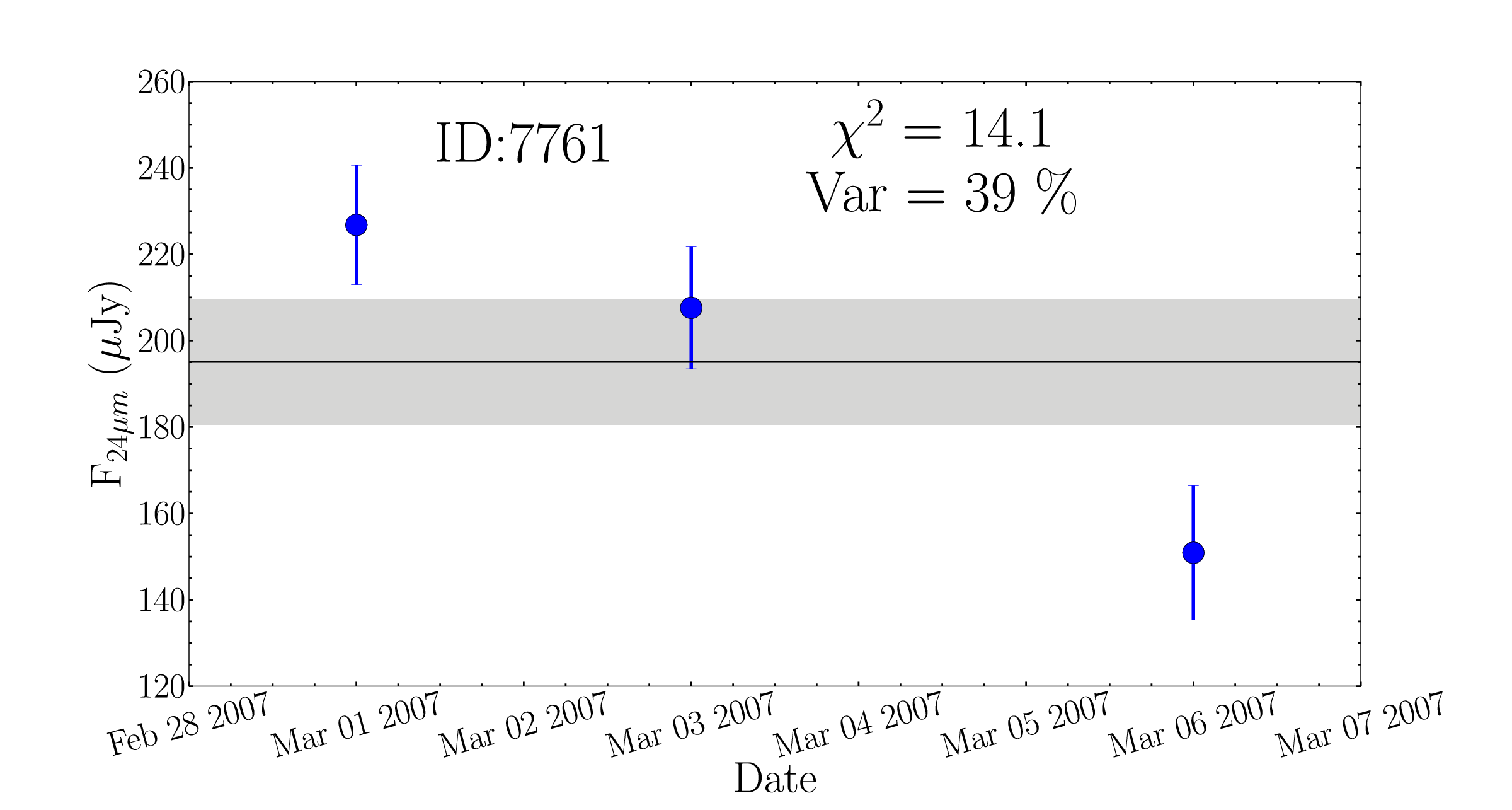}}{\hspace{0cm}}
    \caption{Examples of light curves of two MIPS 24~$\umu$m variable
      sources in GOODS-South. The left panel corresponds to a
      long-term variable source (four epochs) and the right
      panel to a short-term variable candidate (three epochs). The
      flux for each epoch is plotted with its corresponding
      photometric error. The solid line is the 24~$\umu$m mean flux of
      the source and the gray shaded area is the average of the errors
      of the source. Each plot lists the name of the source, the
      $\chi^2$ value, and Var. Light curves for the entire sample
      of MIPS $24\,\umu$m long-term and short-term variable sources are shown in
      Appendices E and F, respectively.}  
    \label{light-curves}
    \end{center}
  \end{minipage}
\end{figure*}

In Figure \ref{light-curves} we show two example light curves, one of
long-term and the other of short-term variable sources. Each plot
shows the name of the source, the $\chi^2$ value and the  measure
  of the variability $Var$ (see below, Equation
\ref{Var-eq}). The light curves for all the long and
short-term 24~$\umu$m variable sources are presented in Appendices
\ref{curvas-luz-long} and \ref{curvas-luz-short}, respectively. 

\begin{figure}
  \subfigure {\includegraphics[width=0.5\textwidth]{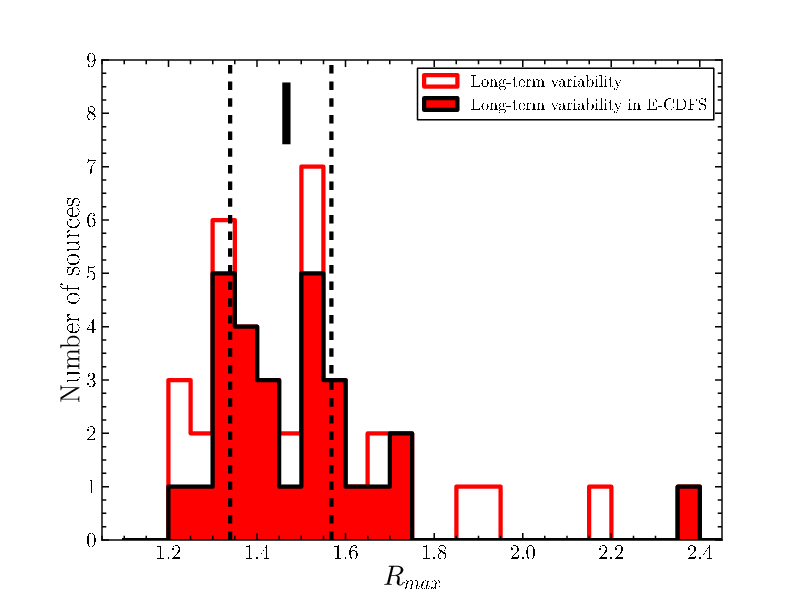}}
  \subfigure {\includegraphics[width=0.5\textwidth]{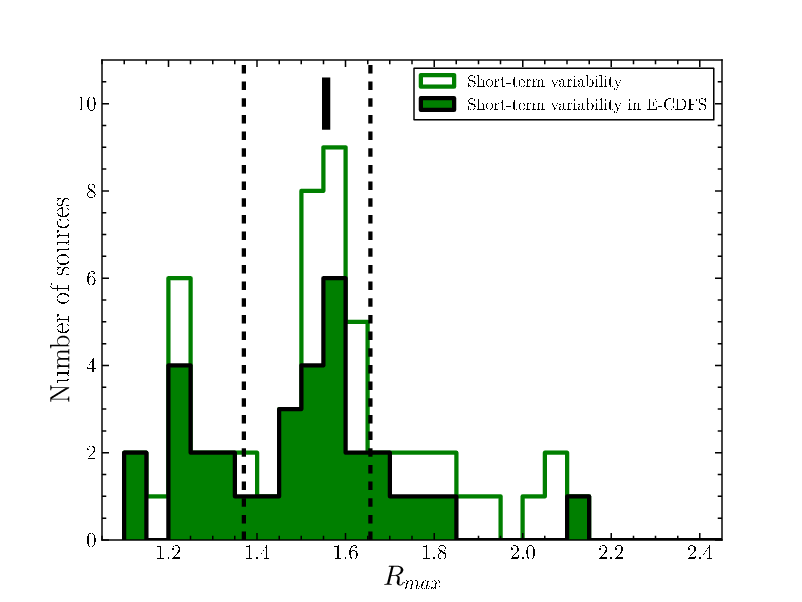}} 
  \caption{Distributions (open histograms) of the maximum to minimum flux ratio,
    $R_{max}$, for the MIPS $24\,\umu$m long-term variable sources (top panel) and
    short-term variable sources (bottom panel). The black lines
    correspond to the median and the dashed 
    lines to the quartiles of the distributions. In both
      panels the filled histograms are the distributions for variable
      sources in the E-CDFS} 
  \label{distribucion-Rmax}
\end{figure}

As a first measure of the variability, we calculated the maximum
to minimum flux ratio, $R_{max}$, as:  $R_{max}=f_{\rm max}/f_{\rm min}$. The
  long-term and short-term 
  variable sources show similar values of the average and median
  $R_{\rm max}$ of approximately $1.5-1.6$ (see
  Table~\ref{properties-candidates} and
Figure~\ref{distribucion-Rmax}).

Another estimate of the variability is the ratio between the maximum
and minimum values and the mean flux $\bar f$ measured as a \%. 

\begin{equation}
  \label{Var-eq}
  Var={{f_{\rm max}-f_{\rm min}}\over{\bar f}} \times 100
\end{equation}
 
\begin{figure}
\centering
 \includegraphics[width=0.52\textwidth]{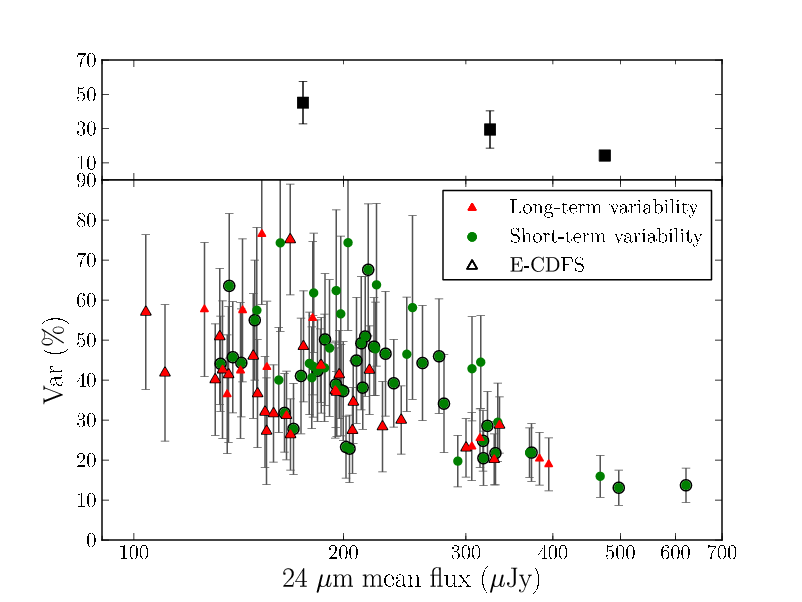}
  \caption{$Var={{f_{\rm max}-f_{\rm min}}\over{\bar f}} \times 100$ as a function of the MIPS $24\,\umu$m mean flux for
    long-term variable sources (red filled triangles) and short-term
    variable sources (green filled circles). The black marked triangles/circles
    correspond to the variable sources in the E-CDFS. The top
    panel shows the average value of  $Var$ with its error for three flux
    intervals (100 $\umu$Jy\textless F\textless 250 $\umu$Jy, 250 $\umu$Jy\textless F\textless 400 $\umu$Jy, F\textgreater 400 $\umu$Jy).}  
  \label{var-vs-flux}
\end{figure}

As can be seen from Table~\ref{properties-candidates}, the typical
24~$\umu$m $Var$ values of the long-term and short-term variable sources are
37-43\%, with typical errors of 12-13\%. In Figure \ref{var-vs-flux}
we show $Var$ against the mean $24\,\umu$m
flux for each candidate (lower panel) with the typical errors (upper
panel). The apparent lack of small values of  
$Var$ at low $24\,\umu$m mean fluxes is because these sources have 
lower S/N detections and therefore higher errors in their photometry,
and for the same variability they do not meet our $\chi^2$
criterion. There is also a lack of large values of $Var$ at high $24\,\umu$m mean fluxes.
 It is due to a statistical effect because the number of sources at high $24\,\umu$m mean fluxes is small,
and the fraction of variable sources with low values of $Var$ is higher than the fraction with large values of $Var$.

Finally, the reduced value of $\chi^2$, which is defined as
  $\chi^2/n$ with $n$ being the number of epochs, has
  also been used as a measure of the $10\,\umu$m variability of local
  quasars by \cite{Neu}. For the long and short-term variable sources
  we find median $\chi^2/n$ values of 3.2 and 3.7, respectively. These are
  slightly higher
than the values measured for local quasars at $10\,\umu$m. We note,
however that our variability criterion in both cases is more restrictive than
that used for the local quasars ($\chi^2/n >1.5$).  

It is not straightforward to compare our measures of the MIPS
  $24\,\umu$m variability
in GOODS-South with studies done in the optical. The optical studies
(e.g., \citealt{Sa}; \citealt{VKG}) used the variability significance
and the variability strength as a measure of the variability. These
parameters are defined 
in a different way than our $Var$ and it is not appropriate to
calculate them for our sources because they assume equal errors for all the
  sources, which is not the case for our epochs as shown in
  Section~2. 

We can compare the MIPS $24\,\umu$m $R_{\rm max}$ values with those
  measured in X-rays. \cite{Yo}  detected X-ray variable 
sources with  maximum-to-minimum flux ratios $R_{\rm max} = 1.5-9.3$
with a median value of 4.1 over a period of 10.8 years. These values are noticeably
higher than those measured at $24\,\umu$m both in short-term and long-term time scales. There are two explanations
for this. First, as pointed out by \cite{Yo}, the limited photon
statistics of their X-ray observations means that sources must be strongly
variable to be identified as such. The second reason is due to the
reprocessed nature of the AGN mid-IR emission. Indeed, in
the context of the AGN dusty torus, the
IR variability of AGN is predicted to be only a fraction of the
AGN intrinsic luminosity variation, to depend of the dust
distribution,  to be delayed with respect to optical variations, and
to depend on the IR wavelength used
(see \citealt{HK} and references therein). This is because the dust is
further away from the central engine than the accretion disk. This has
been confirmed observationally for local quasars \citep{Neu} 
  and Seyfert galaxies \citep{Glass04}.

\begin{table*}
   \begin{minipage}{180mm}
     \begin{center}
    \caption{Properties of the MIPS 24~$\umu$m variable sources.}
    \label{properties-candidates}
    \begin{tabular}{@{}ccccccccc}
      \hline
      Variability & \multicolumn{2}{c}{$f_\nu$  24~$\umu$m ($\umu$Jy)}& \multicolumn{2}{c}{$\chi^2$}& \multicolumn{2}{c}{Var (\%)}& \multicolumn{2}{c}{$R_{max}$}\\
      & Average & Median$^1$ & Average & Median$^1$ & Average & Median$^1$ & Average & Median$^1$  \\
      \hline \hline
      Long-term & 196 & $168^{143}_{218}$& 14.75  & $12.90^{11.94}_{15.46}$& 39.0 & $37.41^{28.3}_{43.7}$& 1.51  & $1.47^{1.33}_{1.57}$ \\
      \hline
      Short-term &  239 & $209^{181}_{279}$ & 12.14  & $11.25^{10.23}_{13.67}$ & 41.7  & $43.4^{29.5}_{49.2}$ & 1.55  & $1.56^{1.36}_{1.66}$\\
      \hline
    \end{tabular}
    \end{center}
    $^1$ Median and quartiles of the distribution.
  \end{minipage}
\end{table*}

\subsection{IRAC colours}
\label{color-color}

In this subsection we investigate the Spitzer-IRAC mid-IR (3.6, 4.5, 5.8, and 8.0 $\umu$m) properties of 
the MIPS $24\,\umu$m variable sources as the IRAC emission has also been
used to select AGN candidates (e.g., \citealt{La}; \citealt{Ste};
\citealt{Almu}; \citealt{Do}; \citealt{La13}).

\citet{La} defined a wedge in an IRAC colour-colour diagram to
 identify AGN via their IR emission, based on the locus of the diagram occupied by quasars.
 \citet{Do} defined a more restrictive IRAC wegde based on the IR power-law criterion of \citet{Almu} 
and the typical errors of the IRAC photometry. 
 This IR power-law wedge was specifically designed
to avoid contamination from high-redshift star
forming galaxies. To do so,  Donley et al. applied a colour cut of $\log(S_{8.0}/S_{4.5})>0.15$ to
avoid high-redshift ($z\ge 2$) star-forming galaxies. They also
applied a vertical cut of $\log(S_{5.8}/S_{3.6})>0.08$ to prevent
contamination due to low-redshift star-forming galaxies  and required that the
IRAC SED of the source rises monotonically. \citet{Do} showed that this power-law wedge selects the majority of luminous X-ray identified 
AGN and therefore it is highly reliable at the expense of losing the
least luminous AGN.  

Finally, we note that recently \citet{La13} put forward a  
new expanded AGN selection criteria with a broader wedge when compared
to that of \cite{La} and imposed a 24~$\umu$m limit of \textgreater
600~$\umu$Jy.  We do not use this new wedge as only 1 short-term 24~$\umu$m variable source is above this limit
(see Figure~\ref{distribucion-flujos}).

To obtain the IRAC data for our sources, we used
the {\it Rainbow} Cosmological Surveys Database, which contains
multi-wavelength  
photometric data as well as spectroscopic information for sources in different cosmological fields, including GOODS-South (see \citealt{PG,PG2}).  
We cross-correlated the MIPS $24\,\umu$m catalogs with the {\it Rainbow}
IRAC sources using a search radius of
2.5$\arcsec$. Of the 39 long-term variable sources, 26 (67\%)
have a single counterpart and the remaining 13 (33\%) have
more than one 
counterpart within a radius of 2.5$\arcsec$. Of the 55 short-term
variable candidates, 44 (80\%) have a single counterpart and the
remaining 11 (20\%) have more than one counterpart within a radius of
2.5$\arcsec$. A visual inspection of the images at different wavelengths allowed us to identify the counterpart of the majority of the variable sources\footnote{In the majority of the sources, inspection of the IRAC images is enough to determine which source dominates in the IR.}. For the rest we used
the data from the nearest source. All the long-term variable
sources have fluxes in all four IRAC bands, whereas only 43
(78\%) of 55 short-term variable candidates do. This is
  because the entire Epoch 7 region is not fully covered by the IRAC
  observations. The flux limits of the $24\,\umu$m variable
    sources are approximately $5\,\umu$Jy at $3.6\,\umu$m, $4\,\umu$Jy
    at $4.5\,\umu$m, $4\,\umu$Jy at $5.8\,\umu$m, and $6\,\umu$Jy at $8\,\umu$m.

\begin{figure}
\centering
  \includegraphics[width=0.52\textwidth]{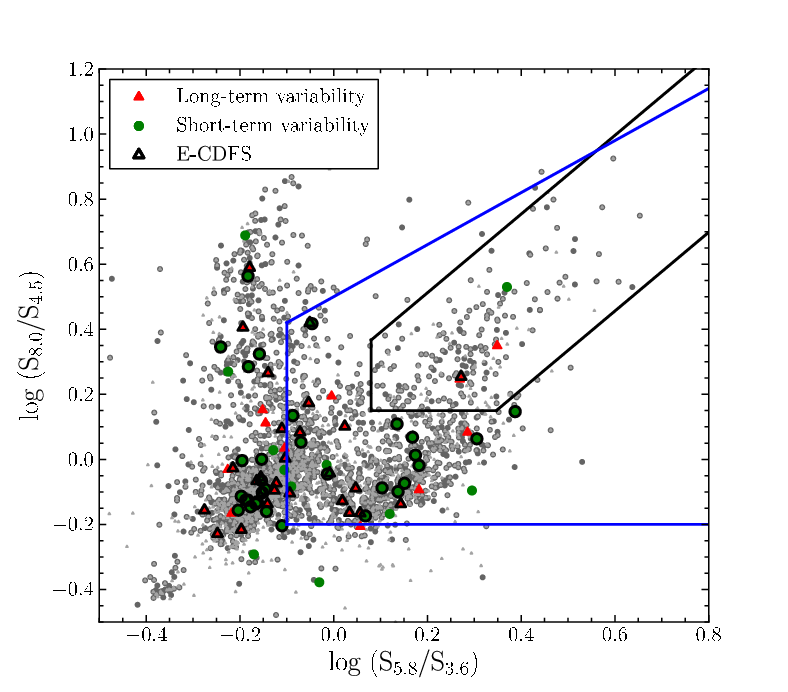}

  \caption{IRAC colour-colour plot of MIPS $24\,\umu$m sources in GOODS-South 
from the {\it Rainbow} database. The red filled triangles and green filled
circles are the long-term and  short-term MIPS $24\,\umu$m variable
sources. The black marked triangles/circles mark the variable
sources in the E-CDFS. The small grey dots are IRAC colours of the non-variable
sources in the studied region. The different AGN wedges are shown as blue solid line for
Lacy et al. (2004) and black solid line for Donley et al. (2012).} 

  \label{diagrama-color-color}
\end{figure}

Figure~\ref{diagrama-color-color} shows the IRAC colour-colour plot for
all the MIPS $24\,\umu$m variable sources detected in the four IRAC bands
together 
with the \citet{La} and \citet{Do} AGN wedges.  
For comparison we also plot the IRAC colours of the
  full (non-variable) MIPS $24\,\umu$m sample in
  the common area of GOODS-South as grey dots. 
Of the 39 long-term variable sources, only 8\% fall in the \citet{Do}
AGN region and 44\% fall in the \citet{La} AGN region. Of the 43
short-term variable sources with IRAC fluxes, 2\% fall in the
\citet{Do} AGN region and 44\% fall in the \citet{La} AGN
region. These fractions of mid-IR variable sources falling inside the
\cite{La} 
  wedge are similar to those found for variable optical sources in
  GOODS-South \citep*[see][]{Villforth12}. In an IRAC
    variability study \cite{Ko}  found a higher
    fraction (approximately 75\%) of near-IR variable objects within an AGN
    wedge similar to that defined by \cite{Ste}. This is most likely
    due to the relatively shallow IRAC observations of their study.

It is also worth noting that the
fraction of objects in the parent MIPS 24 $\umu$m population in GOODS-South that
  are in the \cite{La} wedge is  50\%. However, the \cite{La}
AGN  selection criteria was based on the IRAC colours of bright
SDSS quasars with IRAC $8\,\umu$m fluxes greater than 1\,mJy. In our general
population, only about 1\% of sources are above this
limit. Alternatively, only 2\% of the general sources in GOODS-South
located inside the \cite{La} wedge are found to be variable at
$24\,\umu$m. Such a small fraction of mid-IR variable sources is expected,
   because the probability of detecting mid-IR variability of an AGN
   over the time scales probed and with three/four epochs  is low. Additionally, it is likely that the AGN emission does not have a dominant contribution to the observed 24 $\umu$m emission.

\section{Candidates in the Extended Chandra Deep Field South}
\label{combo-section}

In this section we investigate the  multi-wavelength properties of the MIPS 24~$\umu$m
variable sources located in  the Extended Chandra Deep Field South (E-CDFS), which is located in 
the central area covered by our study (see Figure \ref{candidatos}). 
This field covers an area\footnote{The approximate location of this
    region compared to the region studied here 
can be seen from Figure~\ref{candidatos} where we marked the MIPS
    $24\,\umu$m variable sources in the region of  the E-CDFS}  $\sim$1100~arcmin$^2$. 

The E-CDFS was
observed by COMBO-17 (Classifying Objects by Medium-Band Observations,
a spectrophotometric 17-filter survey) survey \citep{Wo}. The COMBO-17
object catalog contains, in addition to the 17 optical medium-band photometry, the
broad-band $R_{\rm COMBO-17}$ magnitude ($\lambda_c=658$~nm) and
photometric 
redshifts for 63501 objects. We chose the E-CDFS because
  it has been  observed with the deepest multi-wavelength data. We note that
although the COMBO-17 catalog gives a 
photometric redshift,  for the objects without spectroscopic redshift we use the photometric redshifts provided by the
{\it Rainbow} 
database because they are calculated using both optical and IR data (see \citealt{PG2}). We only have spectroscopic redshifts for 8 long-term and 6 short-term variable objects.

For the cross-correlation we used again a search radius of 2.5$\arcsec$. All
the objects within this radius are possible counterparts. Since
  now we are looking at the same area we find relatively similar numbers of
  long (28) and short term (33) variable sources, although still the number of
  short term variable sources is higher due to the deepest central
  area of Epoch 7. Of the
39 long-term variable candidates, 28 are in the E-CDFS and 27 have a detection in the COMBO-17 catalog. Only
19 (70\%) have a single counterpart, whereas the remaining
8 (30\%) have more than one counterpart within a radius of
2.5$\arcsec$. Of the 55 short-time 
variable candidates, 33 are in the E-CDFS and 28 are detected in the COMBO-17 catalog. Only 21 (75\%) have
a single counterpart and the other 7 (25\%) have more than one
counterpart in a radius of 2.5$\arcsec$. In the following discussion
in the case of multiple counterparts we associate the
  MIPS $24\,\,\umu$m source to the nearest
object in the COMBO-17 catalog. 

 The COMBO-17 $R$-band magnitudes
  of the $24\,\umu$m variable sources are given in Tables~C1 and D1 in
  the Appendices.  
The median values are  $R$-band$=22.6\,$mag and $R$-band$=22.3\,$mag  for the long and short-term variable sources,
respectively (see Table~\ref{properties-combo}).  These values are
similar to those of X-ray selected non-broad line AGN in deep cosmological
fields whose optical luminosities are dominated by the host galaxy
\citep[][]{Bauer04}. This is probably the
case as well for 
the MIPS $24\,\umu$m variable sources as they are not dominated by the
AGN (see next section).

We also searched for counterparts in the MUSYC (Multiwavelength Survey by Yale-Chile) catalog \citep{Cardamone10}. See also Section 6.1. This catalog covers all the E-CDFS in the optical and near-IR. We used again a radius of 2.5$\arcsec$ for the cross-correlation. We found 28 long-term variable sources detected in the MUSYC catalog, 23 (82\%) of them with a single counterpart and 33 short-term variable sources, 31 (94\%) with a single counterpart.

As a sanity check, we compared the variable MIPS 24 $\umu$m sources in the E-CDFS with the full variable catalog. We confirmed that their properties in terms of mean $24\,\umu$m fluxes
 (see  Figure~\ref{distribucion-flujos} and Tables~\ref{properties-candidates} and
\ref{properties-combo}) and
  the variability measures $R_{\rm max}$ and $Var$ (see
  Figures~\ref{distribucion-Rmax} and \ref{var-vs-flux}) behave as the general  24 $\umu$m variable population. We therefore expect that the
 properties of the 24 $\umu$m variable sources in the E-CDFS might be extrapolated to the entire variable population.

\begin{table*}
   \begin{minipage}{190mm}
     \begin{center}
    \caption{Properties of MIPS 24~$\umu$m variable candidates in  the E-CDFS.}
    \label{properties-combo}
    \begin{tabular}{@{}cccccccccccccc}
      \hline
      \tiny Variability & \tiny No. & \multicolumn{2}{c}{\tiny{$f_\nu$
          24~$\umu$m ($\umu$Jy)}}&
      \multicolumn{2}{c}{\tiny{$\chi^2$}}&
      \multicolumn{2}{c}{\tiny{Var (\%)}}&
      \multicolumn{2}{c}{\tiny{$R_{max}$}}&
      \multicolumn{2}{c}{\tiny{$z$}}&
      \multicolumn{2}{c}{\tiny{$R_{\rm mag}$}}\\ 
      & & \tiny Average &\tiny Median$^1$ & \tiny Average &\tiny  Median$^1$ &\tiny Average &\tiny  Median$^1$ & \tiny Average &\tiny  Median$^1$ & \tiny Average &\tiny  Median$^1$ & \tiny Average &\tiny  Median$^1$ \\
      \hline \hline
      \tiny Long-term &\tiny 28 &\tiny  186 &\tiny $175^{151}_{207}$ &\tiny 14.98 &\tiny  $12.90^{12.01}_{15.46}$ &\tiny 38.1 &\tiny  $37.4^{30.0}_{43.6}$ &\tiny 1.49 &\tiny  $1.46^{1.36}_{1.57}$ &\tiny 0.94 &\tiny  $0.90^{0.63}_{1.27}$ &\tiny 22.66 &\tiny  $22.63^{21.38}_{23.98}$\\
      \hline
      \tiny  Short-term &\tiny 33&\tiny 240 & \tiny  $212^{184}_{275}$ &\tiny 11.87 &\tiny  $11.10^{10.23}_{12.52}$ &\tiny 38.1 &\tiny  $39.2^{27.8}_{45.9}$ &\tiny 1.49 &\tiny  $1.51^{1.33}_{1.60}$ & \tiny 0.96 &\tiny  $1.00^{0.53}_{1.28}$ &\tiny 22.19 &\tiny  $22.29^{21.01}_{23.49}$\\
      \hline
    \end{tabular}
    \end{center}
     $^1$ Median and quartiles of the distribution.
  \end{minipage}
\end{table*}

\subsection{Photometric redshifts and IRAC properties}\label{photo-z-IRAC} 

\begin{figure*}
  \includegraphics[width=87mm]{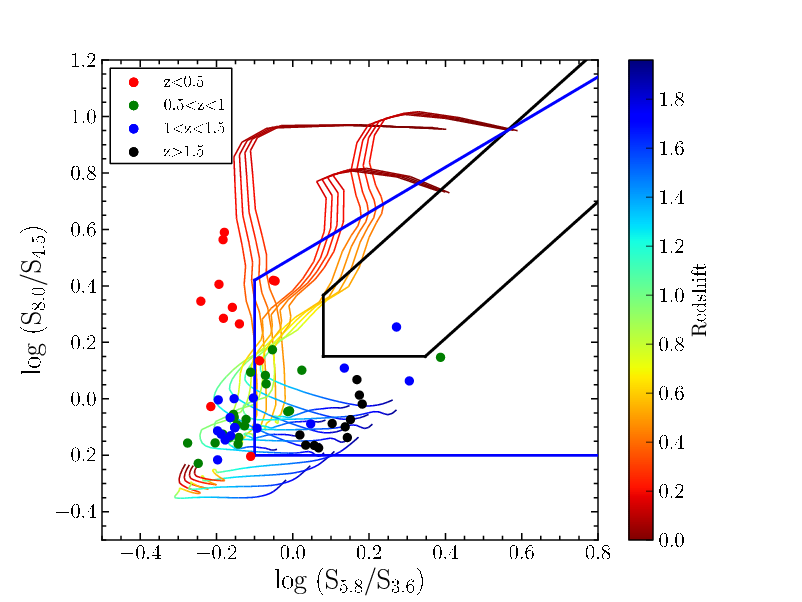}
  \includegraphics[width=87mm]{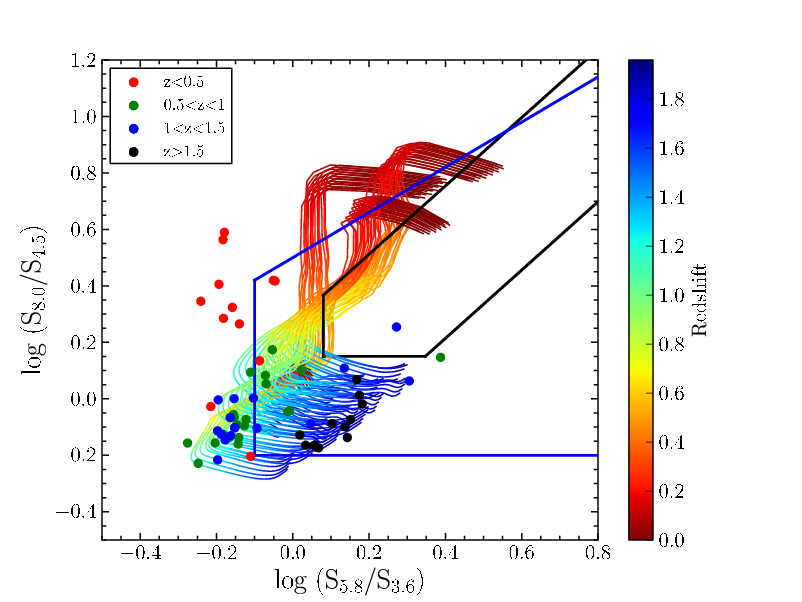}
  \caption{IRAC colour-colour plot of MIPS 24 $\umu$m variable sources
    in the E-CDFS, plotted according to their redshift. The multicoloured
    lines are the predicted IRAC colours of the star-forming templates from
    \citet{Do} with a 20\% AGN contribution (right panel) and
    no AGN contribution (left panel). The four sets of curves 
are for four different templates representing the galaxy contribution
(see Donley et al. 2012 for more details). The redshift
evolution from 0 (top of the curves) up to 2 (bottom of the curves)
plotted (scale on the right hand side of the plots) is
chosen to match the {\it Rainbow} $z$ distribution
of the MIPS $24\,\mu$m variable sources. 
The different AGN wedges  are as in Figure~\ref{diagrama-color-color}. } 

  \label{color-color-z}
\end{figure*}

In this section we study the distribution of the \textit{Rainbow} redshifts for the MIPS $24\,\umu$m variable sources in the E-CDFS. For the long-term 
variable sources the redshifts are between 0.18 and 1.88 and for
the short-term variable source between 0.12 and 1.94.  This redshifts are in accordance  with the redshifts of  the MUSYC catalog. The
average (median) redshifts are similar for the long-term
$z=0.94 (0.90)$ and 
the short term $z=0.96 (1.00)$ variable sources. The average (median) redshifts for the sources in the MUSYC catalog are $z=0.92 (0.85)$ for the long-term and $z=0.98 (0.97)$ for the short-term variable candidates. We are
therefore probing typically variable 
emission at 12~$\umu$m rest-frame. For comparison, the redshift
  distribution of the optical variable sources in GOODS-South has a
  mean value 0.94 for i-band selected sources \citep{Villforth12} and 1.14 for v-band selected sources \citep{Sa}.
 For the X-ray selected AGN in our parent 24 $\umu$m
catalogs the average (median) redshift is $z=1.08 (0.78)$.

Figure~\ref{diagrama-color-color} shows with  black open symbols the IRAC colours of  the MIPS $24\,\umu$m variable sources,
both long and short term. As with the variability  properties, the
variable  sources located in the E-CDFS behave as the general
variable population, with approximately 50\% being in the \citet{La} wedge,
and a small fraction in the \cite{Do} wedge.  As discussed above, the
\cite{Do} criteria are more restrictive to avoid contamination from
star forming galaxies at different redshifts entering the AGN
selection wedge but it misses a large fraction of low-luminosity AGN.
 In consequence, only one of the MIPS $24\,\umu$m
  variable sources in the E-CDFS would be classified as IR
  power-law galaxies according to \cite{Do}. Its ID is given in
  Table~\ref{tabla-comparacion-catalogos}.

We can now use the redshift information in the IRAC colour-colour diagram 
 to investigate whether variable sources in the \cite{La} wedge
  can be  classified as AGN, since the IRAC colours have a strong
  dependence with redshift. As shown in Figure~\ref{color-color-z}, a large
  fraction of the MIPS $24\,\umu$m sources 
falling in the \cite{La} wedge are at $z>1$ (62\% of the objects).

In Figure~\ref{color-color-z} we also
plotted the tracks for an AGN/galaxy
  composite SED with a 20\% AGN contribution (right panel) and no AGN
  contribution (left panel) with redshifts varying
  from 0 up to 2, and different degrees of extinction from
  \cite{Do}. As can be seen from these figures, the colours of
  approximately half  of the MIPS $24\,\umu$m variable sources inside the
  \cite{La} wedge agree with having
  a relatively small AGN fraction and are not 
compatible with a zero AGN fraction for their redshifts. This is similar to what is found for 
optical variable AGN in cosmological fields, where the AGN component is expected to be $\sim$10\% or less
of the total galaxy flux in most cases \citep{Sa}. Those outside this wedge might
  be compatible with just being normal galaxies (although see
  Section~\ref{X-ray}).

\subsection{X-ray properties}\label{X-ray}

One of the main goals of this work is to investigate whether variability at 24 $\umu$m is able to select AGN otherwise missed by deep X-ray exposures.

\subsubsection{Fraction of 24$\umu$m variable sources detected in X-rays}

All the MIPS $24\,\umu$m variable sources, both
  long and short term, were matched against the deepest X-ray
  observations in the CDFS, 
  that is, the 4\,Ms {\it Chandra} catalog presented in
  \cite{Xue}, using again a search radius of 2.5$\arcsec$. These deep observations cover only the central part of
  the E-CDFS. The rest of the area is covered by shallower
  X-ray observations that are part of the E-CDFS observations
  (see 
 references listed in
  Table~\ref{tabla-comparacion-catalogos}). We also cross-matched our variable sources
with these shallower X-ray catalogs with a search radius of 2.5$\arcsec$. We find 7 (25\%) and 4 (12\%) of
  the MIPS $24\,\umu$m long-term and short-term variable sources
  respectively in the E-CDFS are  detected in X-rays (see
  Table~\ref{tabla-comparacion-catalogos} for the ID of the sources).
The lower fraction of X-ray detections among the short-term variable
sources is because a large fraction of these are located outside the
deepest 4\,Ms X-ray region. 
 These fractions of 24 $\umu$m
 variable sources detected in X-rays are in general smaller than for AGN
 candidates selected by optical variability (30-50\%) \citep{Tre8, Sa,
   Villforth12}. 

In the central part of the E-CDFS (CDFS; $\sim$465 arcmin$^2$), which is covered by the deepest X-ray data (\citealt{Xue} catalog),
 30\% of the 24 $\umu$m variable sources are detected in X-rays. Using the catalogs of \citet{Xue} and \citet{Lehmer05, Le},
 and the redshifts provided by the {\it Rainbow} Database, we find that the $24\,\umu$m
  variable sources detected in X-rays have $0.5-8\,$keV luminosities
  ranging from $ \sim  1\times 10^{40}\,{\rm erg \,s}^{-1}$ to $ 
\sim 1\times 10^{44}\,{\rm erg \,s}^{-1}$ (See Table 5).
Although some of our 24~$\umu$m variable sources with an X-ray counterpart
are low luminosity X-ray sources and would be below the limit X-ray
luminosity for the AGN definition, this does not imply that these sources are not
AGN. \citet{Yo} studied sources classified as galaxies in X-rays and
found the 22\% of them presented variability in X-rays, confirming that 
variability selects AGN that might not be selected by other methods. 
There are also many optical variables that are not X-ray detected or that have low X-ray 
to optical flux ratios (see Fig 6 in \citealt{Sa}).

\begin{table*}
 
  \begin{minipage}{180mm}
  
    \begin{center}
    \caption{MIPS $24\,\mu$m variable sources identified with other AGN criteria.}
    \label{tabla-comparacion-catalogos}    
    \begin{tabular}{@{}ccccc}
      
      \hline
      ID& Catalogs& Ref&\citet{La}& X-ray Luminosity\\
      &&&&(0.5-8 KeV) (erg s$^{-1}$)\\
      \hline
      \multicolumn{5}{c}{Long-term variable sources }\\
      \hline \hline

      2552& X-ray, compilation AGN, optical-variable, IR power-law &(2), (3)*, (4), (5), (11), (13)&YES& $1.20 \times 10^{44}$ \\
      13601&X-ray, Chandra 4\,Ms, Radio excess&(2), (6), (9)*, (12)&NO& $1.17 \times 10^{42}$ \\
      12099& X-ray & (3)*, (11)& NO& $1.08 \times 10^{43}$ \\
      10015&relative IR SFR excess, X-ray &(7), (10)*&NO& $8.88 \times 10^{41}$\\
      2226&optical-variable, Chandra 4\,Ms&(8), (9)*&NO& $1.38 \times 10^{40}$ \\
      11976&Chandra 4\,Ms& (9)*&YES& $3.47 \times 10^{42}$\\
      14779&Chandra 4\,Ms& (9)*&NO& $2.87 \times 10^{40}$ \\
      9796&Radio excess& (12)&YES\\
      9579& IR power-law& (13)&YES\footnotemark[1]\\
      4679& IR power-law& (13)&YES\footnotemark[1]\\
      \hline
      \multicolumn{5}{c}{Short-term variable sources}\\
      \hline \hline
      8766& IR power-law, Chandra 4\,Ms &(1), (9)* &YES& $1.52 \times 10^{41}$ \\
      6827& X-ray, optical-variable, IR power-law&(2), (3)*, (5), (11), (13) &YES\footnotemark[1]& $1.41 \times 10^{42}$\\
      7761& X-ray, Chandra 4\,Ms, Radio excess&(2), (6), (9)*, (12)&NO& $1.17 \times 10^{42}$ \\
      7513&optical-variable&(8)&YES\\
      917& Chandra 4\,Ms& (9)*&NO& $2.36 \times 10^{40}$ \\
      2614& Chandra 4\,Ms& (9)*&NO& $3.66 \times 10^{42}$ \\
      540& Radio excess& (12)&YES\footnotemark[1]\\
      \hline
      
    \end{tabular}
      
    \end{center}
      
Notes.- References for the catalogs. (1)~\citet{Almu}; (2)~\citet{Ca}; (3)~\citet{Lehmer05}
(4)~\citet{Ve}; (5)~\citet{Tre8}; (6)~\citet{To}; (7)~\citet{Lu};
(8)~\citet{VKG}; (9)~\citet{Xue}; (10)~\citet{Le}; (11)~\citet{Sil};
(12)~This work (radio excess); (13)~This work (IR power-law according
to \cite{Do}). 

* Reference for the X-ray luminosity.\\
$^1$ Not in the E-CDFS.\\

  \end{minipage}

\end{table*}

\subsubsection{Fraction of X-ray selected AGN found variable at 24$\umu$m}

As explained in the introduction, all AGN are expected to vary over a large range of timescales.
However, the probability of 
detecting AGN variability in the mid-IR is always lower than the optical and near-IR because mid-IR variability is predicted
 to be only a fraction of the AGN  intrinsic luminosity variation. This is
 because the dust responsible for the bulk of the mid-IR emission is further away from the central engine than the accretion disk, and the variability signal is 
expected to be smoothed for large dust distribution (see \citealt{Neu,Glass04,HK}).

Before we compute the fraction of X-ray selected AGN found variable at 24 $\umu$m, we need to calculate
the number of X-ray sources detected in 24 $\umu$m satisfying our criteria in the parent catalogs as explained in 
Section 5.1. We found 211 X-ray sources in the central part of the E-CDFS (classified as AGN and galaxies in 
the \citet{Xue} catalog) satisfying the properties of our parent MIPS 24 $\umu$m catalogs. Of these, only $\sim$4\% are found to be
  variable at $24\,\umu$m on the timescales probed by our study. This fraction is smaller than the
 fraction found in the optical ($\sim$25\%, see \citealt{KS,Sa}). This is explained by model simulations, which predict a more smothered variable signal and longer time scales in the mid-IR
 than in the optical \citep{HK}.  In addition, for low luminosity AGN  most of the mid-IR emission might come from the
host galaxy, which would make it difficult to detect variability (see Section 6.1).

 There are 149 sources classified as AGN in the \cite{Xue} catalog in our parent 24 $\umu$m merged catalogs (see Section 5.1). If we assume that deep X-ray exposures provide the majority of the AGN in the field the total AGN population in this field would be 149 AGN. Assuming the 24 $\umu$m variable sources in the region covered by the \cite{Xue} catalog not detected in X-ray are also AGN, they would only account for a small fraction ($\leq$ 13 \%)
of the total AGN population in this field.

 \subsubsection{Candidates in the deepest X-ray region of the E-CDFS}

As explained in Section 6.2.1, only the central area of the E-CDFS is covered by the deepest X-ray data \citep{Xue}. Since the effective exposure of the {\it Chandra} 4\,Ms survey is not homogeneous (see Fig. 2 of \citealt{Xue}), we selected a central region of $\sim$115 arcmin$^2$ with the deepest and most homogeneous X-ray coverage. In this region we can compare the sources in the parent catalog with the selected variable sources.

In this deepest X-ray region there are 189 sources in the parent 24 $\umu$m long-term catalog and 181 in the parent 24 $\umu$m short-term  catalog. There are only 8 long-term variable sources and 5 short-term variable sources. Tables~\ref{central-region} and \ref{resumen-candidatos-otros-metodos} summarize the results for the deepest X-ray region within the CDFS. As expected, the percentage of 24 $\umu$m variable sources with an X-ray detection (63\% for long-term and 60\% for short-term variable sources) is higher than the percentage in our parent 24 $\umu$m catalog sources detected in X-ray (48\% for long-term and 55\% for short-term variable sources). This is expected because the fraction of X-ray detection is higher in AGN than in not AGN.  Since the number of variable sources in this region is small, the percentages given at Tables~\ref{central-region} and \ref{resumen-candidatos-otros-metodos} suffer from small number statistics.

\begin{table*}
  \begin{minipage}{180mm}
    \begin{center}
    \caption{ Summary of fractions in the deepest X-ray region within the CDFS ($\sim$115 arcmin$^2$).}
    \label{central-region}    
    \begin{tabular}{@{}ccccc}
      
      \hline
      & No. sources& parent catalog& parent catalog& X-ray sources\\
      & parent 24 $\umu$m& with X-ray& with variability& with variability\\
      & catalog& No. (\%)& No. (\%)& No. (\%)\\

      \hline \hline
       Long-term&189&90 (48)&8 (4)&5 (6)\\
       Short-term&181&99 (55)&5 (3)&3 (3)\\
      \hline

    \end{tabular}
    \end{center}

  \end{minipage} 
\end{table*}

\subsection{Monochromatic IR luminosities}

  From the {\it Rainbow} Database we obtained the rest-frame $24\,\umu$m monochromatic luminosities for the MIPS
  $24\,\umu$m variable sources. As the contribution of the AGN to the total luminosity in the 24 $\umu$m variable sources is expected to be low (See Section 6.1), the luminosities were obtained by fitting  the star forming galaxy templates from \citet{Chary01}.  Therefore, the fitted templates provide a reasonable approximation to the rest-frame 24 $\umu$m luminosities of the sources, which arise from star formation in the host galaxy and the putative AGN. For each source we used all the available photometric mid-to-far IR data points to fit the SEDs. Apart from the MIPS 24 $\umu$m flux, the Rainbow catalogs include photometry in the four IRAC bands, MIPS 70 $\umu$m and {\it Herschel}/PACS 100 and 160 $\umu$m. For our sample of 24 $\umu$m variable sources, 32\% have 70 $\umu$m photometry, 23\% have 100 $\umu$m photometry, and 18\% have 160 $\umu$m photometry.

\begin{figure}
   \begin{minipage}{84mm}
    \begin{center}
\includegraphics[width=1.1\textwidth]{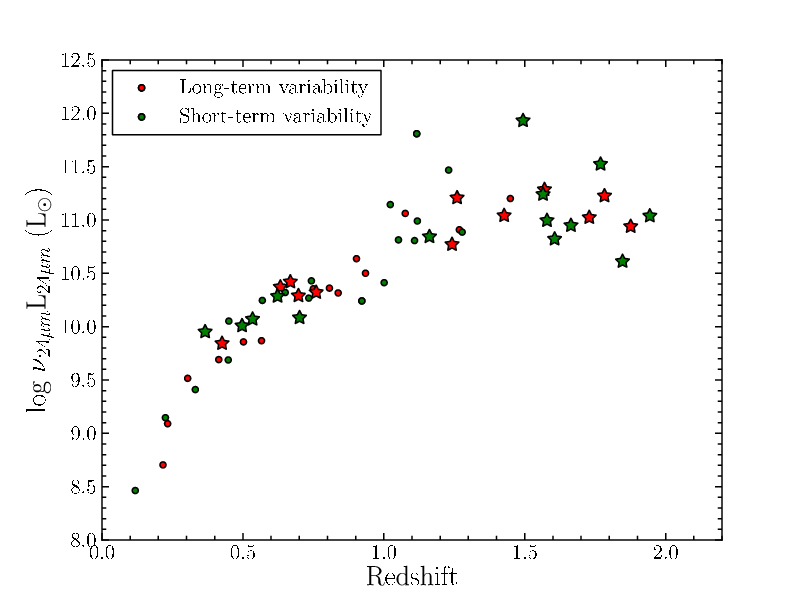}
      \caption{Rest-frame monochromatic 24~$\umu$m luminosity as a
        function of the redshift for the sources in
        the E-CDFS for long-term (red) and short-term (green)
        variable sources. The stars correspond to the
        variable sources satisfying the \citet{La} AGN selection
        criteria} 
      \label{luminosity-z}
      \end{center}
    \end{minipage}
\end{figure}

 Figure \ref{luminosity-z} shows these
  luminosities against the redshift
  for the long-term (red) and short-term (green) variable candidates in the E-CDFS.
The  mean value of rest-frame $\log (\nu L_{24\mu{m}} / L_{\sun})$ is 10.5 for both, the long-term
  and the short-term variable sources. For those
  candidates satisfying the \citet{La} AGN selection criteria the mean
  values are $\log (\nu L_{24\mu{m}} / L_{\sun}) =10.7$ for both, the
  long-term and the short-term 
  variable candidates. Conversely, the candidates not
  satisfying the \citet{La} criteria have mean values of $\log (\nu L_{24\mu{m}} / L_{\sun})= 10.3$ and 
  10.4 for the long-term and the short-term variable candidates, respectively.  This is expected as galaxies in the
    \cite{La} wedge tend to have a
    higher AGN fraction 
    contributing to their IR emission than those outside (see
    previous section).

\begin{table*}
  \begin{minipage}{180mm}
    \begin{center}
    \caption{Summary of fractions of  MIPS $24\,\umu$m variable sources 
selected as AGN by other criteria.}
    \label{resumen-candidatos-otros-metodos}    
    \begin{tabular}{@{}ccccccccc}
      
      \hline
      &No. & X-ray$^1$& radio$^2$ &  other AGN$^3$  &IR$^4$ & Combined$^5$ & \citet{La}$^6$  &  Combined$^7$ \\
      &variable& & excess&catalogs& power law& &criteria&criteria\\
      &sources &No. (\%)&No. (\%)&No. (\%)&No. (\%)&No. (\%) &No. (\%)&No.(\%)\\
      \hline
      \multicolumn{9}{c}{Long-term variable sources}\\
      \hline \hline
       Deepest X-ray region& 8 & 5 (63) & 1 (13) & 3 (38) & 0 (0) & 5 (63) & 3 (38) & 7 (88)\\
      In  the E-CDFS & 28 & 7 (25) & 2 (7) & 4 (14) & 1 (4) & 8 (29) & 12 (43) & 17 (61)\\
      Outside the E-CDFS & 11 & 0 (0) & 0 (0) & 0 (0) & 2 (18) & 2 (18) & 5 (45) & 5 (45)\\
      In IRAC & 39 & 7 (18) & 2 (5) & 4 (10) & 3 (8) & 10 (26) & 17 (44) & 22 (56) \\
      All & 39 & 7 (18) & 2 (5) & 4 (10) & 3 (8) & 10 (26) & 17 (44) & 22 (56) \\
     \hline
     \multicolumn{9}{c}{Short-term variable sources}\\
      \hline \hline
       Deepest X-ray region& 5 & 3 (60) & 1 (20) & 2 (40) & 0 (0) & 3 (60) & 2 (40) & 4 (80)\\
      In the E-CDFS & 33 & 4 (12) & 1 (3) & 3 (9) & 0 (0) & 5 (15) & 14 (42) & 17 (52)\\
      Outside the E-CDFS& 22 & 1 (5) & 1 (5) & 1 (5) & 1 (5) & 2 (9) & 5 (23) & 5 (23)\\
      In IRAC& 43 & 5 (12) & 2 (5) & 4 (9) & 1 (2) & 7 (16) & 19 (44) & 22 (51)\\
      All& 55 & 5 (9) & 2 (4) & 4 (7) & 1 (2) & 7 (13) & 19 (35) & 22 (40)\\
      \hline

    \end{tabular}
    \end{center}

$^1$ Variable MIPS 24 $\umu$m sources detected in X-rays.

$^2$ Variable MIPS 24 $\umu$m sources with radio excess.

$^3$ Variable MIPS 24 $\umu$m sources in other AGN catalogs. (See notes in Table 5). 

$^4$ Variable MIPS 24 $\umu$m sources detected as IR power-law AGN.

$^5$ Combined 1$^{st}$, 2$^{nd}$, 3$^{rd}$, and 4$^{th}$ criteria.

$^6$ Variable MIPS 24 $\umu$m sources satisfying the Lacy et al. 2004 criteria.

$^7$ All the criteria combined.

  \end{minipage} 
\end{table*}

 \subsection{Radio properties}
\label{radio-section}

 We investigate the radio properties of the MIPS
  $24\,\umu$m variable sources, since radio observations are in 
  principle not biased against obscured AGN. Since star-forming
  galaxies also emit at radio frequencies and show a tight correlation between the
  IR and the radio emission \citep[e.g.][]{Helou85,Condon92,Ivison10}, it is also possible to select AGN in cosmological fields by 
  looking for radio excess sources \citep[see e.g.][]{Do05, DelMoro}. 
We cross-correlated our 24 $\umu$m variable sources with the
\citet{Mi} radio 1.4\,GHz source catalog and found that 2 long-term
and 7 (one of which is outside the E-CDFS) short-term variable sources had a radio counterpart within a search
radius of 2.5$\arcsec$.  That is, 7\% and 18\% of the MIPS
  $24\,\umu$m long and short term variable sources.

\begin{figure}
\includegraphics[width=0.52\textwidth]{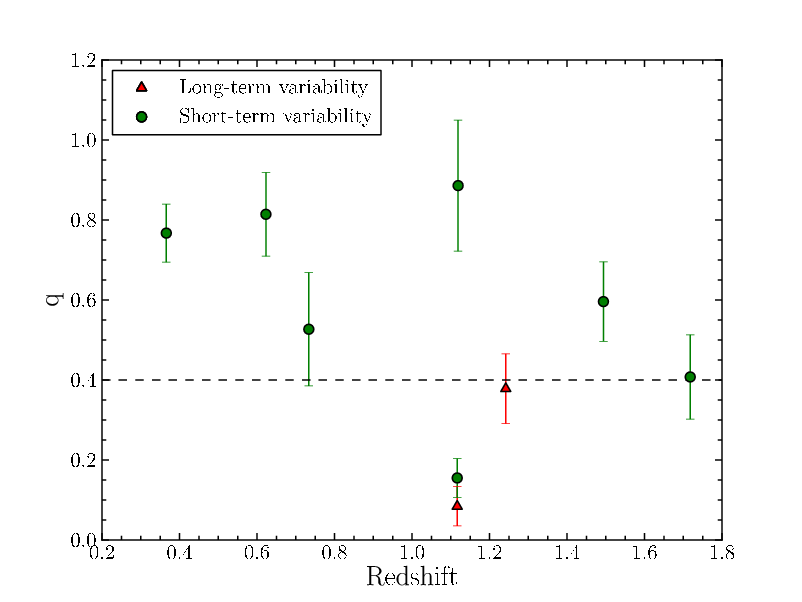}
  \caption{The $q=log(f_{24~\umu {\rm m}}/f_{\rm 1.4~GHz})$ ratio
    versus the redshift for long-term variable
    sources (red triangles) 
      and short-term variable sources (green circles). Sources
        with $q<0.4$ are considered to have a 
      radio excess, as shown by Del Moro et al. (2013). Note we include a source out of the E-CDFS. } 
  \label{radio-q}
\end{figure}

We calculated the $q$ ratio
defined as 
$q=log(f_{24~\umu{\rm m}}/f_{\rm 1.4~GHz})$ (see \citealt{Appleton04}) to determine if any of
these sources present a radio excess.
Figure \ref{radio-q} shows $q$ versus the redshift
for the 14 MIPS $24\,\mu$m variable sources with radio detections at
1.4\,GHz. \cite{Do05} considered radio excess sources those
  having  
q\textless0 for non-K-corrected fluxes. On the other hand,
\cite{DelMoro} demonstrated that this limit misses a large fraction of
sources with radio excesses based on the ratio between the far-IR 
luminosity and the radio emission. From Figure~5 
of \cite{DelMoro}, we can see that sources with $q<0.4$ can be
considered radio excess sources, and therefore AGN candidates. Among
the MIPS $24\,\umu$m variable sources we find only  2 ($7\%$)
long-term and  2 (one of them out of the E-CDFS) ($3\%$) 
short-term variable sources are radio excess sources,  all of them
  at $ z>1.1$. Note that we include the short-term source just above the line (see Figure \ref{radio-q}).  For reference, their IDs are
  given in Table~\ref{tabla-comparacion-catalogos}. These small
  fractions of radio excess sources, if taken as AGN candidates among the
  MIPS $24\,\umu$m variable sources is generally consistent with the
  little overlap between radio selected AGN and AGN selected via their 
  X-ray and/or IR emission in cosmological fields \citep{Do05,
    Hickox09, Villforth12}. 

\subsection{Comparison with other variability studies}
Finally, we cross-correlated our 24~$\umu$m variable sources with sources 
found to be variable in other studies in the E-CDFS, using a search radius of 2.5$\arcsec$. We found two long
term and two short term variable sources at $24\,\umu$m in common with
the optical variability studies of \cite{Tre8} and \cite{VKG}, and 
none with those of  \citet{Co}, \cite{KS}, and \citet{Sa}. Three of these four
sources are also detected in X-rays, as can be seen from
Table~\ref{tabla-comparacion-catalogos}. The low 
correspondence between optical and mid-IR variable sources is expected
given the lags and lower variation of amplitude observed and predicted
in the mid-IR 
for local AGN when compared with those observed in the optical
\citep{Neu,Glass04,HK}.

\section{Discussion and Summary}
\label{summary}
In this work we have used multi-epoch deep {\it Spitzer}/MIPS $24\,\umu$m
observations in GOODS-South to look for mid-IR variable sources. 
  The goal was to identify low luminosity and possibly obscured AGN
  candidates 
  not identified by other methods. To select variable sources
  we used a $\chi^2$-statistics method to take into account 
 the different photometric
errors due to the different depths between epochs and
varying depth within a given mosaic. 
By combining $24\,\umu$m data taken over three years and four epochs
we studied long-term variability over time scales of
  months-years.  Additionally we subdivided the longest duration
  epoch in three subepochs that allowed us to study the short-term
variability in time scales of days over a period of seven days. In each epoch and subepoch,
we restricted the analysis to sources above the 5$\sigma$ detection limit and without neighbours 
at distances of less than 10$\arcsec$ to minimize crowding effects in the photometry. We used 
a 2$\arcsec$ cross-matching radius, imposing additionally 
that the 2$\arcsec$ criterion was fulfilled in each pair of epochs. 

After discarding problematic sources, our sample
contains 39 long-term and 55 short-term mid-IR
variability-selected AGN candidates over  the GOODS-South areas of
1360\,arcmin$^2$ and 1960\,arcmin$^2$, respectively, covered by the different epochs. The expected
fraction of false positives in our sample of variable mid-IR sources 
is estimated to be about 40\%.
 The mid-IR
long and short-term  variable sources comprise approximately 1.7\% and
2.2\% of the parent MIPS 24 $\umu$m samples, respectively. After removing the expected number of false positives the estimated percentages are 1.0\% and 1.4\%. These fractions of variable
sources are typical of optical and near-IR variability studies in 
cosmological fields.
 The typical variability at $24\,\umu$m 
of the sources is 40\%, both
  for the long and short-term variable sources.

In Section~\ref{combo-section} we studied the
properties of these variability selected AGN candidates restricting
the region to the E-CDFS, as it contains the deepest and largest
 multi-wavelength coverage.  We also made use of the {\it Rainbow}
  photometric redshifts that are calculated using optical and IR data for the objects without spectroscopic redshift.
In the E-CDFS, we found 28 and 33 long
and short term mid-IR variable sources, respectively, typically
  at $z=1$  which implies our work is sensitive to variable 
emission at 12~$\umu$m rest-frame. 

We cross-correlated our AGN
candidates with other AGN catalogs including X-ray, radio, and
variable catalogs in the E-CDFS. In the region with
  the best coverage by the  deepest X-ray 
  data, the {\it Chandra} 4\,Ms catalog of \cite{Xue} (CDFS; $\sim$465 arcmin$^2$), 30\%  of the
  variable sources (both short and long term) are also detected in X-rays. However, their 
$0.5-8\,$keV  luminosities are typically $2\times 10^{42}\,{\rm
  erg\,s}^{-1}$, with a few sources with X-ray luminosities of 
$\sim 10^{40}\,{\rm
  erg\,s}^{-1}$. In the {\it Chandra} 4\,Ms catalog of \cite{Xue} there
 are 149 sources identified as AGN due to their X-ray luminosity in our parent MIPS 24 $\umu$m catalogs (i.e., after removing close neighbours and merging individual catalogs), see Section 5.1. If we
 assume that the 24 $\umu$m variable sources  in the region covered by the \cite{Xue} catalog not detected in X-rays are AGN,
 they would only account for a small fraction ($\leq 13\%$) of the total AGN
 population in this field. 

As expected, the fraction of
  $24\,\umu$m variable sources with a radio excess
  ($q=log(f_{24~\umu{\rm m}}/f_{\rm 1.4~GHz})<0.4$)  is small, 
as is the case with variable sources identified in other wavelengths. 
Table~\ref{resumen-candidatos-otros-metodos} summarizes the
results. 

We also  investigated the IRAC properties of the $24\,\umu$m
  variable sources. The fraction of mid-IR variable selected AGN
  candidates  meeting  the \cite{Do} IR power
  law criteria for AGN is small. This is not surprising, as this 
method has been proven to be a
  very reliable method to select luminous AGN, although it is highly incomplete for low luminosity X-ray selected
AGN. Combining the AGN selected by the IR power law criteria with
the above X-ray, radio, and variability criteria, 
we find that 29\% and 15\%
 of the long-term and short-term MIPS $24\,\umu$m
variable sources would be also identified as AGN using other methods 
(see Table~\ref{resumen-candidatos-otros-metodos}). 
The lower fraction for the short-term
variable candidates is because a larger fraction of them lie in the
area with the shallowest X-ray coverage (that is, the
 E-CDFS, see  Section 6.2). 

In Table~\ref{resumen-candidatos-otros-metodos} we also included the
number of MIPS $24\,\umu$m variable sources that fall in the \cite{La}
IRAC colour-colour wedge. Approximately 44\% of the $24\,\umu$m
variable selected AGN candidates are located in this wedge.  
However, using their redshifts we concluded that 
of these only half of them would have colours compatible with a
small ($\sim 20\%$) AGN contribution (see
Section~\ref{photo-z-IRAC}). 

If we combined all these
criteria together we would obtain an upper limit of 
$\sim 56\%$ to the fraction of MIPS $24\,\umu$m variable sources  that
would be identified as AGN by other methods. 
For reference in Table~\ref{resumen-candidatos-otros-metodos} we also
listed these AGN fractions for 
  sources outside the E-CDFS. However, as noted before the
  multi-wavelength coverage and depth of the observations outside this
  region are not as good, so these fractions should be taken as lower
  limits. 

As explained in Section 6.2, only the central area of the E-CDFS is covered by the deepest X-ray data \citep{Xue}. We selected the region with the deepest and most homogeneous X-ray data ($\sim$115 arcmin$^2$), which is also covered by other AGN variability studies \citep{Co,KS,Tre8,VKG,Sa}. Combining all the criteria together, we obtained that $\sim$85\% of the 24 $\umu$m variable sources in this region would be identified as AGN by other methods (see Table~\ref{resumen-candidatos-otros-metodos}). The percentage in this region is higher than when considering  the entire E-CDFS due to the deepest X-ray data and because other AGN catalogs do not cover all the E-CDFS. In this 115 arcmin$^2$ region we compared the parent catalog with the variable sources (see Table~\ref{central-region}). As expected, in this region the fraction of 24 $\umu$m variable sources with an X-ray detection is higher (63\% for long-term and 60\% for short-term variable sources) than that of sources in the parent 24 $\umu$m catalog with X-ray detections (48\% for long-term and 55\% for short-term variable sources). Since the number of variable sources in this region is small, the percentages given at Tables~\ref{central-region} and \ref{resumen-candidatos-otros-metodos} suffer from small number statistics.

 In summary, we have shown that MIPS 24~$\umu$m variability provides a new method to
  identify AGN in cosmological fields. We find, however, that the 24 $\umu$m variable sources only  account for a small fraction  ($\leq$ 13 \%) of the general AGN population.
This is expected because model simulations predict a more
smothered variable signal and longer timescales in the mid-IR
than in the optical \citep{HK}.  Moreover, we found that the AGN contribution to the mid-IR emission 
of these 24 $\umu$m variable sources is low (typically less than 20\%). Since our method is only sensitive to high amplitude variability (see Section 4) then these 24 $\umu$m variable sources are likely to host low-luminosity AGN where the variability is expected to be stronger \citep{Tre}.

\subsection*{Acknowledgements}

 We thank the referee for valuable comments that helped improve the paper.

J.G.-G., A.A.-H., and A.H.-C. acknowledge support from the Augusto
G. Linares research program of the Universidad de Cantabria and from
the Spanish Plan Nacional through grant AYA2012-31447.

 P.G.P.-G. acknowledges support from MINECO grant AYA2012-31277. 

This work has made use of the Rainbow Cosmological Surveys Database,
which is operated by the Universidad Complutense de Madrid (UCM),
partnered with the University of California Observatories at Santa
Cruz (UCO/Lick,UCSC). This research has made use of the NASA/IPAC
Extragalactic Database (NED) which is operated by the Jet Propulsion
Laboratory, California Institute of Technology, under contract with
the National Aeronautics and Space Administration.  

This work is based on observations made with the Spitzer Space Telescope, obtained from the NASA/ IPAC Infrared Science Archive, both of which are operated by the Jet Propulsion Laboratory, California Institute of Technology under a contract with the National Aeronautics and Space Administration.

\appendix

\section[]{Photometric errors}
\label{photometry-errors}

 As we explained in Section~\ref{selection-variable-sources} errors are essential in
the $\chi^2$ method, and therefore  we checked them in each epoch. For every
source in each epoch, we calculated the ratio $F_{epoch}$/$\bar{F}$
and $\Delta$$F_{epoch}/F_{epoch}$, where $F_{epoch}$ is the flux in
each epoch, $\bar{F}$ is the mean flux for the source  from
  measurements in all epochs, and $\Delta
F_{epoch}$ is the error of the flux in each epoch. We separated the
values in bins according to their mean flux value, so each bin
contained 200 sources. For each bin we calculated the median of the
$F_i$/$\bar{F}$ values and +1$\sigma$ and -1$\sigma$  (26$^{th}$ and 84$^{th}$ percentiles) so that between
the median and $\sigma+$ there were the 34\% of the data in the bin,
and the same between the median and $\sigma-$. This is a measure of
the scatter of the fluxes and should be consistent with the
photometric errors.  We also calculated the median of the
$\Delta$$F_{epoch}/F_{epoch}$ for the sources of each bin.  

\begin{figure}
 \includegraphics[width=0.52\textwidth]{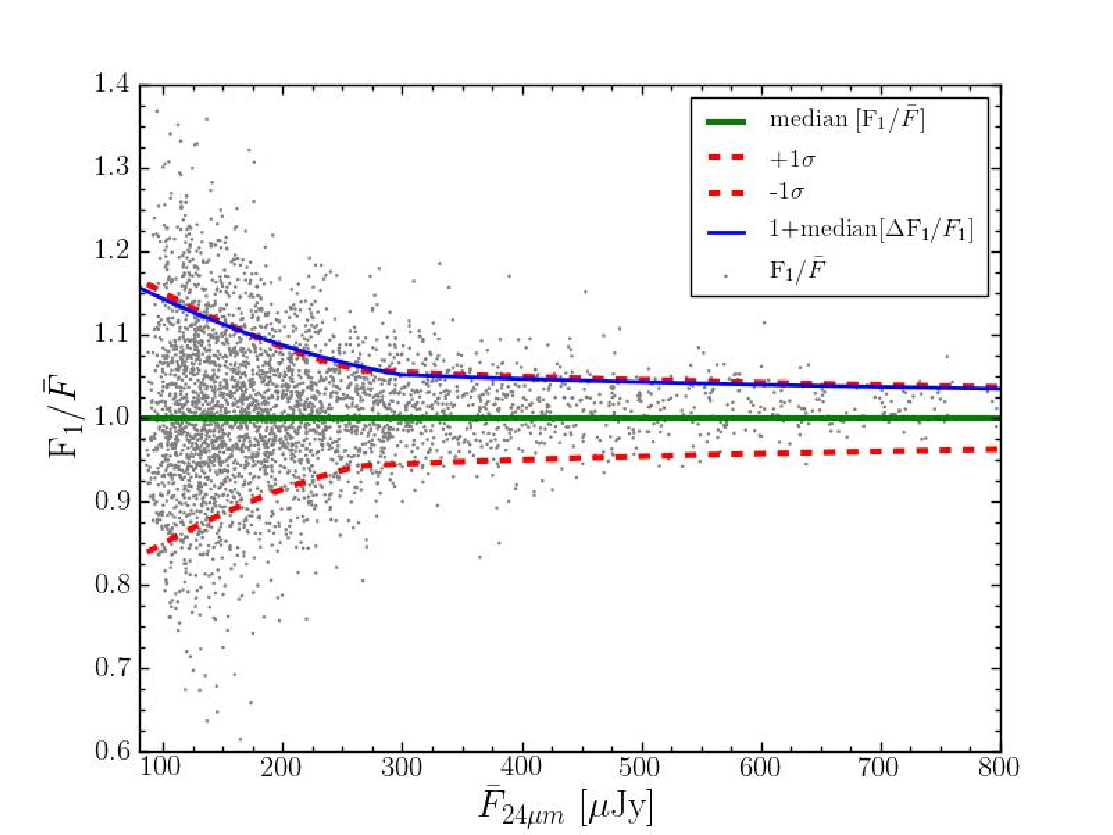}
  \caption{Scatter of the fluxes of the sources in epoch 1 (grey points). The  black line represents the fit of the median in each bin and the red lines represent the fit of the $\pm$1$\sigma$ (26$^{th}$ and 84$^{th}$ percentiles). The fit of the median of the photometric errors for each bin correspond to the blue line.}
  \label{trompeta}
\end{figure}

Figure \ref{trompeta}  plots the scatter in fluxes  as a
  function of the $24\,\umu$m median flux for Epoch
1,  as an example. As can be seen from the figure, the
median of the errors (blue line) is 
consistent with the dispersion of the fluxes (red line). This 
 confirms the validity of the estimated photometric errors used 
to calculate the
$\chi^2$ value. For all the epochs and subepochs these figures are
similar and the photometric errors are consistent with the dispersion
of the fluxes. 

\section[]{Kolmogorov-Smirnov test for $\chi^2$ distribution}
\label{KS-test-appendix}

We performed a Kolmogorov-Smirnov test (hereafter KS-test) in order to determine
 whether our observed distributions  of $\chi^2$ values differ
significantly 
from the theoretical distribution  based on the assumption of gaussian photometric errors. We did the test for the data
  used to study both the long-term
and short-term variability at $24\,\umu$m.

\begin{figure}
\centering
 \includegraphics[width=84mm]{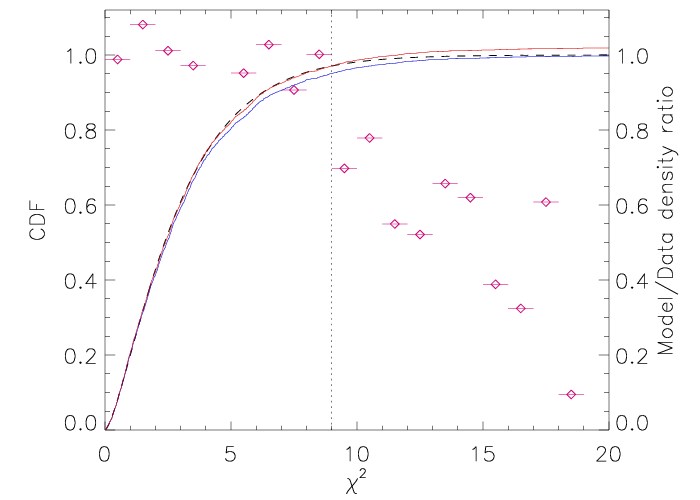}
  \caption{KS-test for the long-term data. The dashed line is the theoretical $\chi^2$ distribution for 3 degrees of freedom corresponding to long-term variability. The blue line is the CDF of the data and the red line is the scaled and truncated CDF in $\chi^2$=9 (vertical dotted line). The magenta symbols (right axis) represent the ratio between the theoretical distribution of $\chi^2$ and the number of sources in the sample, in intervals of $\Delta\chi^2$=1.}
  \label{KS-long}
\end{figure}

Figure \ref{KS-long} shows the cumulative distribution function, CDF,
for the long-term data. As can be seen from the figure  (and confirmed by a KS test), the
theoretical $\chi^2$ 
distribution for 3 degrees of freedom (dashed line) is incompatible
with the CDF of the data (blue line). This is
because the tail of objects with high $\chi^2$ is more populated than
expected from the photometric errors alone. This is shown with
the magenta symbols which are the ratio between the theoretical
distribution of $\chi^2$ and the number of sources in the sample, in
intervals of $\Delta\chi^2$=1. This ratio presents a small deficit of
sources in the range $\chi^2$=1-2 and an increasing excess at higher $\chi^2$
values. We truncated and rescaled the CDF at $\chi^2$=9 (red
line). The rescaled CDF follows well the theoretical distribution for
$\chi^2$\textless~9. There is a small depression around $\chi^2$=5,
but it is not significant. The KS-test found no significant
differences between the theoretical and the observed distributions
below $\chi^2$\textless~9. Cutting the CDF in $\chi^2$=10 the
differences start to be significant. This  indicates there is an excess of $\chi^2$\textgreater~9 sources, which is evidence for variability.

\begin{figure}
  \centering
  \includegraphics[width=84mm]{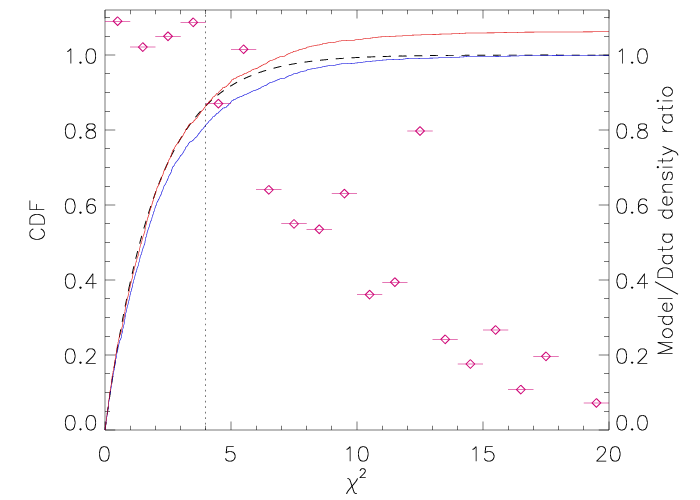}
  \caption{KS-test for the short-term data. The symbols are the same as Figure \ref{KS-long}. In this case the dashed line corresponds to the theoretical $\chi^2$ distribution for 2 degrees of freedom corresponding to short-term variability. The dotted line is the truncation of the CDF in $\chi^2$=4}
  \label{KS-short}
\end{figure}

Figure \ref{KS-short} shows the CDF for the short-term data. The
theoretical distribution corresponds to a $\chi^2$ distribution with 2
degrees of freedom.  The rescaled CDF follows well the theoretical
distribution for $\chi^2$\textless~4. For higher values of $\chi^2$
the difference between the theoretical and the observed distribution
is significant, again indicating that our criterion is valid for
selecting variable sources.

\section[]{Images of the variable candidates in the different epochs}

\begin{figure*}
  \begin{minipage}{180mm}
    \begin{center}

    \subfigure [Epoch 1]{\includegraphics[width=39mm]{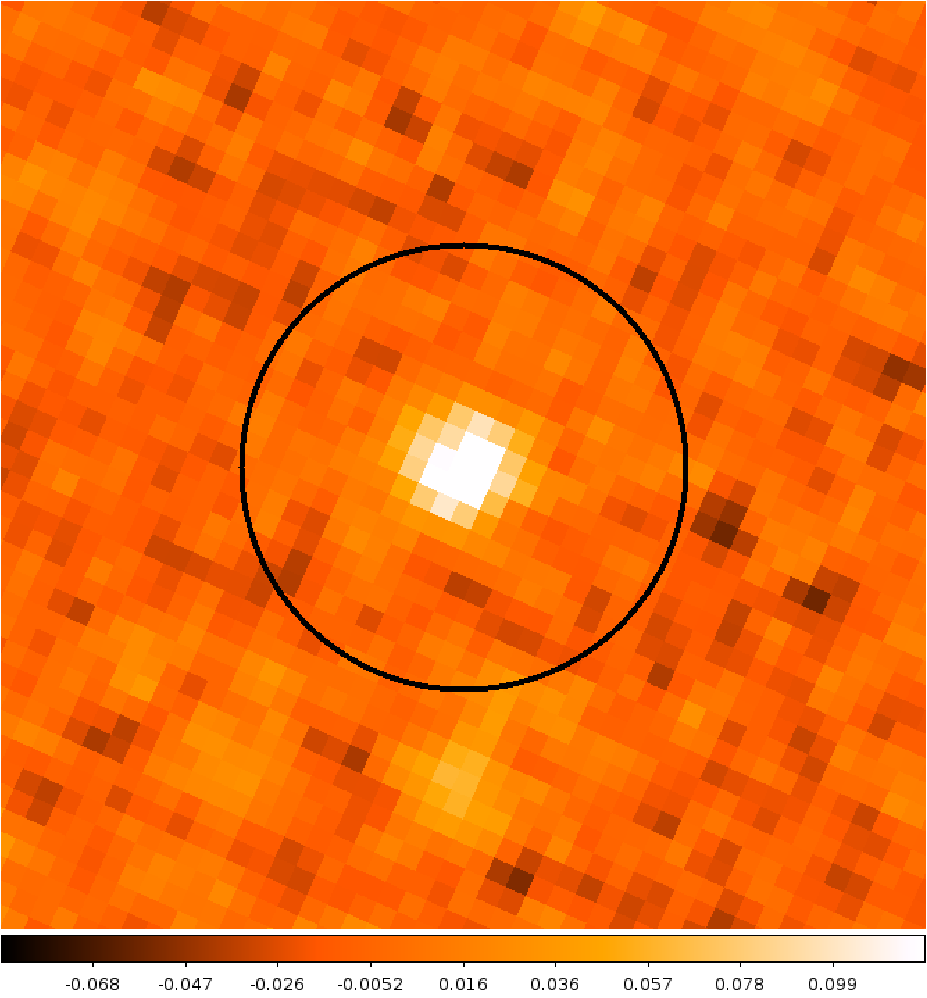}}{\hspace{0cm}}
    \subfigure [Epoch 3]{\includegraphics[width=39mm]{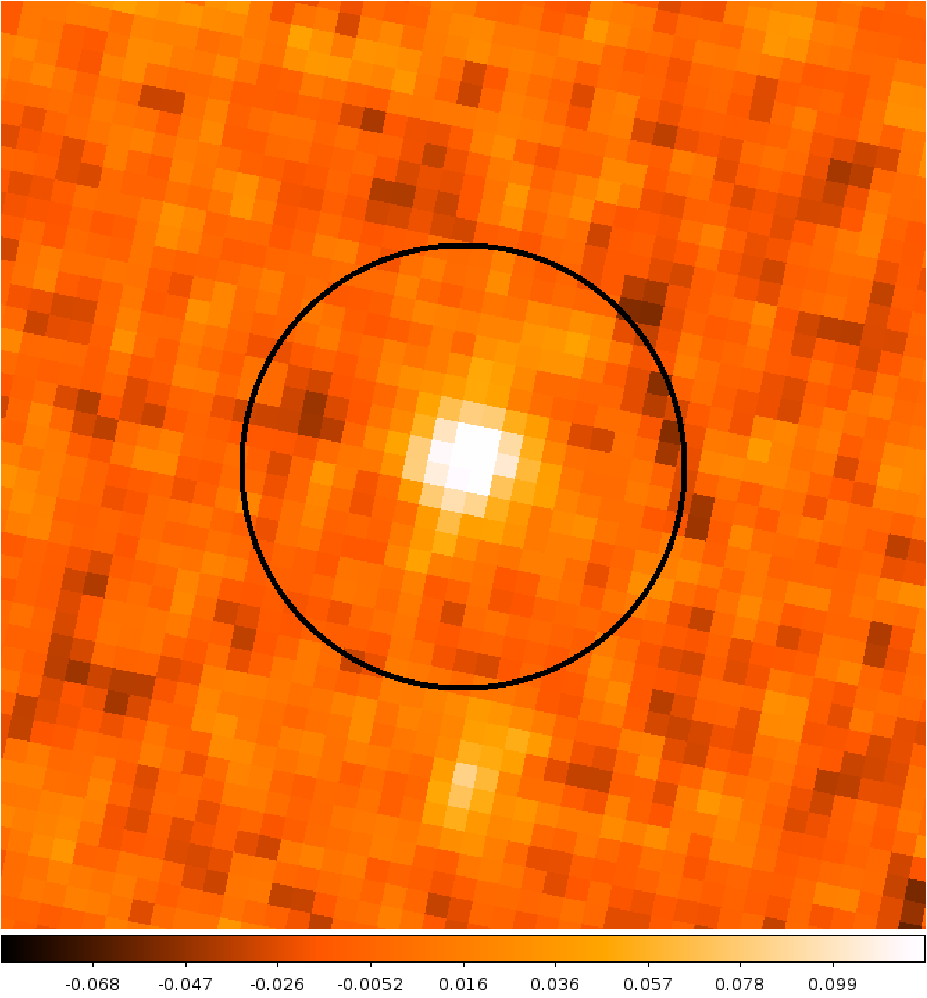}}{\hspace{0cm}}
    \subfigure [Epoch 6]{\includegraphics[width=39mm]{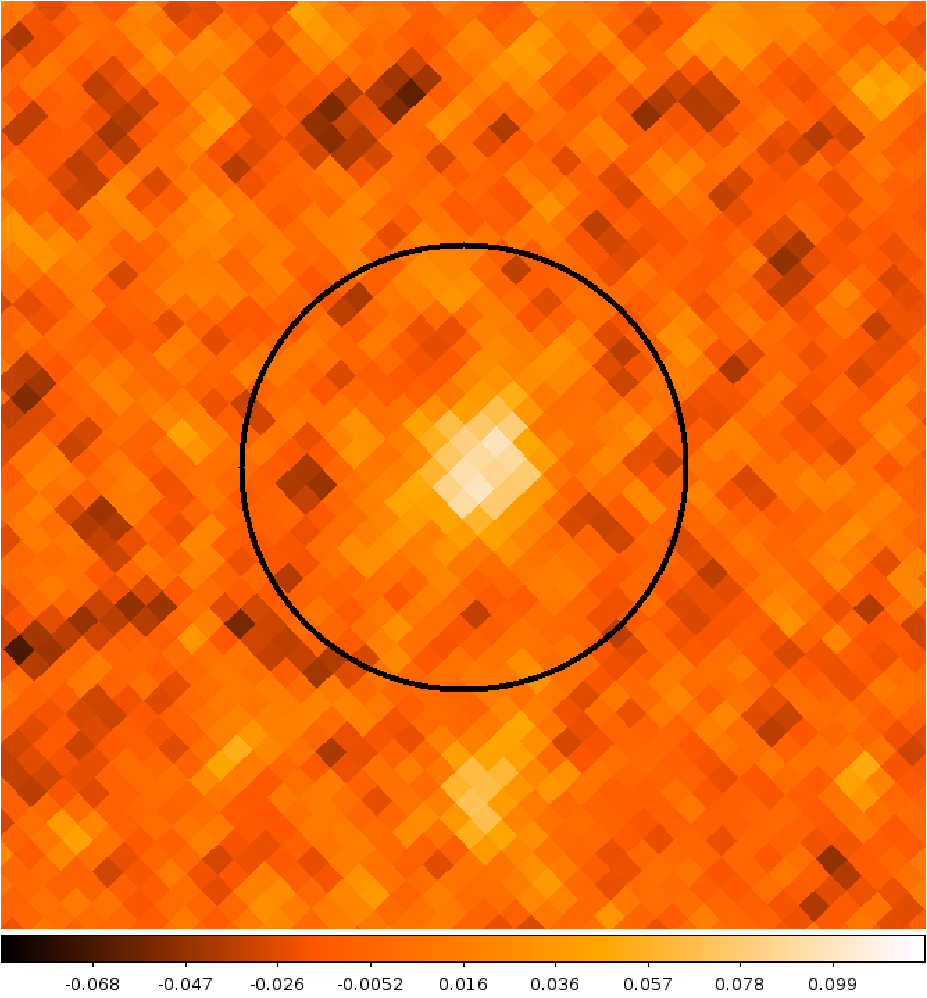}}{\hspace{0cm}}
    \subfigure [Epoch 7]{\includegraphics[width=39mm]{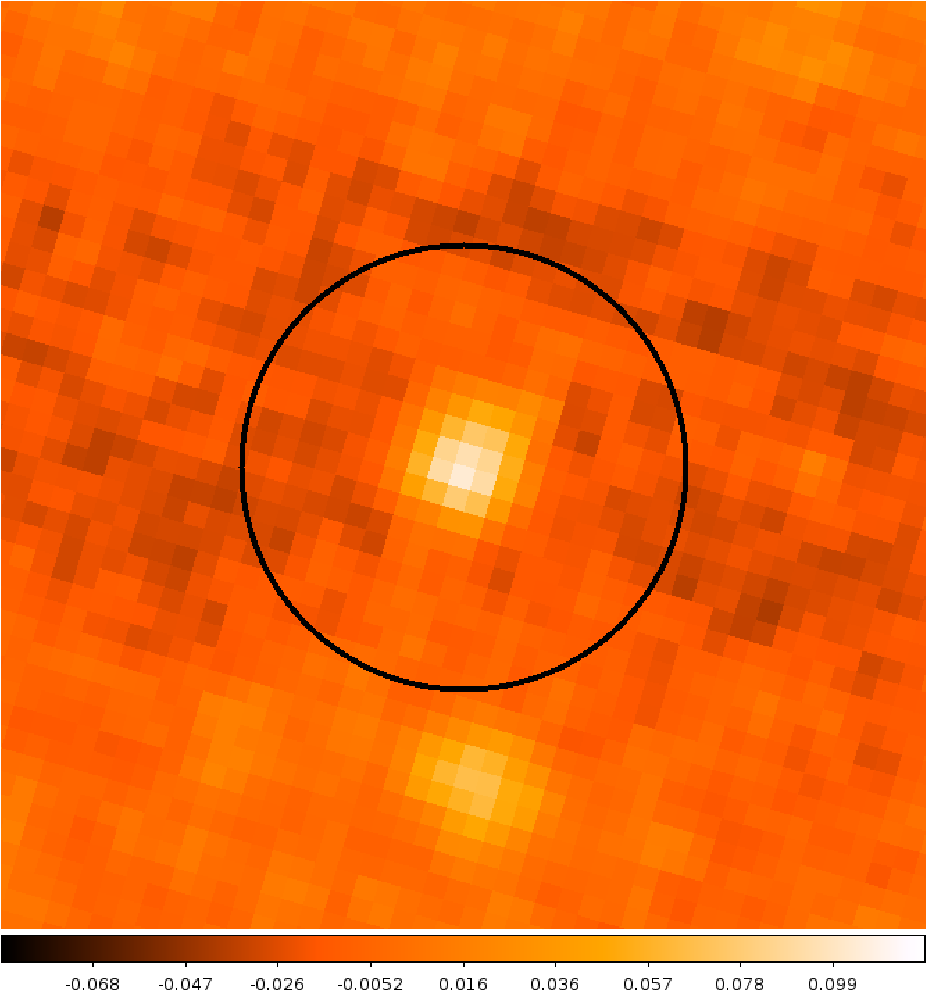}}{\hspace{0cm}}\\
  
    \caption{MIPS 24 $\umu$m images of the four epochs of the long-term variable candidate ID:5109. The FoV of the images is 50$\arcsec$x50$\arcsec$. The black circle represents the source and has a radius of 12$\arcsec$.}
    \end{center}
  \end{minipage}
\end{figure*}

\begin{figure*}
  \begin{minipage}{180mm}
    \begin{center}

    \subfigure [Epoch 1]{\includegraphics[width=39mm]{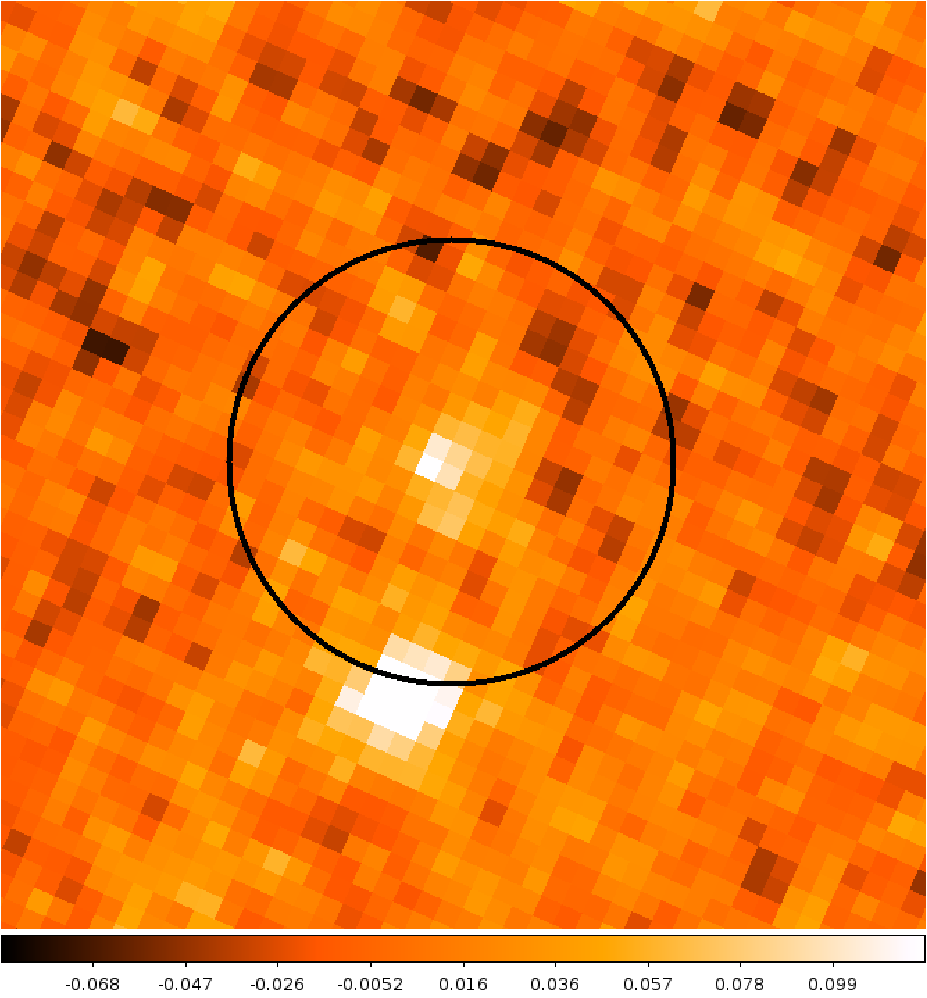}}{\hspace{0cm}}
    \subfigure [Epoch 3]{\includegraphics[width=39mm]{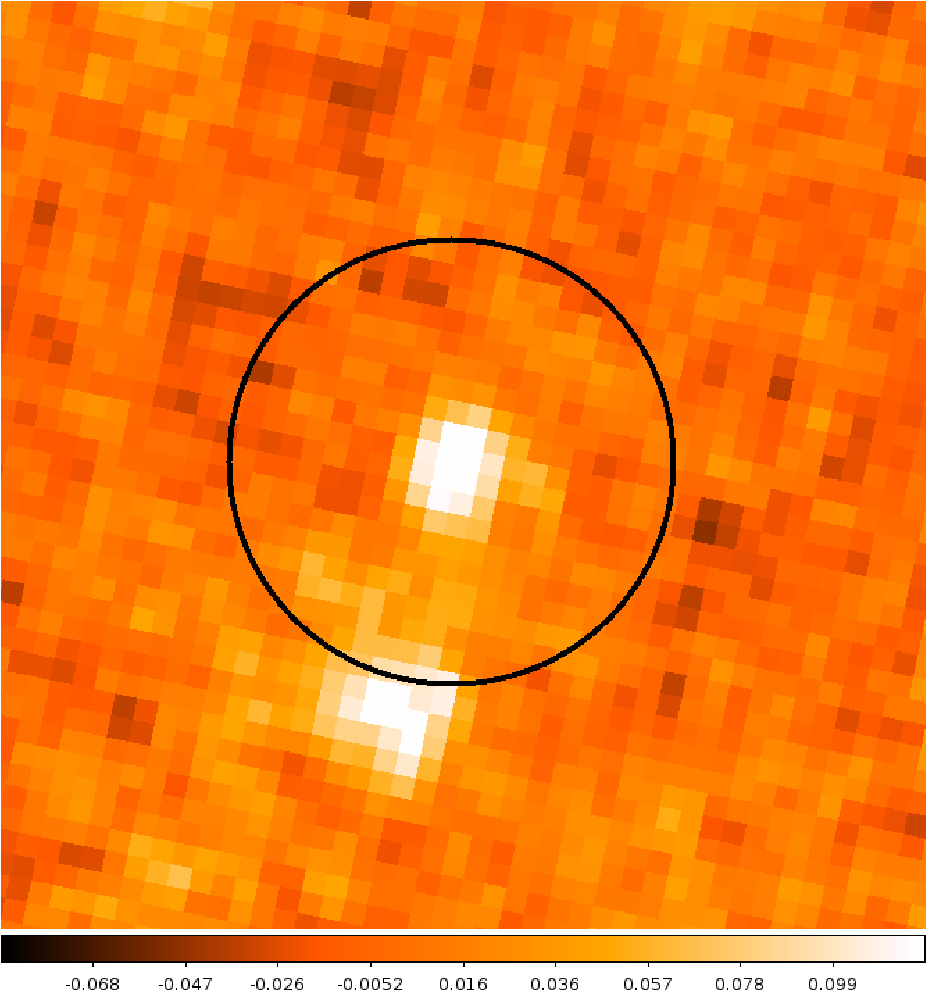}}{\hspace{0cm}}
    \subfigure [Epoch 6]{\includegraphics[width=39mm]{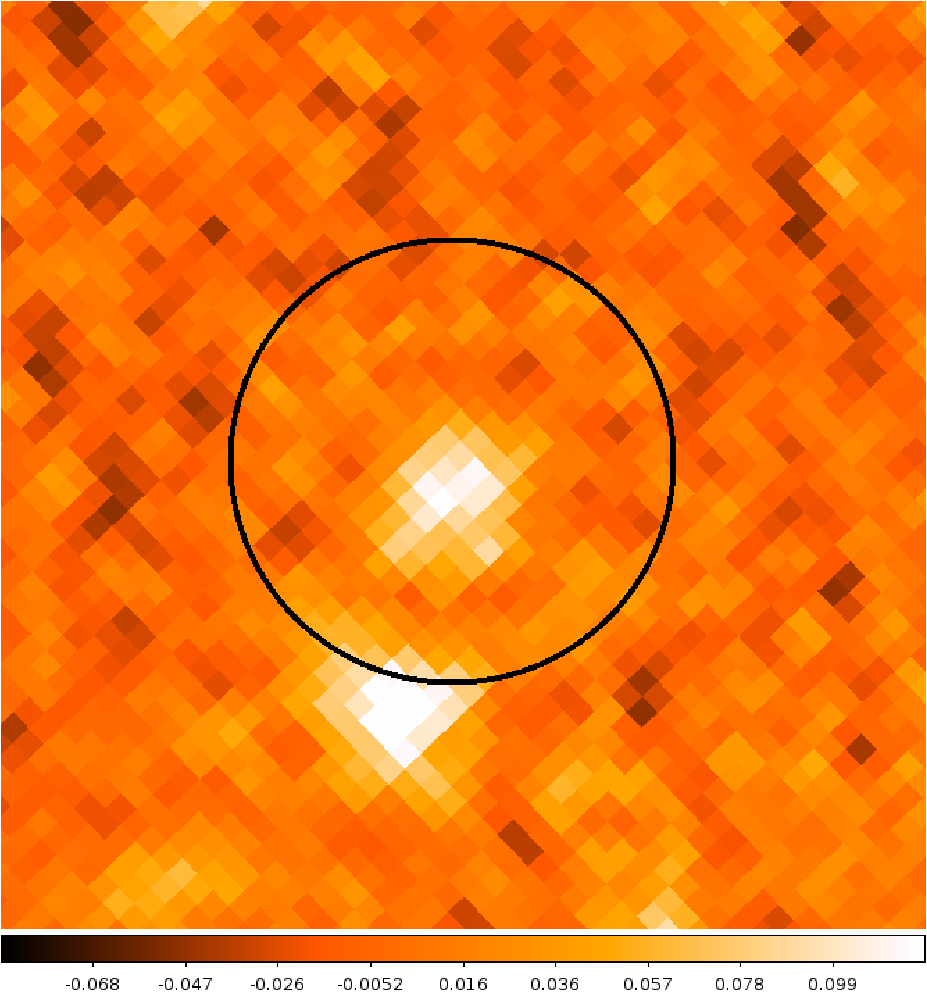}}{\hspace{0cm}}
    \subfigure [Epoch 7]{\includegraphics[width=39mm]{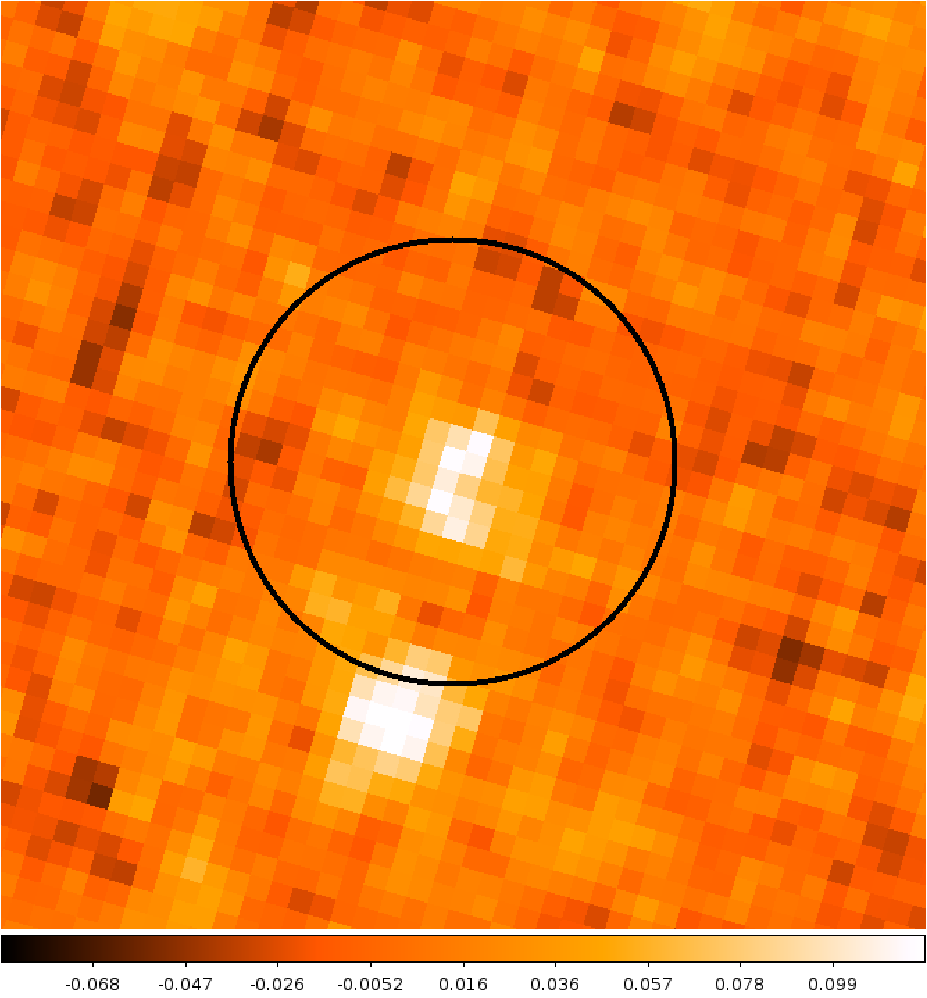}}{\hspace{0cm}}\\
  
    \caption{Same as Figure C1 for the long-term variable candidate ID:5086} 
    \end{center}
  \end{minipage}
\end{figure*}

\begin{figure*}
  \begin{minipage}{180mm}
    \begin{center}

    \subfigure [Epoch 7a]{\includegraphics[width=39mm]{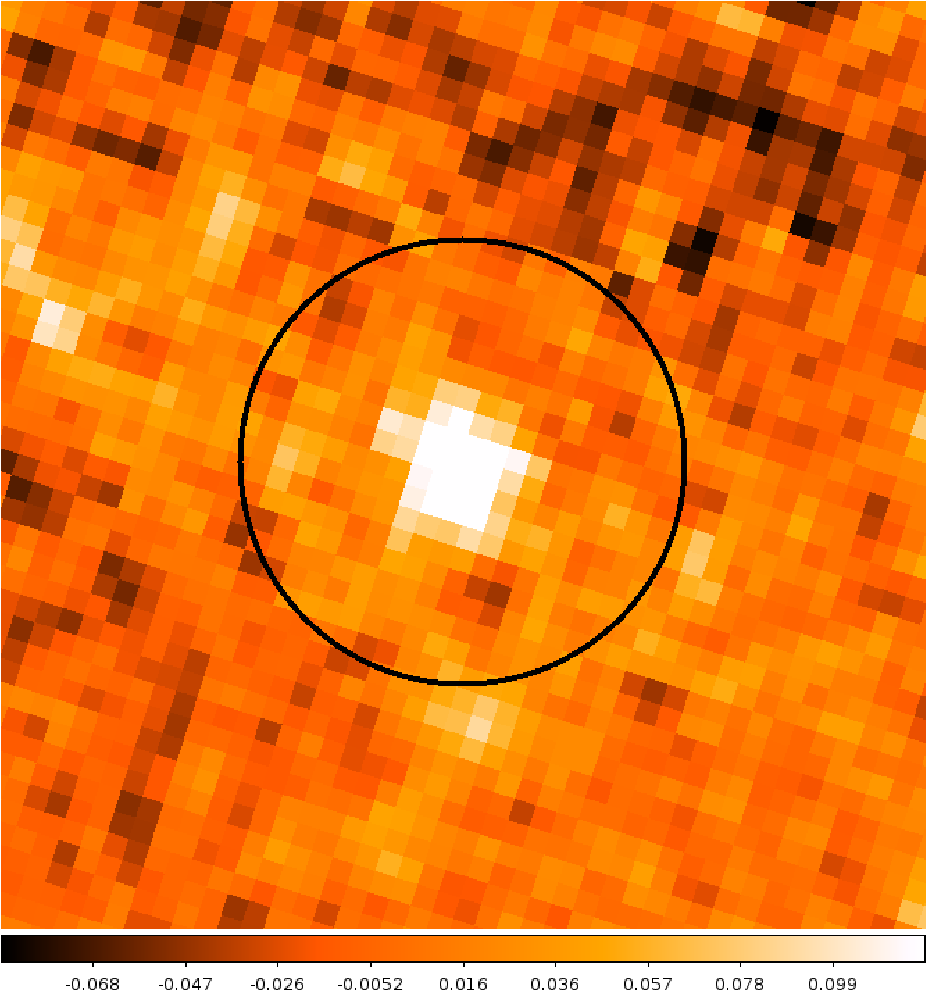}}{\hspace{0cm}}
    \subfigure [Epoch 7b]{\includegraphics[width=39mm]{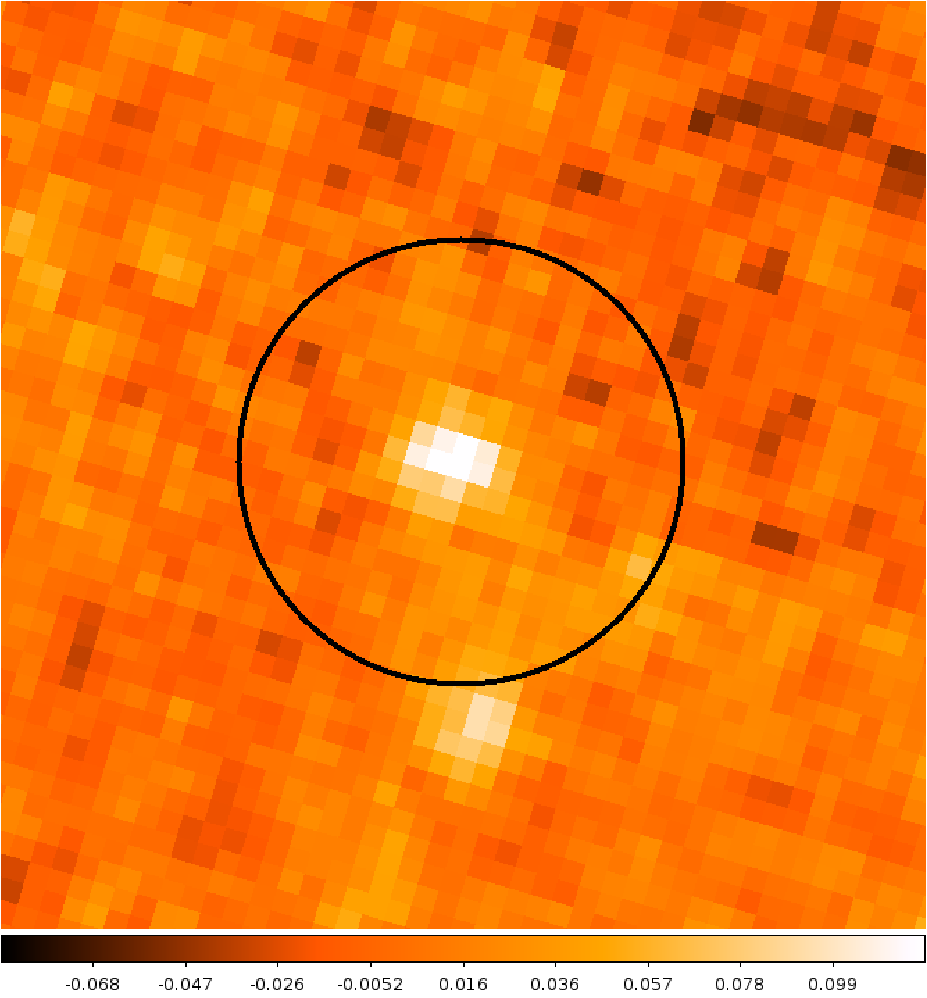}}{\hspace{0cm}}
    \subfigure [Epoch 7c]{\includegraphics[width=39mm]{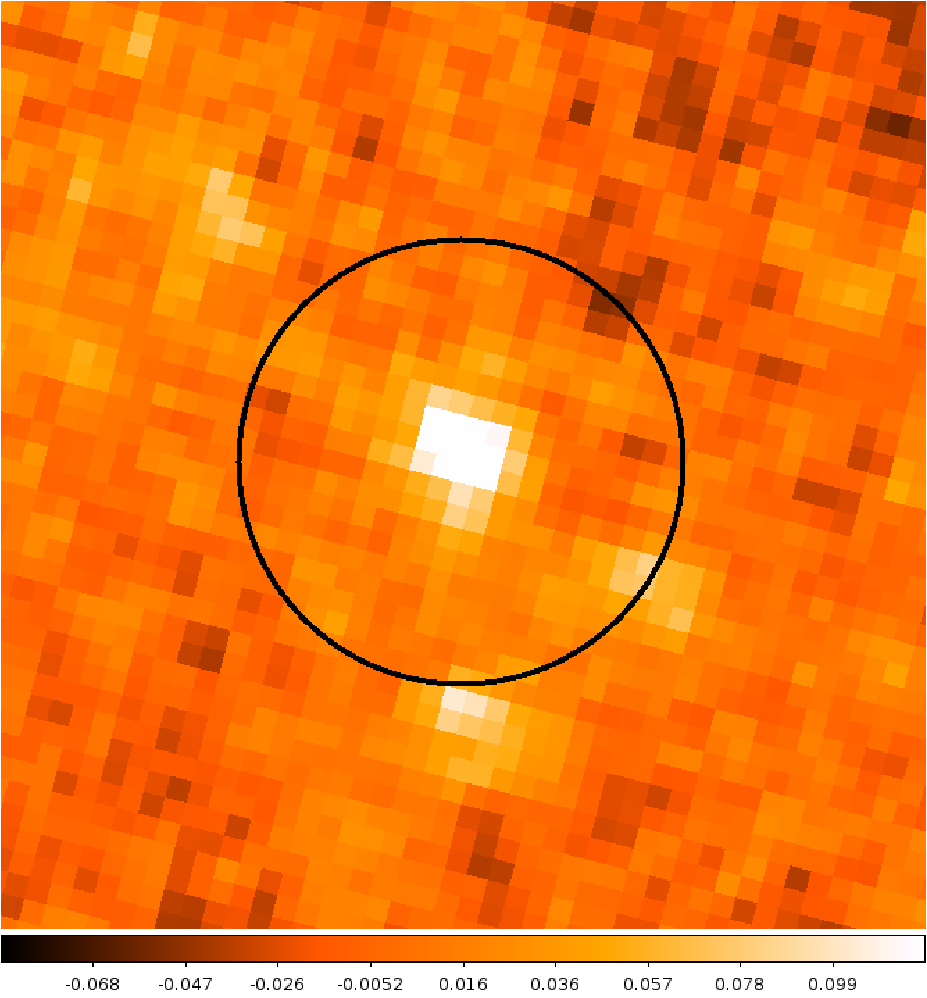}}{\hspace{0cm}}\\
    
    \caption{Same as Figure C1 for the short-term variable candidate ID:7513} 
    \end{center}
  \end{minipage}
\end{figure*}

\begin{figure*}
  \begin{minipage}{180mm}
    \begin{center}

    \subfigure [Epoch 7a]{\includegraphics[width=39mm]{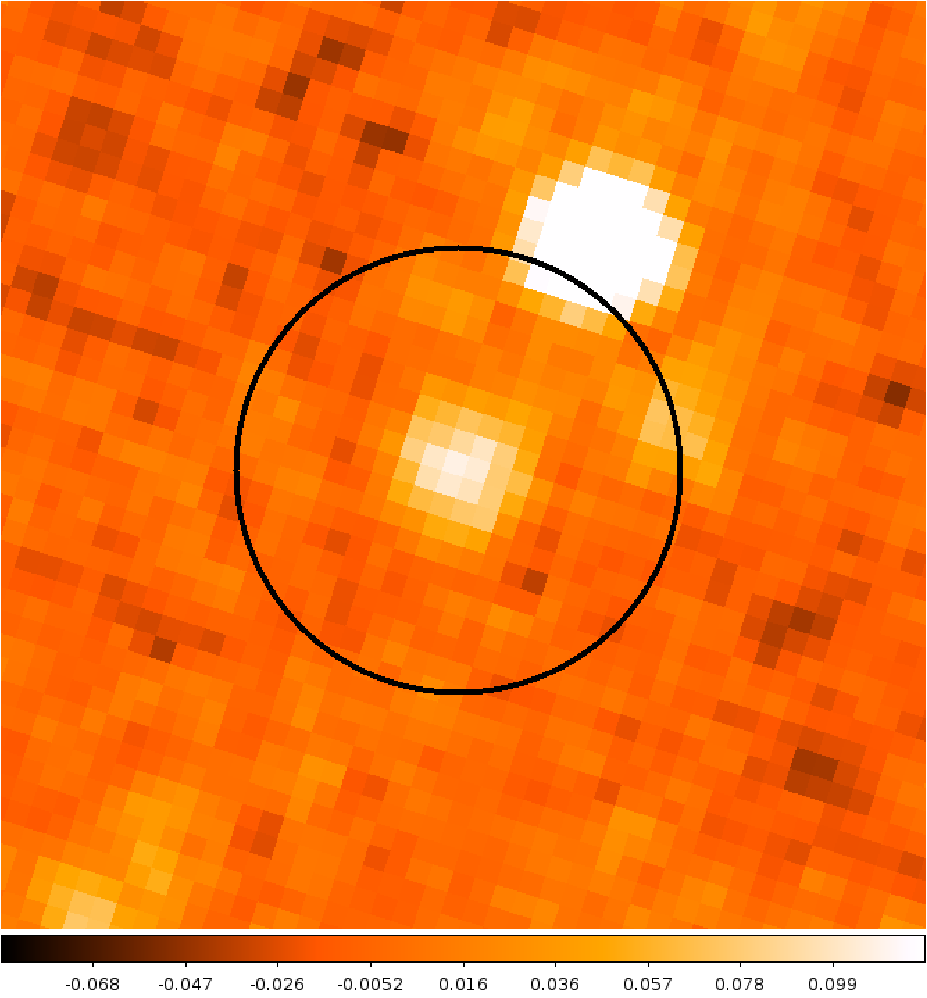}}{\hspace{0cm}}
    \subfigure [Epoch 7b]{\includegraphics[width=39mm]{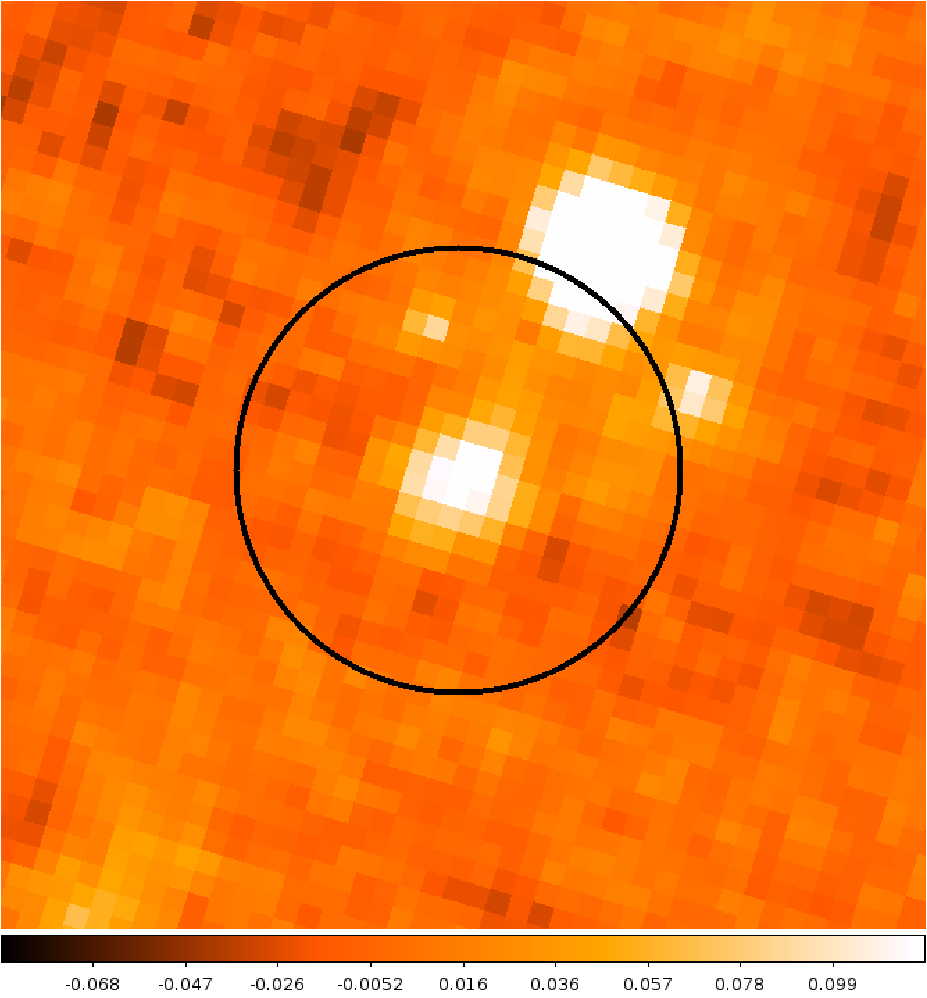}}{\hspace{0cm}}
    \subfigure [Epoch 7c]{\includegraphics[width=39mm]{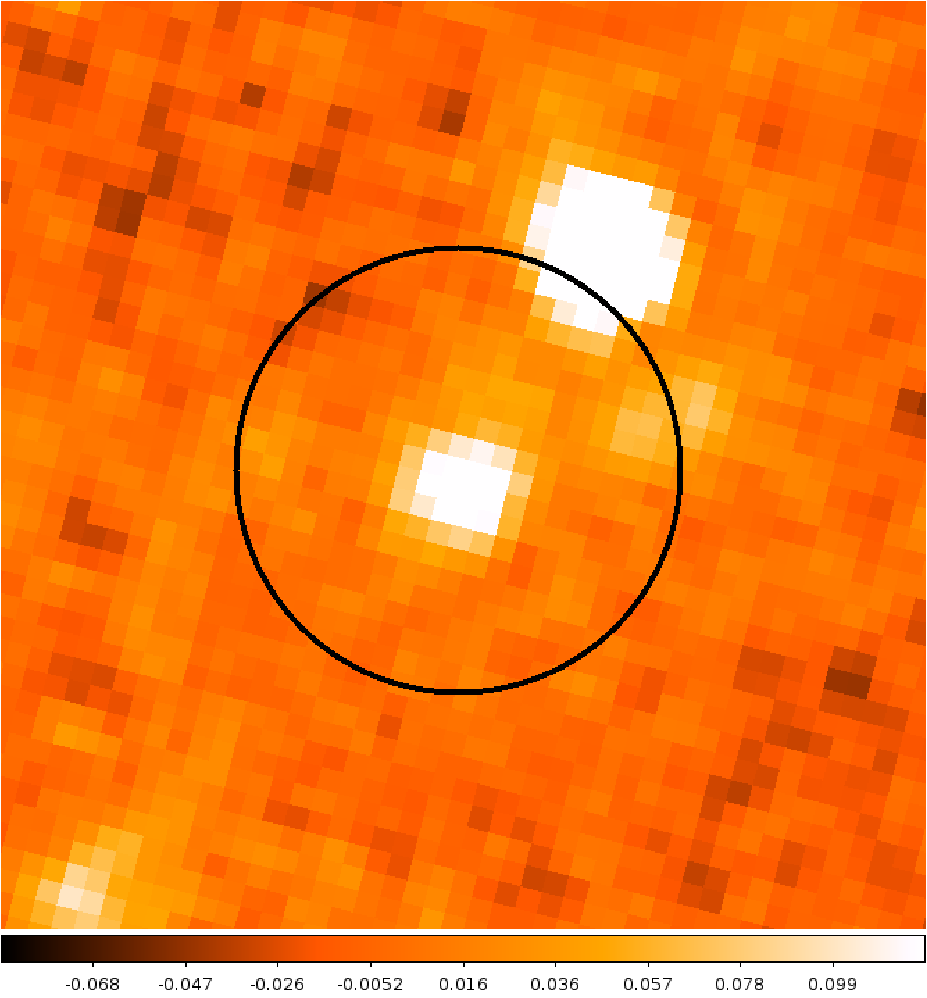}}{\hspace{0cm}}\\
    
    \caption{Same as Figure C1 for the short-term variable candidate ID:7921} 
    \end{center}
  \end{minipage}
\end{figure*}

\section[]{Long-term variable candidates catalog}
\label{tablas-apendice-long}

\begin{table*}
  \begin{minipage}{190mm}
    \begin{center}
      \caption{Catalog of the long-term variable candidates.}
      \label{long-term-candidates}
      \begin{tabular}{@{}ccccccccccccccc}
        \hline
        \tiny ID &\tiny  RA &\tiny  DEC &\tiny  Flux[epoch 1] &\tiny  Flux[epoch 3] &\tiny  Flux[epoch 6] &\tiny  Flux[epoch 7] &\tiny  $\bar{F}$ &\tiny  $\chi^2$ &\tiny  Var &\tiny  $R_{max}$  &\tiny  z$^+$ &\tiny R &\tiny  X-ray? &\tiny  Radio?\\
        &\tiny (J2000)&\tiny  (J2000)&\tiny ($\umu$Jy) &\tiny ($\umu$Jy) &\tiny ($\umu$Jy) &\tiny ($\umu$Jy)&\tiny ($\umu$Jy)&& \tiny  (\%)&&&&&\\
        \hline \hline
        
        \tiny 7348 &\tiny 52.8655903 &\tiny -27.5329591 &\tiny 410$\pm$16 &\tiny 410$\pm$24 &\tiny 416$\pm$20 &\tiny 342$\pm$17 &\tiny 395 &\tiny 12.23&\tiny 18.9&\tiny 1.22 &\tiny 0.71 &\tiny - &\tiny NO &\tiny NO\\
        
        \tiny 13441$^*$\footnotemark[1]$^,$\footnotemark[2] &\tiny 52.8882698 &\tiny -27.7897526 &\tiny 144$\pm$14 &\tiny 156$\pm$20 &\tiny 148$\pm$18 &\tiny 99$\pm$11 &\tiny 137 &\tiny 12.48 &\tiny 41.4&\tiny 1.57 &\tiny 0.57 &\tiny 20.98 &\tiny NO &\tiny NO\\
       
        \tiny 5071$^*$ &\tiny 52.9801950 &\tiny -27.8067295 &\tiny 129$\pm$13 &\tiny 83$\pm$14 &\tiny 104$\pm$21 &\tiny 127$\pm$7 &\tiny 111 &\tiny 11.39&\tiny 41.8 &\tiny 1.56 &\tiny 0.75 &\tiny 23.08 &\tiny NO &\tiny NO\\
        \tiny 11519 &\tiny 53.0214235 &\tiny -27.4722517 &\tiny 178$\pm$16 &\tiny 191$\pm$15 &\tiny 128$\pm$15 &\tiny 124$\pm$21 &\tiny 155 &\tiny 13.61&\tiny 43.2&\tiny 1.54 &\tiny 1.07 &\tiny - &\tiny NO &\tiny NO\\
        \tiny 12269 &\tiny 53.0288333 &\tiny -28.1259252 &\tiny 114$\pm$19 &\tiny 174$\pm$15 &\tiny 163$\pm$19 &\tiny 118$\pm$13 &\tiny 142 &\tiny 11.71&\tiny 42.4&\tiny 1.53 &\tiny 0.29 &\tiny - &\tiny NO &\tiny NO\\
        \tiny 14622$^*$\footnotemark[2] &\tiny 53.0467505 &\tiny -27.5834456 &\tiny 201$\pm$14 &\tiny 225$\pm$14 &\tiny 177$\pm$16 &\tiny 269$\pm$18 &\tiny 218 &\tiny 16.75&\tiny 42.5&\tiny 1.52 &\tiny 1.43 &\tiny 25.22 &\tiny NO &\tiny NO\\
        \tiny 9579 &\tiny 53.0534834 &\tiny -27.4522969 &\tiny 403$\pm$17 &\tiny 418$\pm$17 &\tiny 341$\pm$19 &\tiny 369$\pm$21 &\tiny 383 &\tiny 11.63&\tiny 20.3&\tiny 1.23 &\tiny 0.19 &\tiny - &\tiny NO &\tiny NO\\
        \tiny 16416 &\tiny 53.0543385 &\tiny -27.3906117 &\tiny 159$\pm$17 &\tiny 151$\pm$15 &\tiny 163$\pm$19 &\tiny 251$\pm$31 &\tiny 180 &\tiny 11.87&\tiny 55.5&\tiny 1.67 &\tiny 0.57 &\tiny - &\tiny NO &\tiny NO\\
        \tiny 5359$^*$ &\tiny 53.0719302 &\tiny -27.9225433 &\tiny 112$\pm$14 &\tiny 169$\pm$18 &\tiny 139$\pm$21 &\tiny 116$\pm$7 &\tiny 134 &\tiny 12.39&\tiny 42.6&\tiny 1.51 &\tiny 1.27 &\tiny 24.98 &\tiny NO &\tiny NO\\
       
        \tiny 5080$^*$\footnotemark[1]$^,$\footnotemark[2] &\tiny 53.0779999 &\tiny -27.7740151 &\tiny 247$\pm$17 &\tiny 260$\pm$16 &\tiny 196$\pm$20 &\tiny 208$\pm$10 &\tiny 228 &\tiny 11.73&\tiny 28.3&\tiny 1.33 &\tiny 1.729 &\tiny 24.08 &\tiny NO &\tiny NO\\
        \tiny 11976$^*$ &\tiny 53.0885467 &\tiny -27.8504550 &\tiny 134$\pm$17 &\tiny 161$\pm$14 &\tiny 126$\pm$19 &\tiny 182$\pm$8 &\tiny 151 &\tiny 18.92&\tiny 36.6&\tiny 1.44 &\tiny 1.78 &\tiny 23.28 &\tiny YES &\tiny NO\\
        \tiny 9796$^*$\footnotemark[1]$^,$\footnotemark[2] &\tiny 53.1110983 &\tiny -27.6040756 &\tiny 185$\pm$13 &\tiny 172$\pm$16 &\tiny 234$\pm$16 &\tiny 152$\pm$15 &\tiny 186 &\tiny 14.77&\tiny 43.7&\tiny 1.53 &\tiny 1.24 &\tiny 24.79 &\tiny NO &\tiny YES\\
       
        \tiny 5109$^*$\footnotemark[2] &\tiny 53.1515045 &\tiny -27.7620615 &\tiny 224$\pm$13 &\tiny 214$\pm$19 &\tiny 162$\pm$17 &\tiny 143$\pm$10 &\tiny 186 &\tiny 30.41&\tiny 43.6&\tiny 1.57 &\tiny 0.426 &\tiny 22.16 &\tiny NO &\tiny NO\\
       
        \tiny 7742$^*$\footnotemark[1]$^,$\footnotemark[2] &\tiny 53.1942822 &\tiny -27.6723954 &\tiny 333$\pm$15 &\tiny 358$\pm$25 &\tiny 373$\pm$17 &\tiny 277$\pm$16 &\tiny 335 &\tiny 19.00&\tiny 28.8&\tiny 1.35 &\tiny 0.668 &\tiny 21.84 &\tiny NO &\tiny NO\\
        \tiny 10015$^*$\footnotemark[1]$^,$\footnotemark[2] &\tiny 53.1948840 &\tiny -27.7538449 &\tiny 172$\pm$13 &\tiny 146$\pm$16 &\tiny 184$\pm$17 &\tiny 133$\pm$9 &\tiny  159 &\tiny 11.67&\tiny 31.6&\tiny 1.38 &\tiny 0.838 &\tiny 22.96 &\tiny YES &\tiny NO\\
        \tiny 2226$^*$\footnotemark[1]$^,$\footnotemark[2] &\tiny 53.2001248 &\tiny -27.8155580 &\tiny 146$\pm$13 &\tiny 155$\pm$17 &\tiny 180$\pm$22 &\tiny 191$\pm$8 &\tiny 168 &\tiny 12.37&\tiny 26.4&\tiny 1.30 &\tiny 0.233 &\tiny 20.78 &\tiny YES &\tiny NO\\
        \tiny13601$^*$\footnotemark[2] &\tiny 53.2022070 &\tiny  -27.8263358 &\tiny 149$\pm$19 &\tiny 174$\pm$16 &\tiny 144$\pm$16 &\tiny 196$\pm$8 &\tiny 166 &\tiny 17.41&\tiny 31.1&\tiny 1.36 &\tiny 1.117 &\tiny 25.08 &\tiny YES &\tiny YES\\
        \tiny 14779$^*$\footnotemark[1]$^,$\footnotemark[2] &\tiny 53.2047528 &\tiny -27.7432269 &\tiny 100$\pm$13 &\tiny 81$\pm$17 &\tiny 94$\pm$15 &\tiny 141$\pm$10 &\tiny 104 &\tiny 15.46&\tiny 57.0&\tiny 1.73 &\tiny 0.216 &\tiny 19.75 &\tiny YES &\tiny NO\\
        \tiny2324$^*$\footnotemark[2] &\tiny 53.2064728 &\tiny -27.8675984 &\tiny 174$\pm$16 &\tiny 135$\pm$13 &\tiny 135$\pm$19 &\tiny 177$\pm$8 &\tiny 155 &\tiny 11.94&\tiny 27.2&\tiny 1.31 &\tiny 0.414 &\tiny 21.38 &\tiny NO &\tiny NO\\
        
        \tiny10402 &\tiny 53.2076102 &\tiny -28.0760407 &\tiny 345$\pm$14 &\tiny 335$\pm$19 &\tiny 312$\pm$15 &\tiny 265$\pm$18 &\tiny 314 &\tiny 13.35&\tiny 25.5&\tiny 1.30 &\tiny 1.95 &\tiny - &\tiny NO &\tiny NO\\
        \tiny 4679 &\tiny 53.2103280 &\tiny -27.5243703 &\tiny 340$\pm$23 &\tiny 269$\pm$17 &\tiny 341$\pm$19 &\tiny 274$\pm$18 &\tiny 306 &\tiny 13.33&\tiny 23.4&\tiny 1.27 &\tiny 0.98 &\tiny - &\tiny NO &\tiny NO\\
        \tiny5930 &\tiny 53.2134267 &\tiny -28.1566425 &\tiny 136$\pm$14 &\tiny 178$\pm$16 &\tiny 95$\pm$20 &\tiny 164$\pm$16 &\tiny 143 &\tiny 12.72&\tiny 57.4&\tiny 1.86 &\tiny - &\tiny - &\tiny NO &\tiny NO\\
        \tiny4878$^*$ &\tiny53.2139709 &\tiny -27.6210732 &\tiny 212$\pm$14 &\tiny 273$\pm$15 &\tiny 166$\pm$15 &\tiny 212$\pm$16 &\tiny 207 &\tiny 11.88&\tiny 34.5&\tiny 1.43 &\tiny 1.88 &\tiny - &\tiny NO &\tiny NO\\
        
        \tiny10377$^*$\footnotemark[1] &\tiny 53.2681253 &\tiny -28.0246684 &\tiny 198$\pm$19 &\tiny 169$\pm$16 &\tiny 171$\pm$25 &\tiny 241$\pm$17 &\tiny 195 &\tiny 11.58&\tiny 37.4&\tiny 1.43 &\tiny 0.92 &\tiny 22.83 &\tiny NO &\tiny NO\\
        \tiny2380$^*$\footnotemark[1]$^,$\footnotemark[2] &\tiny 53.2730602 &\tiny -27.8755176 &\tiny 179$\pm$18 &\tiny 111$\pm$15 &\tiny 165$\pm$15 &\tiny 140$\pm$8 &\tiny 149 &\tiny 12.01&\tiny 46.0&\tiny 1.62 &\tiny 0.50 &\tiny 20.99 &\tiny NO &\tiny NO\\
        \tiny12354 &\tiny 53.2756206 &\tiny -28.0988060 &\tiny 149$\pm$13 &\tiny 122$\pm$17 &\tiny 153$\pm$15 &\tiny 81$\pm$15 &\tiny 126  &\tiny 16.58&\tiny 57.6&\tiny 1.90 &\tiny 1.86 &\tiny - &\tiny NO &\tiny NO\\
        \tiny8295$^*$ &\tiny 53.2766409 &\tiny -28.0183704 &\tiny 174$\pm$13 &\tiny 190$\pm$16 &\tiny 217$\pm$15 &\tiny 91$\pm$17 &\tiny 168 &\tiny 34.06&\tiny 75.1&\tiny 2.39 &\tiny 1.57 &\tiny 24.21 &\tiny NO &\tiny NO\\
        
        \tiny13829$^*$\footnotemark[1]$^,$\footnotemark[2] &\tiny 53.2853000 &\tiny -28.0627180 &\tiny 137$\pm$13 &\tiny 161$\pm$18 &\tiny 140$\pm$16 &\tiny 94$\pm$13 &\tiny 133 &\tiny 12.01&\tiny 50.9&\tiny 1.72 &\tiny 0.76 &\tiny 22.88 &\tiny NO &\tiny NO\\
        
        \tiny2552$^*$\footnotemark[1]$^,$\footnotemark[2] &\tiny 53.2943522 &\tiny -27.9635147 &\tiny 250$\pm$14 &\tiny 314$\pm$17 &\tiny 361$\pm$16 &\tiny 348$\pm$12 &\tiny 330 &\tiny 13.59&\tiny 20.2&\tiny 1.23 &\tiny 1.26 &\tiny 21.21 &\tiny YES &\tiny NO\\
        \tiny5086$^*$\footnotemark[1]$^,$\footnotemark[2] &\tiny 53.2980525 &\tiny -27.6902849 &\tiny 130$\pm$19 &\tiny 215$\pm$14 &\tiny 185$\pm$18 &\tiny 171$\pm$23 &\tiny 175 &\tiny 13.65&\tiny 48.4&\tiny 1.65 &\tiny 0.90 &\tiny 22.95 &\tiny NO &\tiny NO\\
        \tiny5451$^*$ &\tiny 53.3000223 &\tiny -27.8779051 &\tiny 283$\pm$18 &\tiny 345$\pm$16 &\tiny 275$\pm$15 &\tiny 298$\pm$10 &\tiny 300 &\tiny 11.82&\tiny 23.1&\tiny 1.25 &\tiny 0.81 &\tiny 22.52 &\tiny NO &\tiny NO\\
        \tiny2253$^*$\footnotemark[1]$^,$\footnotemark[2] &\tiny 53.3011532 &\tiny -27.7885025 &\tiny 220$\pm$17 &\tiny 182$\pm$15 &\tiny 239$\pm$16 &\tiny 184$\pm$10 &\tiny 206 &\tiny 12.90&\tiny 27.4&\tiny 1.31 &\tiny 1.45 &\tiny 23.98 &\tiny NO &\tiny NO\\

        \tiny12132$^*$\footnotemark[1] &\tiny 53.3266480 &\tiny -27.8986874 &\tiny 149$\pm$15 &\tiny 208$\pm$15 &\tiny 231$\pm$15 &\tiny 201$\pm$10 &\tiny 197 &\tiny 15.46&\tiny 41.4&\tiny 1.55 &\tiny 0.63 &\tiny 21.31 &\tiny NO &\tiny NO\\
        \tiny5766$^*$\footnotemark[2] &\tiny 53.3296181 &\tiny -28.0256251 &\tiny 241$\pm$13 &\tiny 201$\pm$13 &\tiny 274$\pm$16 &\tiny 252$\pm$11 &\tiny 242 &\tiny 14.64&\tiny 30.0&\tiny 1.36 &\tiny 0.30 &\tiny 21.58 &\tiny NO &\tiny NO\\
        \tiny8379$^*$\footnotemark[1] &\tiny 53.3585985 &\tiny -28.0450788 &\tiny 203$\pm$15 &\tiny 158$\pm$13 &\tiny 230$\pm$17 &\tiny 192$\pm$9 &\tiny 196 &\tiny 12.77&\tiny 37.0&\tiny 1.46 &\tiny 1.08 &\tiny 22.63 &\tiny NO &\tiny NO\\
        \tiny12099$^*$\footnotemark[1]$^,$\footnotemark[2] &\tiny 53.3617713 &\tiny -27.8518082 &\tiny 100$\pm$16 &\tiny 119$\pm$21 &\tiny 152$\pm$16 &\tiny 153$\pm$8 &\tiny 131 &\tiny 12.56&\tiny 40.1&\tiny 1.53 &\tiny 0.94 &\tiny 22.31 &\tiny YES &\tiny NO\\
        
        \tiny5403$^*$\footnotemark[1]$^,$\footnotemark[2] &\tiny 53.3798560 &\tiny -27.8245098 &\tiny 145$\pm$26 &\tiny 178$\pm$20 &\tiny 166$\pm$18 &\tiny 129$\pm$8 &\tiny 154 &\tiny 13.92&\tiny 32.0&\tiny 1.38 &\tiny 0.70 &\tiny 22.14 &\tiny NO &\tiny NO\\
        \tiny5582 &\tiny 53.4022377 &\tiny -27.9101321 &\tiny 133$\pm$27 &\tiny 150$\pm$24 &\tiny 156$\pm$18 &\tiny 106$\pm$8 &\tiny 136 &\tiny 15.81&\tiny 36.4&\tiny 1.47 &\tiny 1.93 &\tiny - &\tiny NO &\tiny NO\\
        \tiny6061 &\tiny 53.4205850 &\tiny -28.1435063 &\tiny 163$\pm$15 &\tiny 217$\pm$21 &\tiny 101$\pm$16 &\tiny 130$\pm$17 &\tiny 153 &\tiny 22.74&\tiny 76.5&\tiny 2.16 &\tiny 0.81 &\tiny - &\tiny NO &\tiny NO\\

        \hline
      \end{tabular}
    \end{center}
    $^+$ Redshifts with three decimal points are spectroscopic redshifts.\\
    $^*$ In the E-CDFS.\\
    $^1$ In COMBO-17 catalog and only one counterpart in 2.5$\arcsec$ in Rainbow catalog.\\
    $^2$ Only one counterpart in 2.5$\arcsec$ in COMBO-17 catalog.\\
    The references for the spectroscopic redshifts are: 
    ID 5080: \cite{Grazian2006};
    ID 5109: \cite{Mignoli2005};
    ID 7742, 10015, 14779, 2324: \cite{Lefevre2004};
    ID 2226: \cite{Balestra2010};
    ID 13601: \cite{Mainieri2008}.
    
  \end{minipage}
\end{table*}

\section[]{Short-term variable candidates catalog}
\label{tablas-apendice-short}

\begin{table*}
  \begin{minipage}{200mm}
    \begin{center}
    \caption{Catalog of the short-term variable candidates.}
    \label{short-term-candidates}
    \begin{tabular}{@{}cccccccccccccc}
      \hline
      \tiny ID &\tiny  RA &\tiny  DEC &\tiny  Flux [epoch7a] &\tiny  Flux [epoch7b] &\tiny  Flux7 [epoch7c] &\tiny  $\bar{F}$ &\tiny  $\chi^2$ &\tiny  Var &\tiny  $R_{max}$  &\tiny  z$^+$ &\tiny R &\tiny  X-ray? &\tiny  Radio?\\
      &\tiny (J2000)&\tiny  (J2000)&\tiny ($\umu$Jy) &\tiny ($\umu$Jy) &\tiny ($\umu$Jy)&\tiny ($\umu$Jy)&& \tiny  (\%)&&&&&\\
      \hline \hline

      \tiny 13 &\tiny 52.7297992 &\tiny -27.5739281 &\tiny 298$\pm$30 &\tiny 216$\pm$33 &\tiny 156$\pm$33 &\tiny 223 &\tiny 10.80 &\tiny 63.8 &\tiny 1.92&\tiny - &\tiny - &\tiny NO &\tiny NO\\
      
      \tiny 2634$^*$\footnotemark[1]$^,$\footnotemark[2] &\tiny 52.8866973 &\tiny -27.8719379 &\tiny 211$\pm$15 &\tiny 229$\pm$15 &\tiny 155$\pm$17 &\tiny 198 &\tiny 11.81 &\tiny 37.5 &\tiny 1.48&\tiny 0.65 &\tiny 20.30 &\tiny NO &\tiny NO\\
      
      \tiny 6314$^*$\footnotemark[1]$^,$\footnotemark[2] &\tiny 52.8931060 &\tiny -27.8764410 &\tiny 200$\pm$14 &\tiny 194$\pm$15 &\tiny 128$\pm$21 &\tiny 174 &\tiny 10.12 &\tiny 41.0 &\tiny 1.56&\tiny 0.74 &\tiny 21.20 &\tiny NO &\tiny NO\\
      \tiny 937$^*$ &\tiny 52.8942467 &\tiny -27.8840075 &\tiny 326$\pm$14 &\tiny 281$\pm$14 &\tiny 346$\pm$16 &\tiny 318 &\tiny 10.15 &\tiny 20.5 &\tiny 1.23&\tiny 1.16 &\tiny - &\tiny NO &\tiny NO\\
      \tiny 2625$^*$\footnotemark[2] &\tiny 52.8962139 &\tiny -27.8649395 &\tiny 119$\pm$14 &\tiny 103$\pm$18 &\tiny 190$\pm$16 &\tiny 137 &\tiny 16.43 &\tiny 63.5 &\tiny 1.85&\tiny 1.28 &\tiny 23.94 &\tiny NO &\tiny NO\\
     
      \tiny 763$^*$ &\tiny 52.9090009 &\tiny -27.7866579 &\tiny 339$\pm$20 &\tiny 291$\pm$15 &\tiny 363$\pm$21 &\tiny 331 &\tiny 9.37 &\tiny 21.7 &\tiny 1.25&\tiny 1.05 &\tiny 22.07 &\tiny NO &\tiny NO\\
      \tiny 2356$^*$\footnotemark[1]$^,$\footnotemark[2] &\tiny 52.9251368 &\tiny -27.7383352 &\tiny 231$\pm$26 &\tiny 162$\pm$20 &\tiny 152$\pm$24 &\tiny 215 &\tiny 12.52 &\tiny 50.8 &\tiny 1.72&\tiny 1.02 &\tiny 22.38 &\tiny NO &\tiny NO\\ 
   
      \tiny 6827 &\tiny 52.9338732 &\tiny -28.0834027 &\tiny 406$\pm$17 &\tiny 325$\pm$16 &\tiny 380$\pm$19 &\tiny 370 &\tiny 13.22 &\tiny 21.8 &\tiny 1.25&\tiny 0.44 &\tiny - &\tiny YES &\tiny NO\\
      
      \tiny 4648$^*$\footnotemark[1]$^,$\footnotemark[2] &\tiny 52.9379937 &\tiny -27.8746505 &\tiny 163$\pm$11 &\tiny 104$\pm$11 &\tiny 133$\pm$17 &\tiny 133 &\tiny 14.29 &\tiny 44.1 &\tiny 1.56&\tiny 0.734 &\tiny 23.49 &\tiny NO &\tiny YES\\
      
      \tiny 1947 &\tiny 52.9862313 &\tiny -27.5157096 &\tiny 346$\pm$31 &\tiny 208$\pm$25 &\tiny 200$\pm$48 &\tiny 251 &\tiny 13.67 &\tiny 58.1 &\tiny 1.73&\tiny 0.81 &\tiny - &\tiny NO &\tiny NO\\
      \tiny 8181 &\tiny 52.9972222 &\tiny -28.1038541 &\tiny 222$\pm$15 &\tiny 178$\pm$15 &\tiny 143$\pm$17 &\tiny 181 &\tiny 12.25 &\tiny 43.4 &\tiny 1.55&\tiny - &\tiny - &\tiny NO &\tiny NO\\
      \tiny 718$^*$\footnotemark[1]$^,$\footnotemark[2] &\tiny 53.0047558 &\tiny -27.7255025 &\tiny 243$\pm$26 &\tiny 149$\pm$20 &\tiny 235$\pm$31 &\tiny 209 &\tiny 11.87 &\tiny 44.9 &\tiny 1.63&\tiny 1.001 &\tiny 22.29 &\tiny NO &\tiny NO\\
      \tiny 7513$^*$\footnotemark[1]$^,$\footnotemark[2] &\tiny 53.0134300 &\tiny -27.7581000 &\tiny 272$\pm$19 &\tiny 165$\pm$18 &\tiny 227$\pm$18 &\tiny 221 &\tiny 16.47 &\tiny 48.3 &\tiny 1.65&\tiny 0.534 &\tiny 21.01 &\tiny NO &\tiny NO\\
      \tiny 917$^*$\footnotemark[1]$^,$\footnotemark[2] &\tiny 53.0192273 &\tiny -27.8352082 &\tiny 228$\pm$12 &\tiny 197$\pm$13 &\tiny 181$\pm$10 &\tiny 202 &\tiny 9.23 &\tiny 23.2 &\tiny 1.26&\tiny 0.23 &\tiny 19.93 &\tiny YES &\tiny NO\\
      \tiny 8613$^*$\footnotemark[1]$^,$\footnotemark[2] &\tiny 53.0286682 &\tiny -27.6356610 &\tiny 217$\pm$37 &\tiny 158$\pm$20 &\tiny 262$\pm$28 &\tiny 212 &\tiny 10.48 &\tiny 49.2 &\tiny 1.66&\tiny 0.50 &\tiny 20.92 &\tiny NO &\tiny NO\\
      \tiny 1943 &\tiny 53.0316927 &\tiny -27.5017437 &\tiny 389$\pm$32 &\tiny 258$\pm$22 &\tiny 271$\pm$29 &\tiny 306 &\tiny 12.61 &\tiny 42.8 &\tiny 1.51&\tiny 1.91 &\tiny - &\tiny NO &\tiny NO\\
      \tiny 7921$^*$\footnotemark[2] &\tiny 53.0476619 &\tiny -27.9473127 &\tiny 173$\pm$15 &\tiny 212$\pm$12 &\tiny 254$\pm$16 &\tiny 213 &\tiny 13.81 &\tiny 38.1 &\tiny 1.47&\tiny 1.11 &\tiny 22.21 &\tiny NO &\tiny NO\\
      
      \tiny 2277$^*$ &\tiny 53.0555289 &\tiny -27.6597959 &\tiny 238$\pm$40 &\tiny 230$\pm$21 &\tiny 356$\pm$32 &\tiny 275 &\tiny 11.85 &\tiny 45.9 &\tiny 1.55 &\tiny 1.49 &\tiny - &\tiny NO &\tiny YES\\
     
      \tiny 8766$^*$ &\tiny 53.0819932 &\tiny -27.7672103 &\tiny 369$\pm$17 &\tiny 321$\pm$18 &\tiny 277$\pm$22 &\tiny 322 &\tiny 11.94 &\tiny 28.5 &\tiny 1.33&\tiny 0.62 &\tiny - &\tiny YES &\tiny YES\\
      \tiny 517$^*$\footnotemark[1]$^,$\footnotemark[2] &\tiny 53.0839618 &\tiny -27.5734177 &\tiny 179$\pm$29 &\tiny 286$\pm$21 &\tiny 225$\pm$28 &\tiny 230 &\tiny 10.60 &\tiny 46.6 &\tiny 1.60&\tiny 1.12 &\tiny 21.08 &\tiny NO &\tiny YES\\
      \tiny 1503 &\tiny 53.0869545 &\tiny -28.1404091 &\tiny 197$\pm$32 &\tiny 142$\pm$19 &\tiny 234$\pm$26 &\tiny 191 &\tiny 9.59 &\tiny 48.0 &\tiny 1.65&\tiny 0.69 &\tiny - &\tiny NO &\tiny NO\\
      \tiny 10885$^*$ &\tiny 53.0925557 &\tiny -27.9857229 &\tiny 360$\pm$18 &\tiny 311$\pm$15 &\tiny 282$\pm$16 &\tiny 318 &\tiny 11.10 &\tiny 24.8 &\tiny 1.28&\tiny 0.57 &\tiny - &\tiny NO &\tiny NO\\
      
      \tiny 6978 &\tiny 53.1064166 &\tiny -28.1262429 &\tiny 389$\pm$24 &\tiny 307$\pm$19 &\tiny 249$\pm$27 &\tiny 315 &\tiny 15.63 &\tiny 44.5 &\tiny 1.56&\tiny 1.03 &\tiny - &\tiny NO &\tiny NO\\
      \tiny 185$^*$\footnotemark[1]$^,$\footnotemark[2] &\tiny 53.1121892 &\tiny -28.0514201 &\tiny 341$\pm$18 &\tiny 354$\pm$22 &\tiny 423$\pm$20 &\tiny 373 &\tiny 10.06 &\tiny 21.9 &\tiny 1.24&\tiny 1.57 &\tiny 24.00 &\tiny NO &\tiny NO\\
     
      \tiny 3265$^*$ &\tiny 53.1250338 &\tiny -28.0383985 &\tiny 234$\pm$25 &\tiny 191$\pm$16 &\tiny 140$\pm$18 &\tiny 188 &\tiny 10.79 &\tiny 50.2 &\tiny 1.68&\tiny 1.85 &\tiny 23.61 &\tiny NO &\tiny NO\\
      
      \tiny 4501$^*$\footnotemark[1]$^,$\footnotemark[2] &\tiny 53.1367677 &\tiny -27.7688610 &\tiny 610$\pm$20 &\tiny 669$\pm$19 &\tiny 584$\pm$18 &\tiny 621 &\tiny 10.60 &\tiny 13.7 &\tiny 1.15&\tiny 0.366 &\tiny 20.36 &\tiny NO &\tiny YES\\
     
      \tiny 2091 &\tiny 53.1518662 &\tiny -27.5393472 &\tiny 257$\pm$31 &\tiny 145$\pm$21 &\tiny 194$\pm$30 &\tiny 198 &\tiny 9.93 &\tiny 56.5 &\tiny 1.78&\tiny 0.28 &\tiny - &\tiny NO &\tiny NO\\
     
      \tiny 2847$^*$\footnotemark[1]$^,$\footnotemark[2] &\tiny 53.1549571 &\tiny -27.8767845 &\tiny 234$\pm$11 &\tiny 192$\pm$11 &\tiny 187$\pm$13 &\tiny 204 &\tiny 10.23 &\tiny 22.8 &\tiny 1.25&\tiny 0.331 &\tiny 20.49 &\tiny NO &\tiny NO\\
      \tiny 1925 &\tiny 53.1620312 &\tiny -27.4562601 &\tiny 222$\pm$28 &\tiny 243$\pm$25 &\tiny 121$\pm$29 &\tiny 195 &\tiny 11.14 &\tiny 62.4 &\tiny 2.01&\tiny 1.00 &\tiny - &\tiny NO &\tiny NO\\
     
      \tiny 3702 &\tiny 53.1828042 &\tiny -28.2246423 &\tiny 293$\pm$38 &\tiny 142$\pm$20 &\tiny 175$\pm$29 &\tiny 203 &\tiny 15.70 &\tiny 74.3 &\tiny 2.07&\tiny - &\tiny - &\tiny NO &\tiny NO\\
      \tiny 7761$^*$\footnotemark[1]$^,$\footnotemark[2] &\tiny 53.2024886 &\tiny -27.8262335 &\tiny 227$\pm$14 &\tiny 208$\pm$14 &\tiny 151$\pm$16 &\tiny 195 &\tiny 14.12 &\tiny 38.9 &\tiny 1.50&\tiny 1.117 &\tiny 25.08 &\tiny YES &\tiny YES\\
      
      \tiny 4477$^*$\footnotemark[1] &\tiny 53.2235604 &\tiny -27.7345674 &\tiny 253$\pm$18 &\tiny 274$\pm$17 &\tiny 182$\pm$19 &\tiny 236 &\tiny 14.08 &\tiny 39.2 &\tiny 1.51&\tiny 1.58 &\tiny 23.97 &\tiny NO &\tiny NO\\
     
      \tiny 540 &\tiny 53.2480035 &\tiny -27.5404651 &\tiny 139$\pm$27 &\tiny 114$\pm$20 &\tiny 235$\pm$28 &\tiny 162 &\tiny 13.26 &\tiny 74.3 &\tiny 2.06&\tiny 1.72 &\tiny - &\tiny NO &\tiny YES\\

      \tiny 2614$^*$\footnotemark[1]$^,$\footnotemark[2] &\tiny 53.2610999 &\tiny -27.7598248 &\tiny 106$\pm$13 &\tiny 141$\pm$17 &\tiny 169$\pm$13 &\tiny 139 &\tiny 11.25 &\tiny 45.7 &\tiny 1.60&\tiny 1.23 &\tiny 23.34 &\tiny YES &\tiny NO\\
      
      \tiny 6876$^*$ &\tiny 53.2681484 &\tiny -28.0248935 &\tiny 129$\pm$23 &\tiny 246$\pm$20 &\tiny 276$\pm$26 &\tiny 217 &\tiny 21.45&\tiny 67.6 &\tiny 2.13&\tiny 0.92 &\tiny 22.83 &\tiny NO &\tiny NO\\
      \tiny 6017$^*$ &\tiny 53.2758295 &\tiny -27.6029799 &\tiny 206$\pm$28 &\tiny 321$\pm$24 &\tiny 253$\pm$30 &\tiny 260 &\tiny 10.25 &\tiny 44.3 &\tiny 1.56&\tiny 1.77 &\tiny 23.94 &\tiny NO &\tiny NO\\
      
      \tiny 2409$^*$\footnotemark[1]$^,$\footnotemark[2] &\tiny 53.2795549 &\tiny -27.6633534 &\tiny 322$\pm$27 &\tiny 227$\pm$20 &\tiny 288$\pm$32 &\tiny 279 &\tiny 9.23  &\tiny 34.1 &\tiny 1.42&\tiny 0.45  &\tiny 21.35 &\tiny NO &\tiny NO\\
      
      \tiny 10903$^*$ &\tiny 53.2826321 &\tiny -27.9443909 &\tiny 200$\pm$16 &\tiny 237$\pm$17 &\tiny 163$\pm$18 &\tiny 200 &\tiny 9.24 &\tiny 37.2 &\tiny 1.46&\tiny 0.12 &\tiny 20.91 &\tiny NO &\tiny NO\\
      \tiny 1209$^*$\footnotemark[1]$^,$\footnotemark[2] &\tiny 53.2846948 &\tiny -27.9283298 &\tiny 176$\pm$13 &\tiny 139$\pm$15 &\tiny 113$\pm$14 &\tiny 143 &\tiny 11.46 &\tiny 44.3 &\tiny 1.56&\tiny 0.70 &\tiny 21.66 &\tiny NO &\tiny NO\\
      \tiny 7869$^*$\footnotemark[1]$^,$\footnotemark[2] &\tiny 53.3057210 &\tiny -27.8438699 &\tiny 185$\pm$10 &\tiny 186$\pm$15 &\tiny 138$\pm$13 &\tiny 170 &\tiny 9.86 &\tiny 27.8 &\tiny 1.34&\tiny 0.32 &\tiny 19.51 &\tiny NO &\tiny NO\\
      \tiny 8106$^*$ &\tiny 53.3173805 &\tiny -27.9729970 &\tiny 178$\pm$13 &\tiny 225$\pm$17 &\tiny 148$\pm$16 &\tiny 184 &\tiny 11.56 &\tiny 42.3 &\tiny 1.53&\tiny 0.45 &\tiny 22.63 &\tiny NO &\tiny NO\\
     
      \tiny 6918$^*$\footnotemark[1]$^,$\footnotemark[2] &\tiny 53.3352837 &\tiny -28.0254291 &\tiny 108$\pm$17 &\tiny 150$\pm$13 &\tiny 190$\pm$14 &\tiny 149 &\tiny 14.42 &\tiny 55.0 &\tiny 1.76&\tiny 1.61 &\tiny 23.41 &\tiny NO &\tiny NO\\
      
      \tiny 8050$^*$\footnotemark[2] &\tiny 53.3430057 &\tiny -27.9321166 &\tiny 480$\pm$19 &\tiny 539$\pm$16 &\tiny 474$\pm$15 &\tiny 498 &\tiny 10.33 &\tiny 13.1 &\tiny 1.14&\tiny 1.66 &\tiny 23.43 &\tiny NO &\tiny NO\\
      
      \tiny 123$^*$ &\tiny 53.3747793 &\tiny -27.8008735 &\tiny 164$\pm$10 &\tiny 139$\pm$12 &\tiny 191$\pm$11 &\tiny 165 &\tiny 10.80 &\tiny 31.8 &\tiny 1.38&\tiny 1.94 &\tiny - &\tiny NO &\tiny NO\\
      
      \tiny 3632 &\tiny 53.3868526 &\tiny -28.1329288  &\tiny 303$\pm$28 &\tiny 189$\pm$21 &\tiny 249$\pm$26 &\tiny 247 &\tiny 11.93 &\tiny 46.5 &\tiny 1.61&\tiny 0.62 &\tiny - &\tiny NO &\tiny NO\\
      
      \tiny 10053 &\tiny 53.4529616 &\tiny -27.8423718 &\tiny 322$\pm$13 &\tiny 265$\pm$14 &\tiny 290$\pm$12 &\tiny 292 &\tiny 9.62 &\tiny 19.7 &\tiny 1.22&\tiny - &\tiny - &\tiny NO &\tiny NO\\
      \tiny 2734 &\tiny 53.4539570 &\tiny -27.7478151 &\tiny 162$\pm$16 &\tiny 187$\pm$24 &\tiny 101$\pm$19 &\tiny 150 &\tiny 9.70 &\tiny 57.4 &\tiny 1.85&\tiny 0.48 &\tiny - &\tiny NO &\tiny NO\\
      \tiny 1032 &\tiny 53.4584105 &\tiny -27.7783558 &\tiny 168$\pm$17 &\tiny 191$\pm$16 &\tiny 126$\pm$14 &\tiny 162 &\tiny 10.35 &\tiny 40.0 &\tiny 1.51&\tiny 1.05 &\tiny - &\tiny NO &\tiny NO\\
      
      \tiny 6622 &\tiny 53.4639456 &\tiny -27.8334061 &\tiny 507$\pm$18 &\tiny 464$\pm$21 &\tiny 432$\pm$17 &\tiny 468 &\tiny 9.30 &\tiny 15.9&\tiny 1.17&\tiny - &\tiny - &\tiny NO &\tiny NO\\
      \tiny 4767 &\tiny 53.4656628 &\tiny -27.7659261 &\tiny 134$\pm$14 &\tiny 212$\pm$17 &\tiny 190$\pm$16 &\tiny 179 &\tiny 14.28&\tiny 44.1 &\tiny 1.59&\tiny 0.11 &\tiny - &\tiny NO &\tiny NO\\
      \tiny 1039 &\tiny 53.4768023 &\tiny -27.7778086 &\tiny 168$\pm$19 &\tiny 244$\pm$17 &\tiny 132$\pm$20 &\tiny 181 &\tiny 20.08 &\tiny 61.8 &\tiny 1.85&\tiny - &\tiny - &\tiny NO &\tiny NO\\
      
      \tiny 6814 &\tiny 53.4846854 &\tiny -27.9221903 &\tiny 180$\pm$18 &\tiny 144$\pm$14 &\tiny 217$\pm$18 &\tiny 180 &\tiny 10.95 &\tiny 40.6 &\tiny 1.51&\tiny - &\tiny - &\tiny NO &\tiny NO\\
      \tiny 5284 &\tiny 53.4862820 &\tiny -27.9564971 &\tiny 225$\pm$20 &\tiny 144$\pm$14 &\tiny 195$\pm$15 &\tiny 188 &\tiny 13.10 &\tiny 43.0 &\tiny 1.56&\tiny - &\tiny - &\tiny NO &\tiny NO\\
      \tiny 7936 &\tiny 53.4911453 &\tiny -27.8278011 &\tiny 363$\pm$23 &\tiny 269$\pm$26 &\tiny 368$\pm$19 &\tiny 333 &\tiny 10.89 &\tiny 29.5 &\tiny 1.37&\tiny - &\tiny - &\tiny NO &\tiny NO\\
      \tiny 12955 &\tiny 53.4921523 &\tiny -27.8248350 &\tiny 212$\pm$14 &\tiny 175$\pm$16 &\tiny 281$\pm$20 &\tiny 222 &\tiny 17.93 &\tiny 47.8 &\tiny 1.61&\tiny - &\tiny - &\tiny NO &\tiny NO\\

      \hline
    \end{tabular}
    \end{center}
    $^+$ Redshifts with three decimal points are spectroscopic redshifts.\\
    $^*$ In the E-CDFS.\\
    $^1$ In COMBO-17 catalog and only one counterpart in 2.5$\arcsec$ in Rainbow catalog.\\
    $^2$ Only one counterpart in 2.5$\arcsec$ in COMBO-17 catalog.\\
     The references for the spectroscopic redshifts are: 
    ID 4648, 2847: \cite{Lefevre2004};
    ID 718, 7513: \cite{Balestra2010};
    ID 4501: \cite{Mignoli2005};
    ID 7761: \cite{Mainieri2008}.
    
  \end{minipage}
\end{table*}

\section[]{Long-term variable candidates light curves}
\label{curvas-luz-long}

\begin{figure*}
  \begin{minipage}{200mm}
    \begin{center}
    
    \subfigure {\includegraphics[width=47mm]{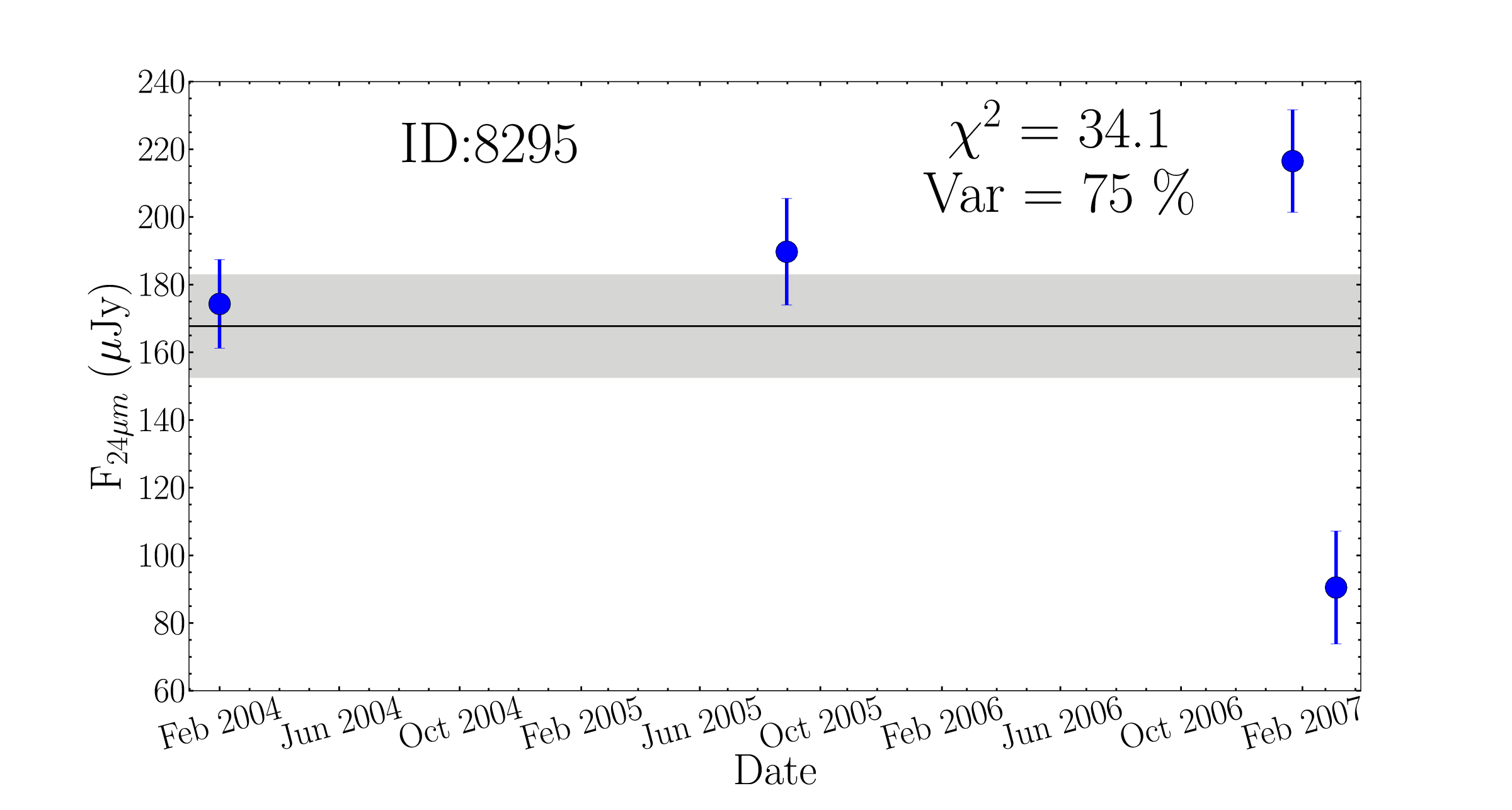}}{\hspace{0cm}}
    \subfigure {\includegraphics[width=47mm]{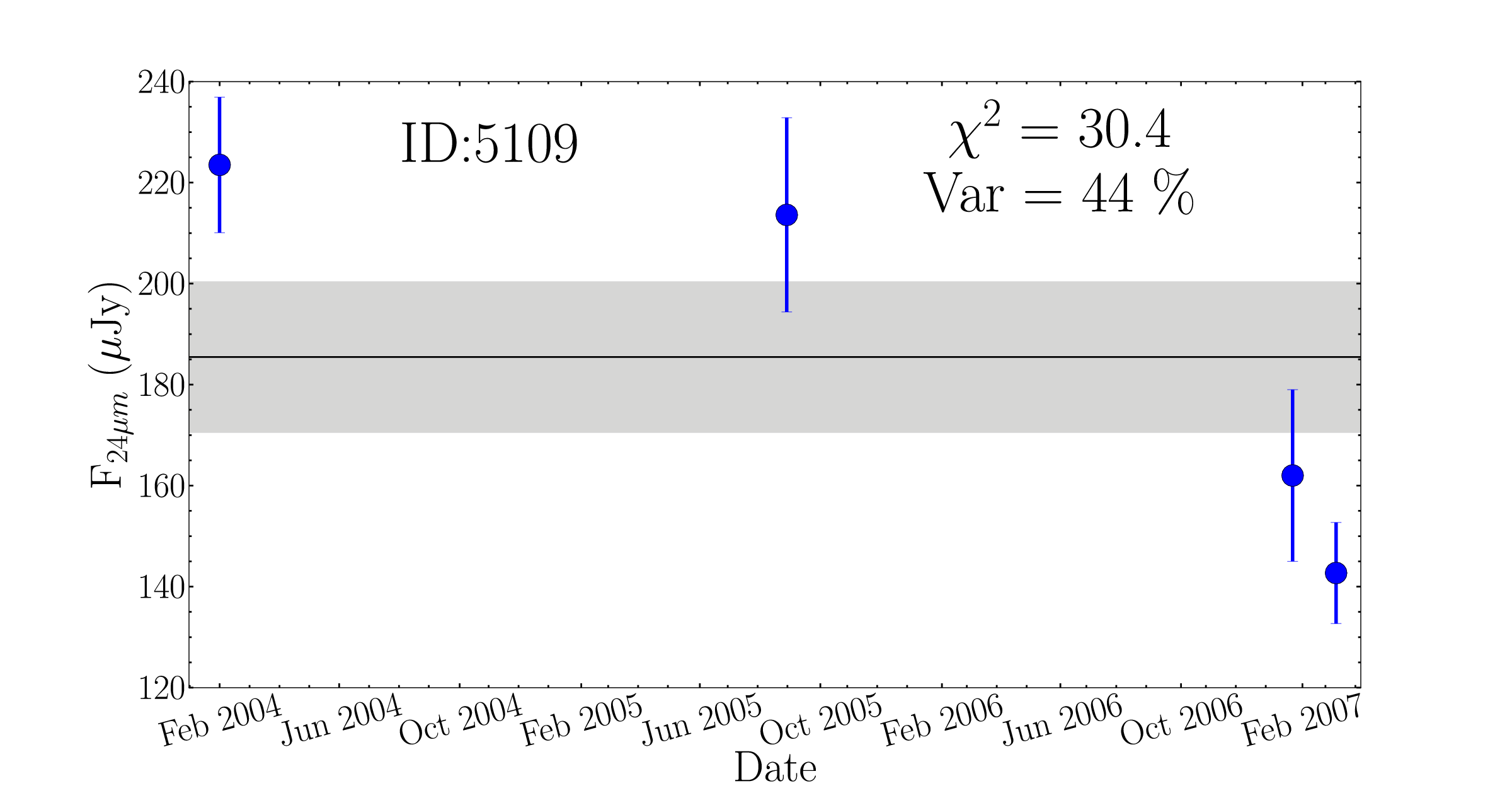}}{\hspace{0cm}}
    \subfigure {\includegraphics[width=47mm]{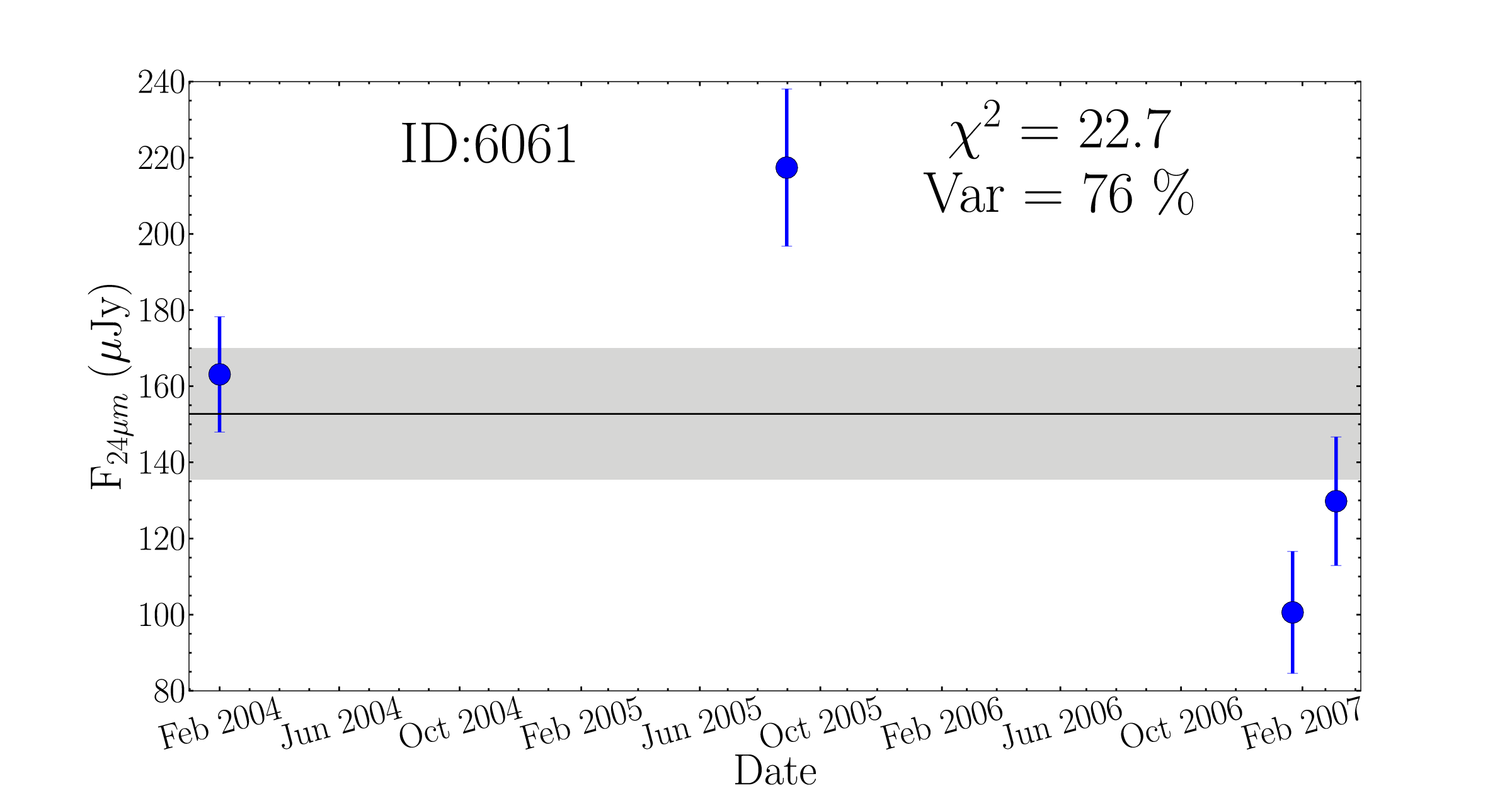}}{\hspace{0cm}}
    \subfigure {\includegraphics[width=47mm]{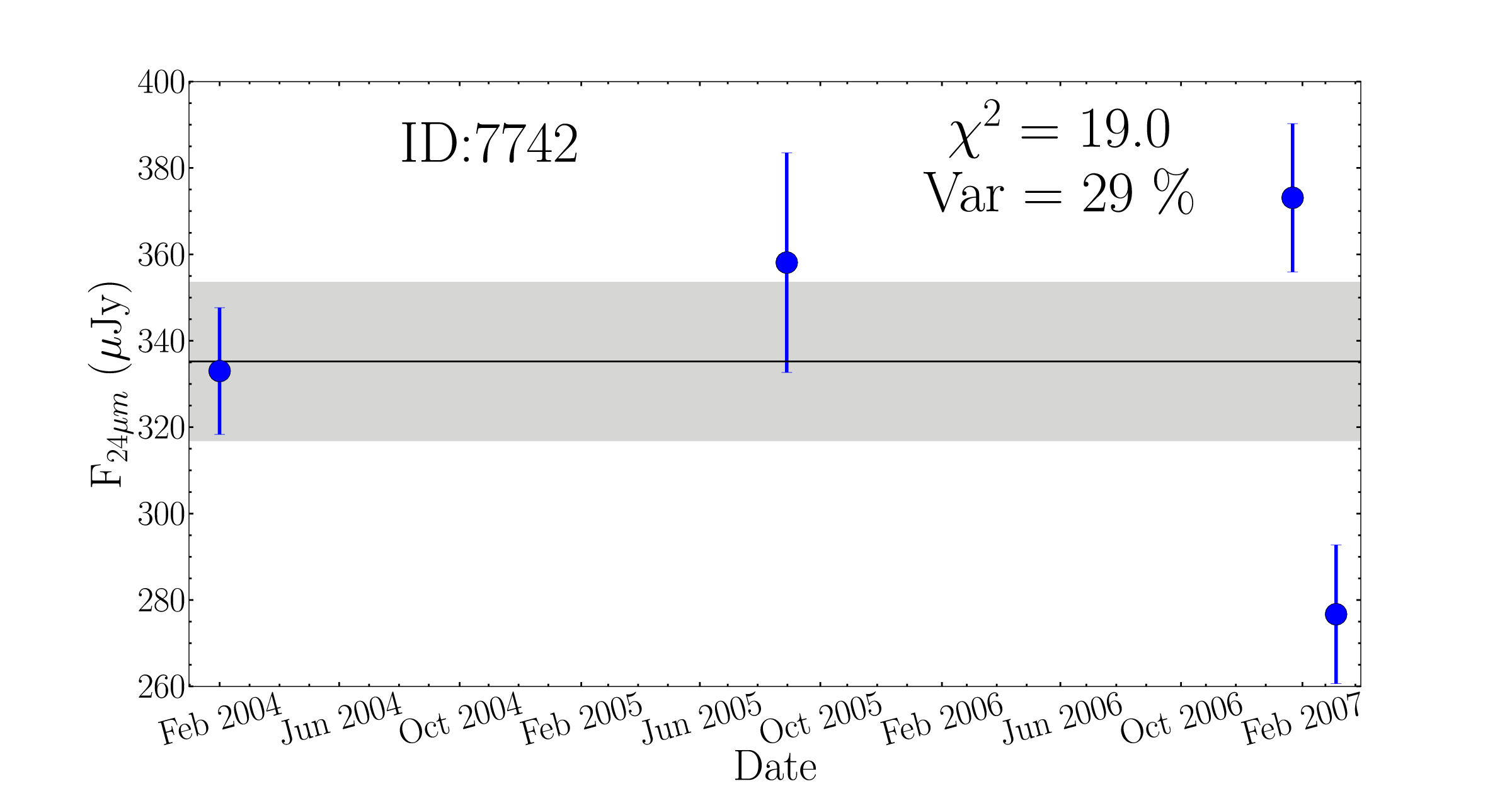}}{\hspace{0cm}}
    \subfigure {\includegraphics[width=47mm]{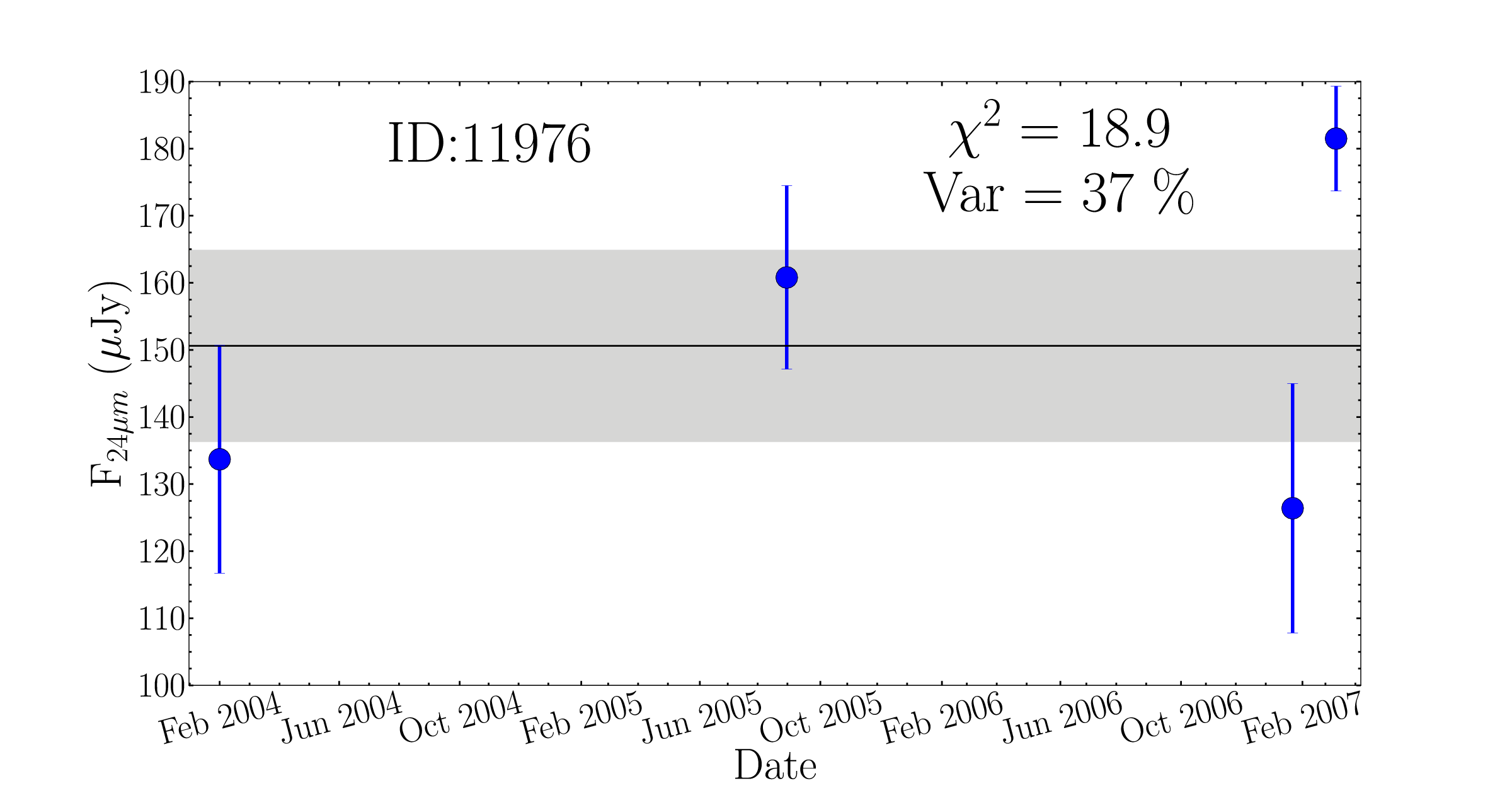}}{\hspace{0cm}}
    \subfigure {\includegraphics[width=47mm]{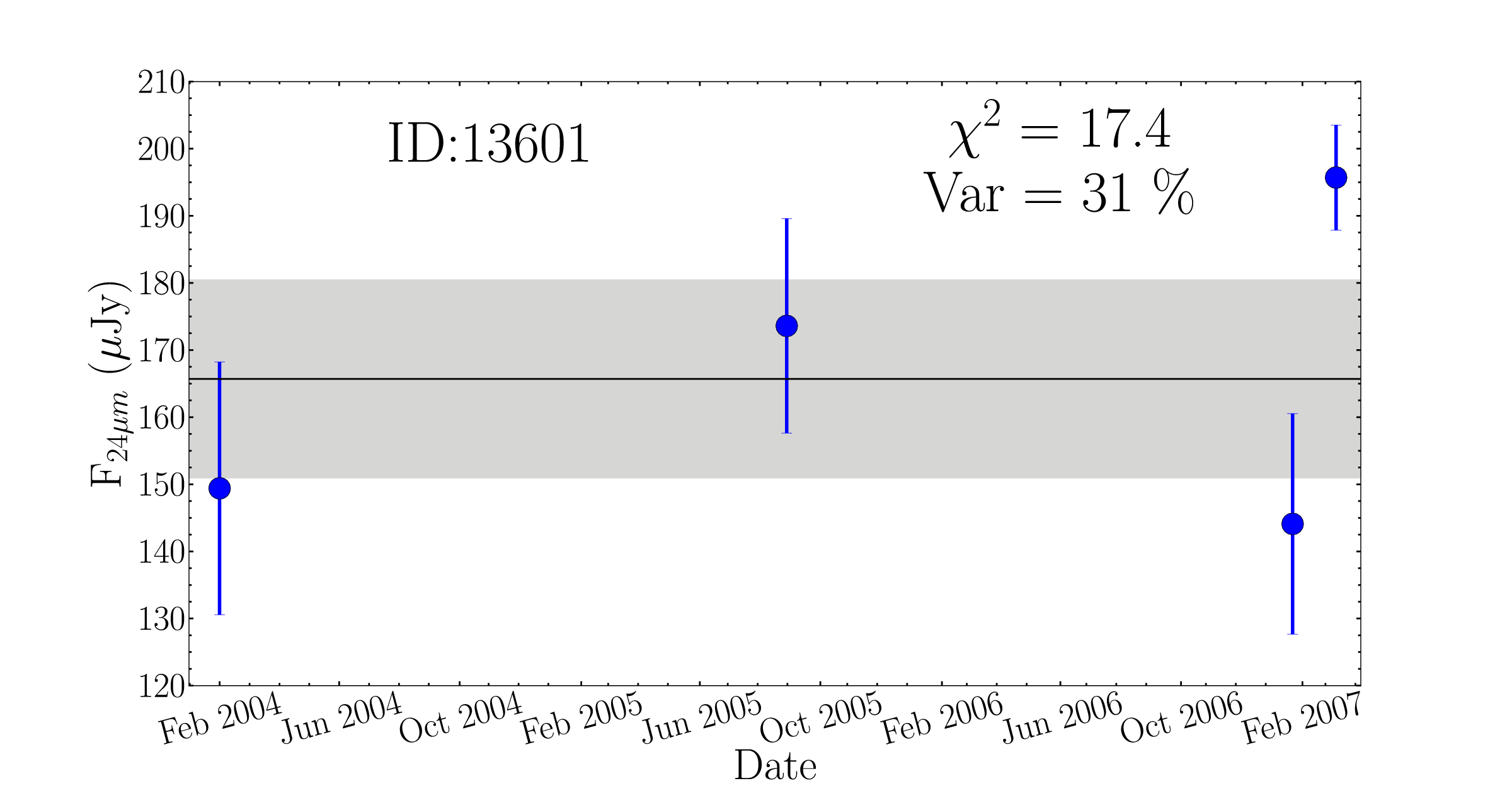}}{\hspace{0cm}}
    \subfigure {\includegraphics[width=47mm]{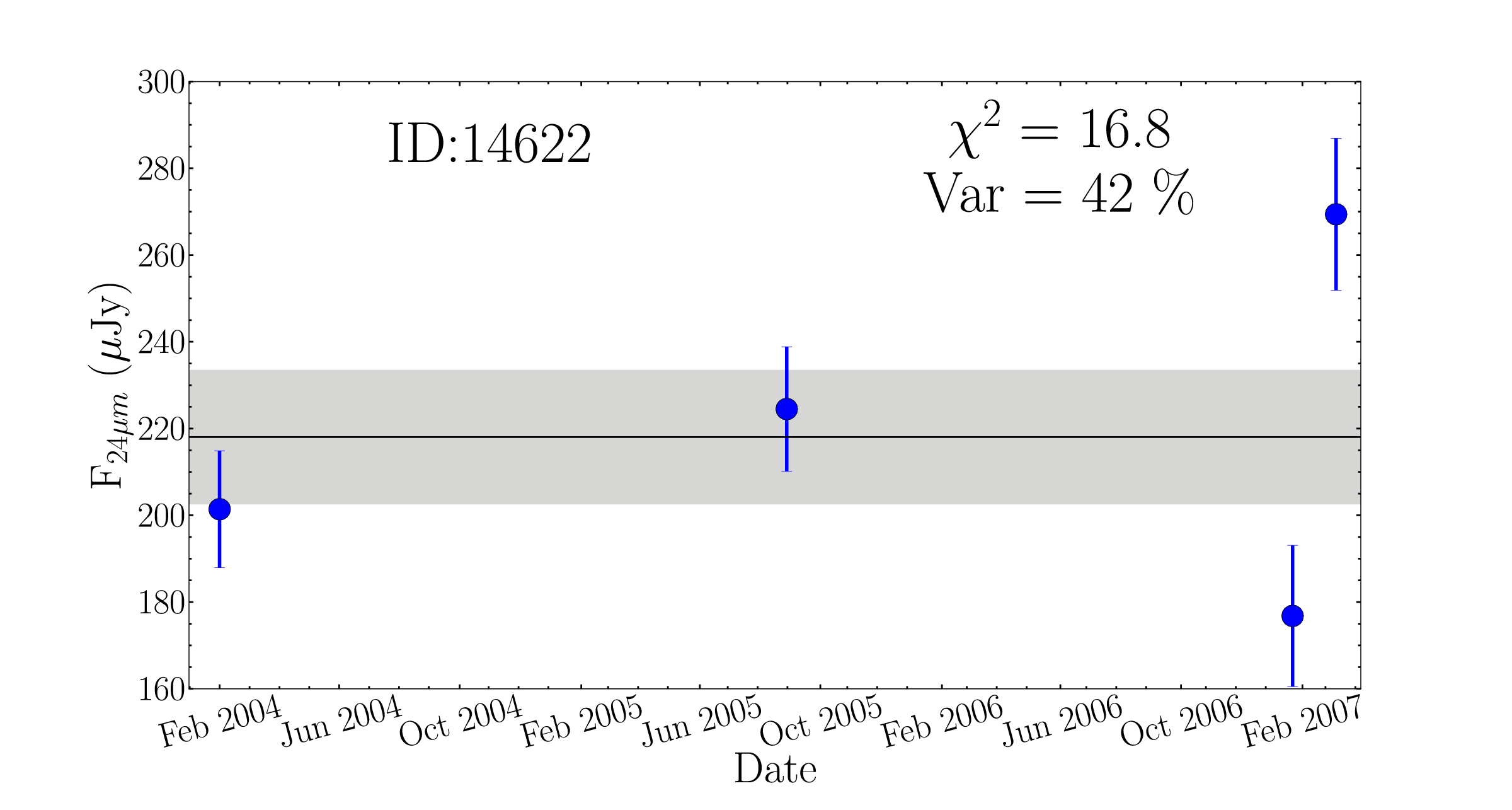}}{\hspace{0cm}}
    \subfigure {\includegraphics[width=47mm]{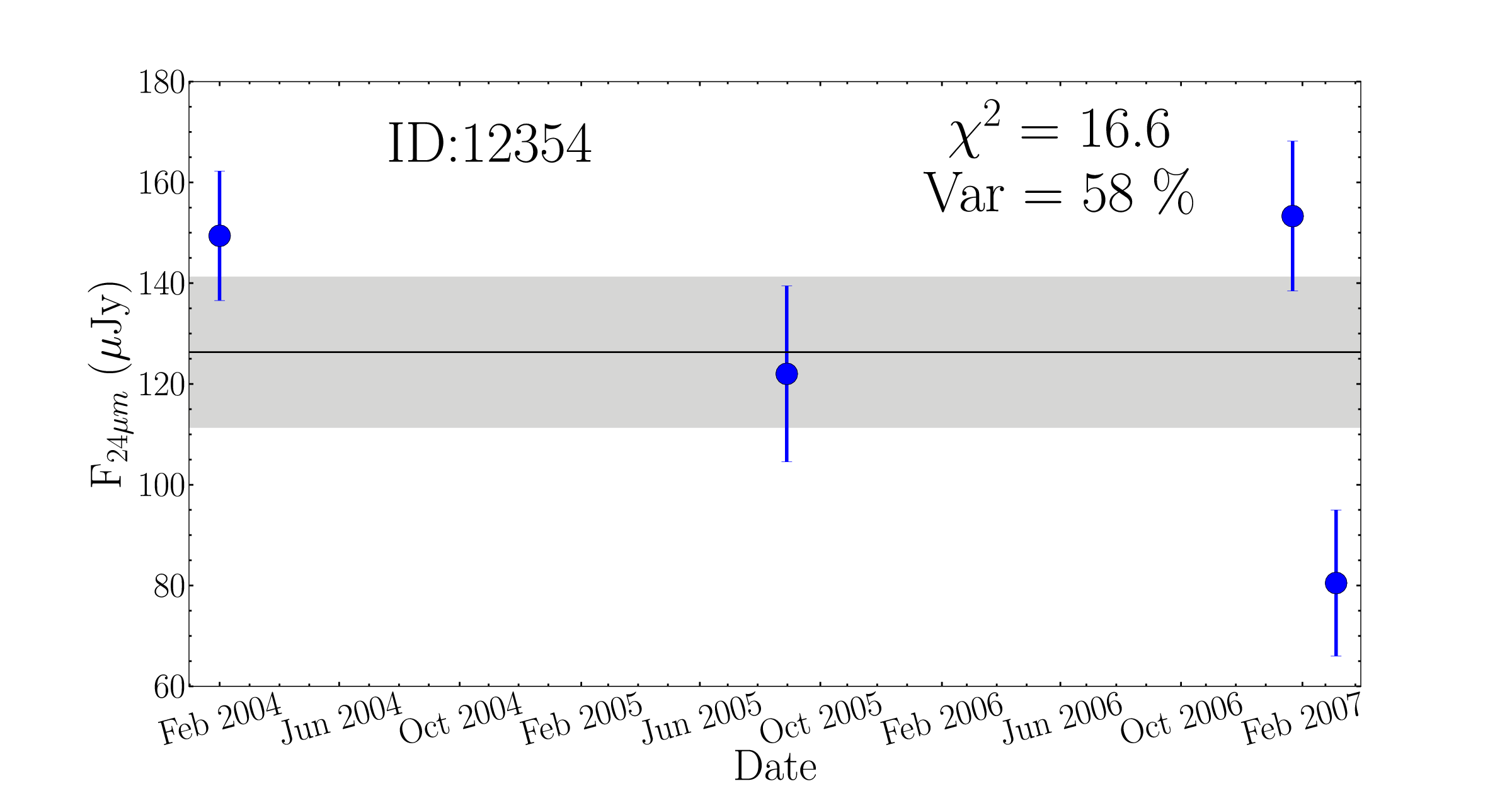}}{\hspace{0cm}}
    \subfigure {\includegraphics[width=47mm]{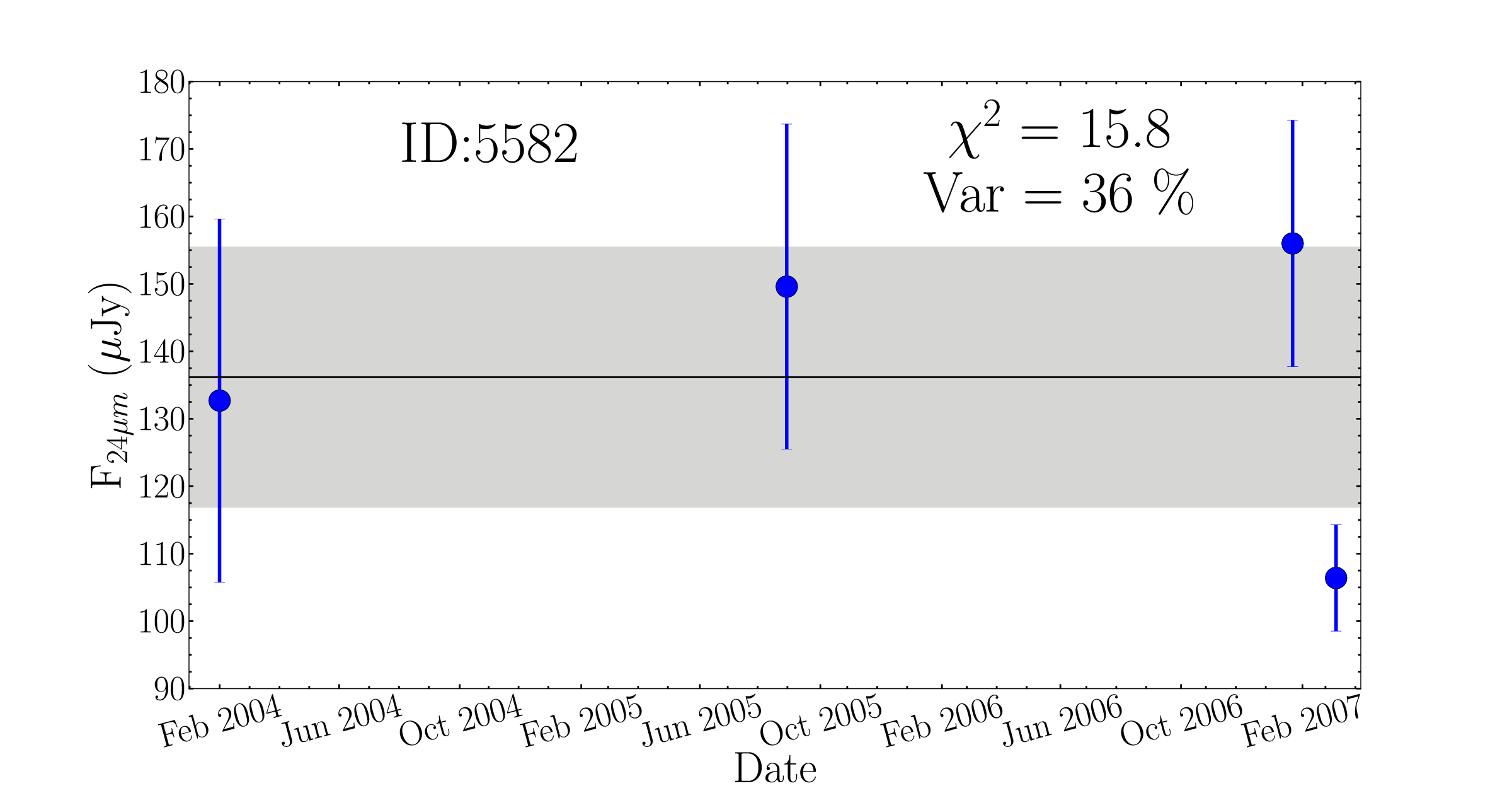}}{\hspace{0cm}}
    \subfigure {\includegraphics[width=47mm]{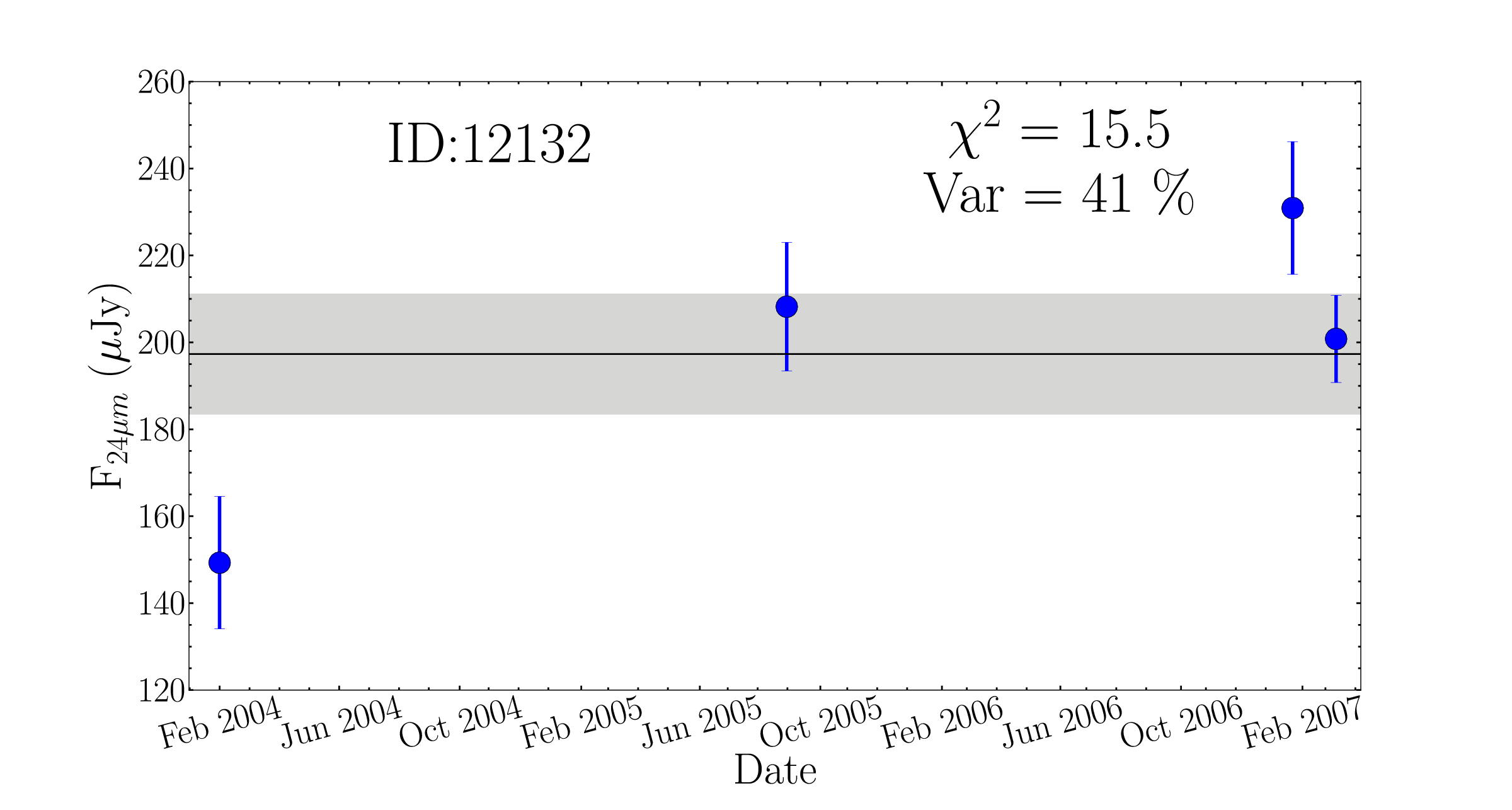}}{\hspace{0cm}}
    \subfigure {\includegraphics[width=47mm]{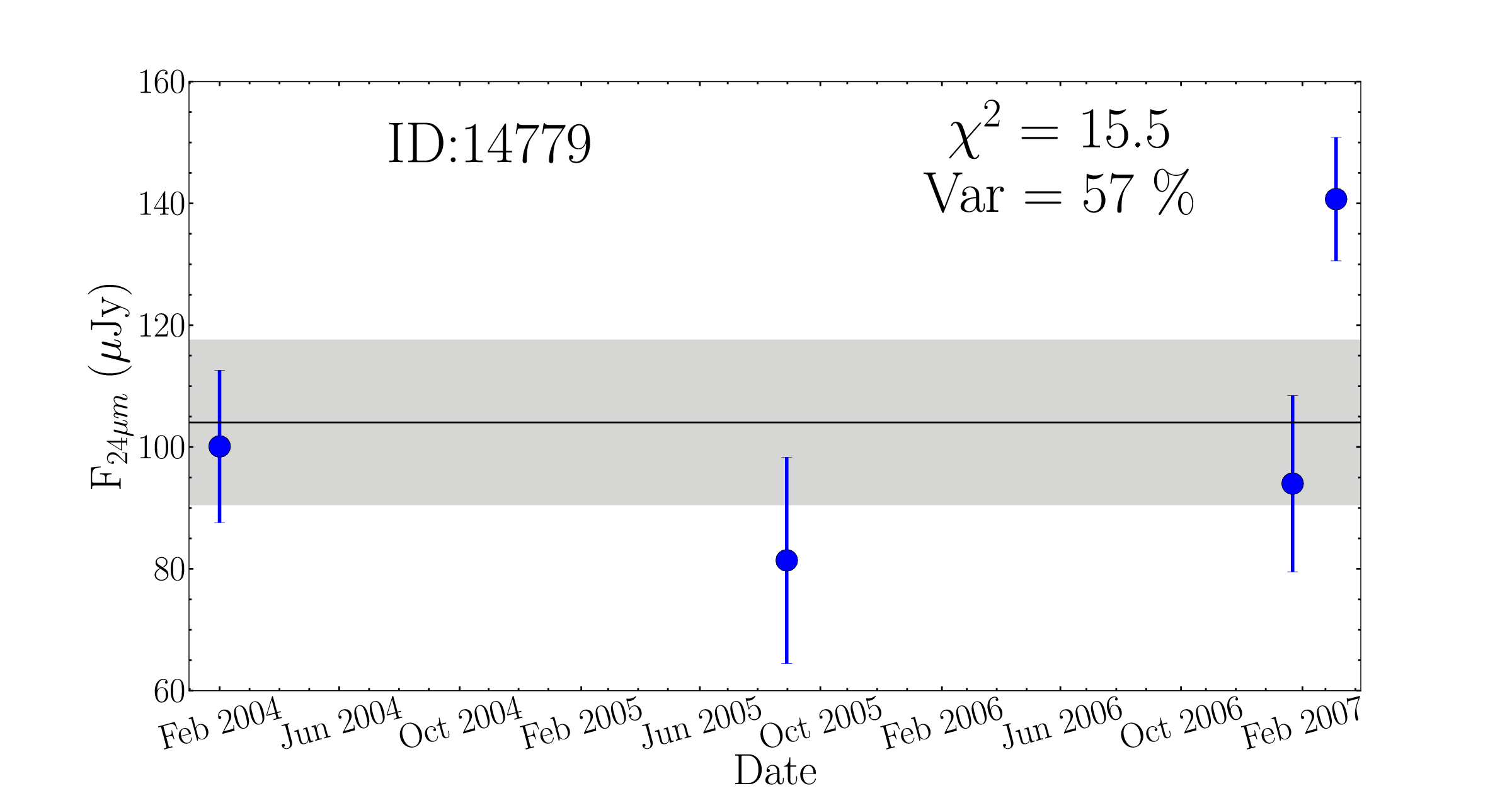}}{\hspace{0cm}}
    \subfigure {\includegraphics[width=47mm]{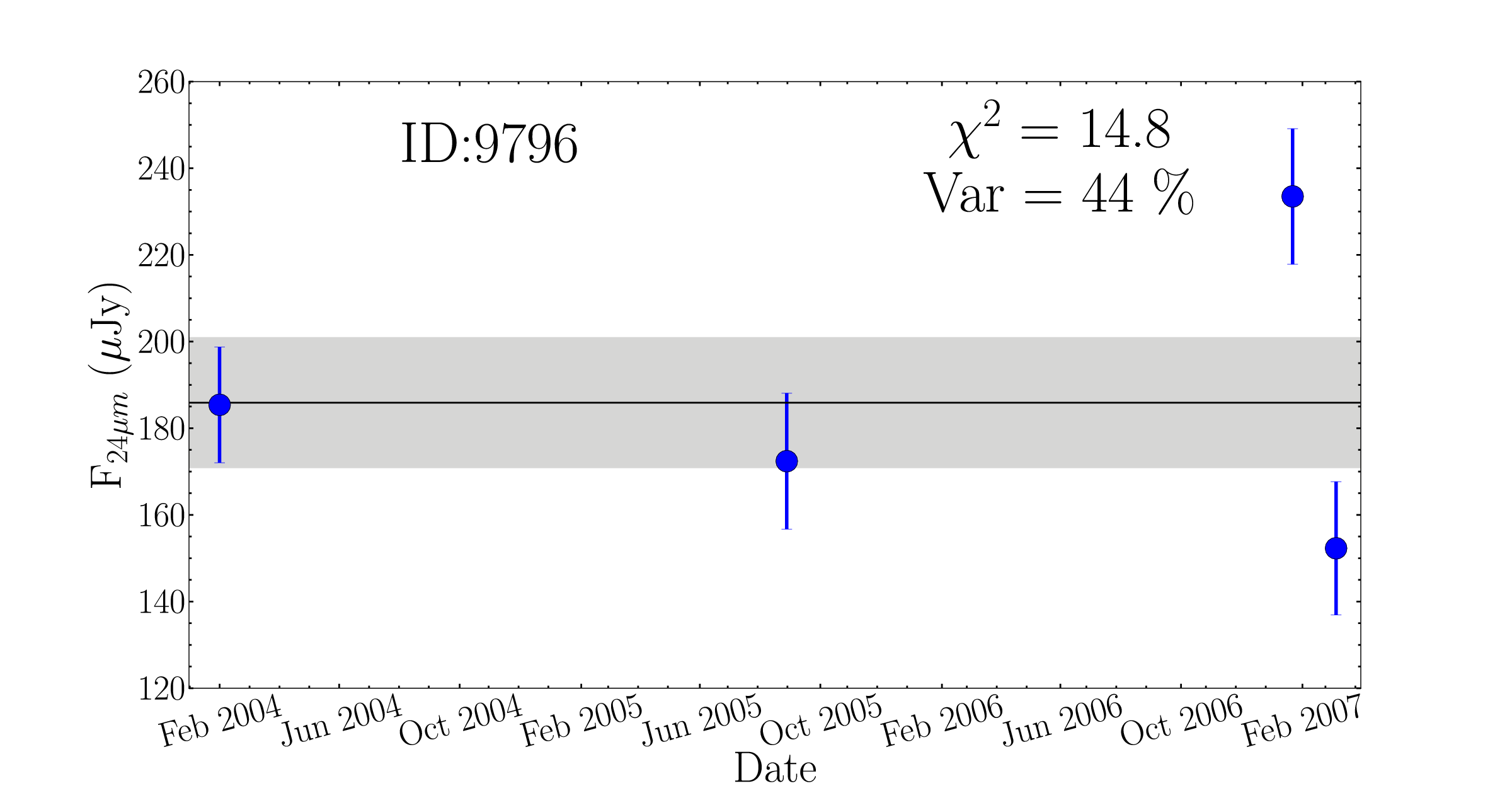}}{\hspace{0cm}}
    \subfigure {\includegraphics[width=47mm]{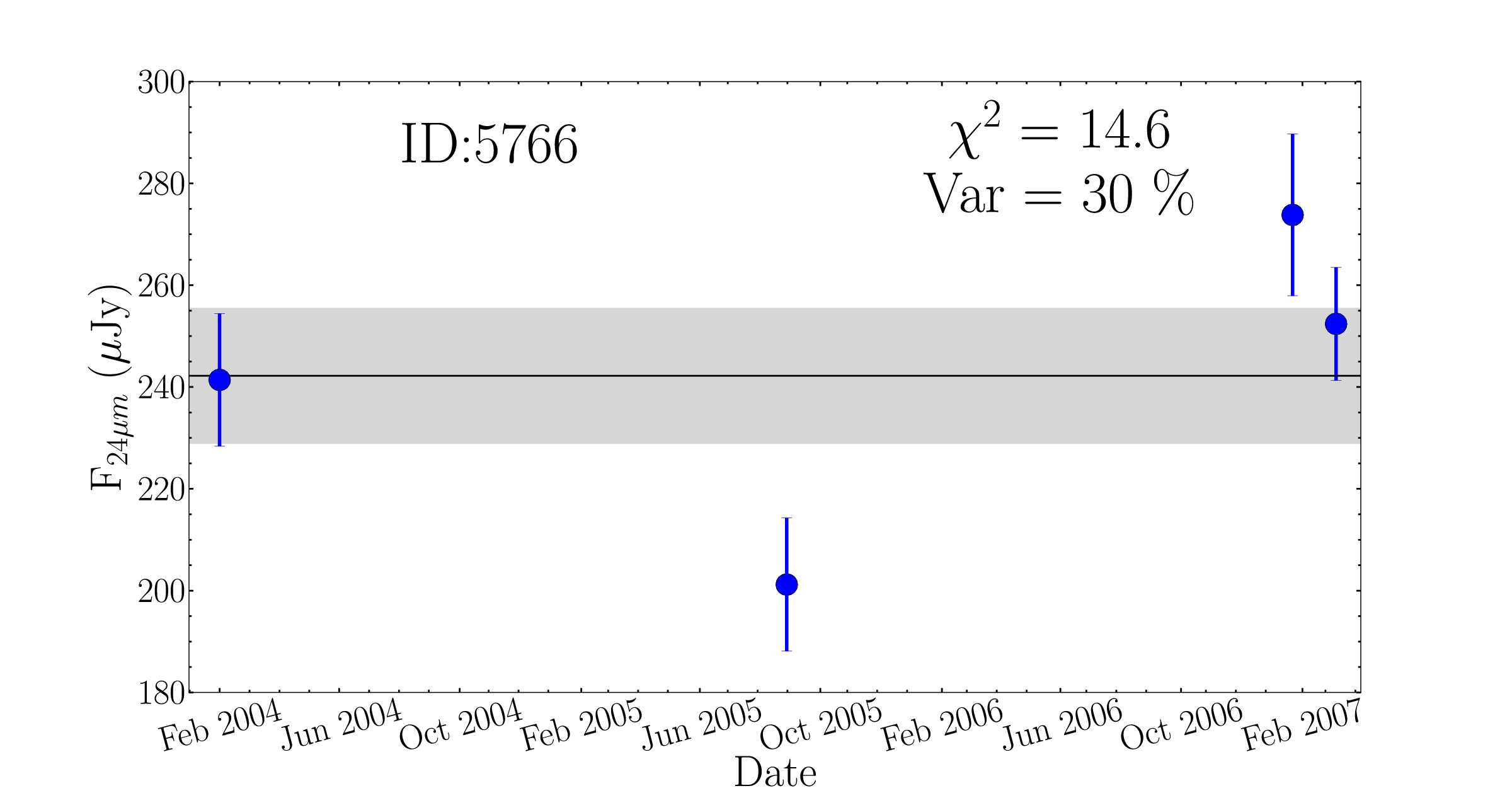}}{\hspace{0cm}}
    \subfigure {\includegraphics[width=47mm]{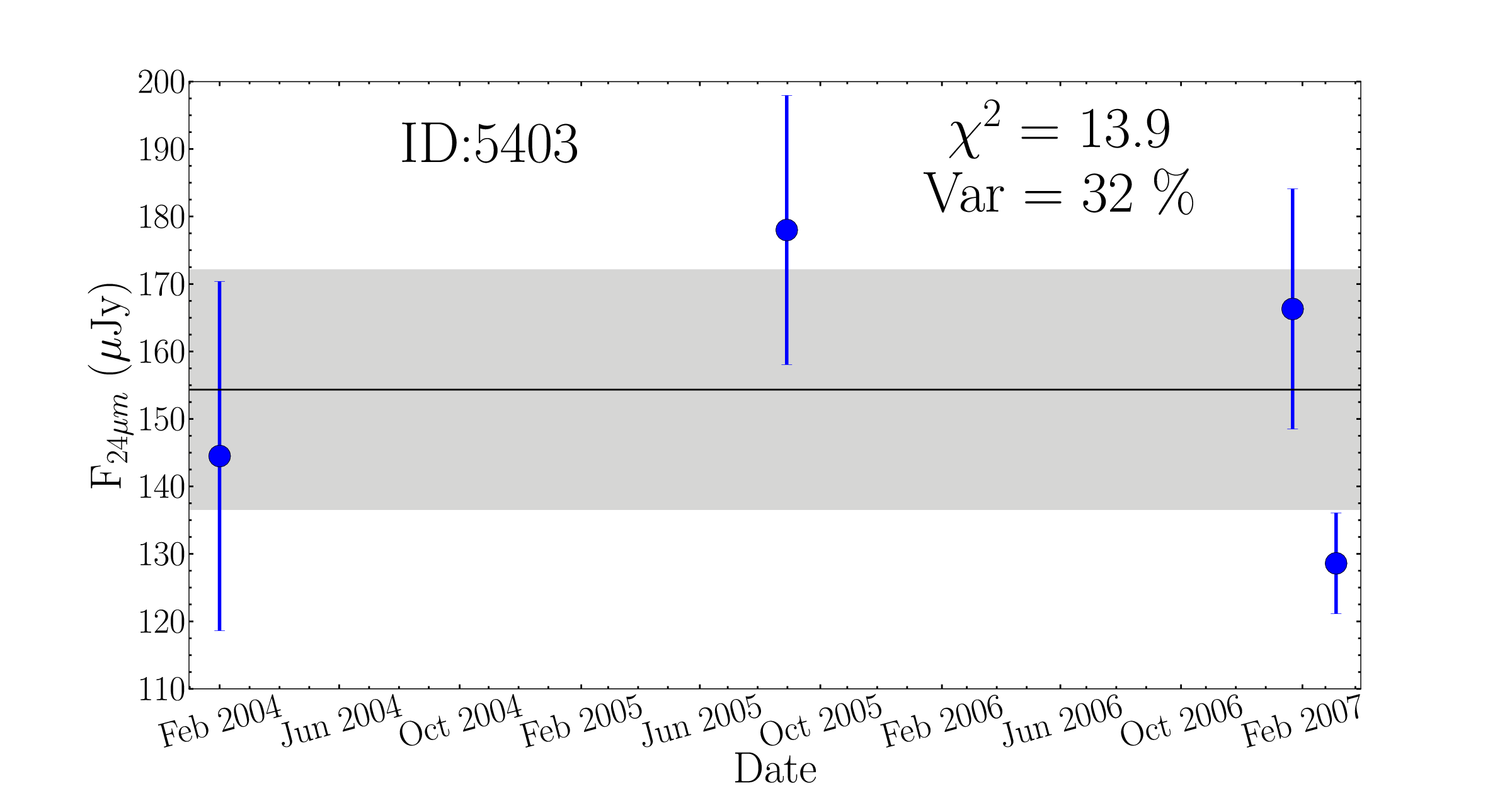}}{\hspace{0cm}}
    \subfigure {\includegraphics[width=47mm]{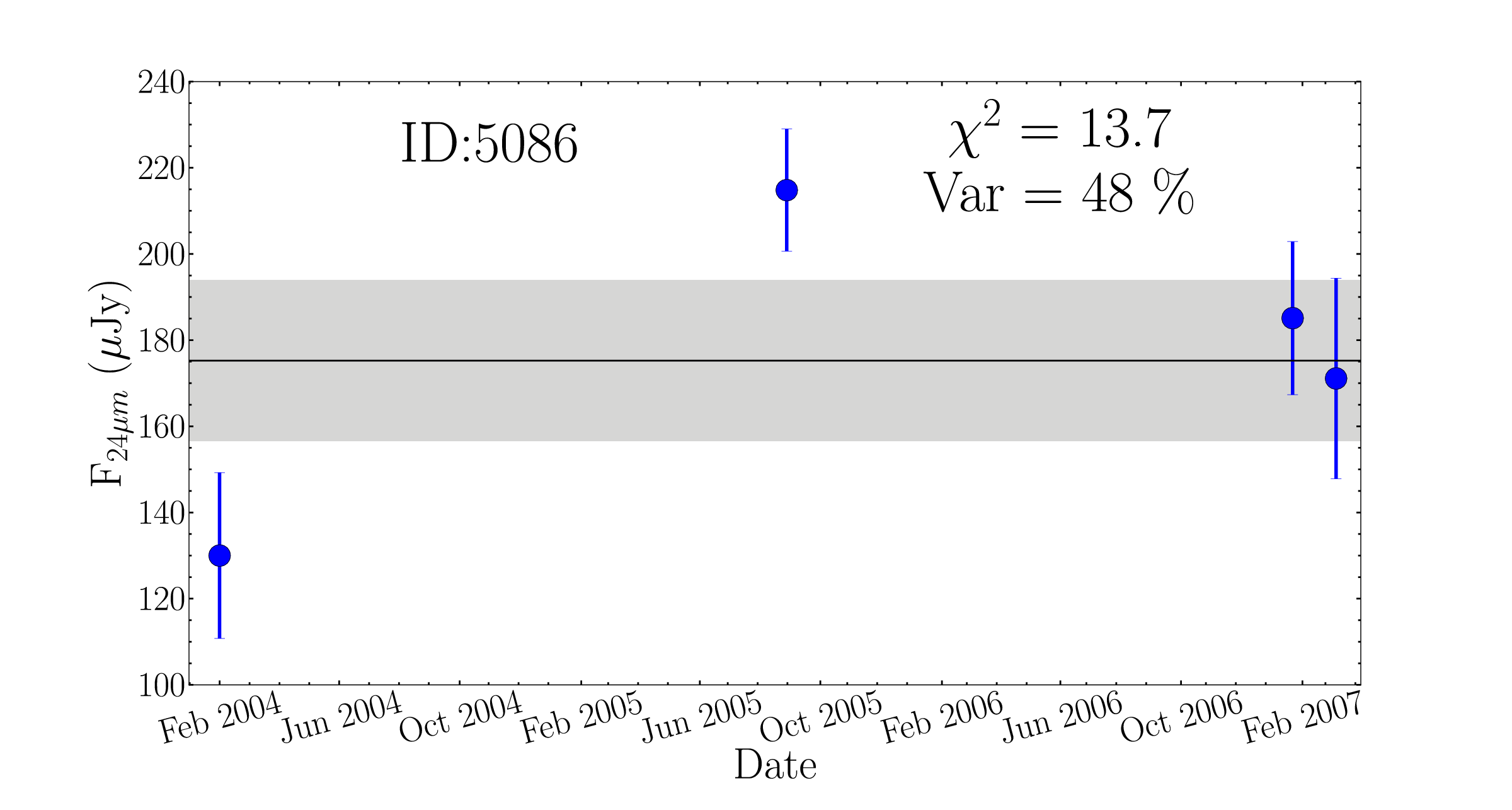}}{\hspace{0cm}}
    \subfigure {\includegraphics[width=47mm]{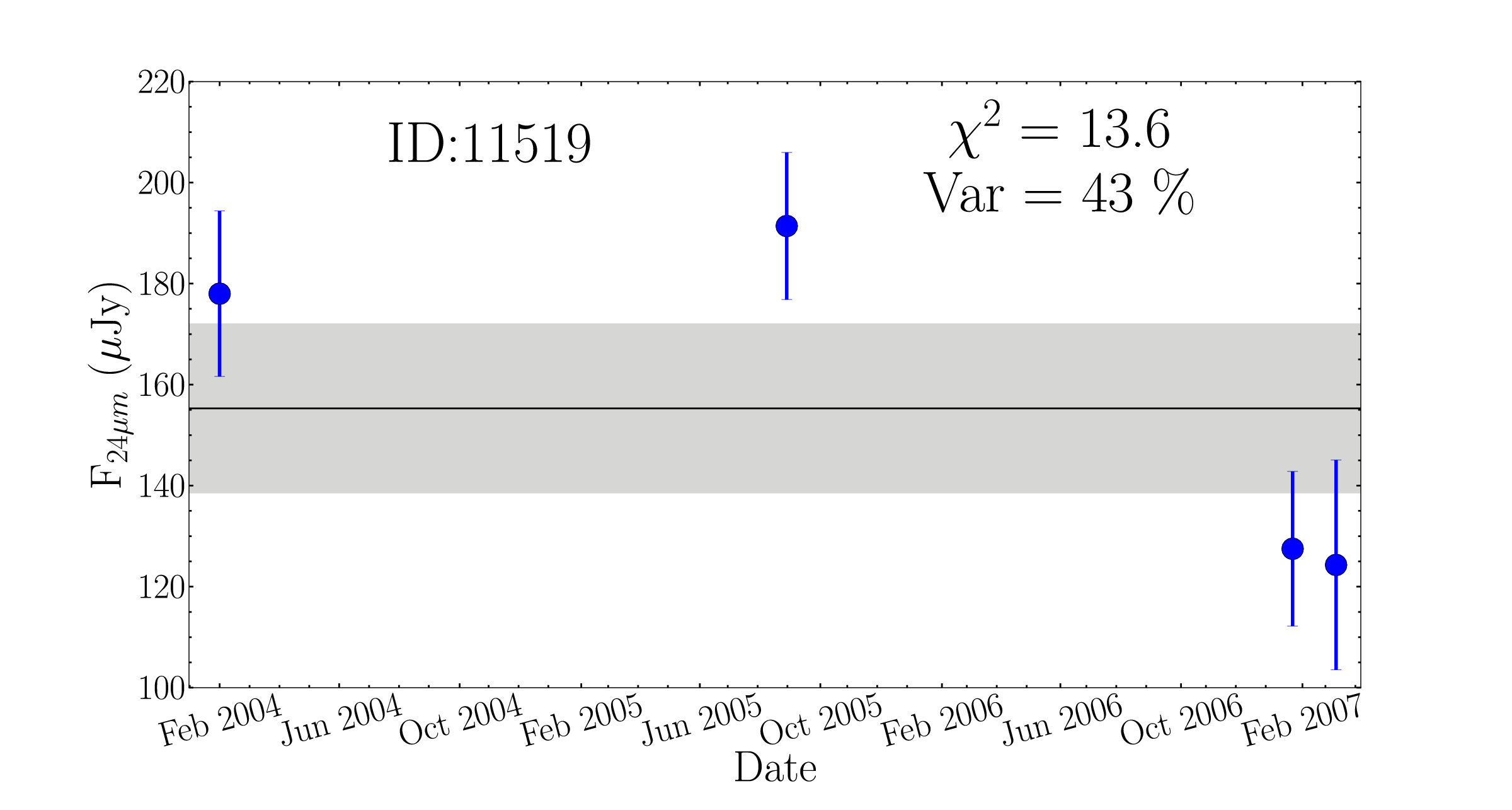}}{\hspace{0cm}}
    \subfigure {\includegraphics[width=47mm]{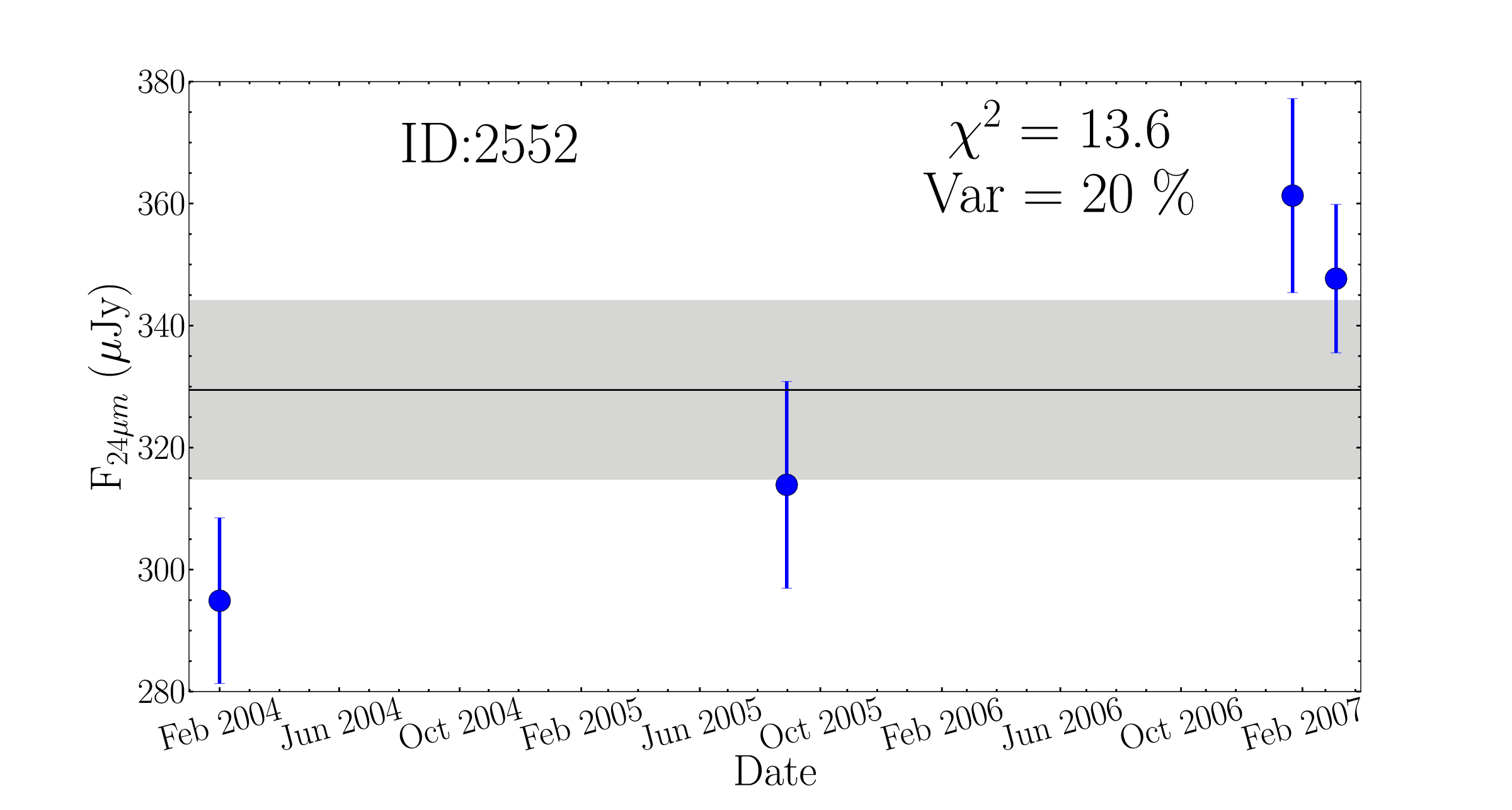}}{\hspace{0cm}}
    \subfigure {\includegraphics[width=47mm]{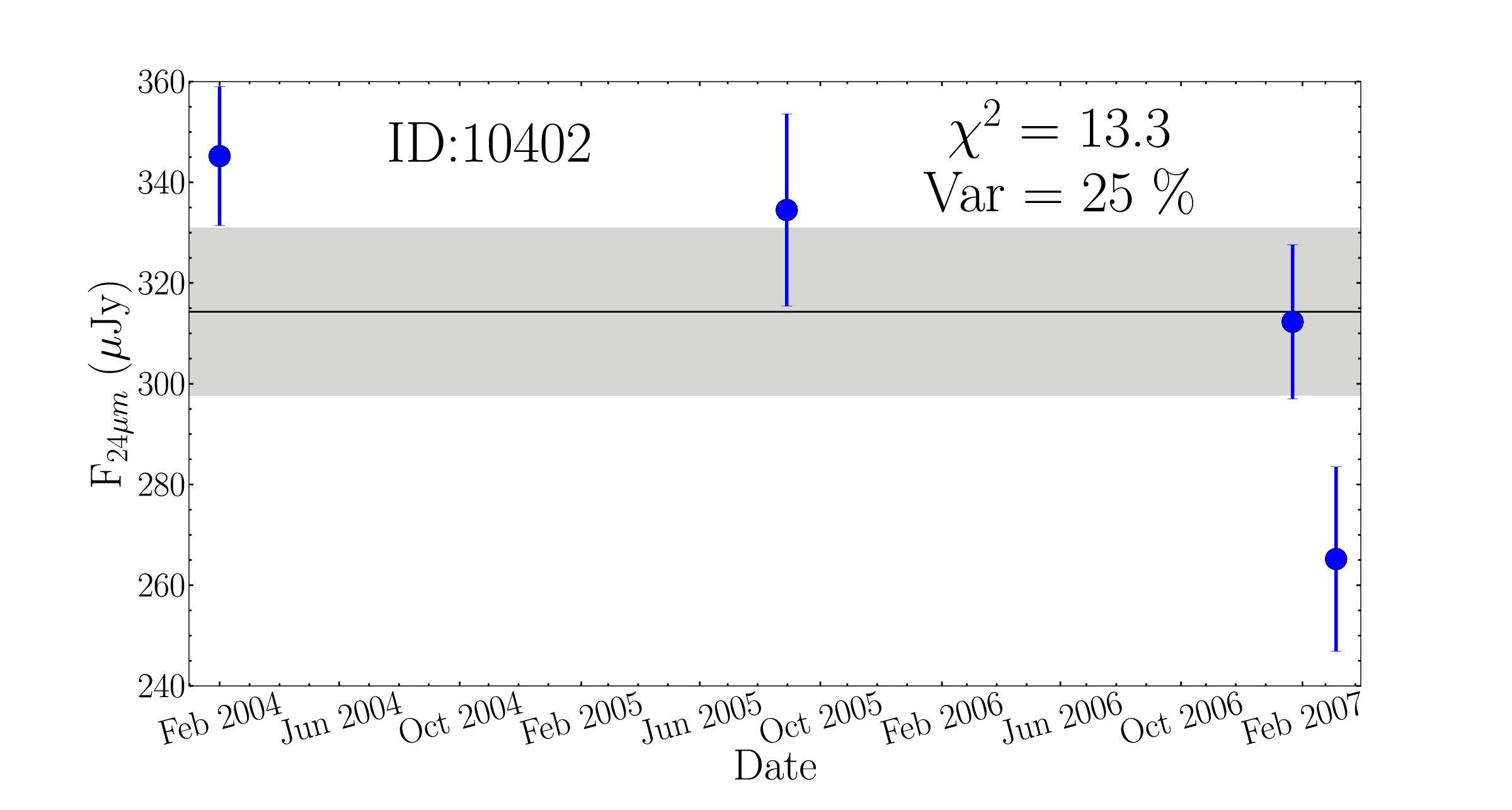}}{\hspace{0cm}}
    \subfigure {\includegraphics[width=47mm]{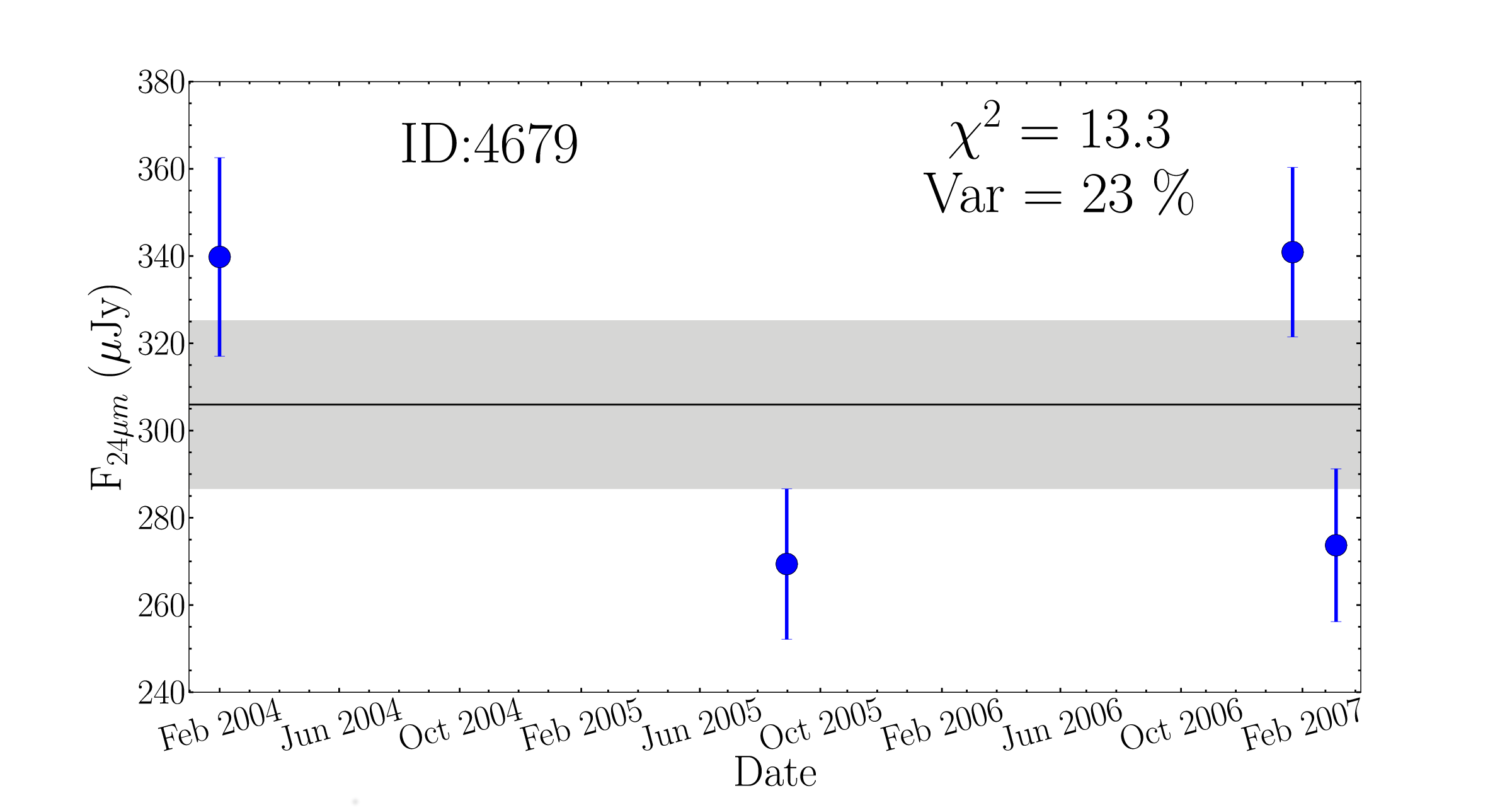}}{\hspace{0cm}}
    \subfigure {\includegraphics[width=47mm]{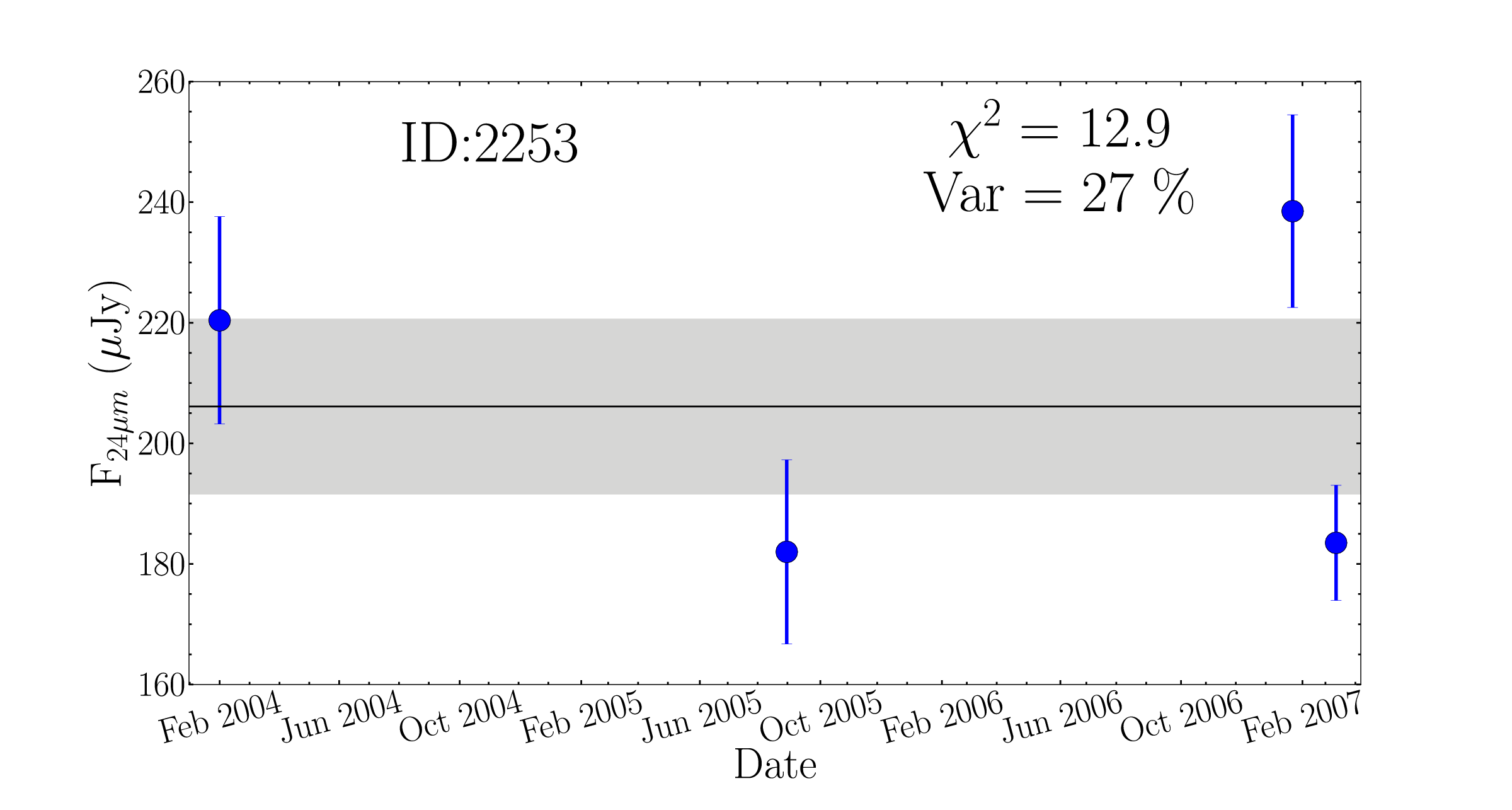}}{\hspace{0cm}}
    \subfigure {\includegraphics[width=47mm]{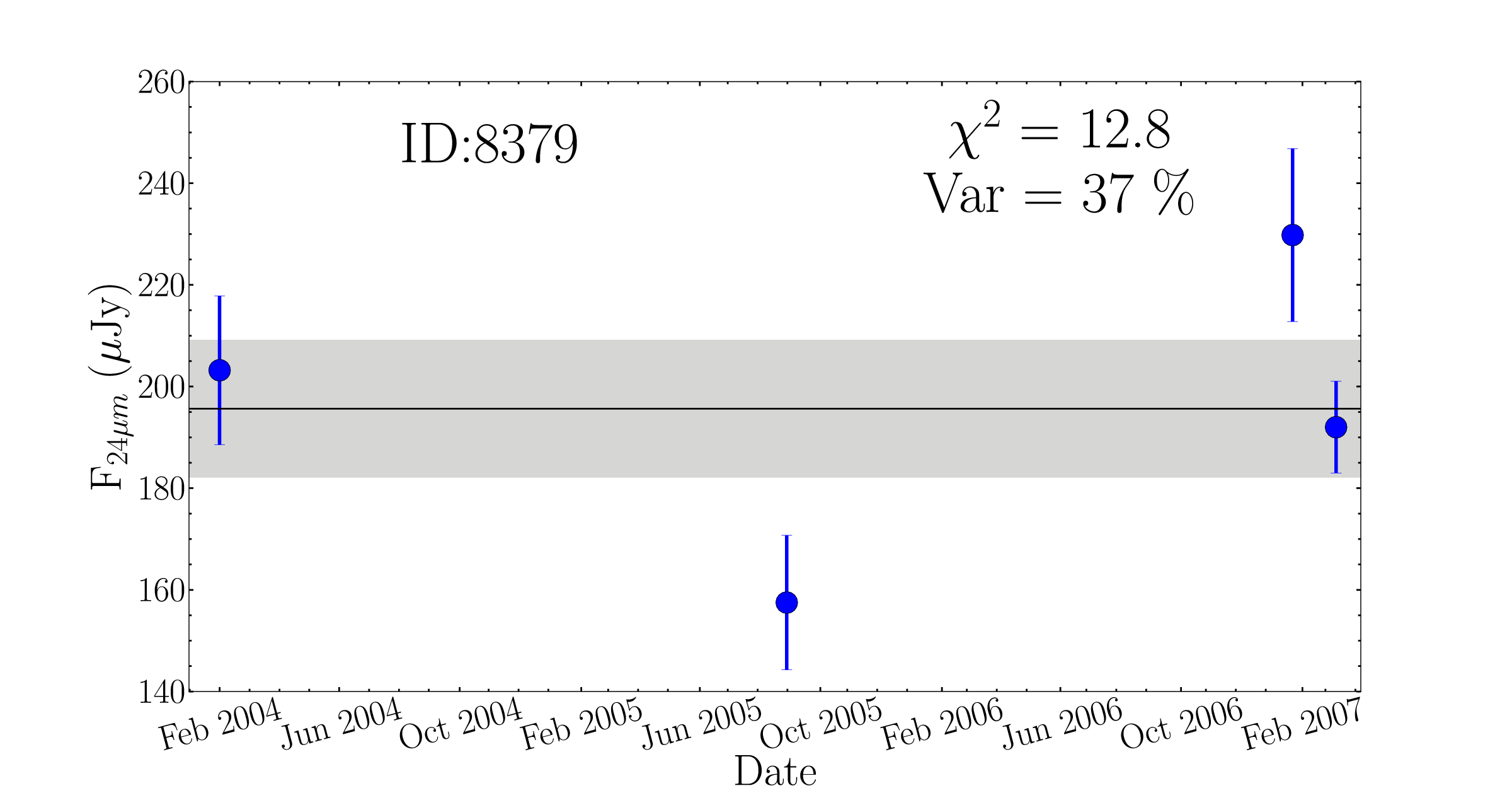}}{\hspace{0cm}}
    \subfigure {\includegraphics[width=47mm]{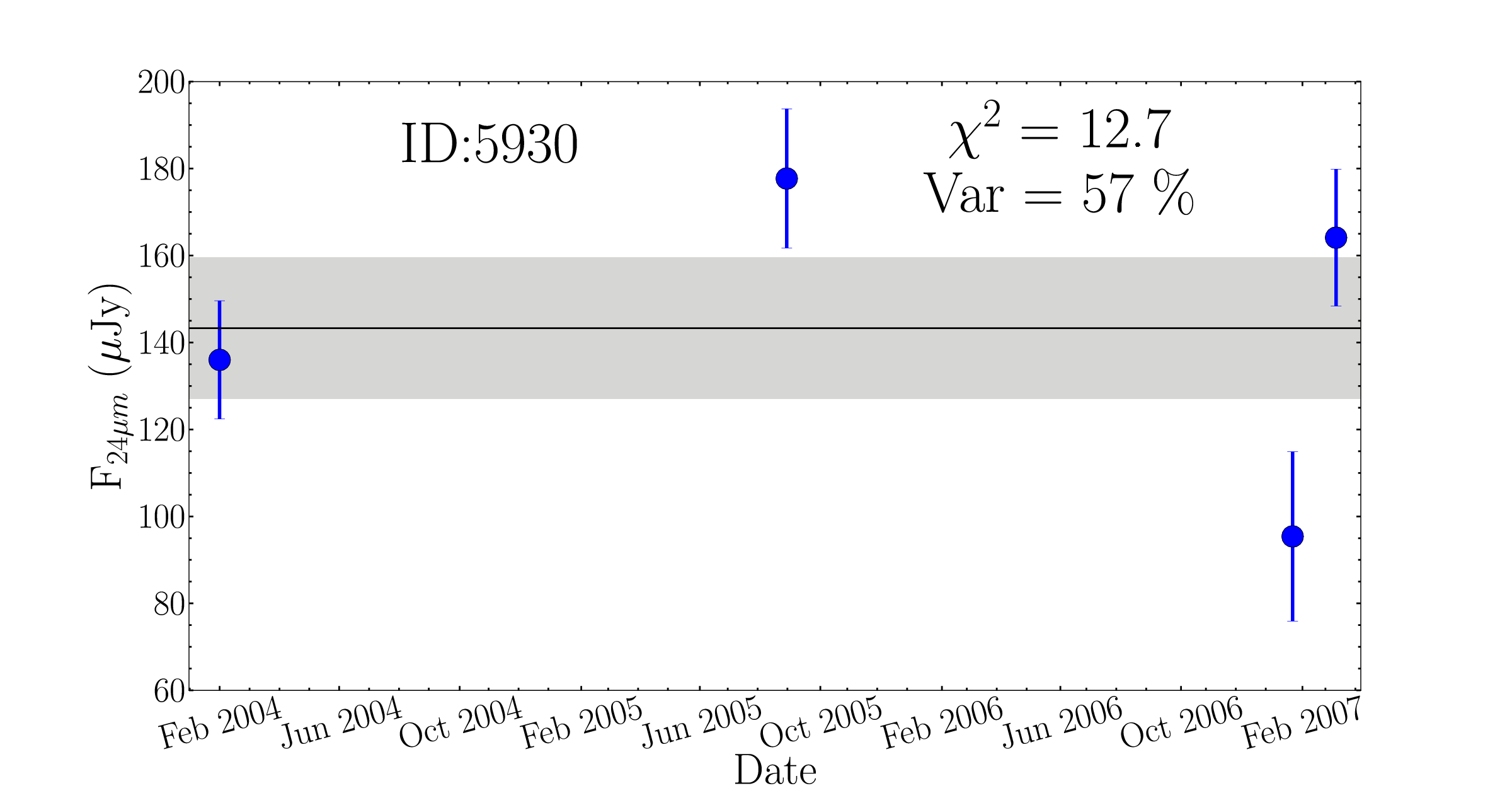}}{\hspace{0cm}}
    \subfigure {\includegraphics[width=47mm]{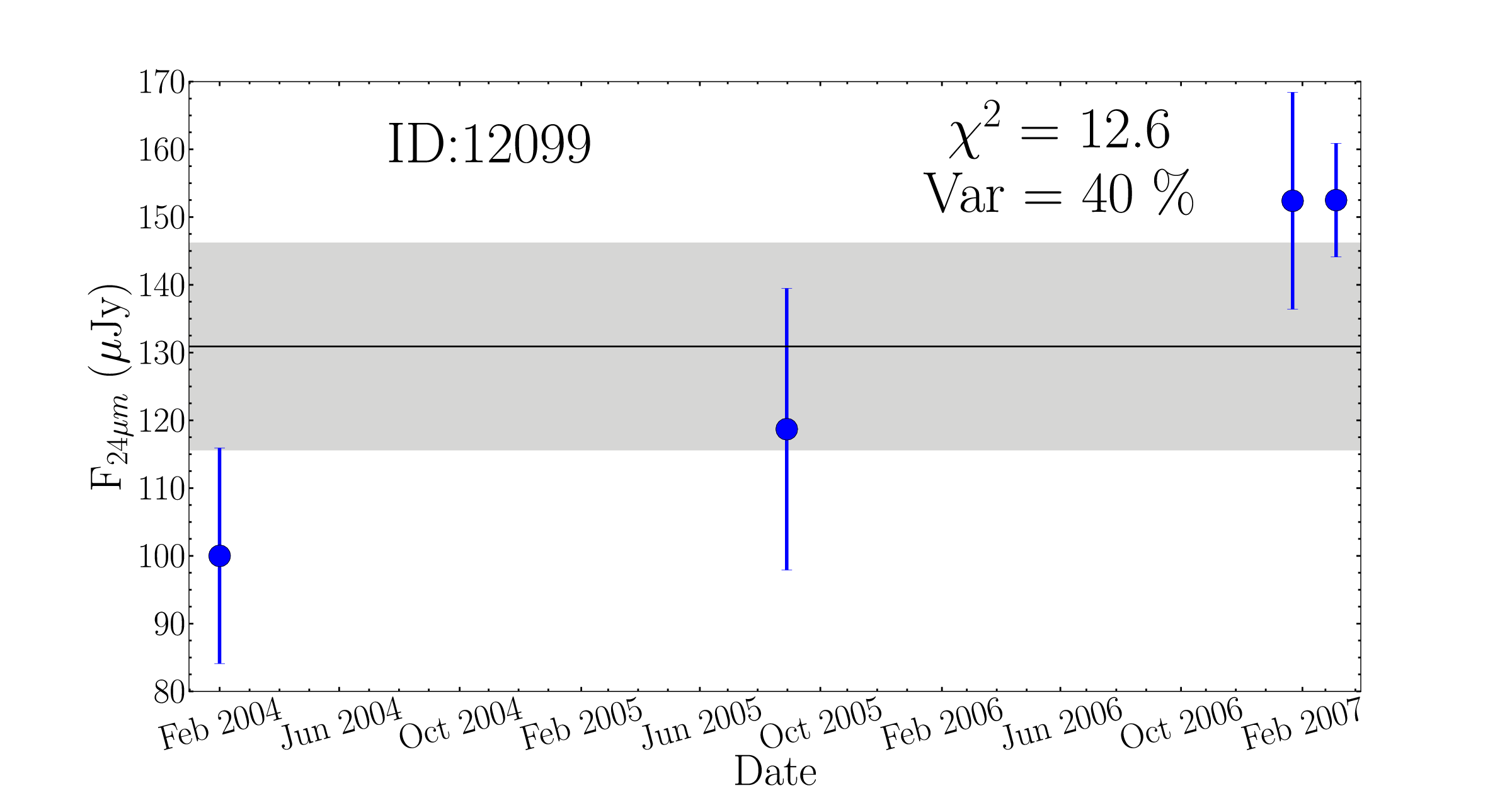}}{\hspace{0cm}}
    \subfigure {\includegraphics[width=47mm]{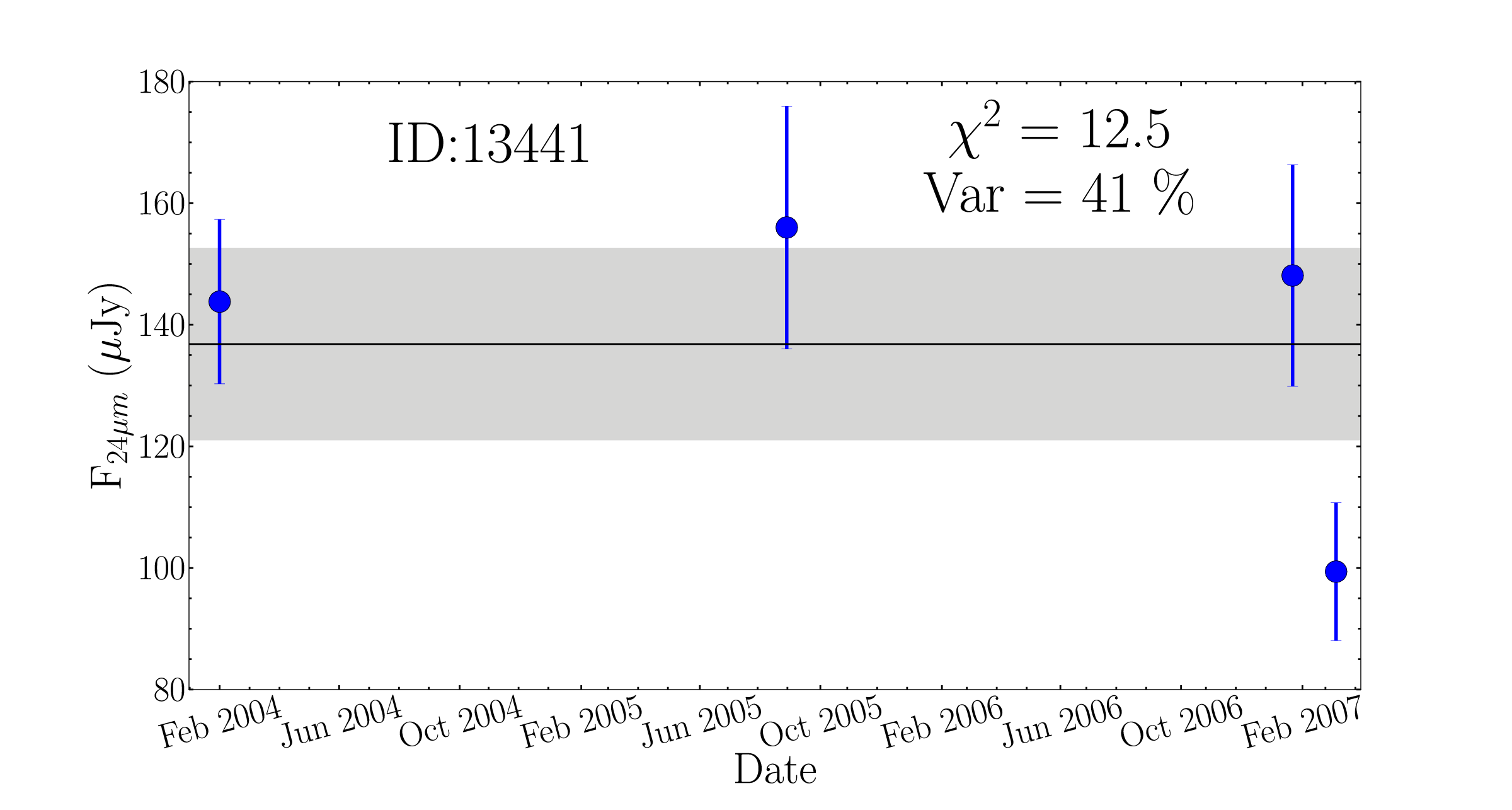}}{\hspace{0cm}}
    \subfigure {\includegraphics[width=47mm]{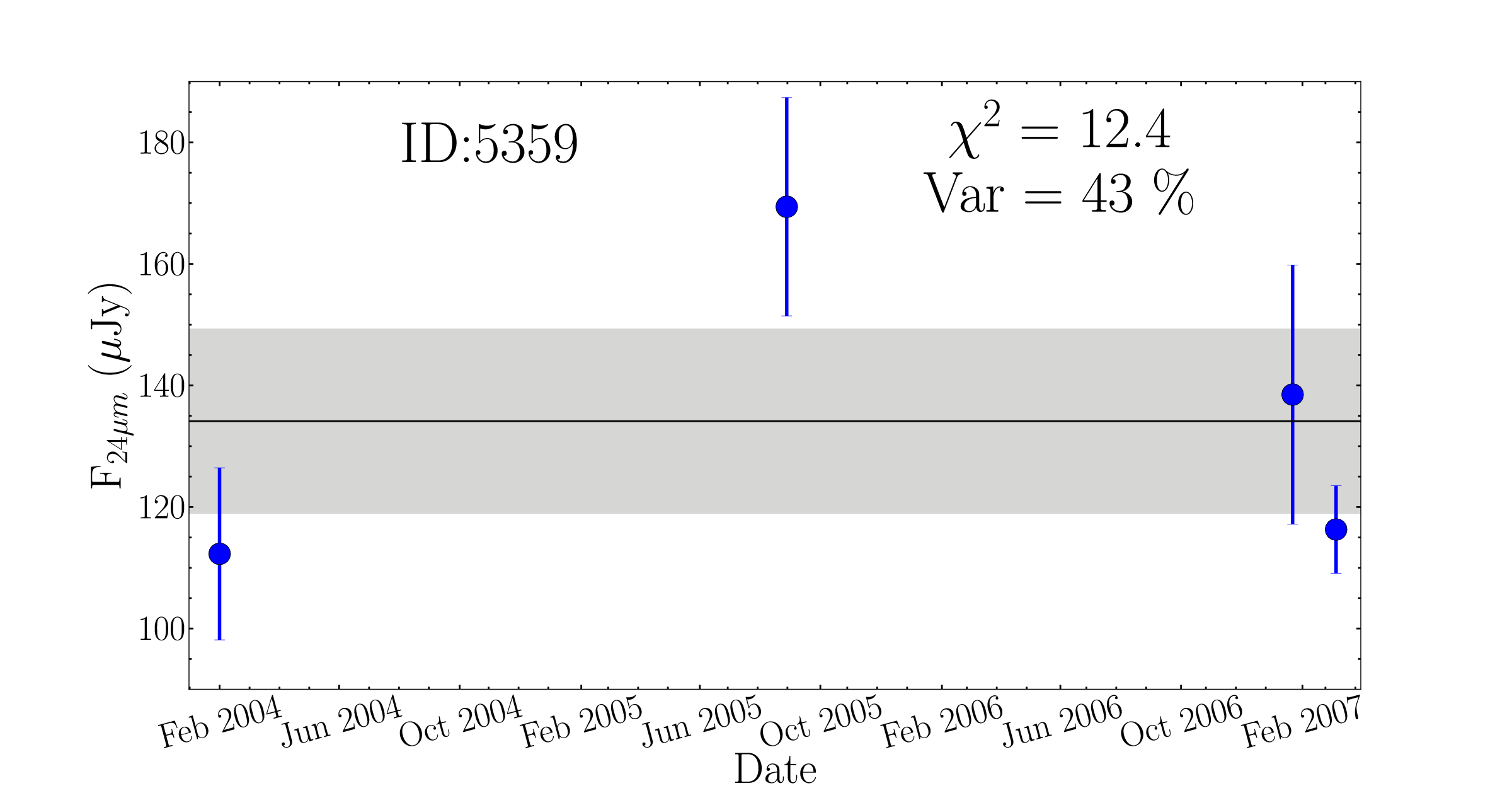}}{\hspace{0cm}}
    \subfigure {\includegraphics[width=47mm]{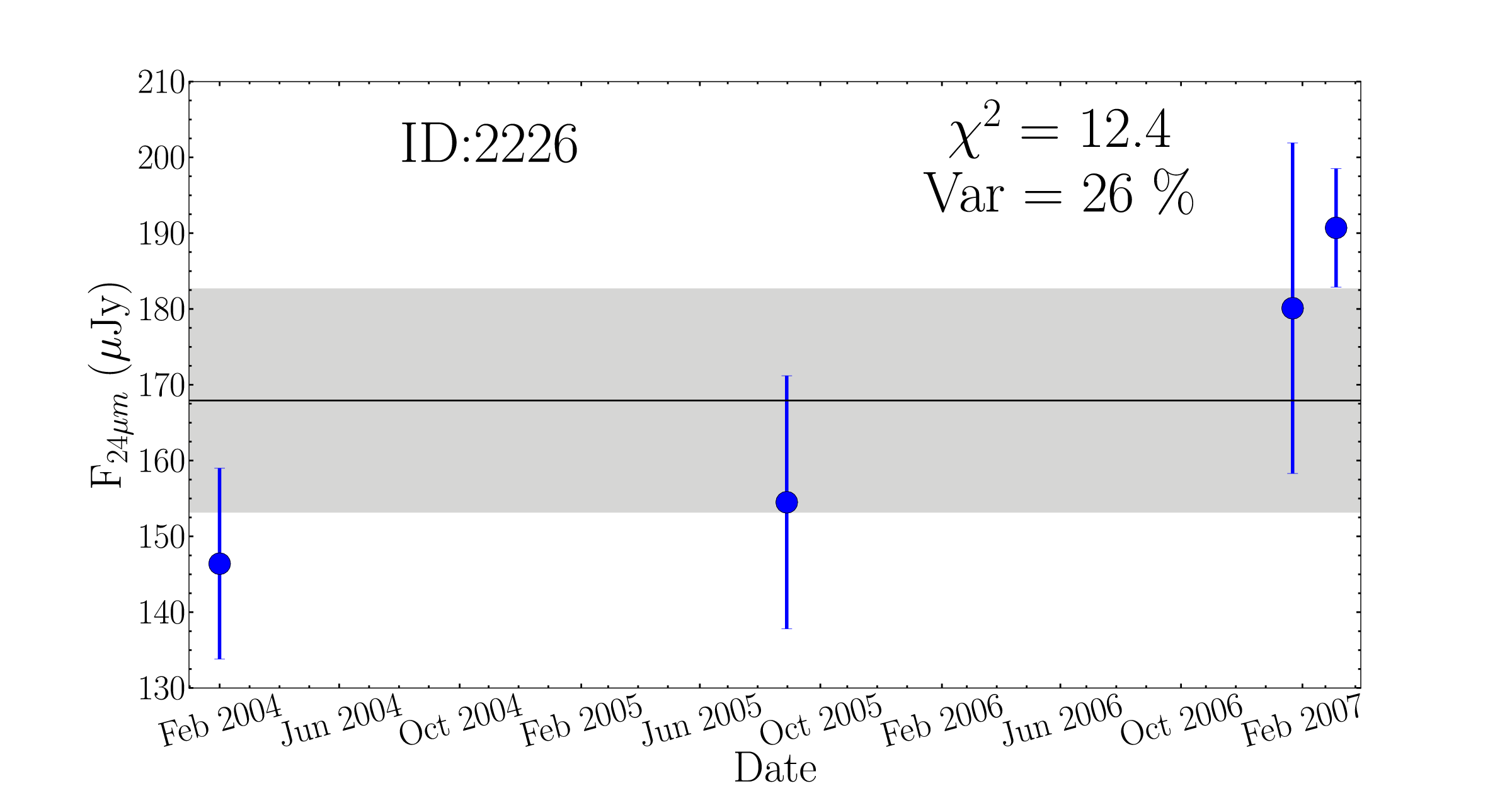}}{\hspace{0cm}}
    \subfigure {\includegraphics[width=47mm]{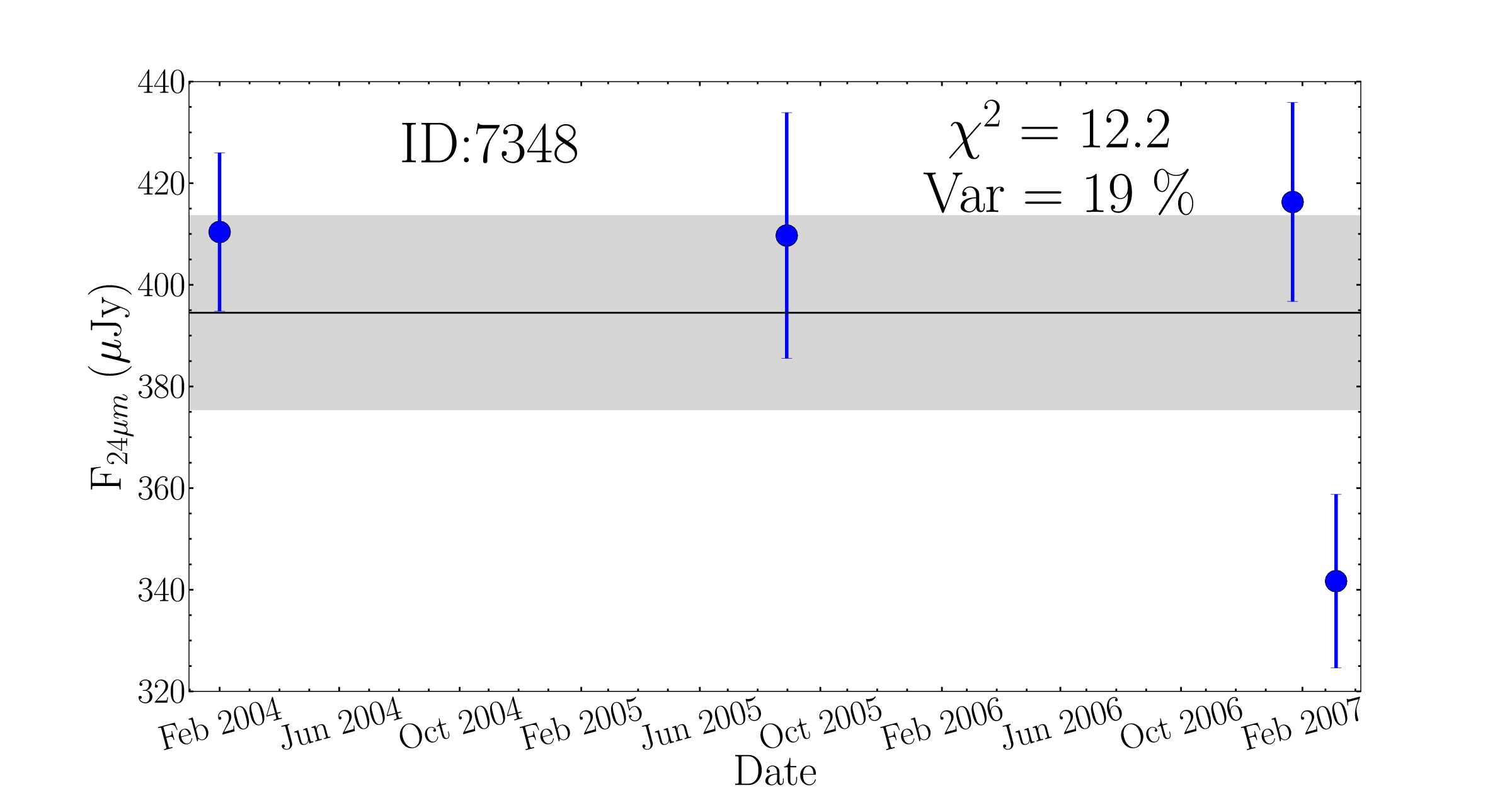}}{\hspace{0cm}}
    \subfigure {\includegraphics[width=47mm]{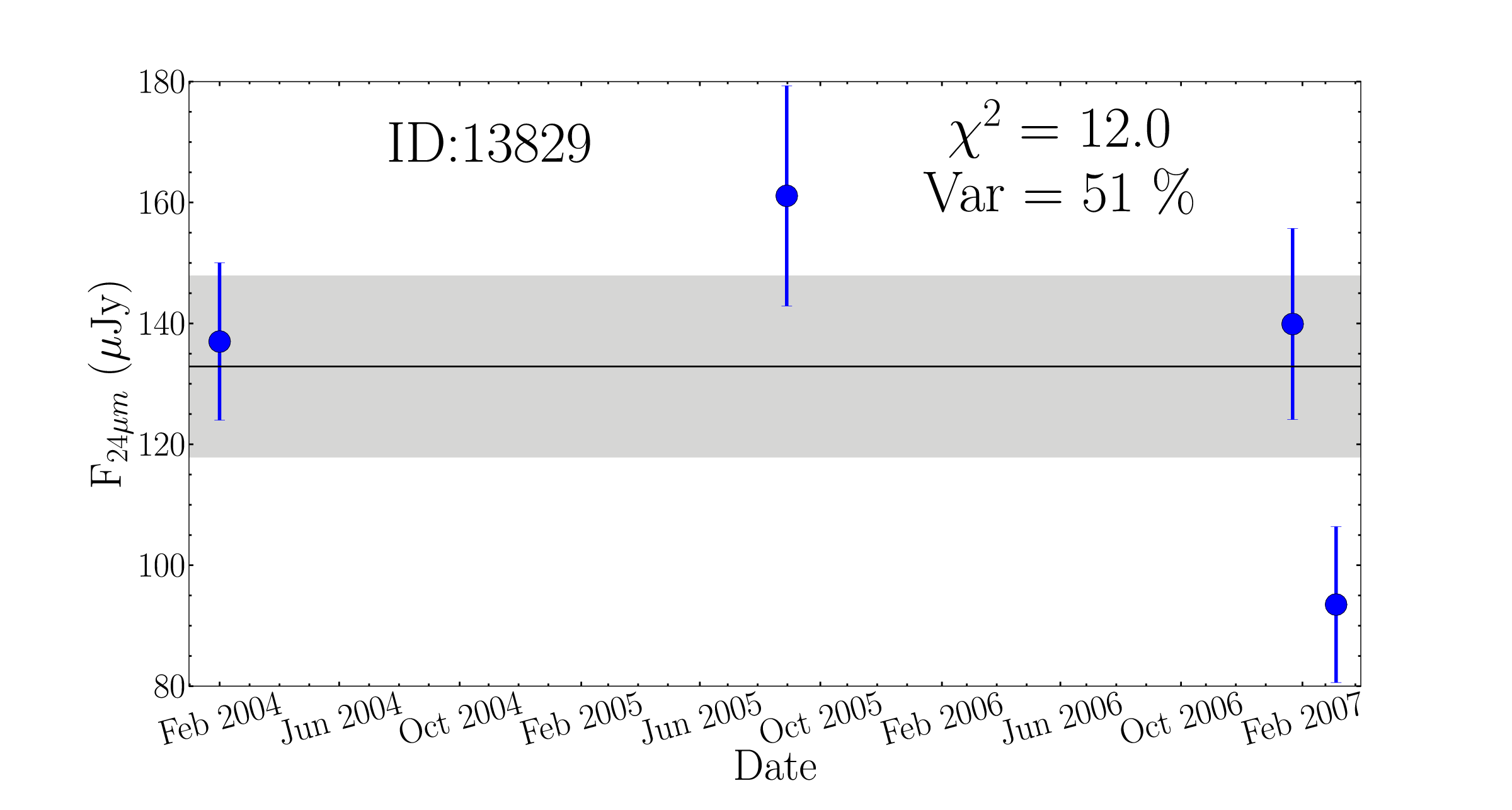}}{\hspace{0cm}}
    \subfigure {\includegraphics[width=47mm]{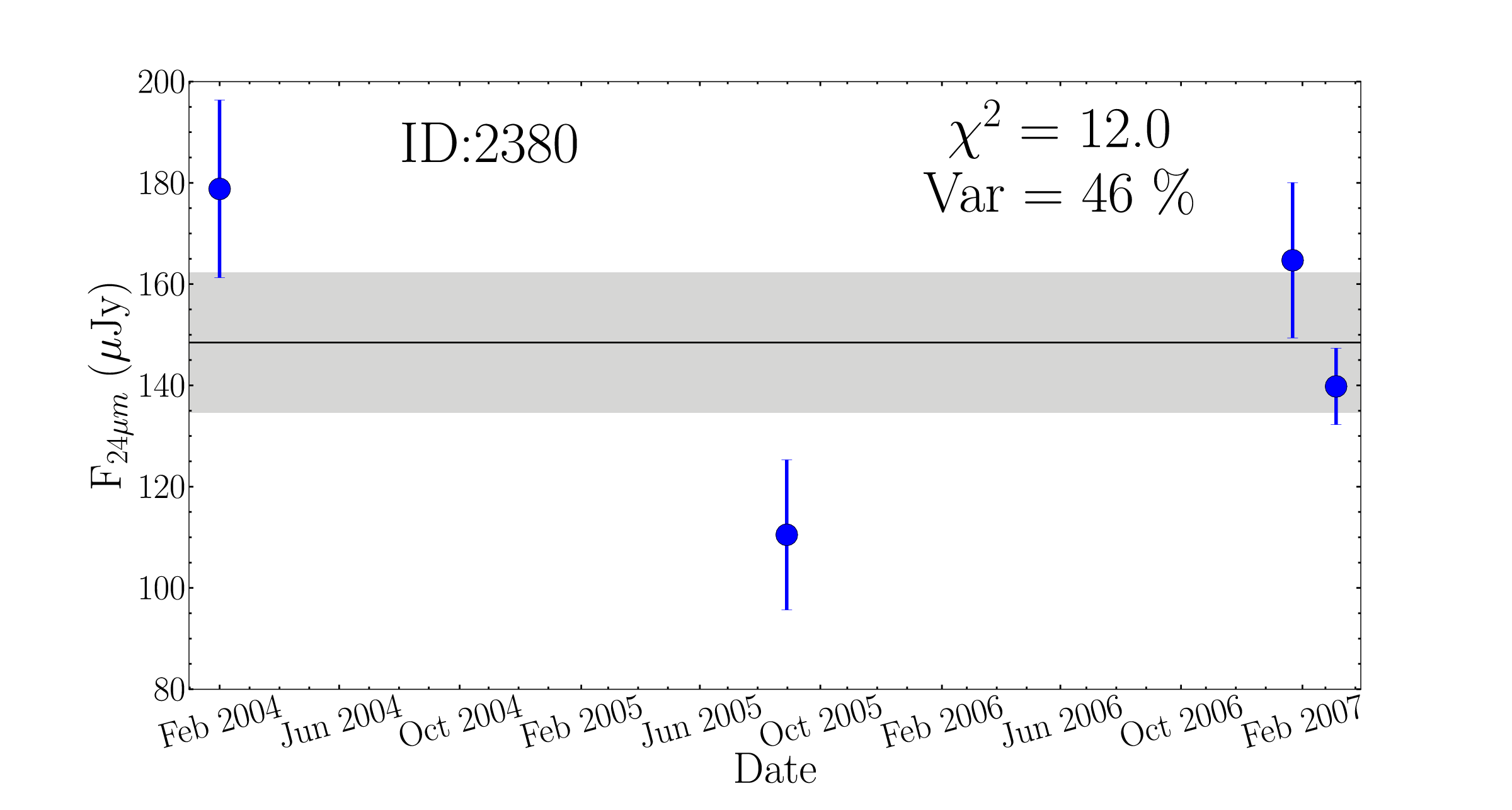}}{\hspace{0cm}}
    \subfigure {\includegraphics[width=47mm]{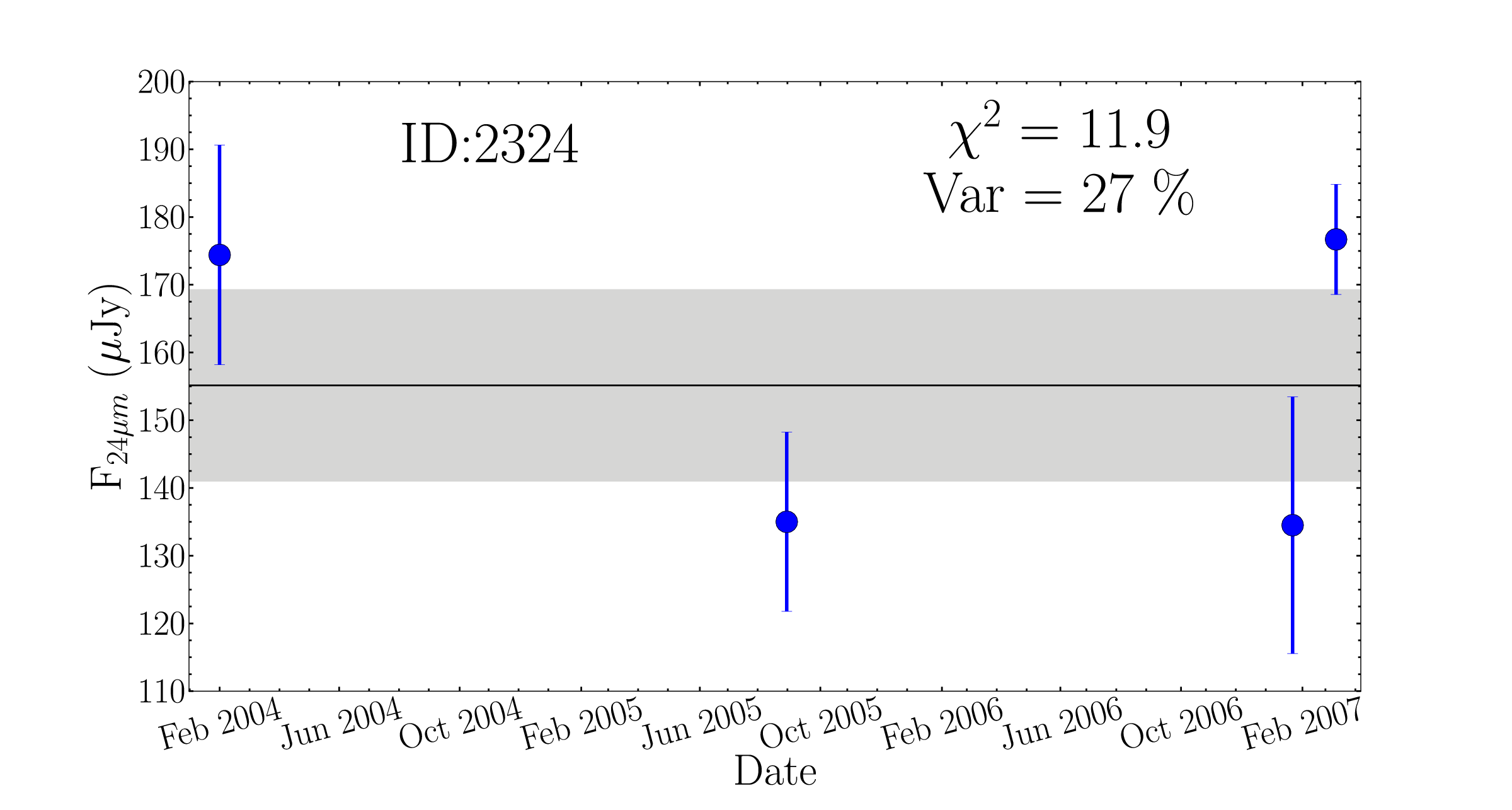}}{\hspace{0cm}}
    \subfigure {\includegraphics[width=47mm]{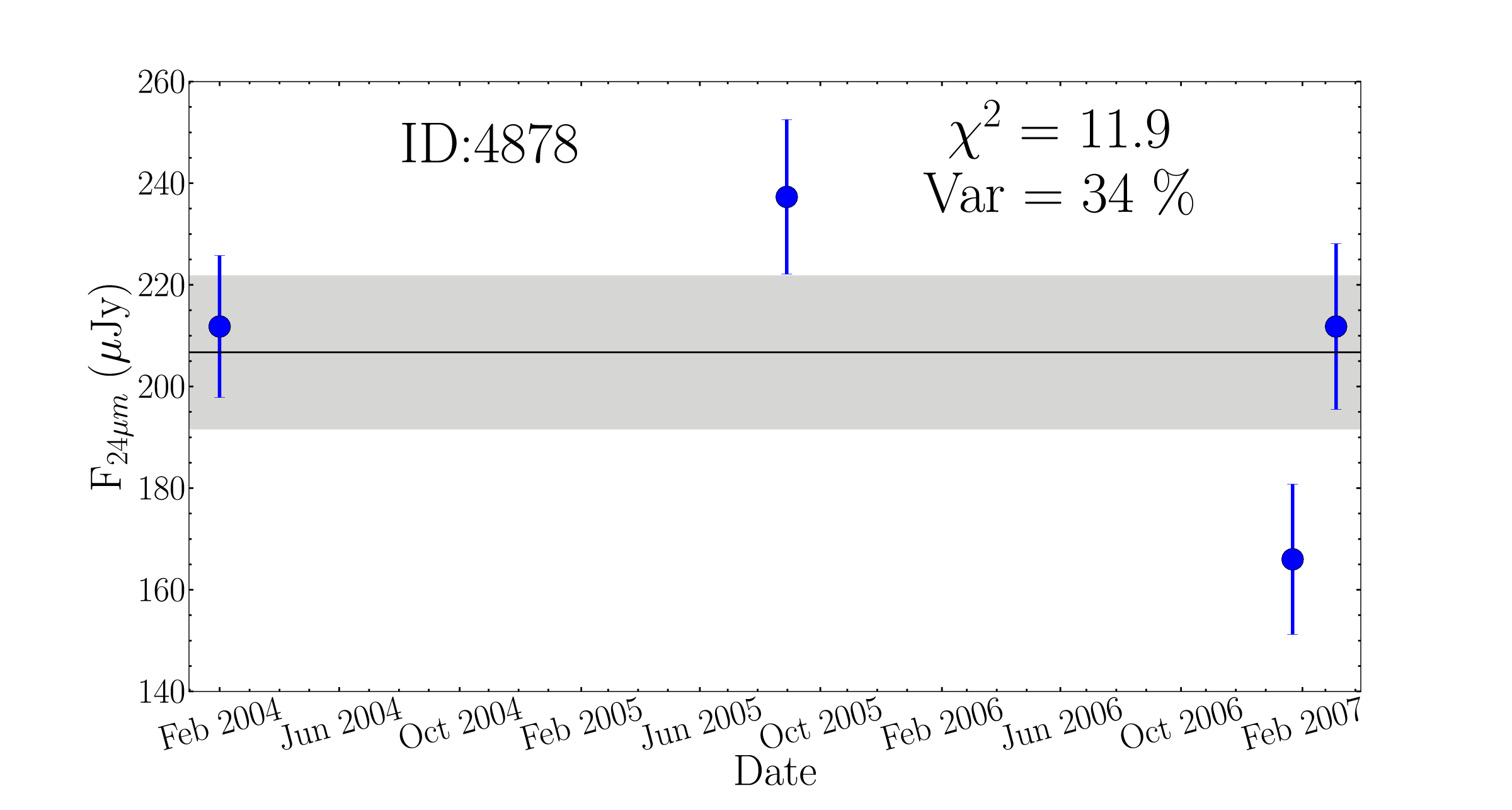}}{\hspace{0cm}}
    \subfigure {\includegraphics[width=47mm]{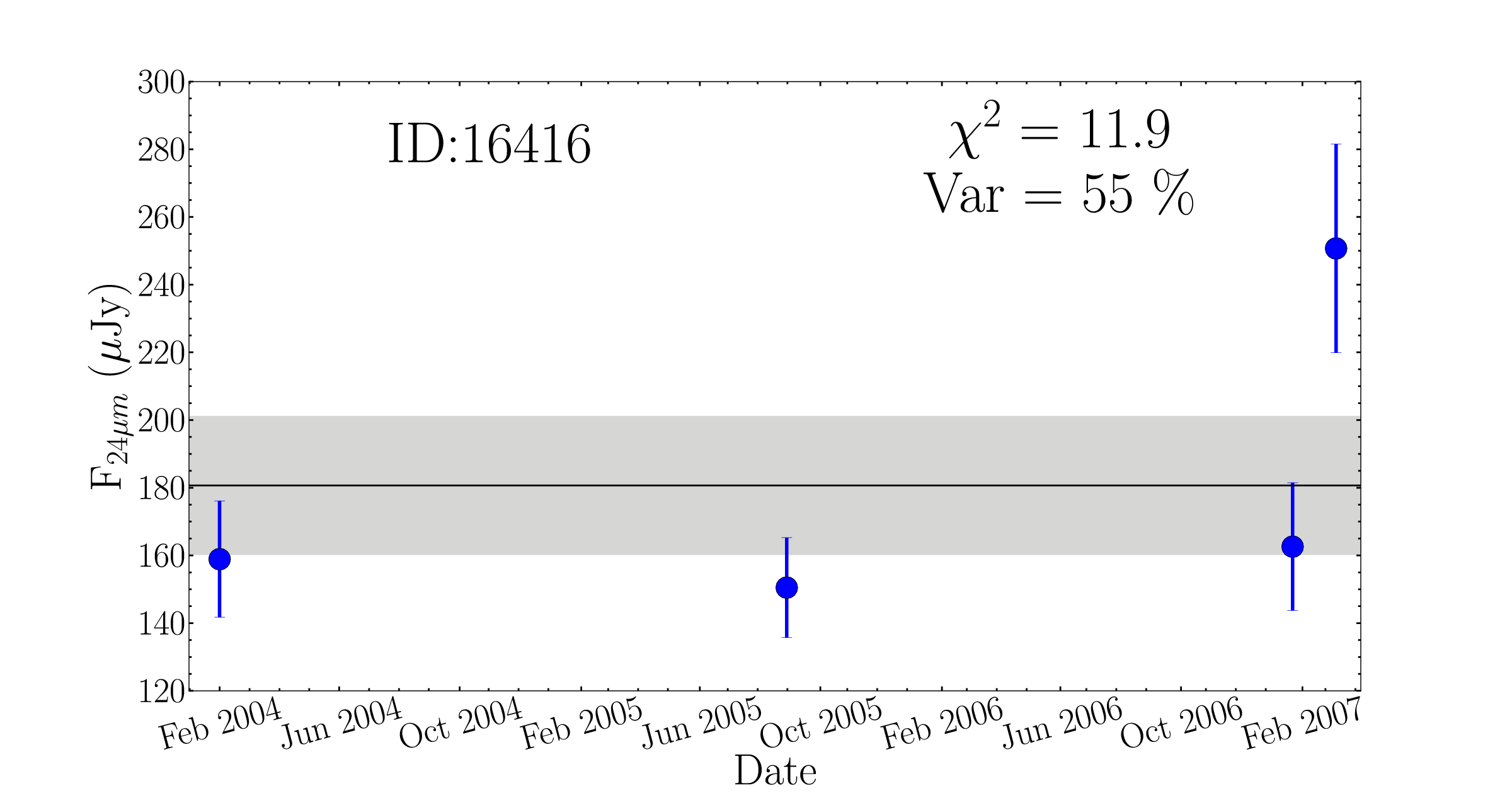}}{\hspace{0cm}}

    \caption{Light curves of MIPS 24 $\umu$m long-term variable candidates.} 
    \label{curvas-luz}
    \end{center}
  \end{minipage}
\end{figure*}

\begin{figure*}
  \begin{minipage}{200mm}
    \begin{center}

    \subfigure {\includegraphics[width=47mm]{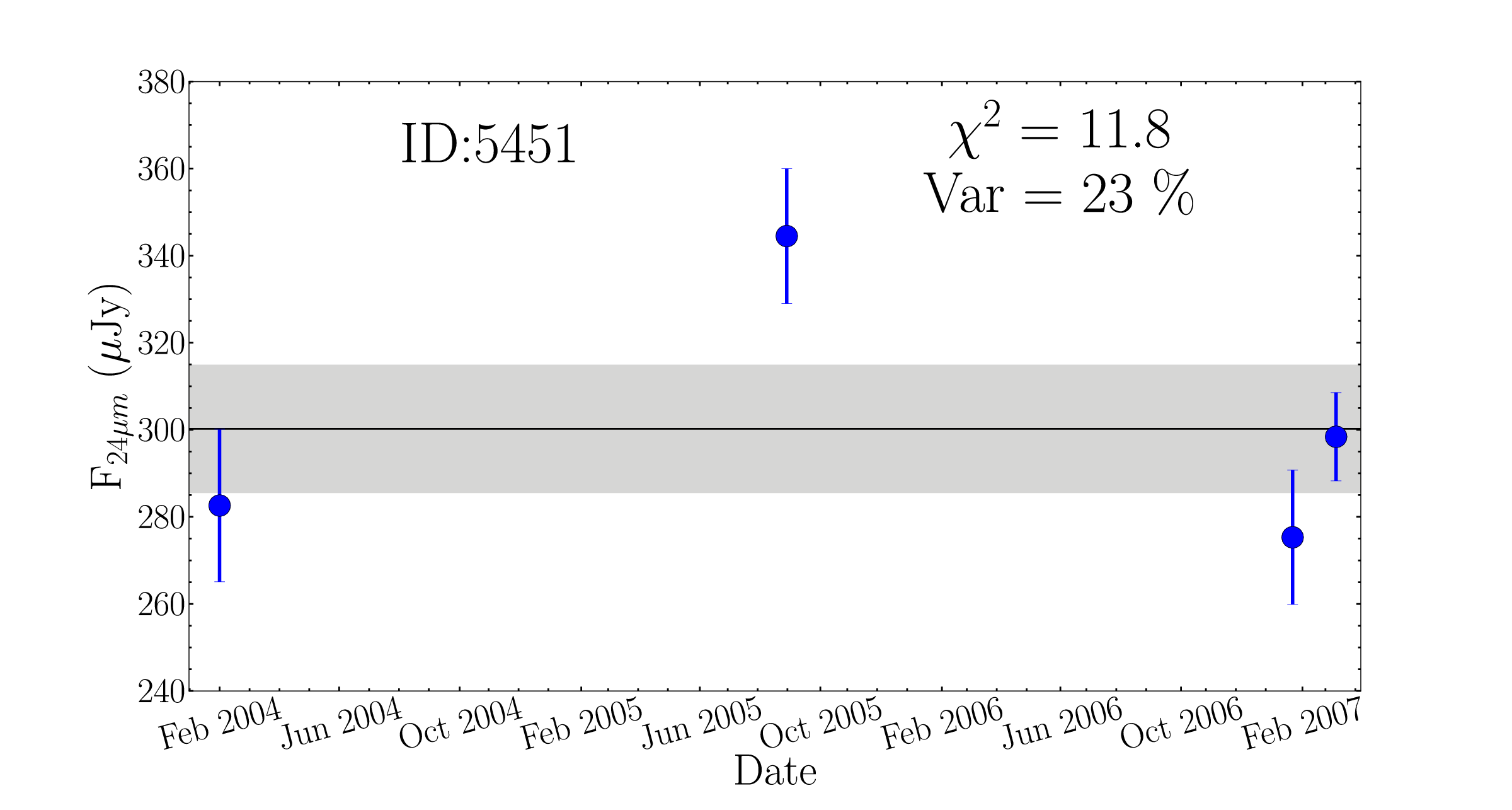}}{\hspace{0cm}}
    \subfigure {\includegraphics[width=47mm]{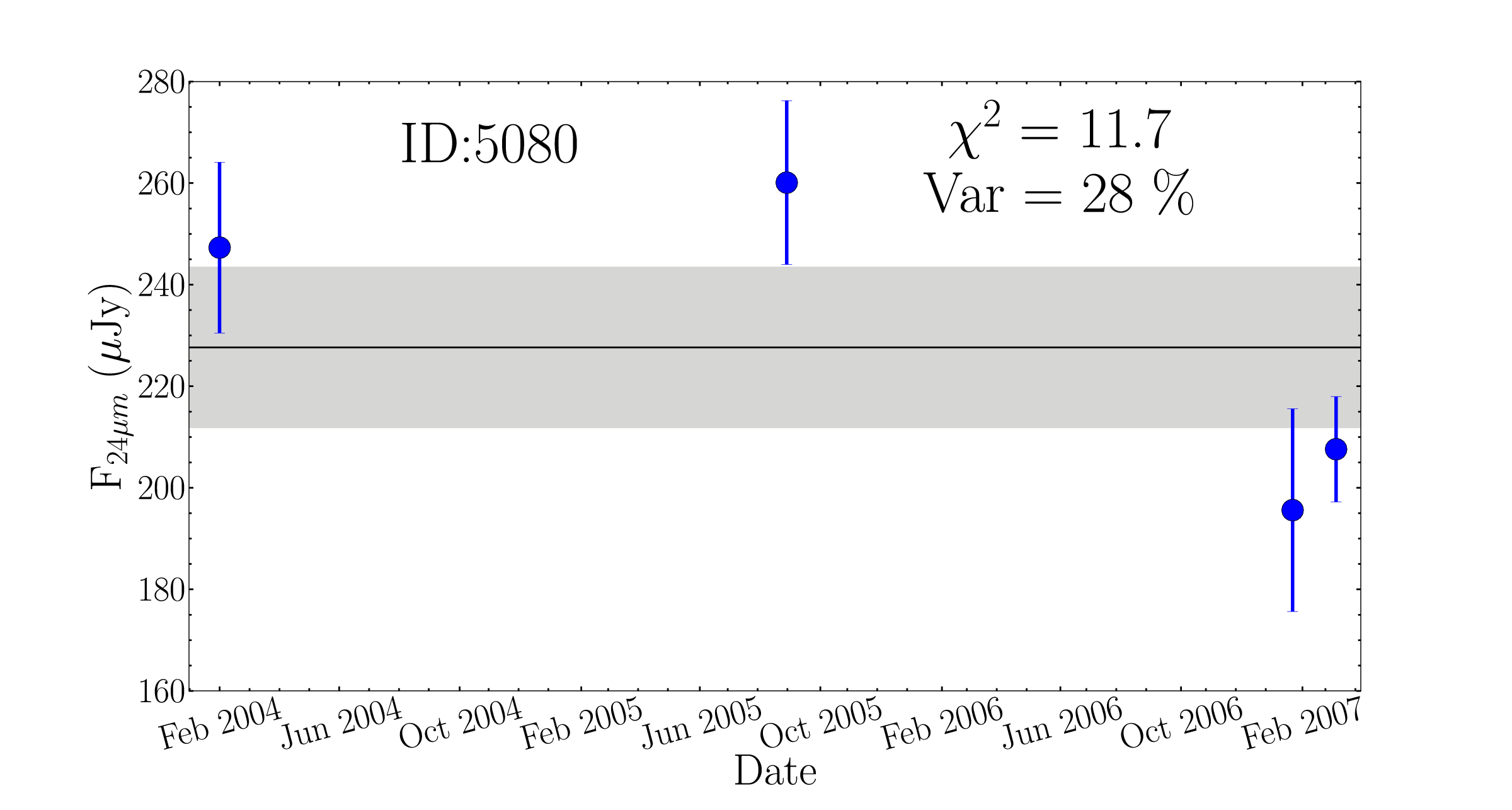}}{\hspace{0cm}}
    \subfigure {\includegraphics[width=47mm]{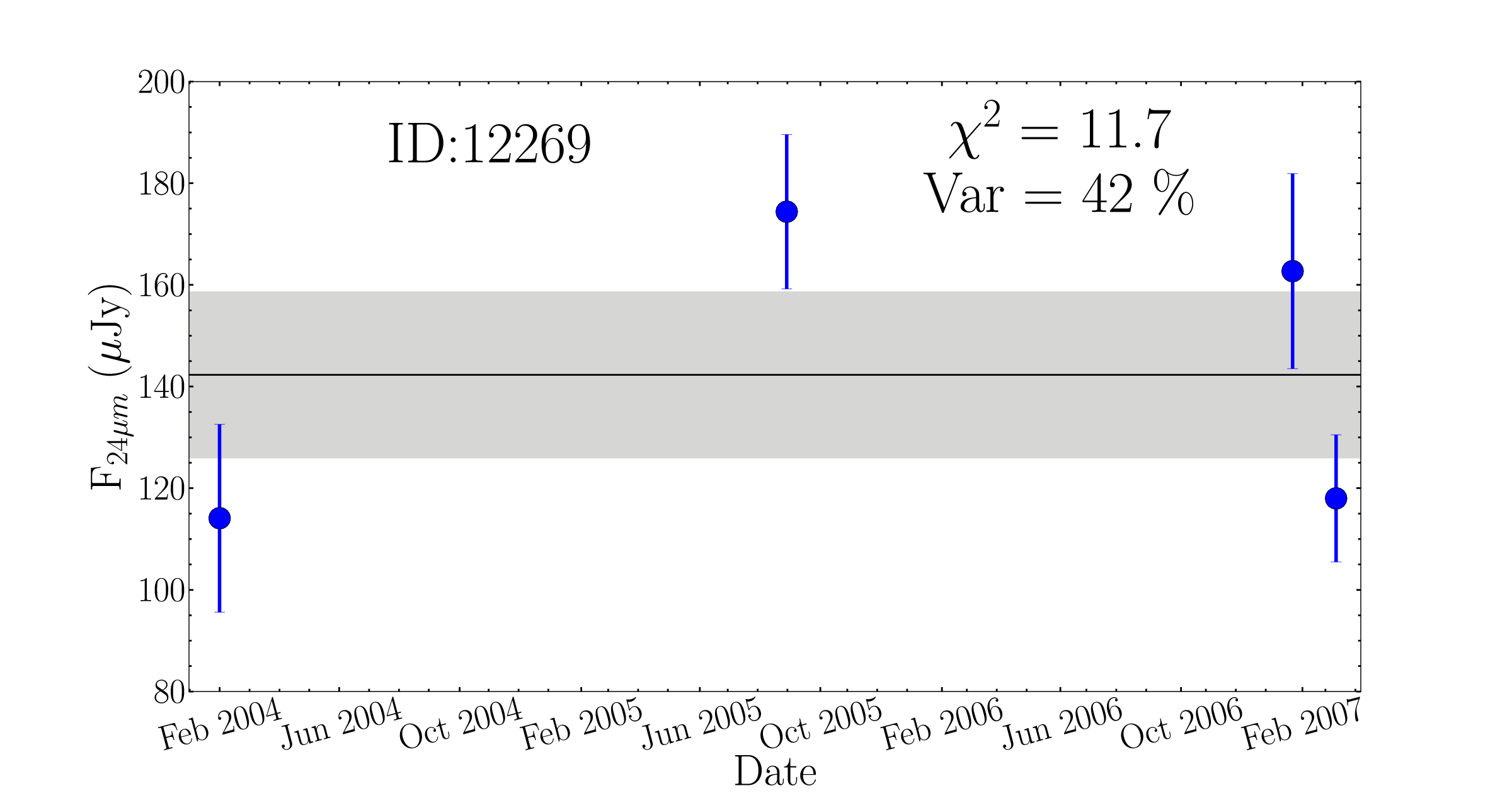}}{\hspace{0cm}}
    \subfigure {\includegraphics[width=47mm]{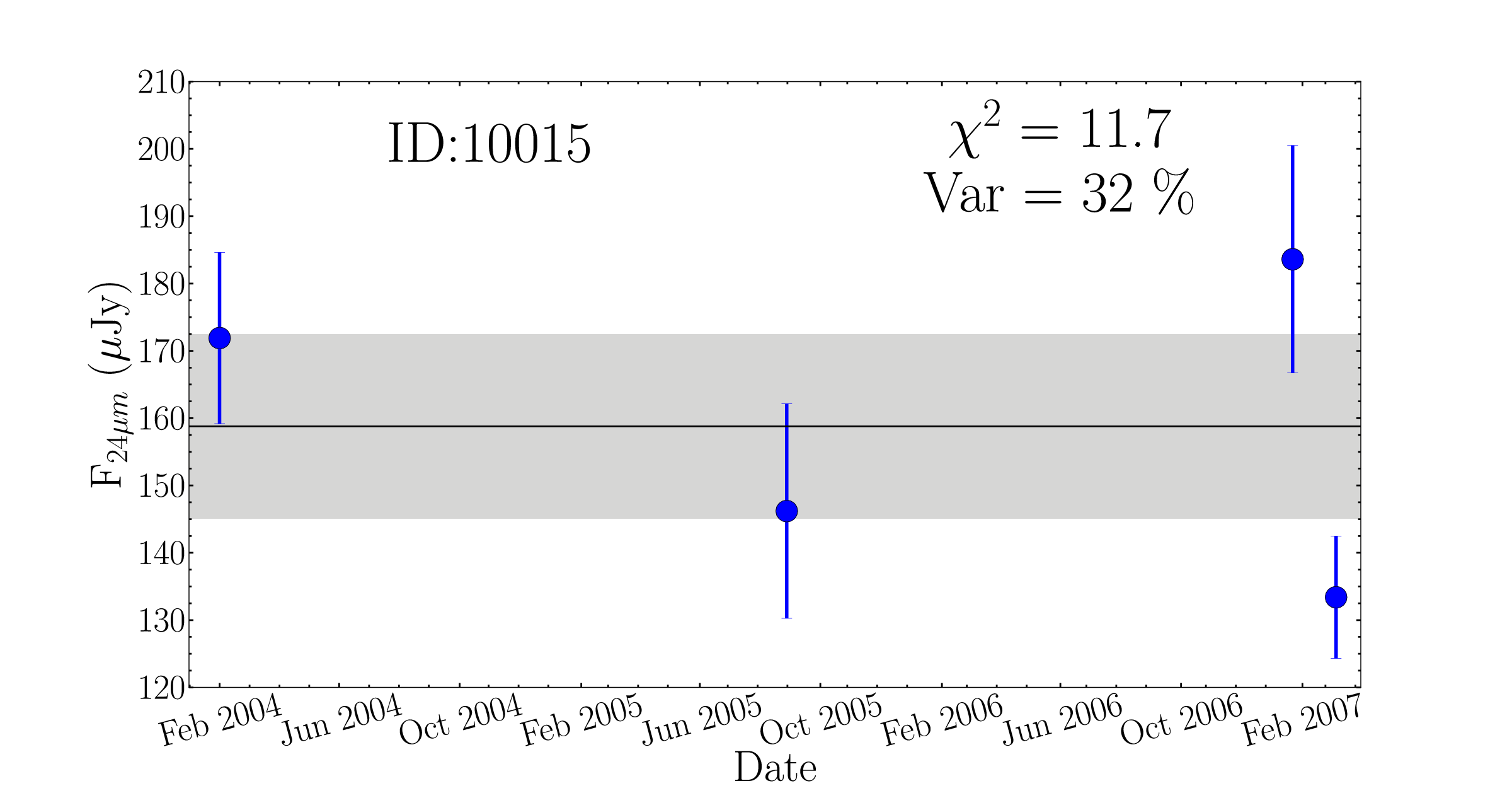}}{\hspace{0cm}}
    \subfigure {\includegraphics[width=47mm]{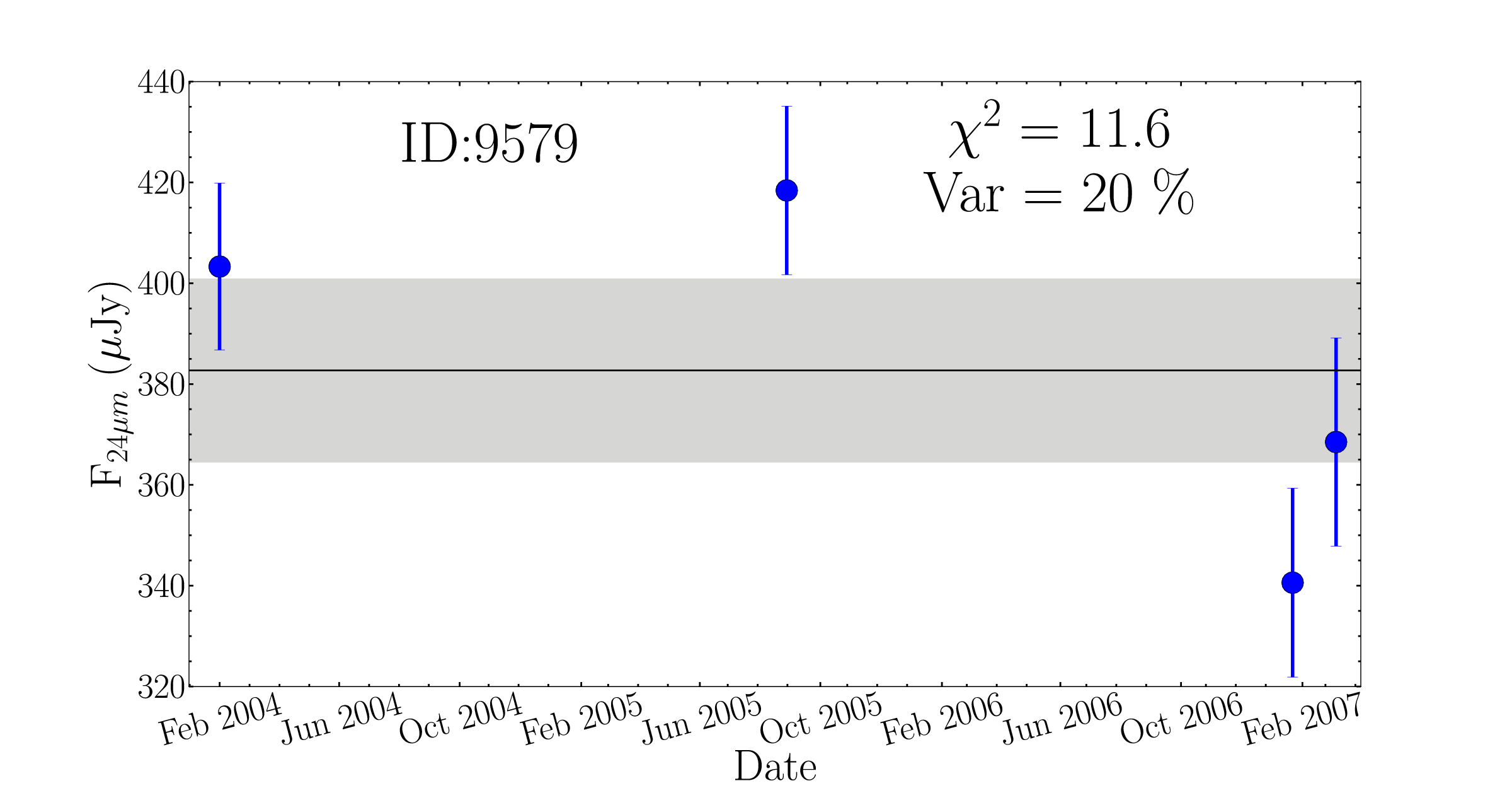}}{\hspace{0cm}}
    \subfigure {\includegraphics[width=47mm]{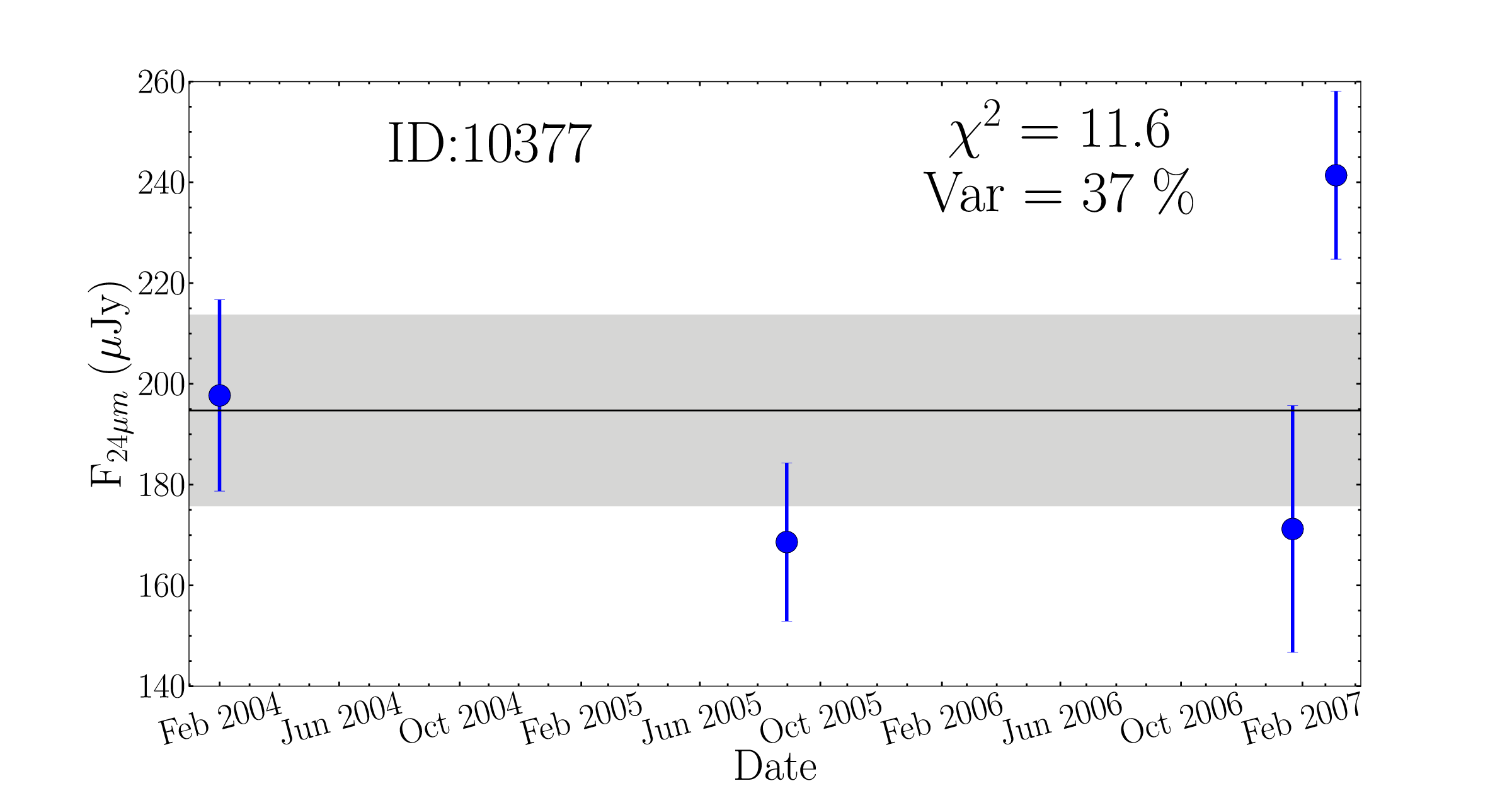}}{\hspace{0cm}}
    \subfigure {\includegraphics[width=47mm]{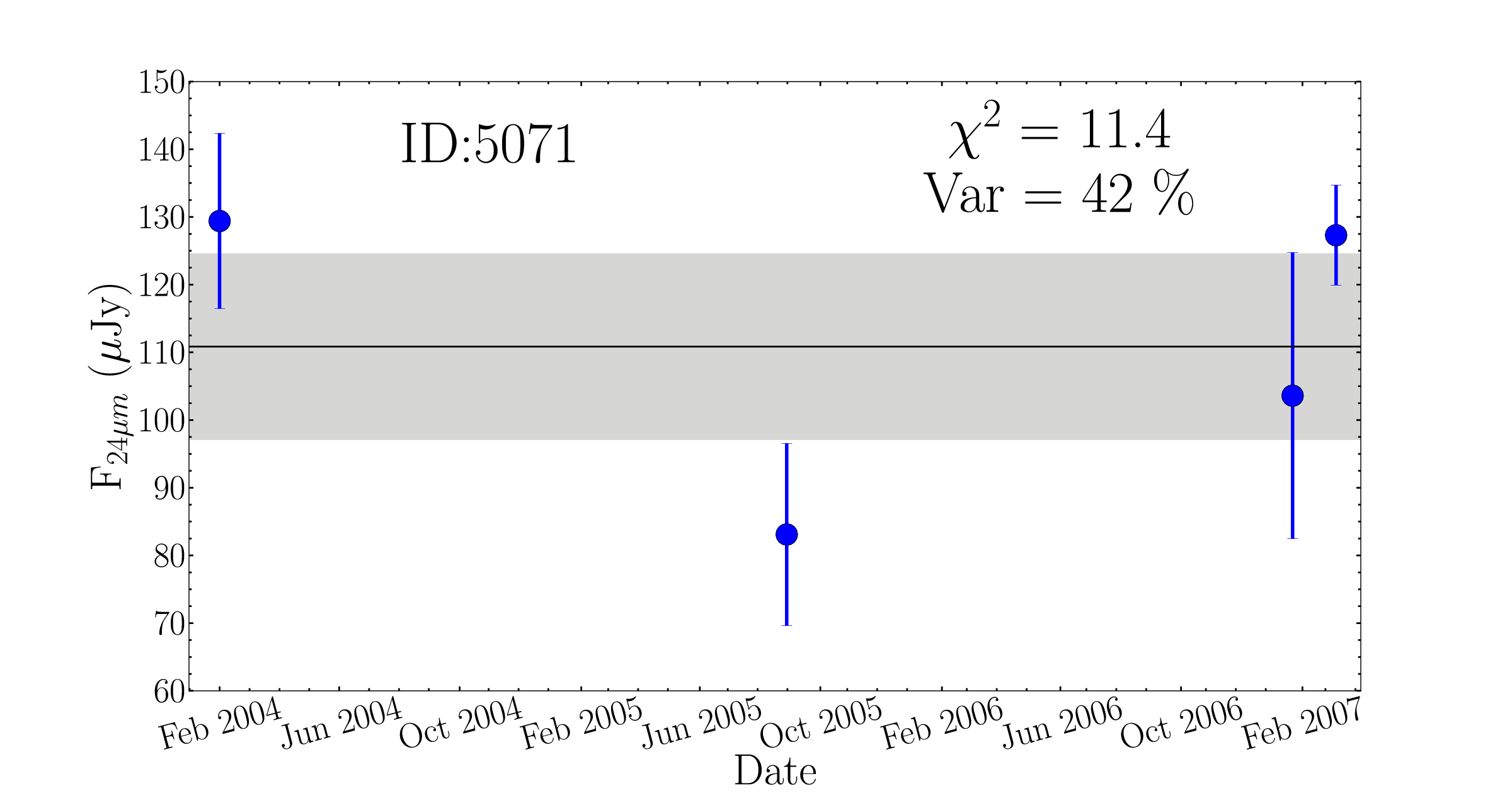}}{\hspace{0cm}}

    \contcaption{} 
    \label{curvas-luz2}
    \end{center}
  \end{minipage}
\end{figure*}

\section[]{Short-term variable candidates light curves}
\label{curvas-luz-short}

\begin{figure*}
  \begin{minipage}{200mm}
    \begin{center}

      \subfigure {\includegraphics[width=47mm]{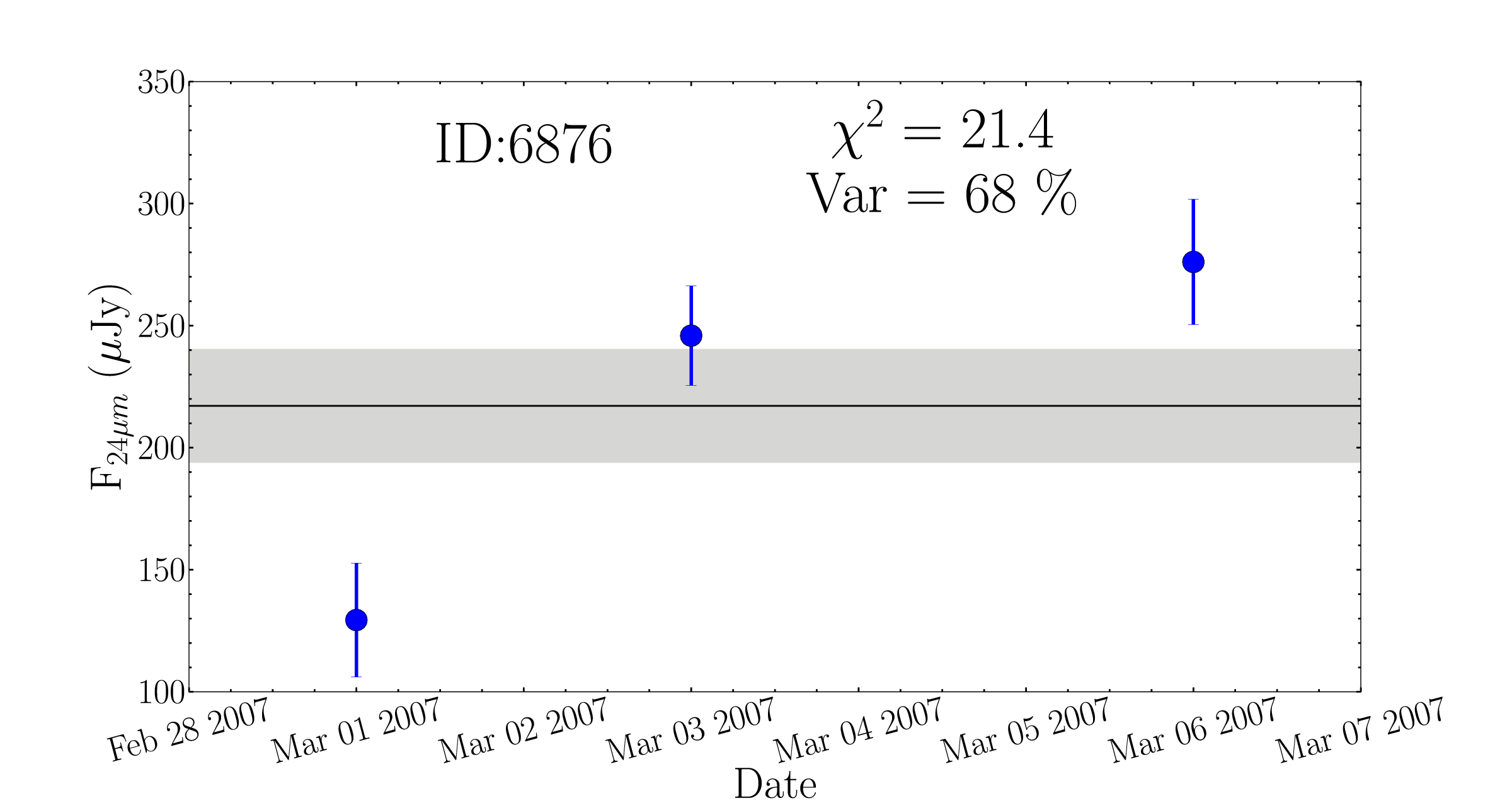}}{\hspace{0cm}}
      \subfigure {\includegraphics[width=47mm]{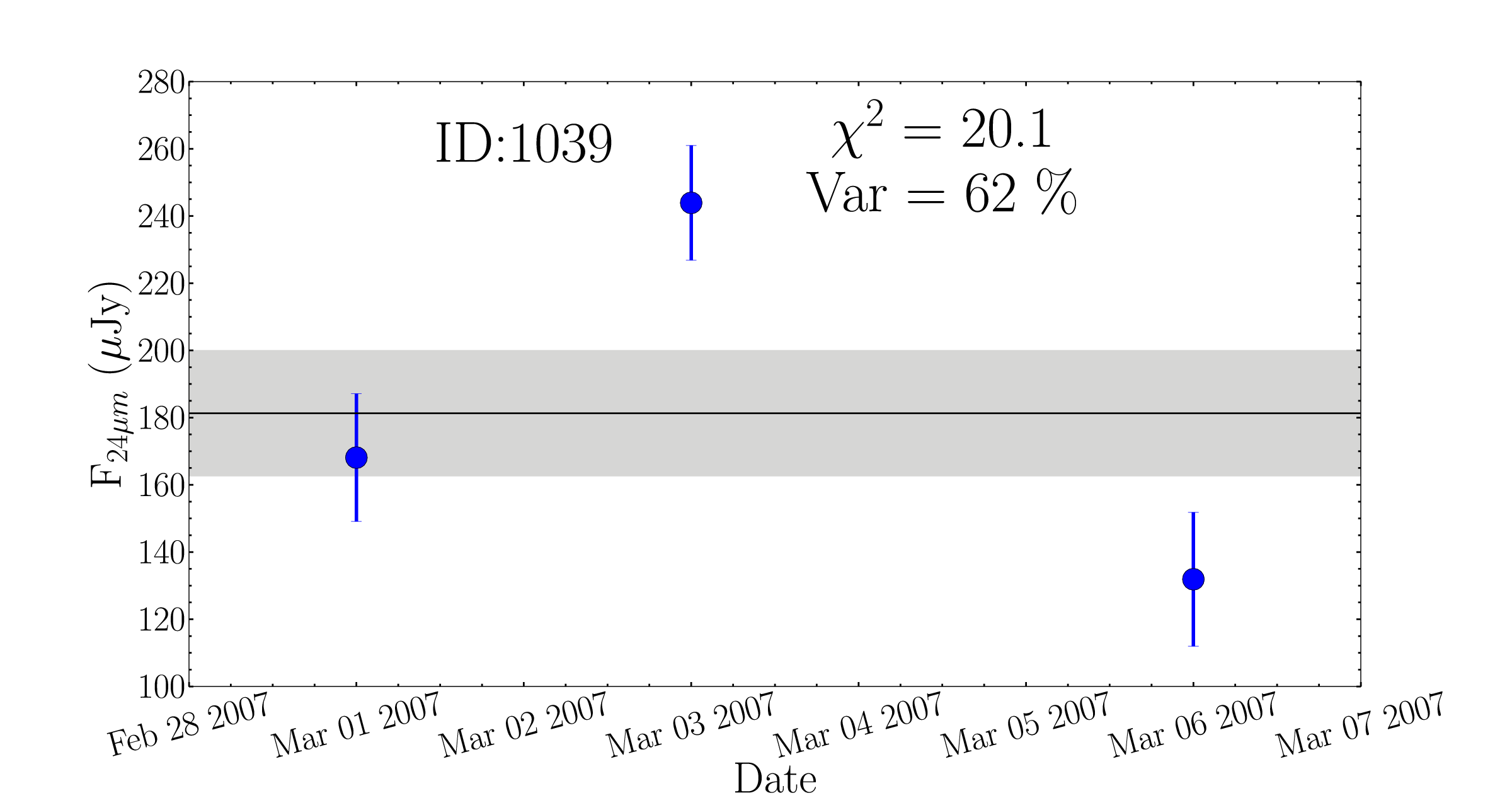}}{\hspace{0cm}}
      \subfigure {\includegraphics[width=47mm]{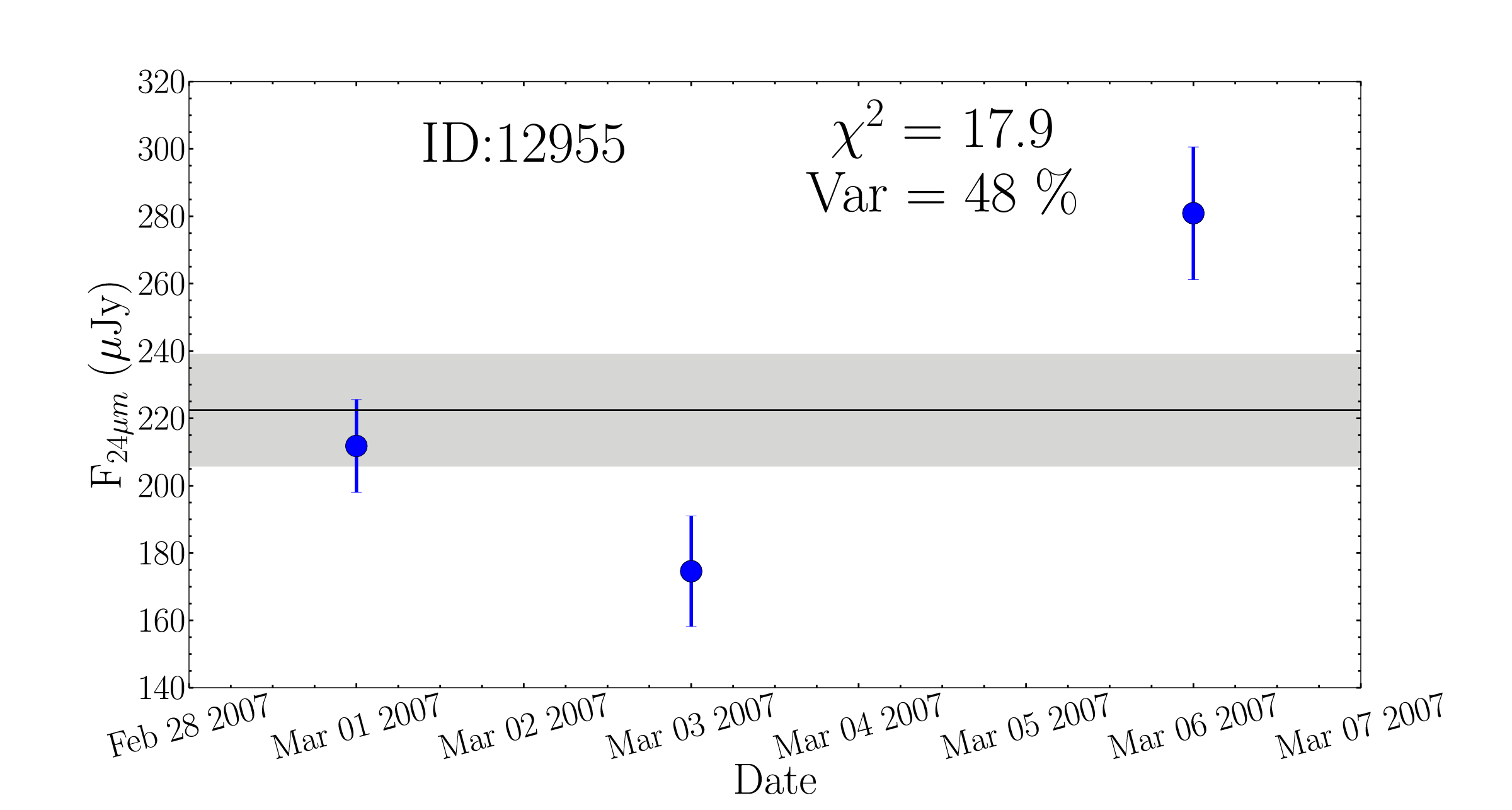}}{\hspace{0cm}}
      \subfigure {\includegraphics[width=47mm]{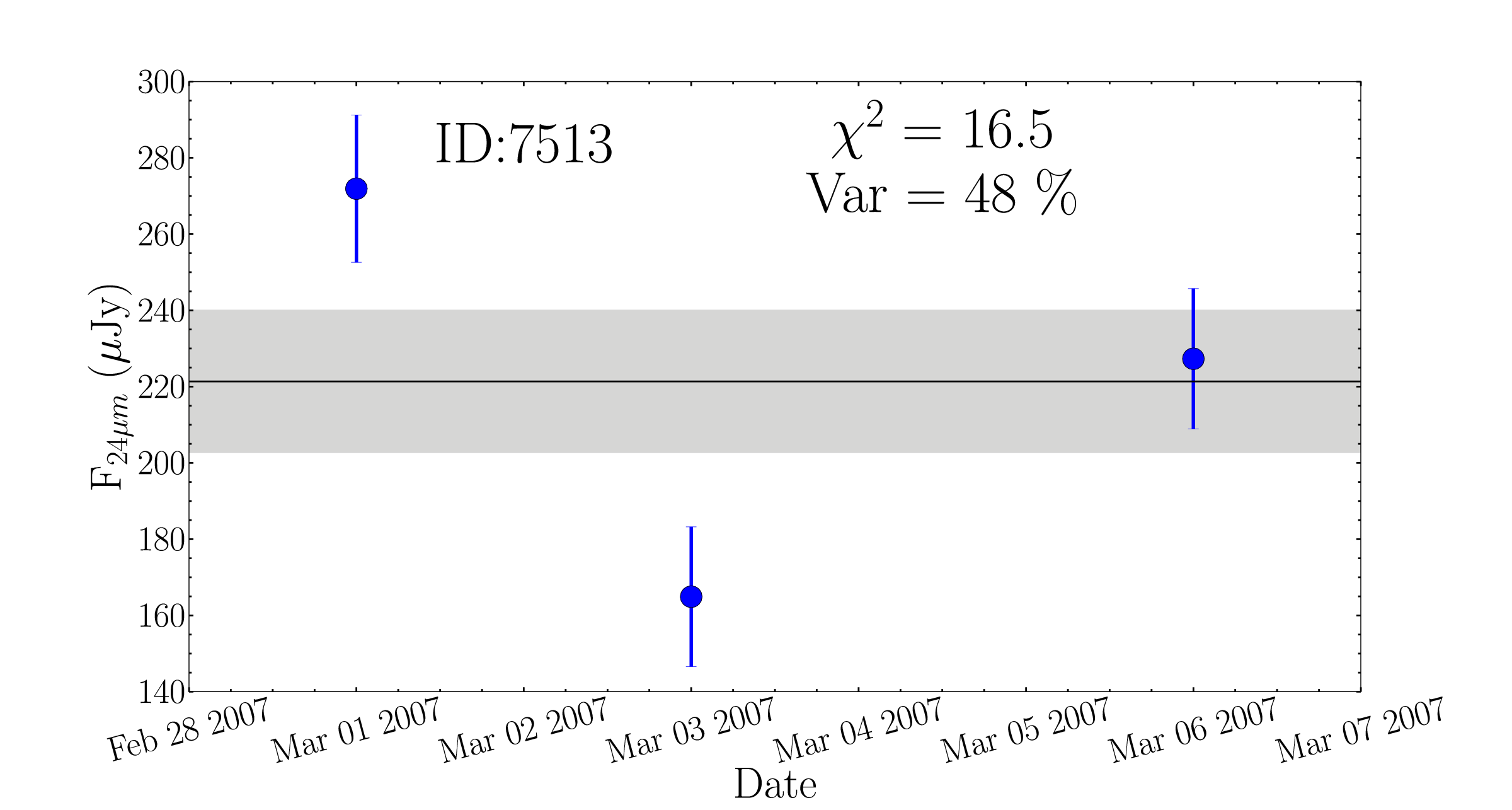}}{\hspace{0cm}}
      \subfigure {\includegraphics[width=47mm]{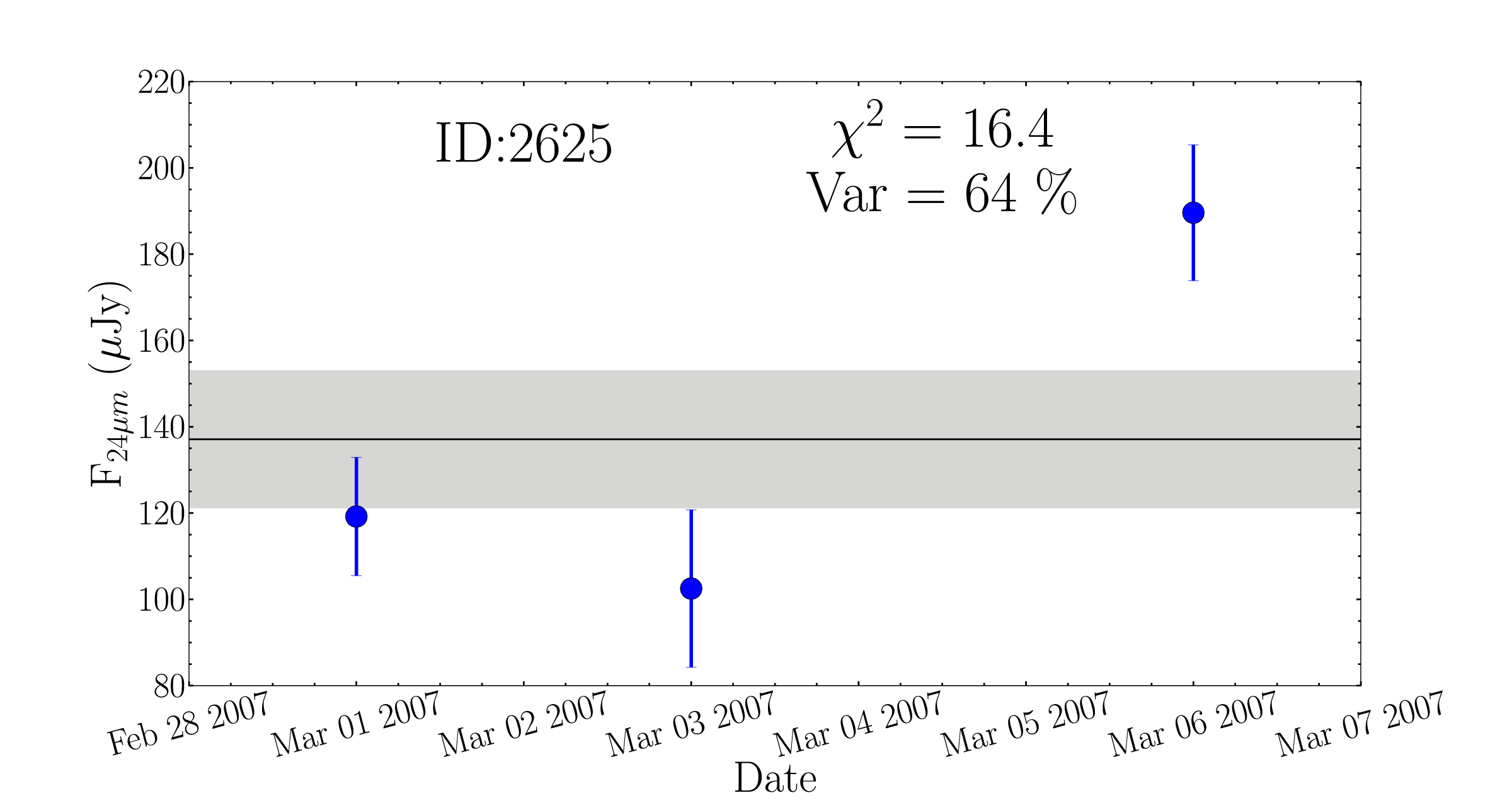}}{\hspace{0cm}}
      \subfigure {\includegraphics[width=47mm]{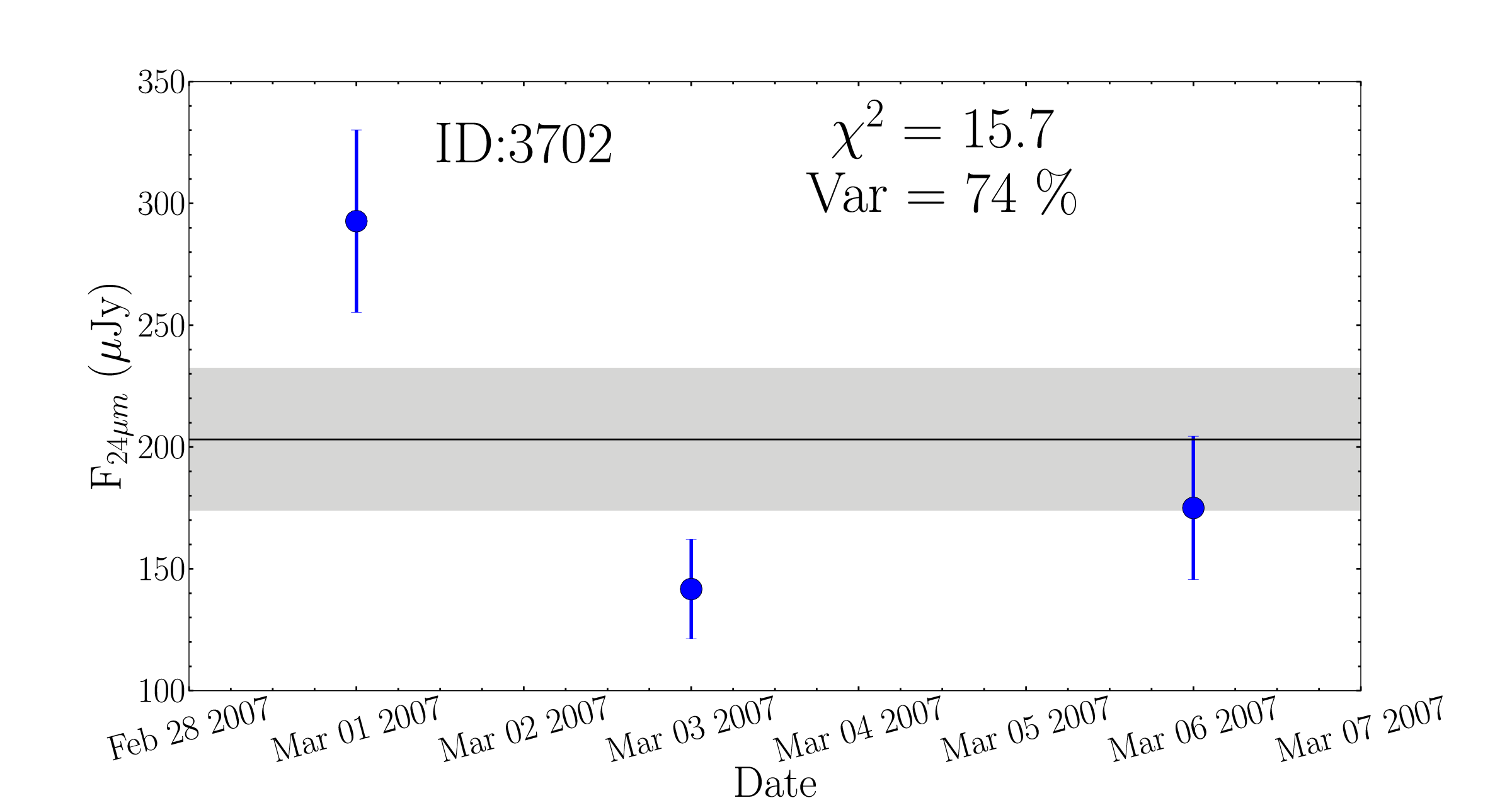}}{\hspace{0cm}}
      \subfigure {\includegraphics[width=47mm]{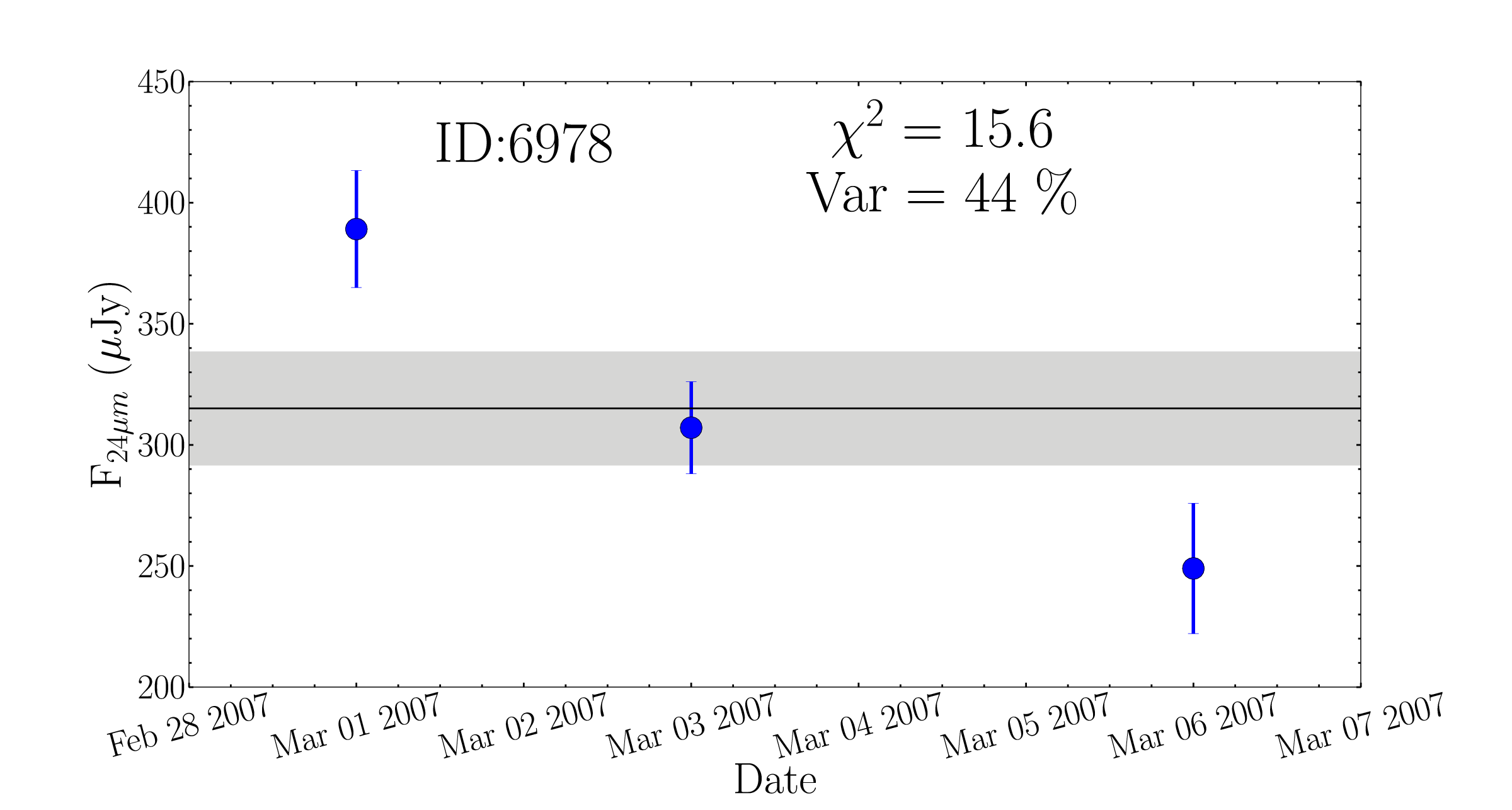}}{\hspace{0cm}}
      \subfigure {\includegraphics[width=47mm]{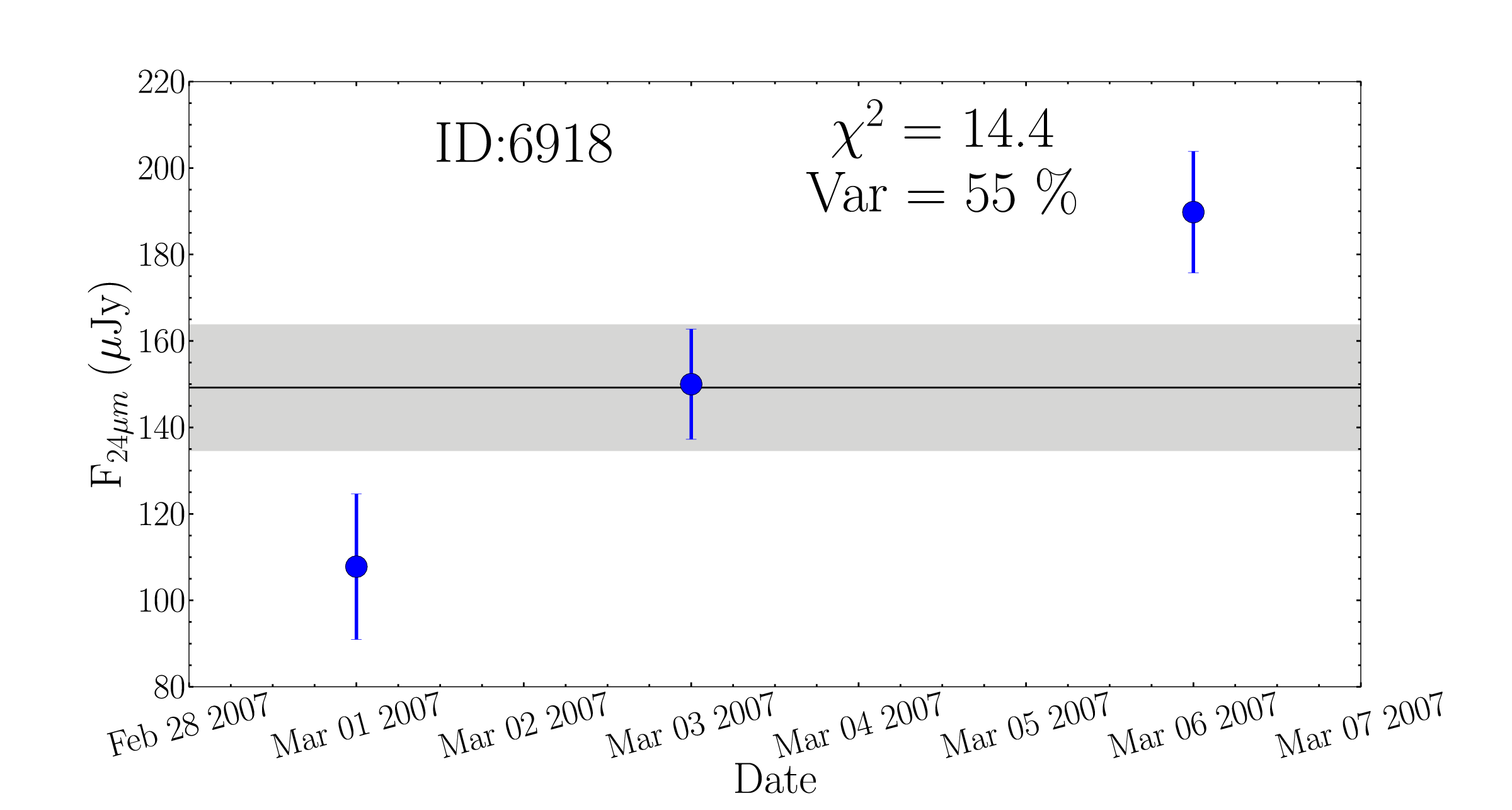}}{\hspace{0cm}}
      \subfigure {\includegraphics[width=47mm]{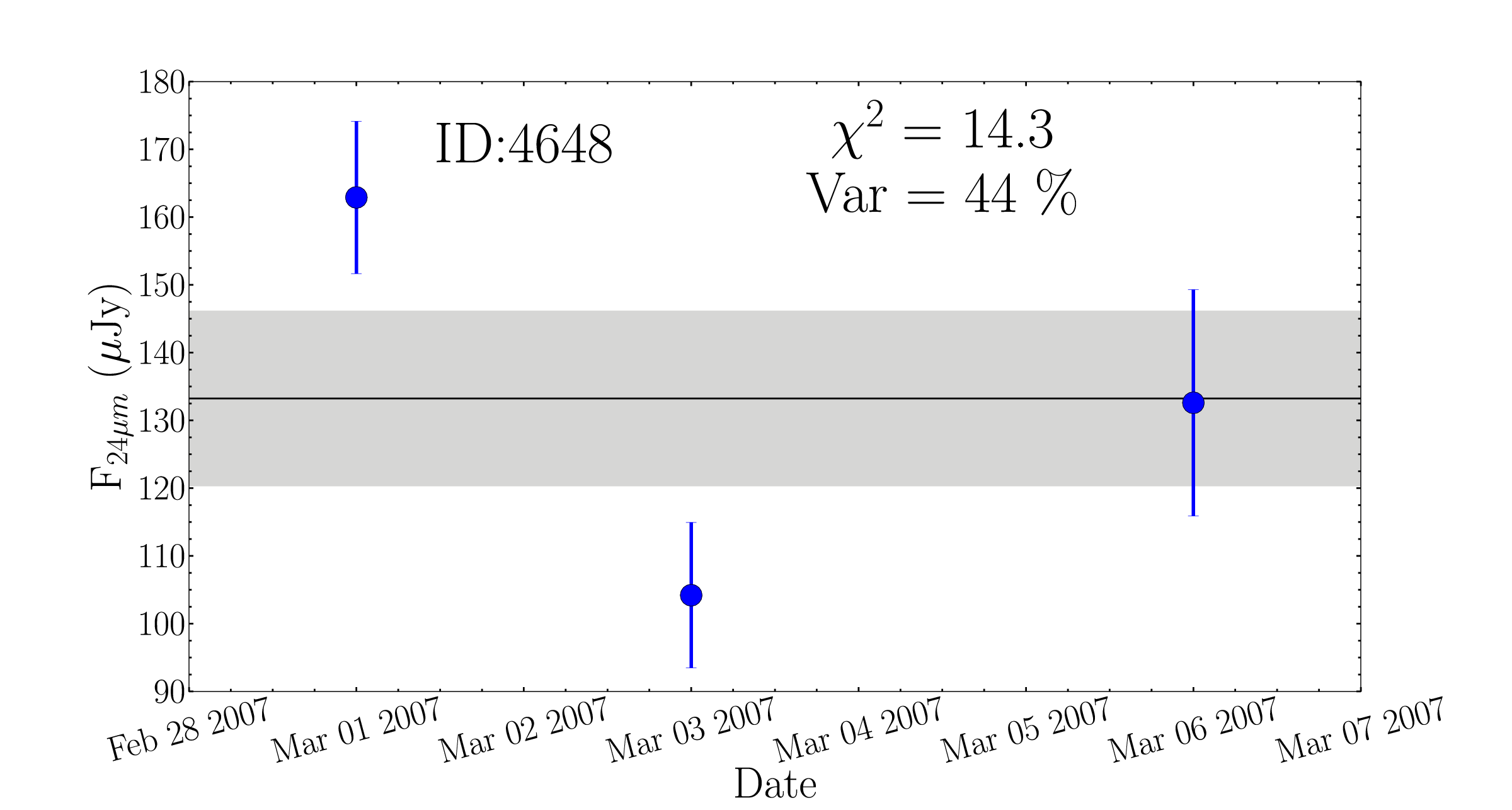}}{\hspace{0cm}}
      \subfigure {\includegraphics[width=47mm]{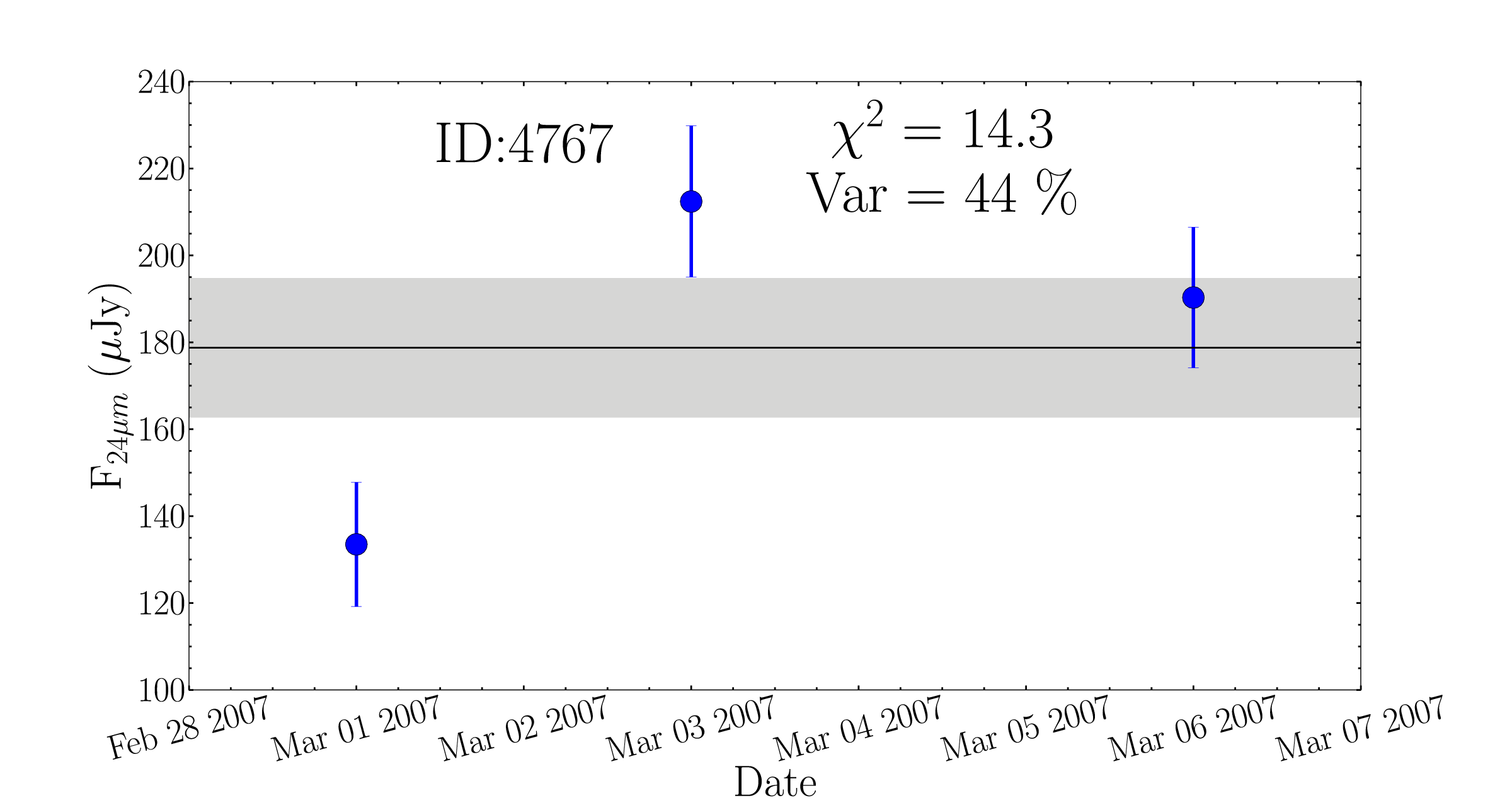}}{\hspace{0cm}}
      \subfigure {\includegraphics[width=47mm]{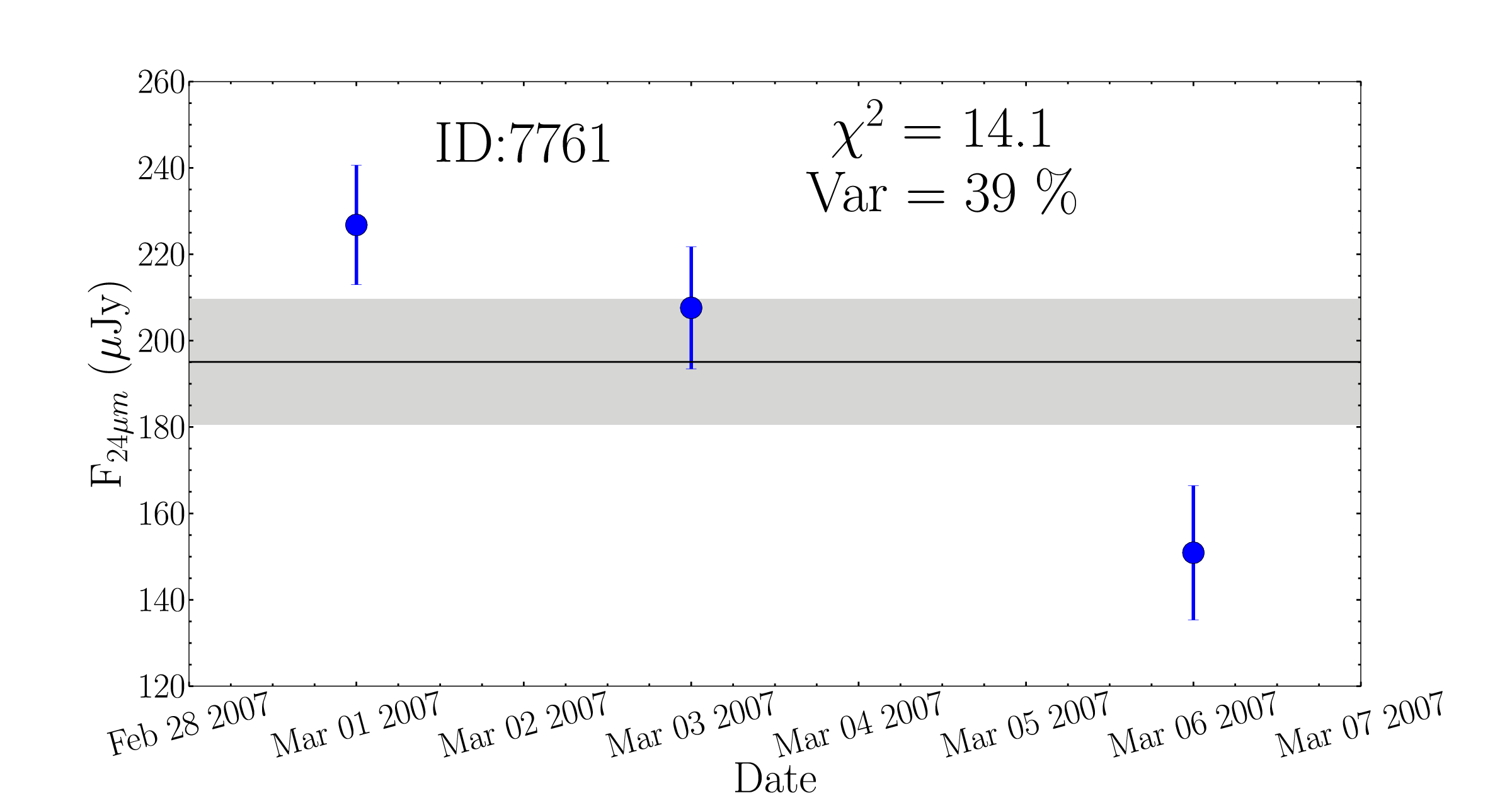}}{\hspace{0cm}}
      \subfigure {\includegraphics[width=47mm]{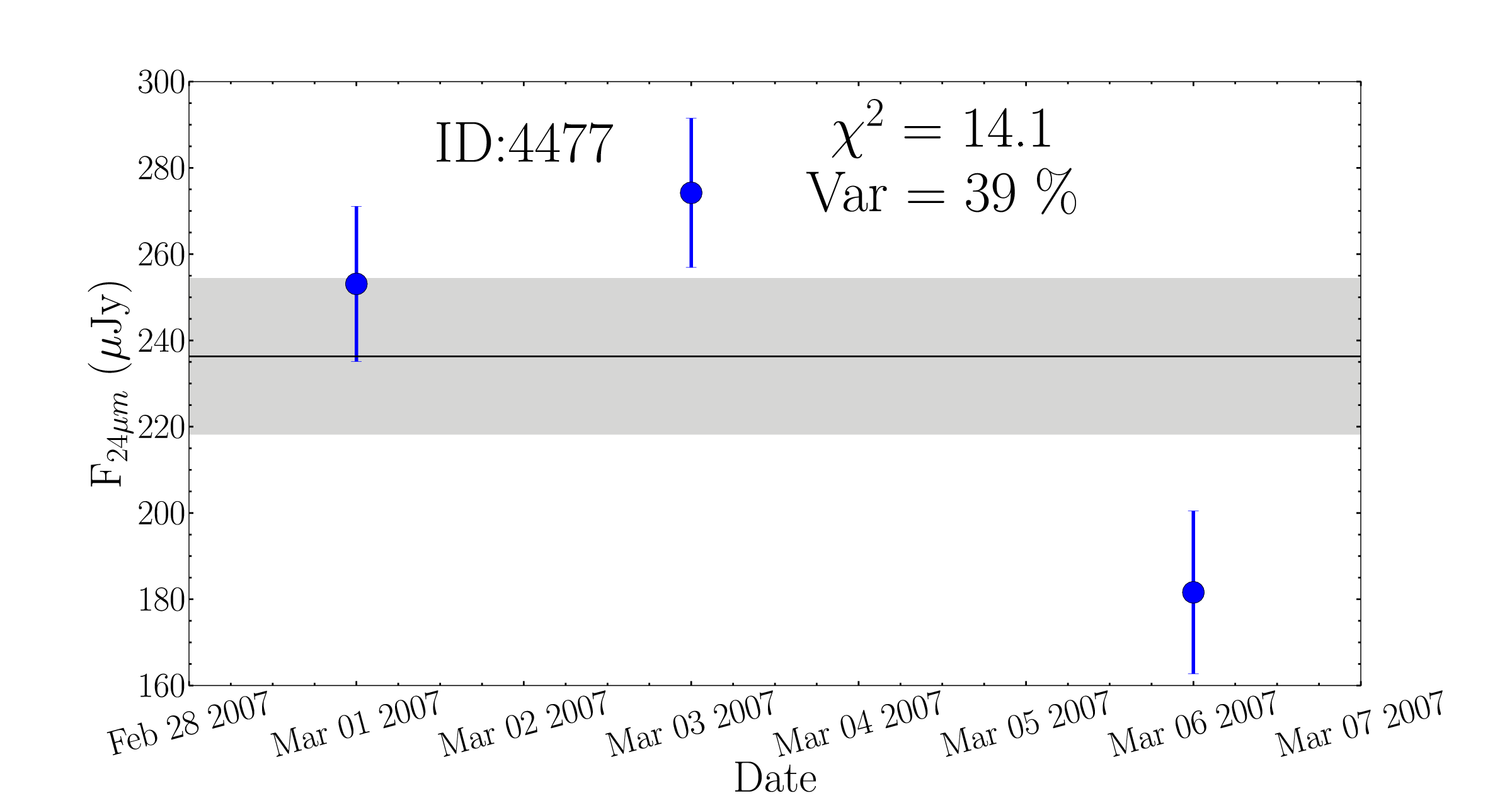}}{\hspace{0cm}}
      \subfigure {\includegraphics[width=47mm]{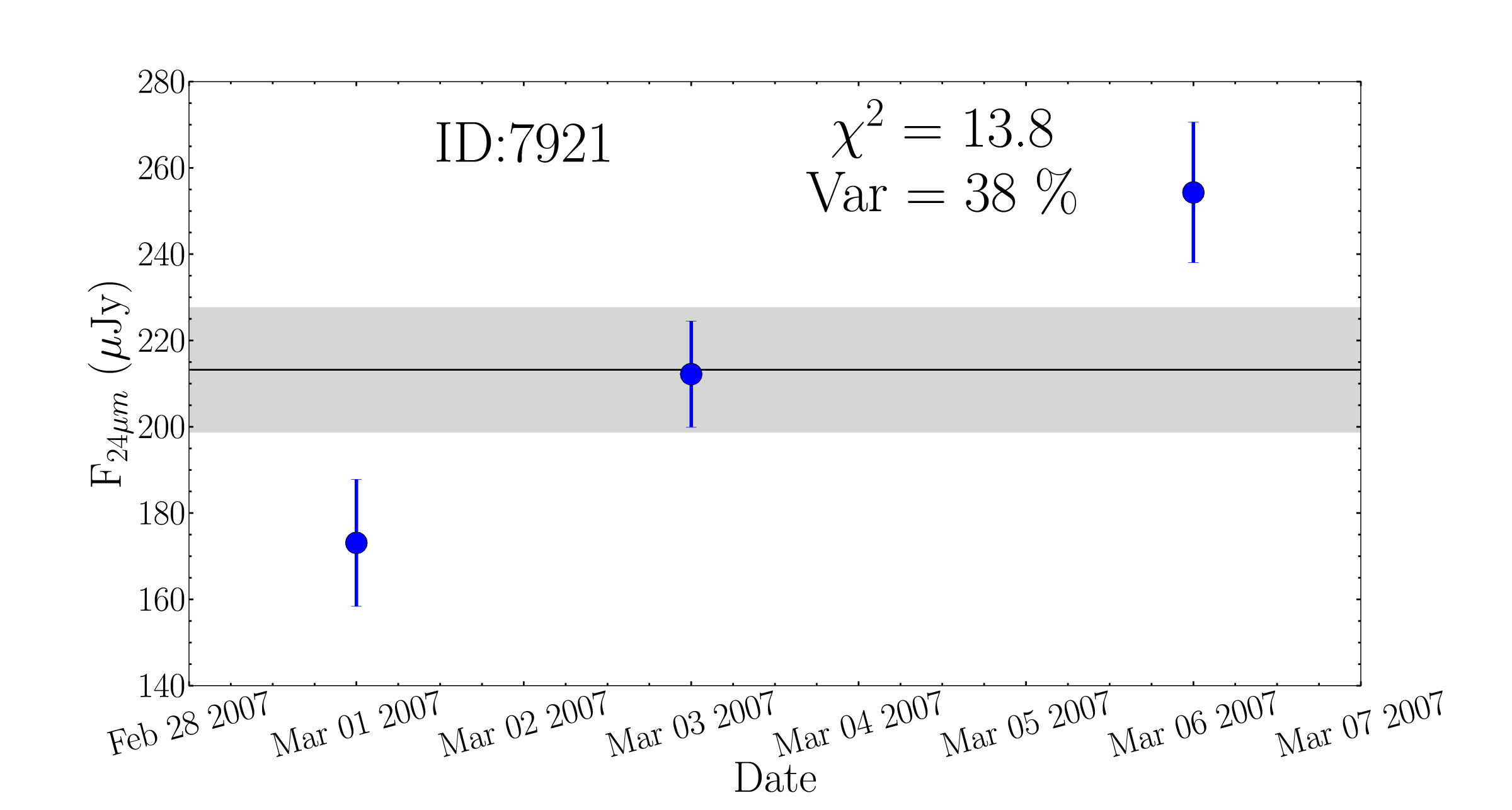}}{\hspace{0cm}}
      \subfigure {\includegraphics[width=47mm]{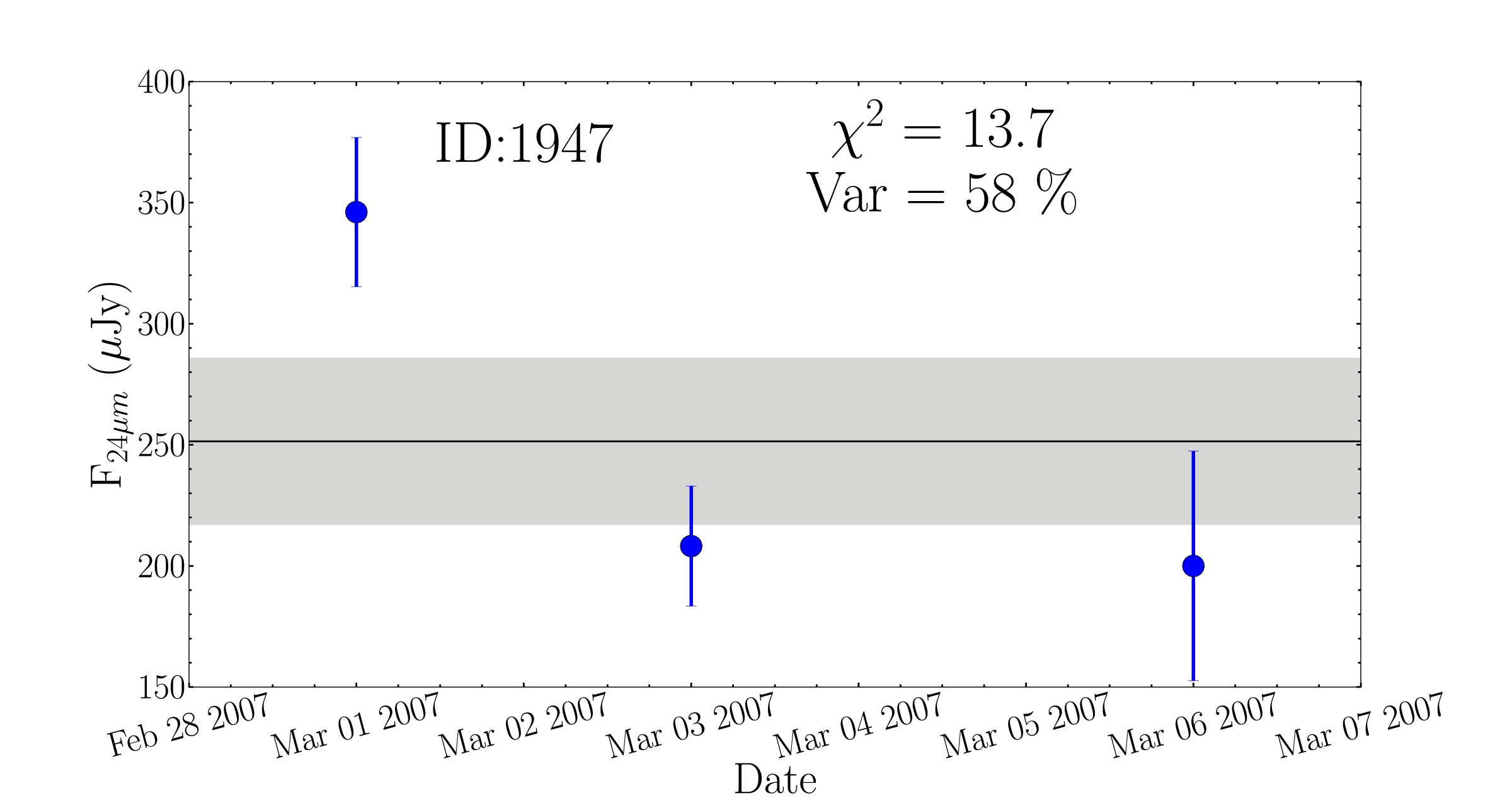}}{\hspace{0cm}}
      \subfigure {\includegraphics[width=47mm]{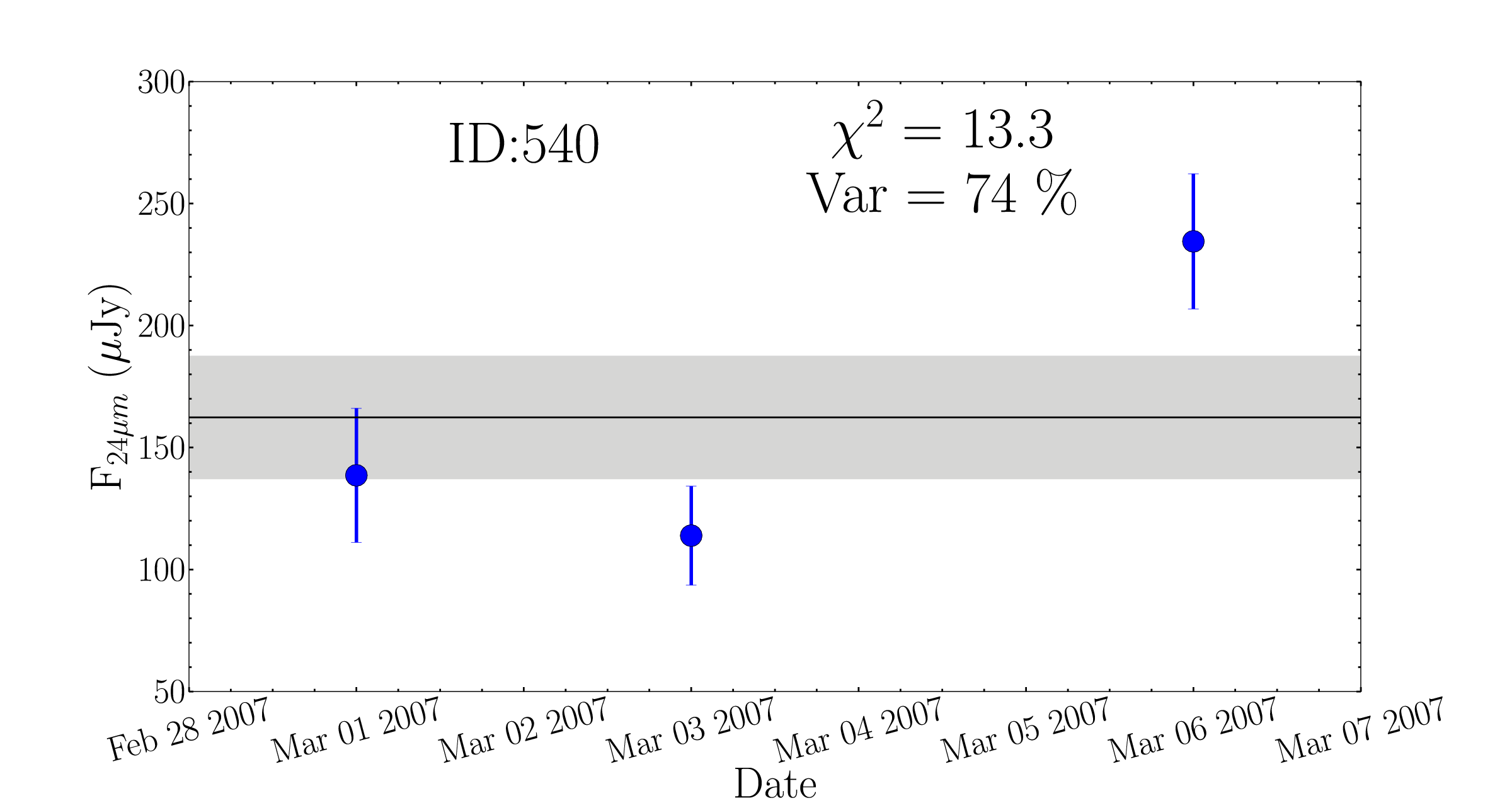}}{\hspace{0cm}}
      \subfigure {\includegraphics[width=47mm]{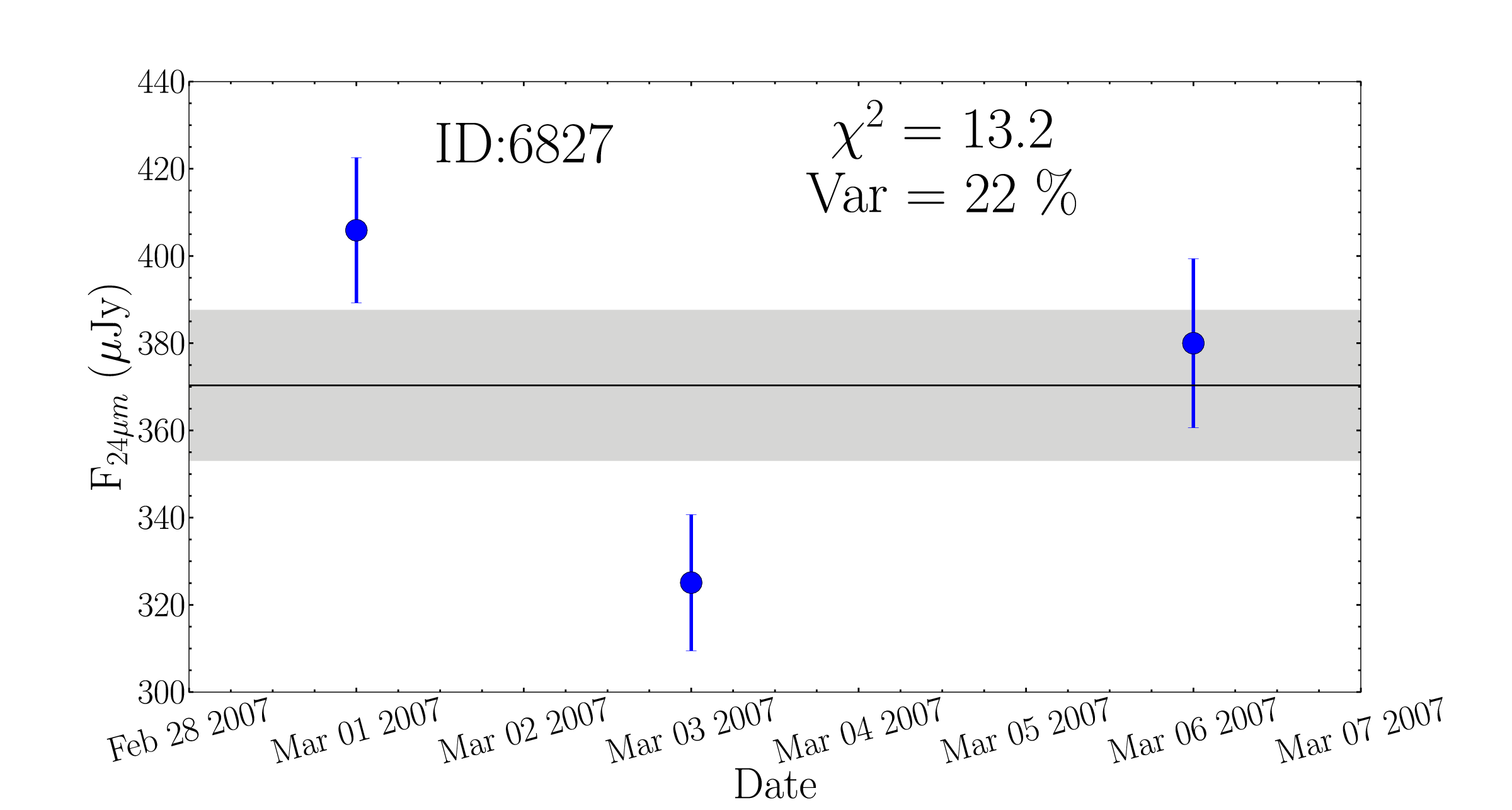}}{\hspace{0cm}}
      \subfigure {\includegraphics[width=47mm]{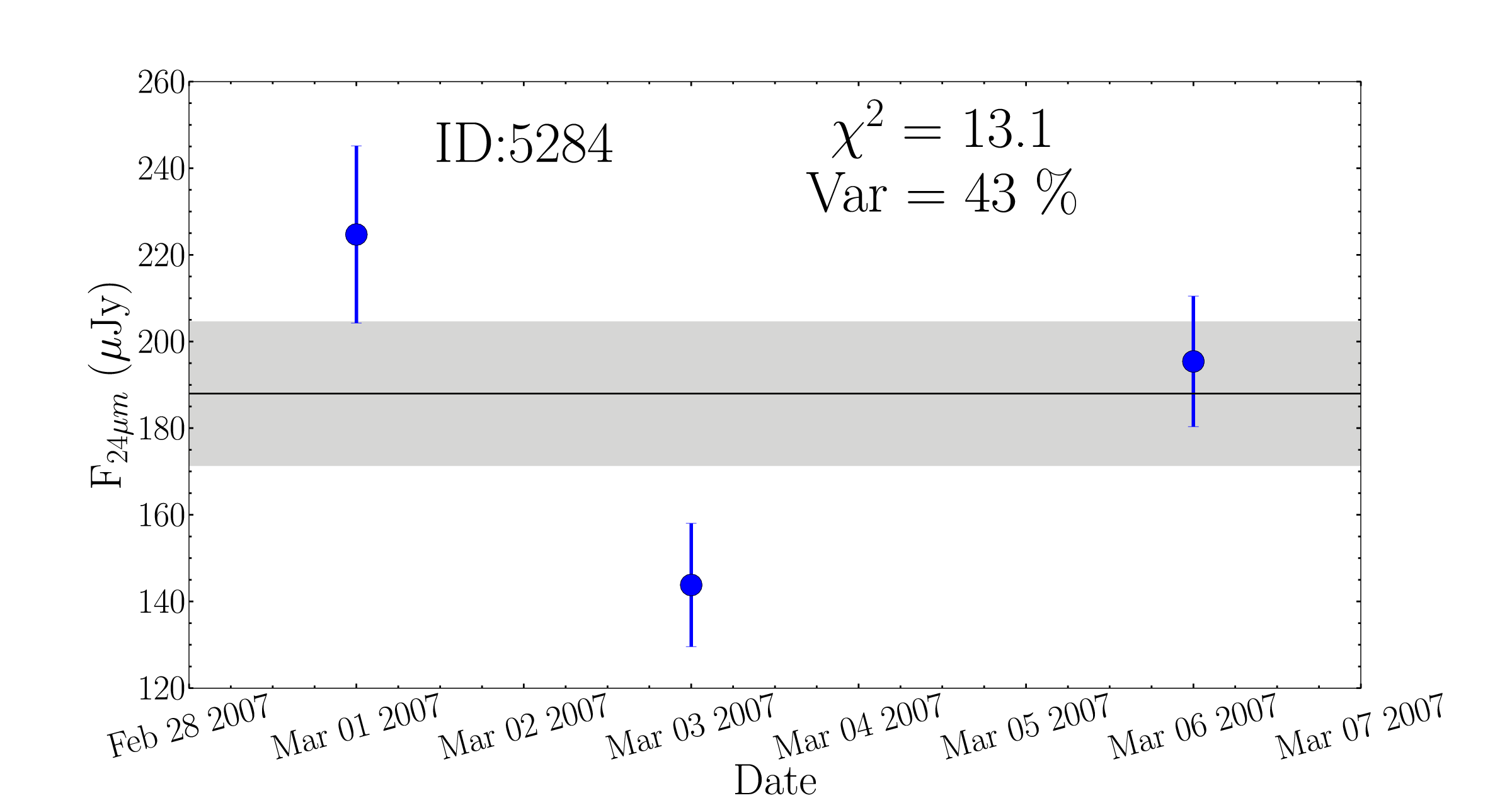}}{\hspace{0cm}}
      \subfigure {\includegraphics[width=47mm]{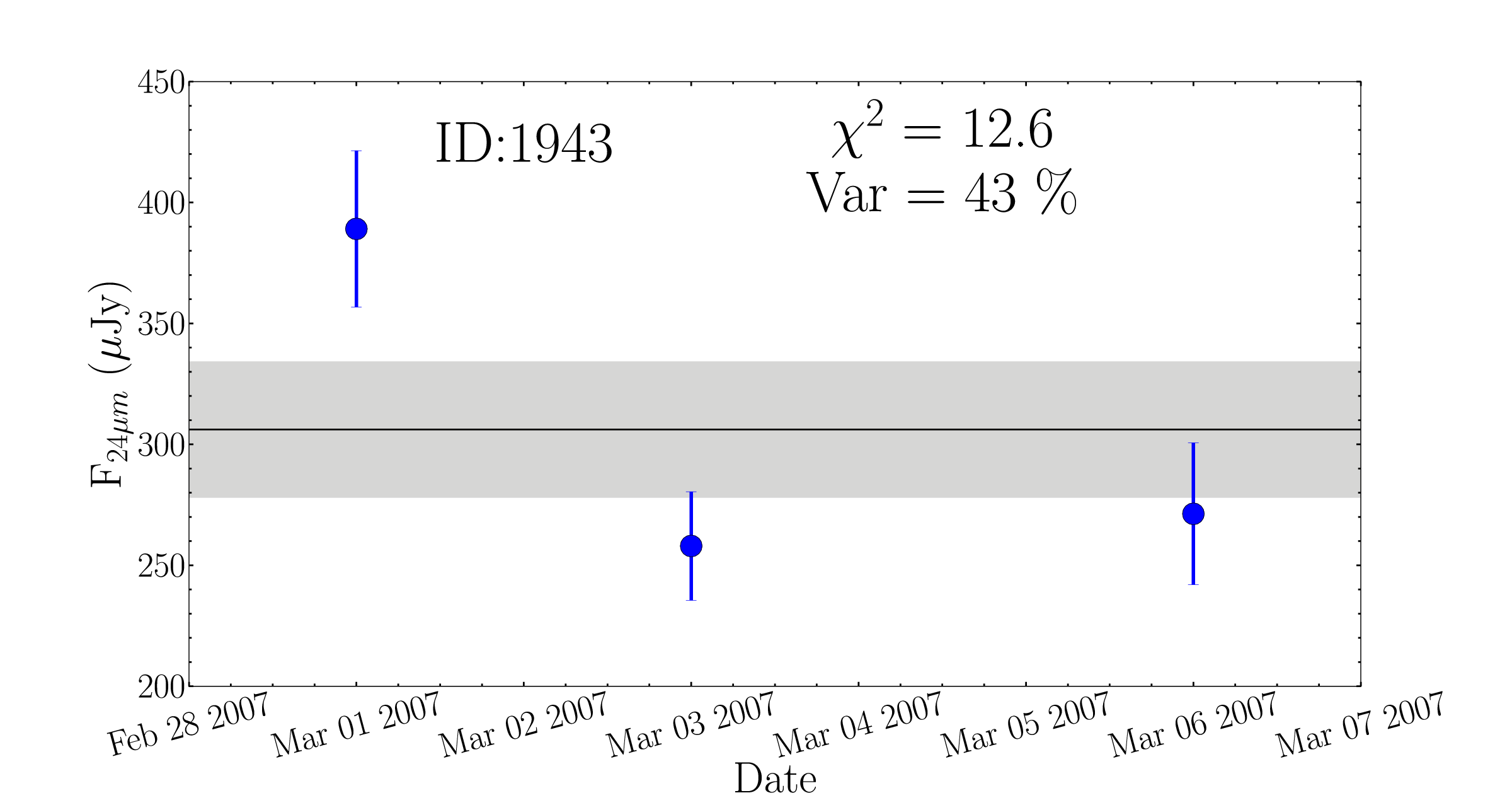}}{\hspace{0cm}}
      \subfigure {\includegraphics[width=47mm]{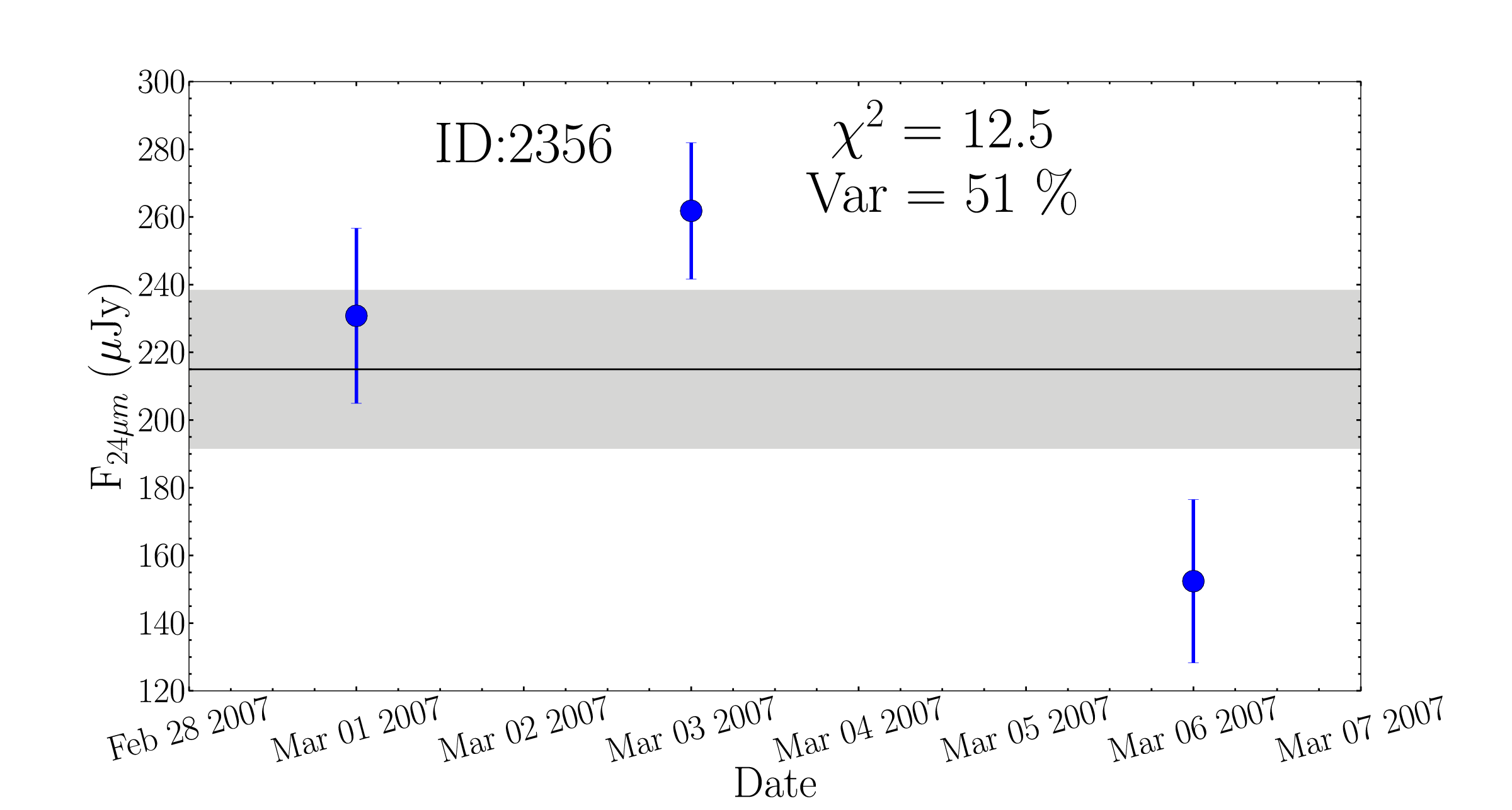}}{\hspace{0cm}}
      \subfigure {\includegraphics[width=47mm]{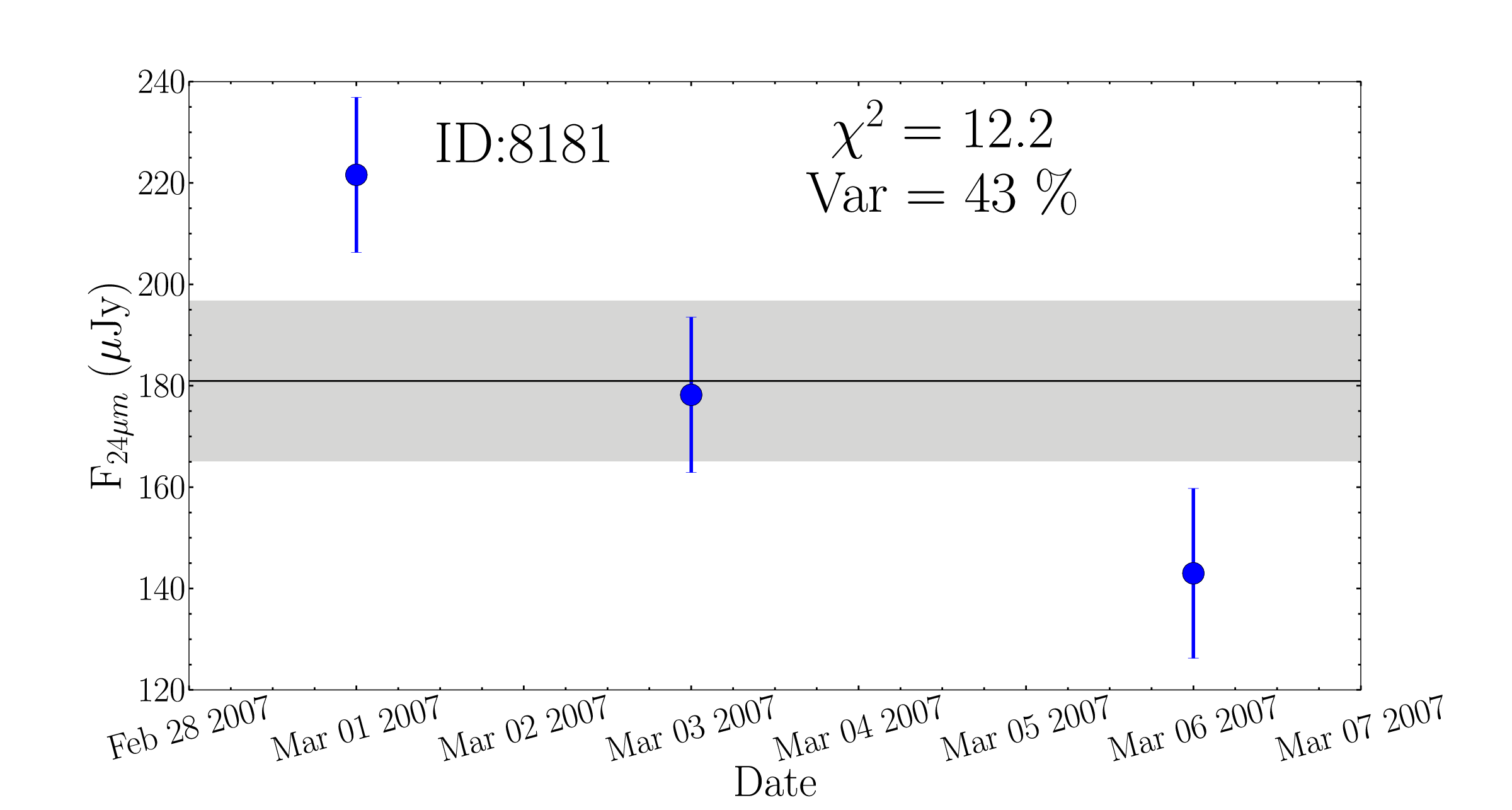}}{\hspace{0cm}}
      \subfigure {\includegraphics[width=47mm]{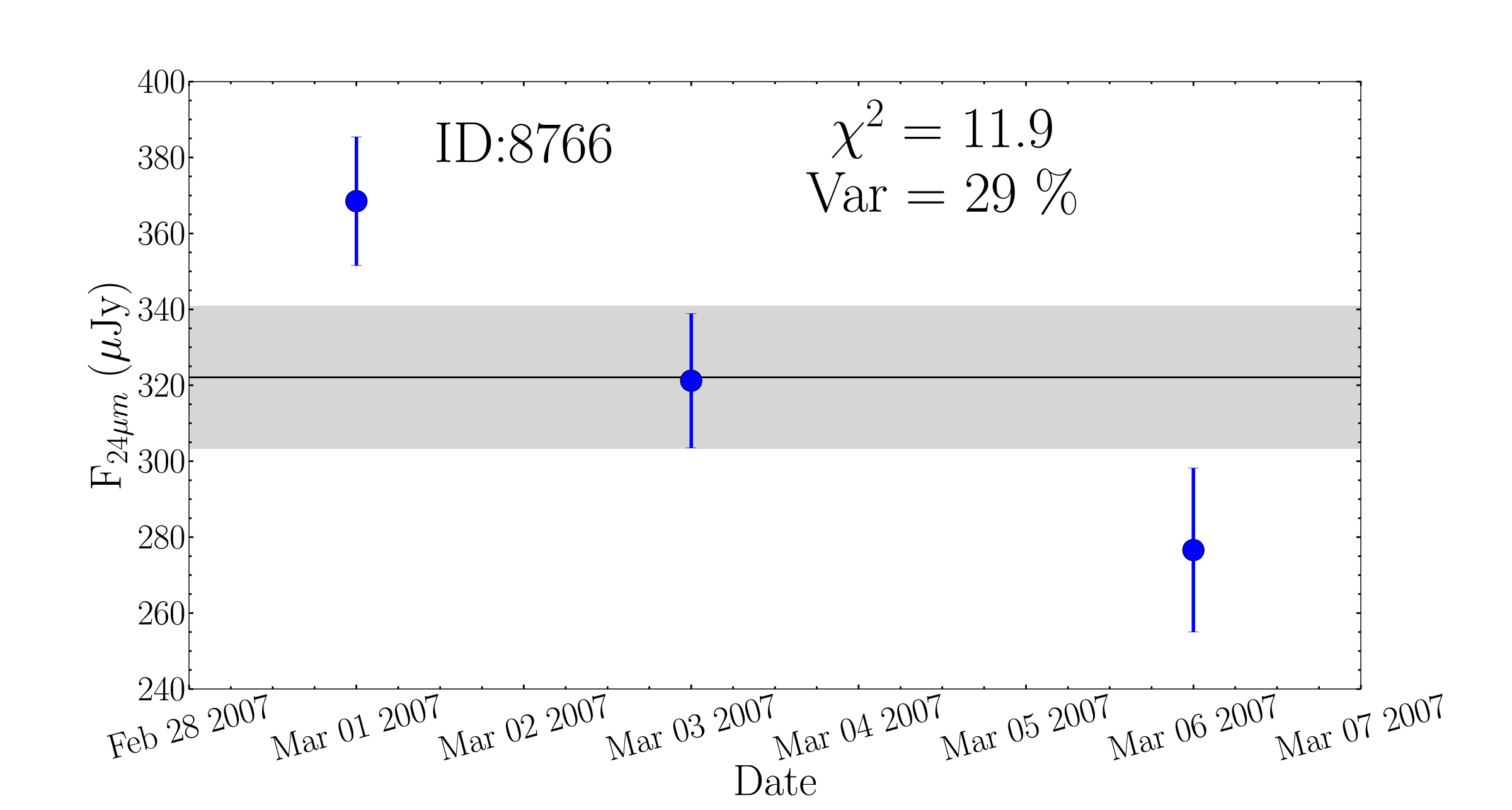}}{\hspace{0cm}}
      \subfigure {\includegraphics[width=47mm]{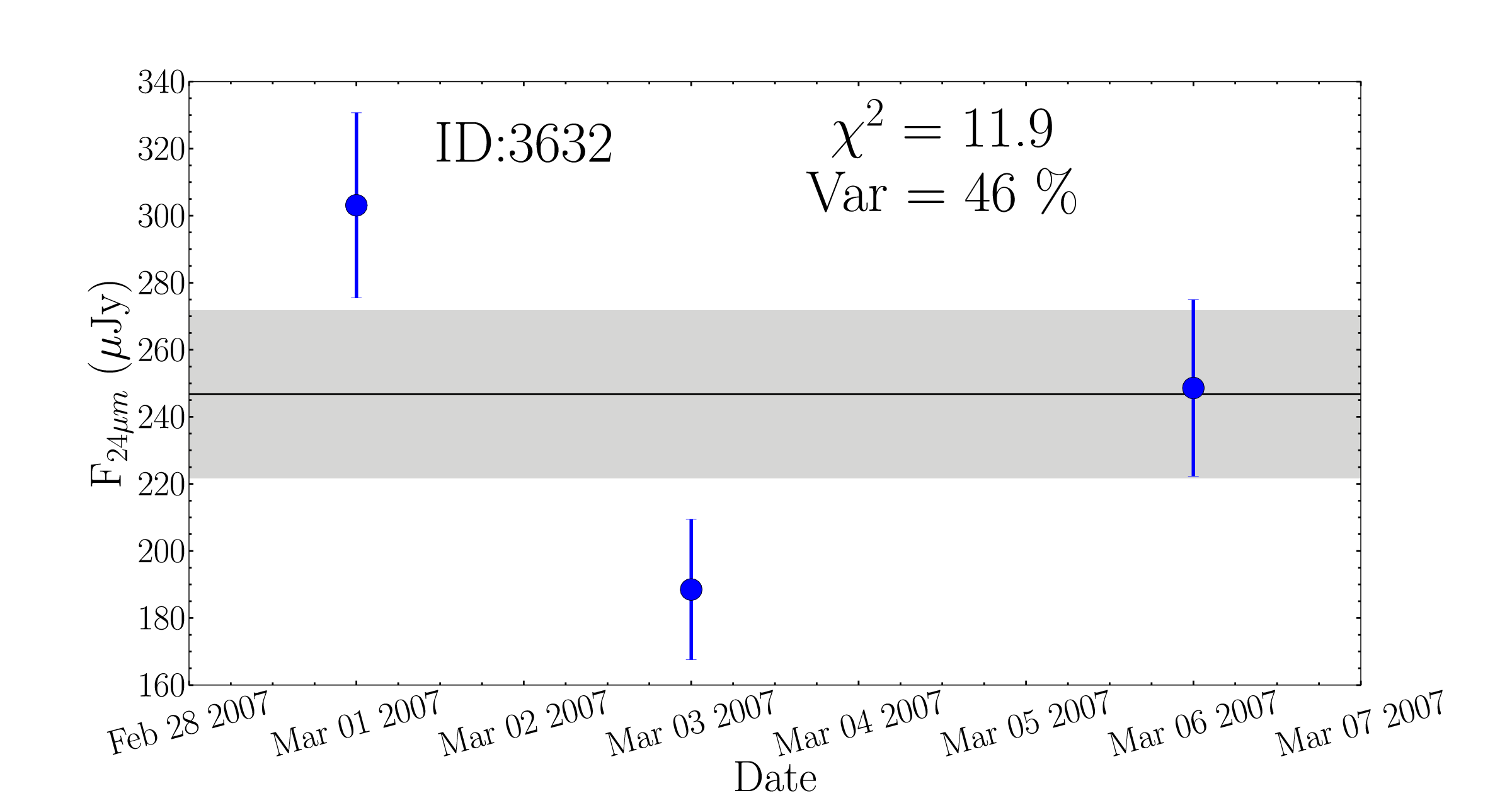}}{\hspace{0cm}}
      \subfigure {\includegraphics[width=47mm]{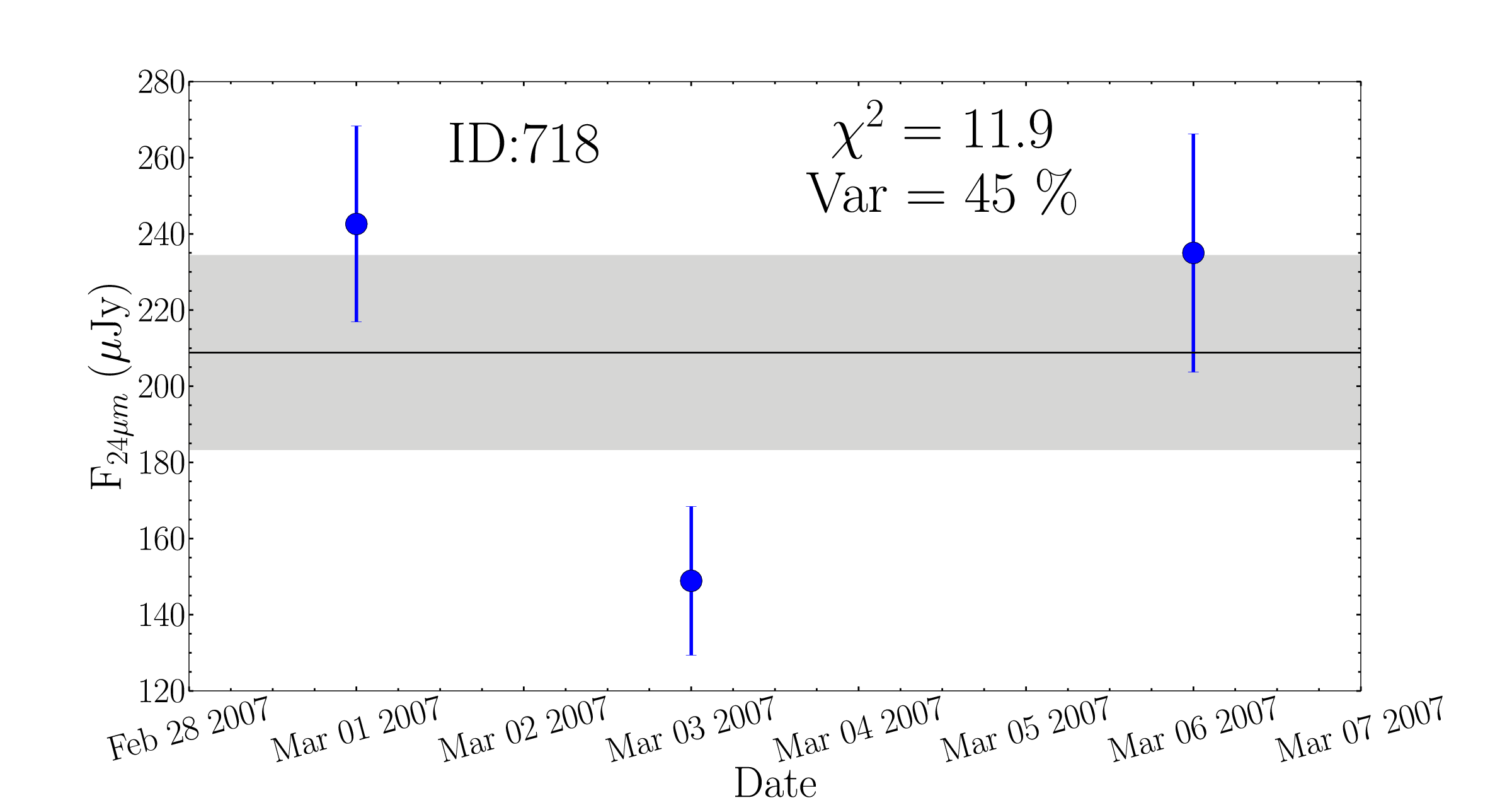}}{\hspace{0cm}}
      \subfigure {\includegraphics[width=47mm]{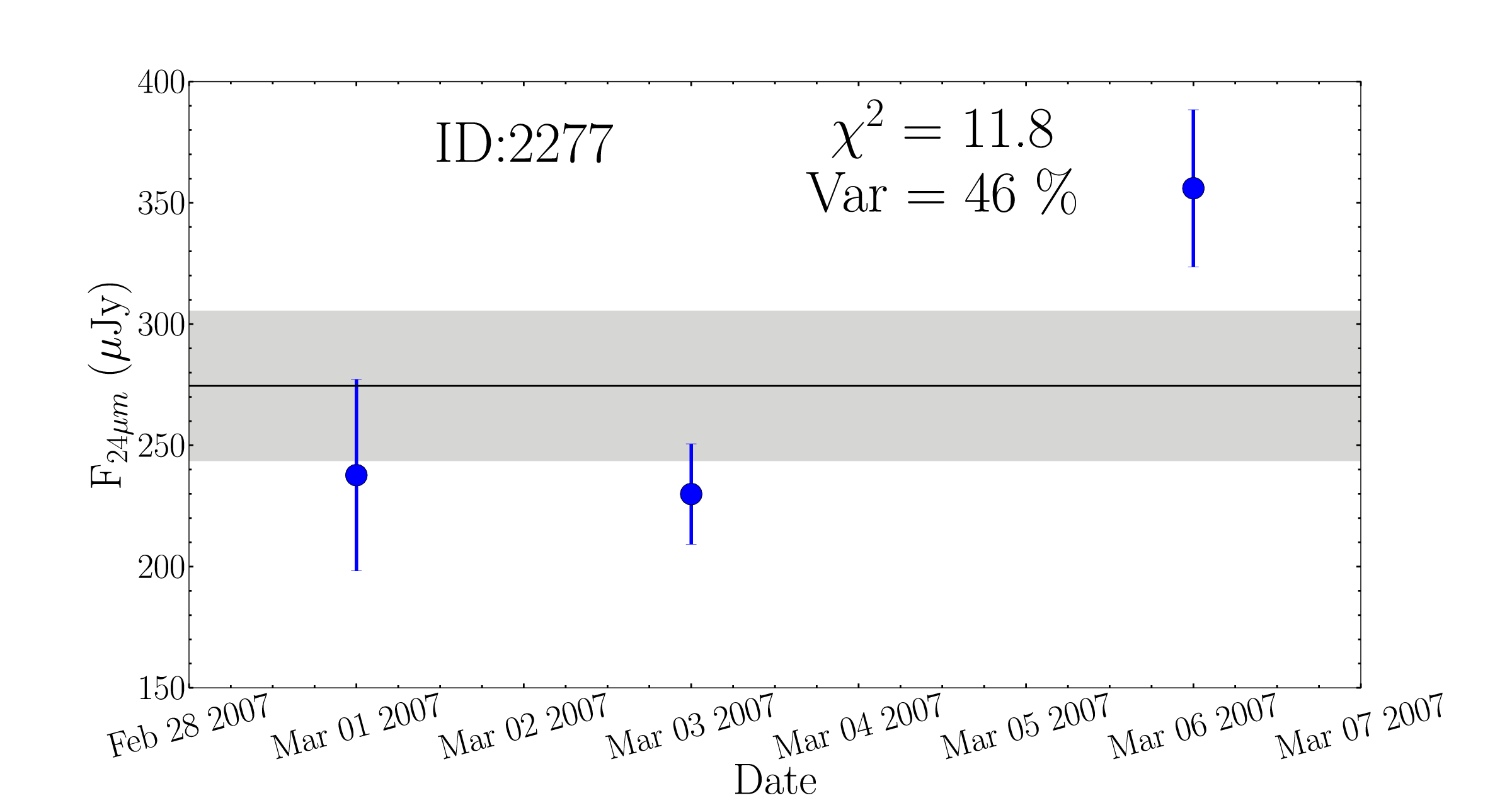}}{\hspace{0cm}}
      \subfigure {\includegraphics[width=47mm]{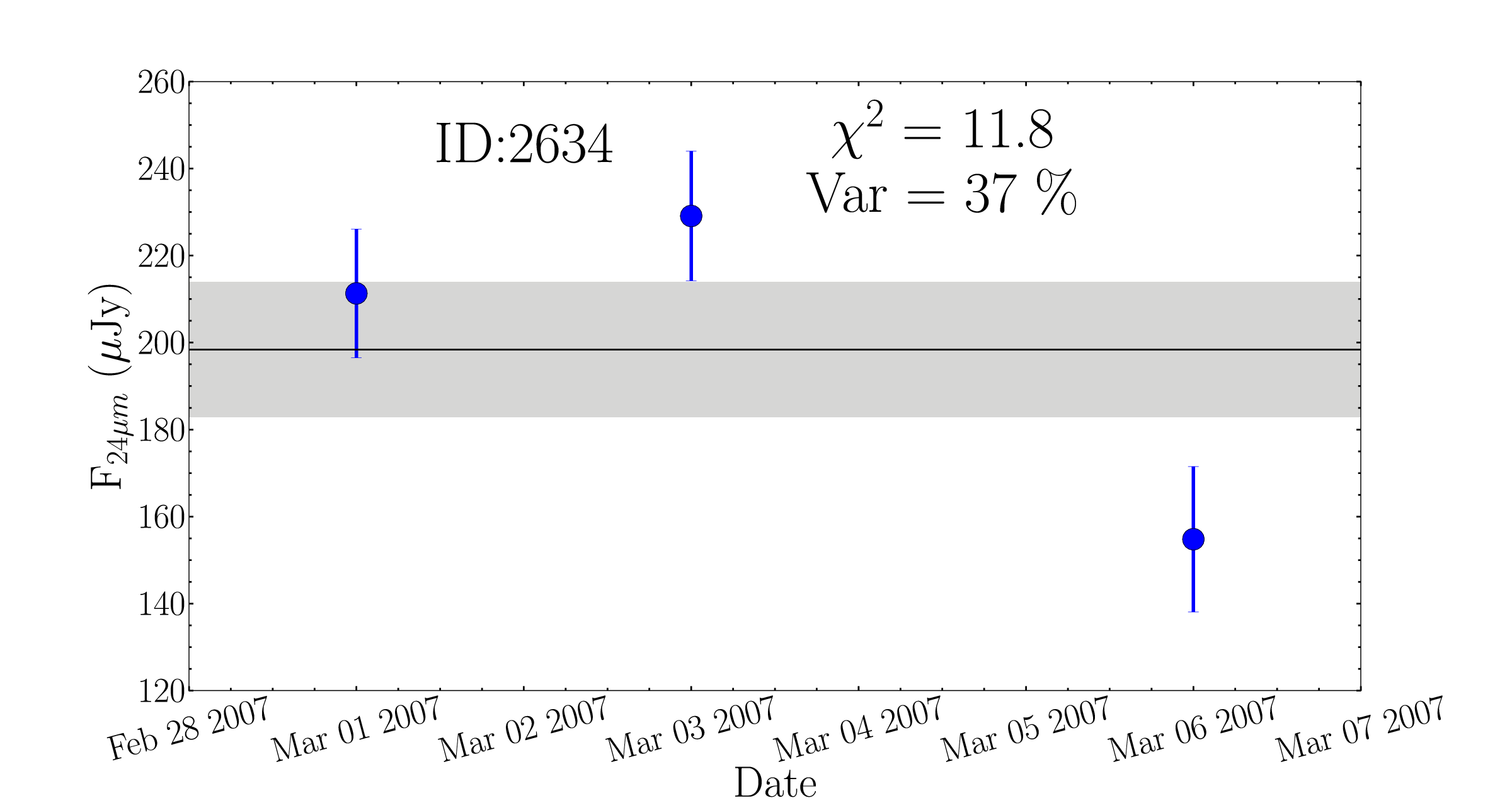}}{\hspace{0cm}}
      \subfigure {\includegraphics[width=47mm]{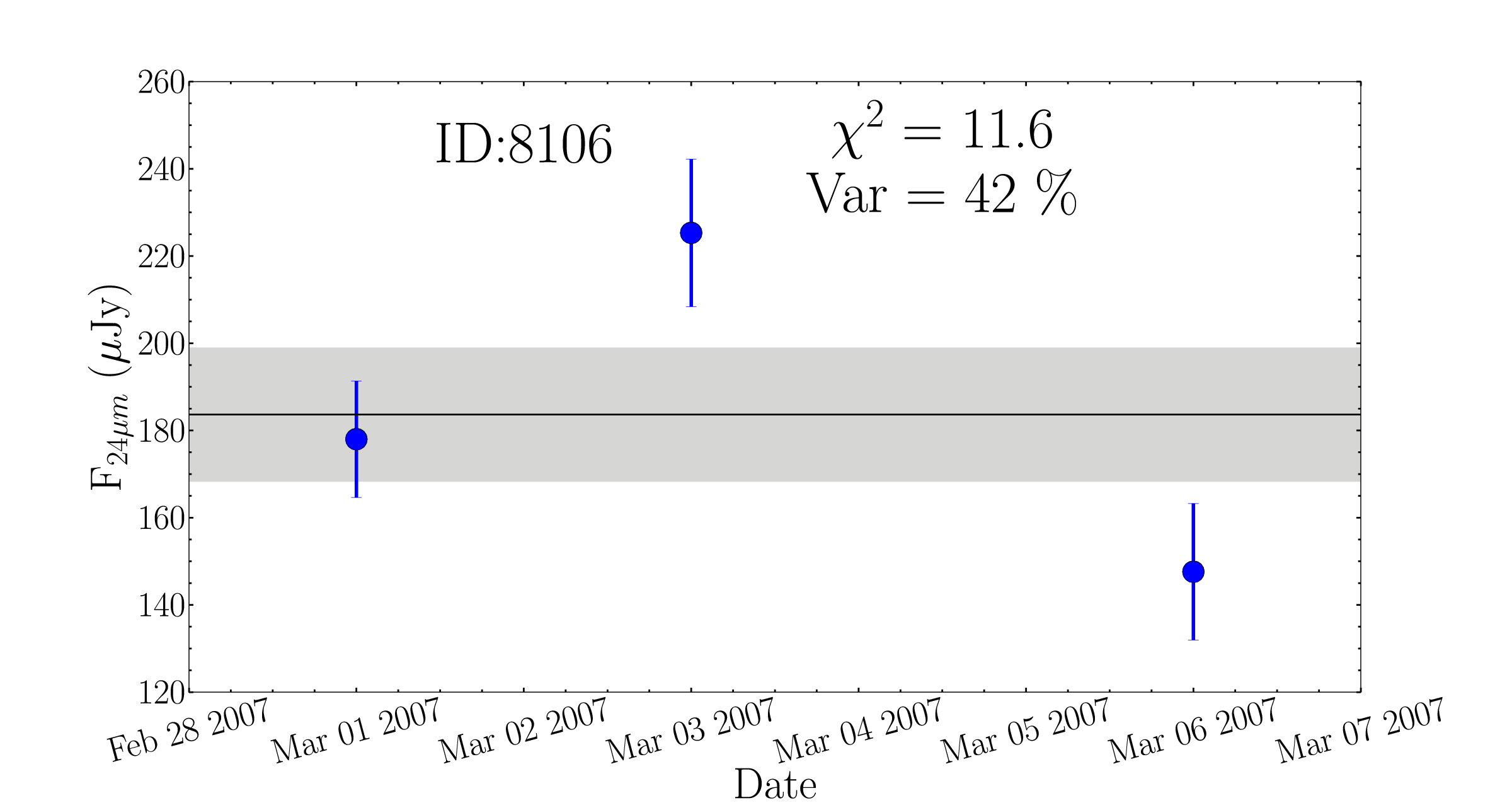}}{\hspace{0cm}}
      \subfigure {\includegraphics[width=47mm]{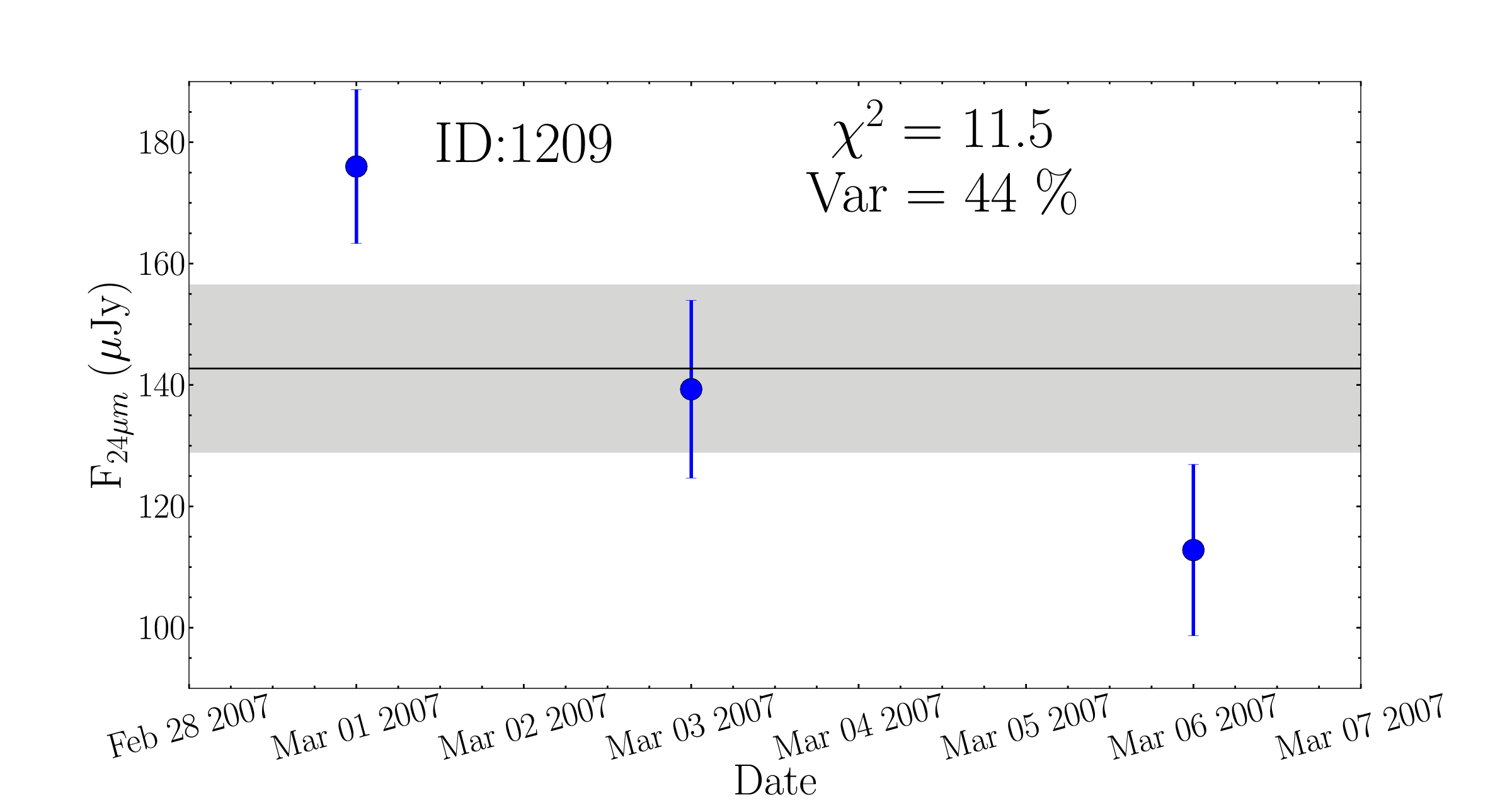}}{\hspace{0cm}}
      \subfigure {\includegraphics[width=47mm]{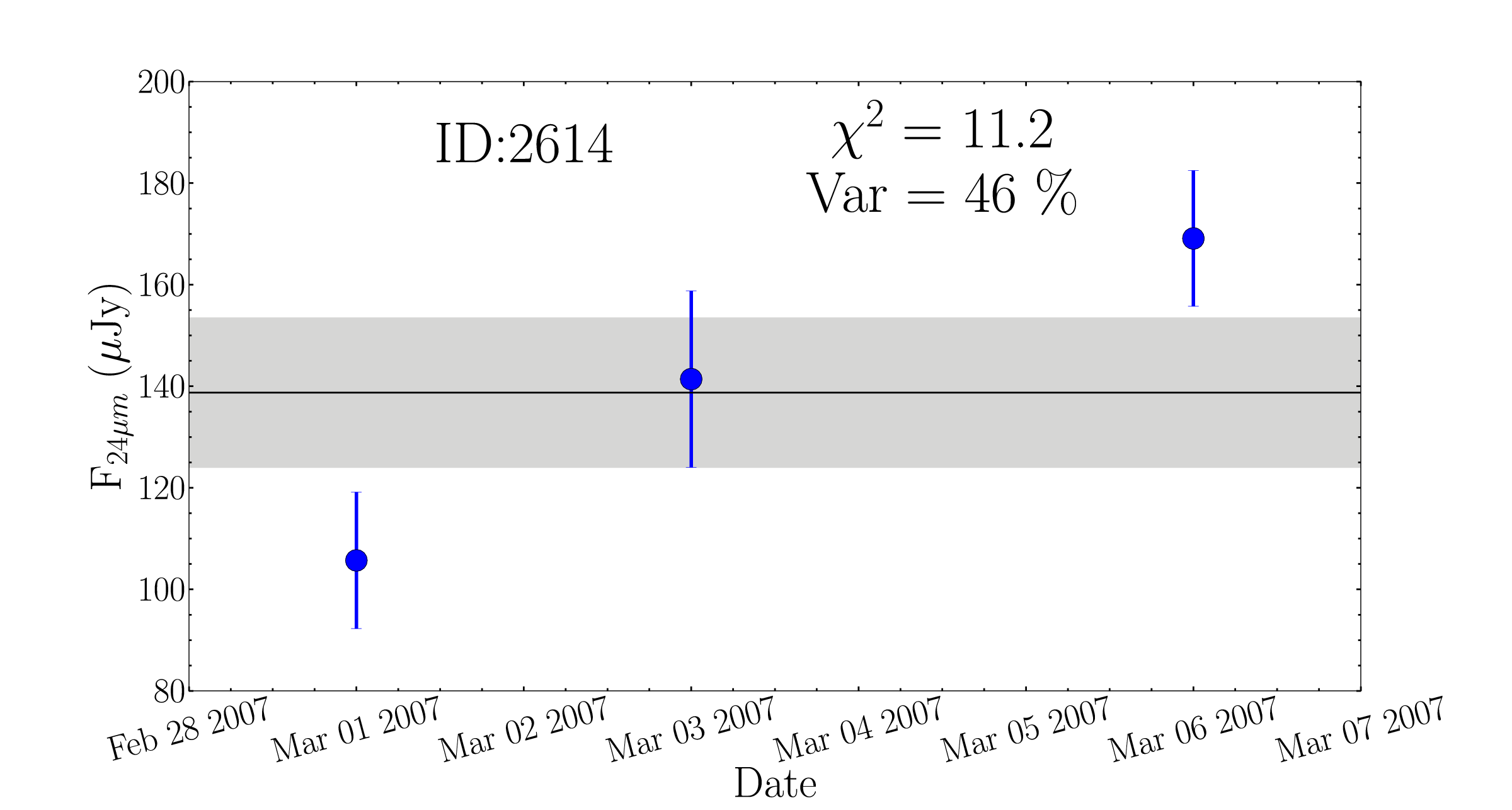}}{\hspace{0cm}}
      \subfigure {\includegraphics[width=47mm]{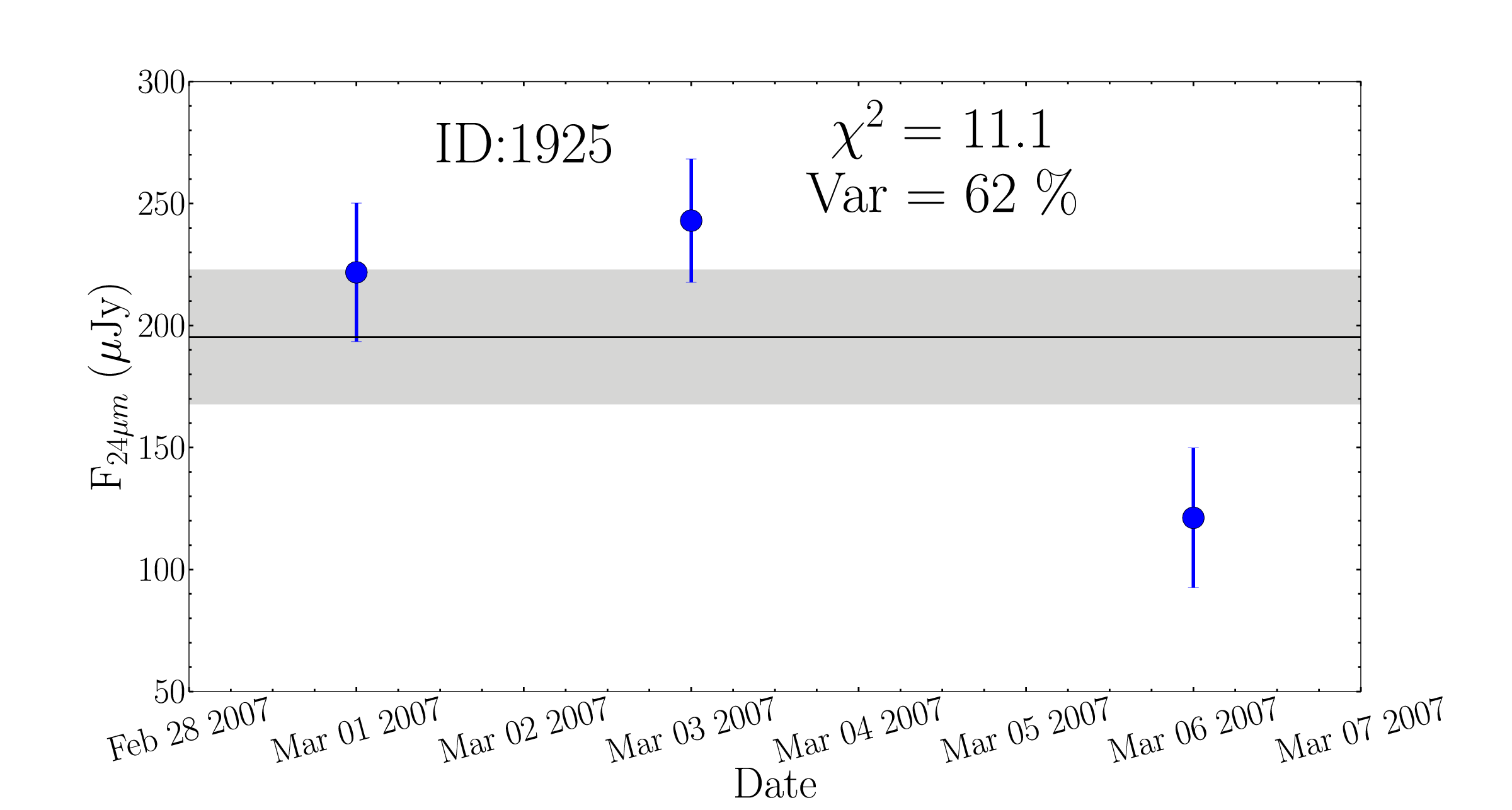}}{\hspace{0cm}}
      \subfigure {\includegraphics[width=47mm]{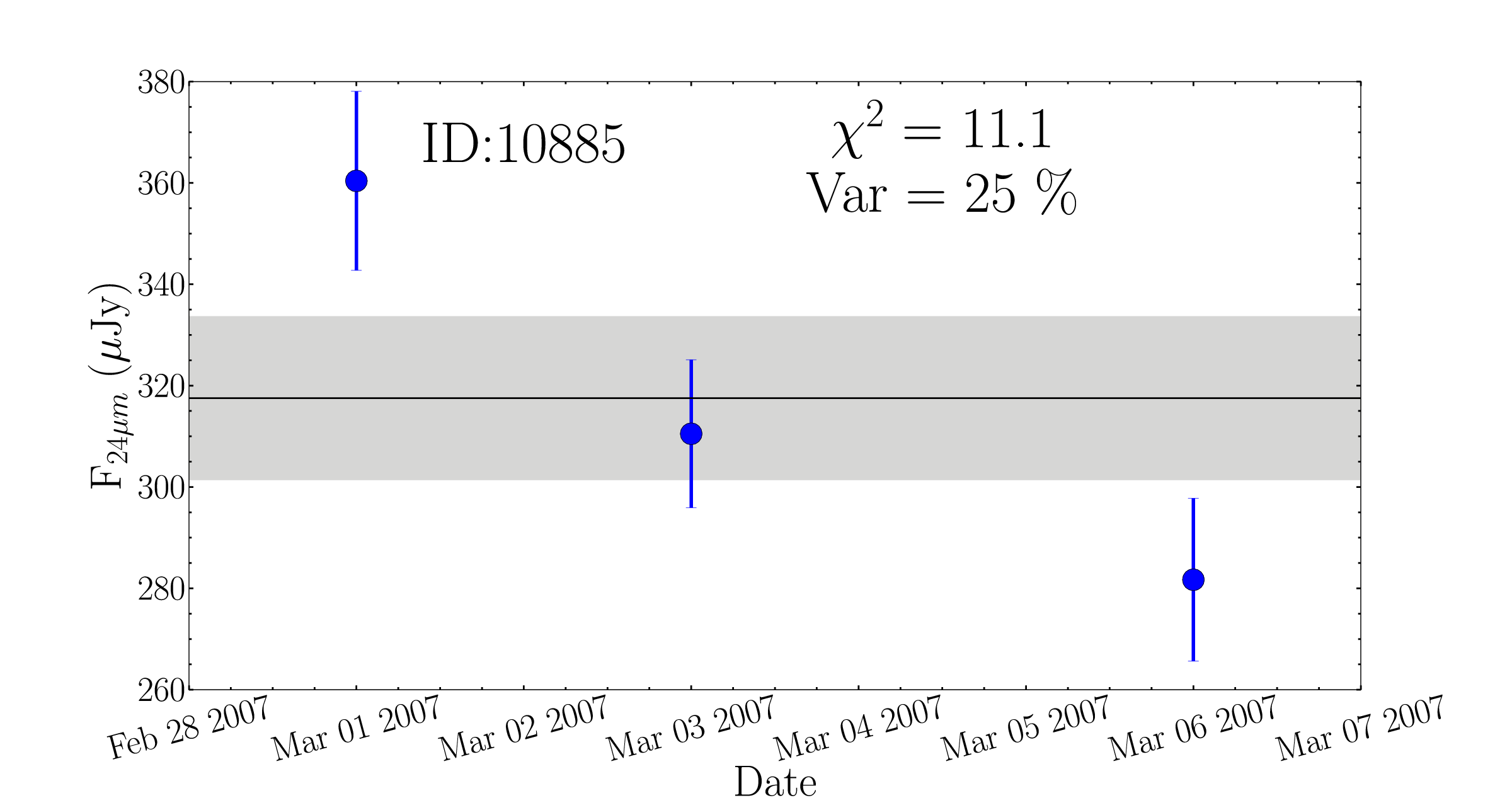}}{\hspace{0cm}}
      \subfigure {\includegraphics[width=47mm]{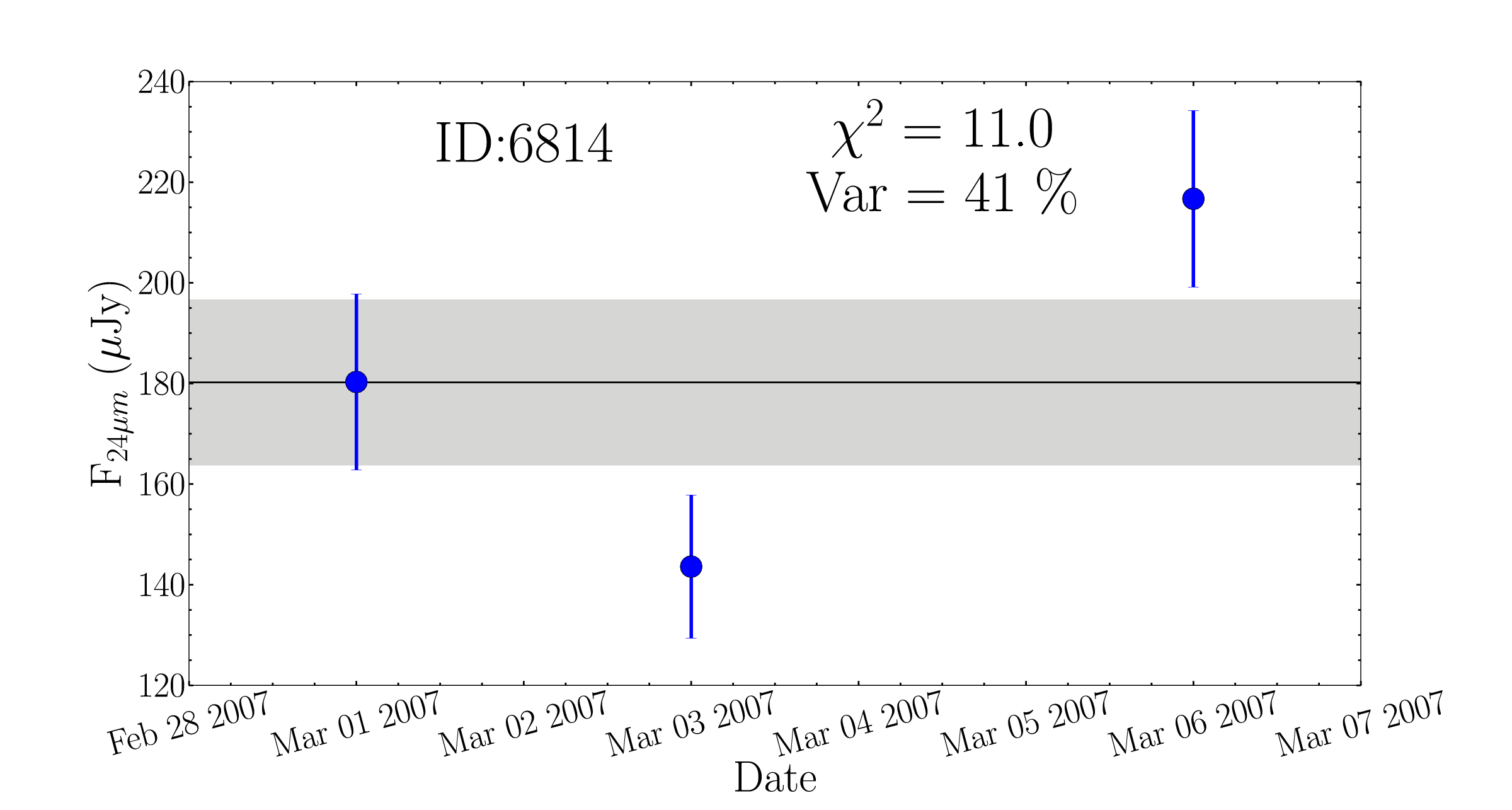}}{\hspace{0cm}}
      \subfigure {\includegraphics[width=47mm]{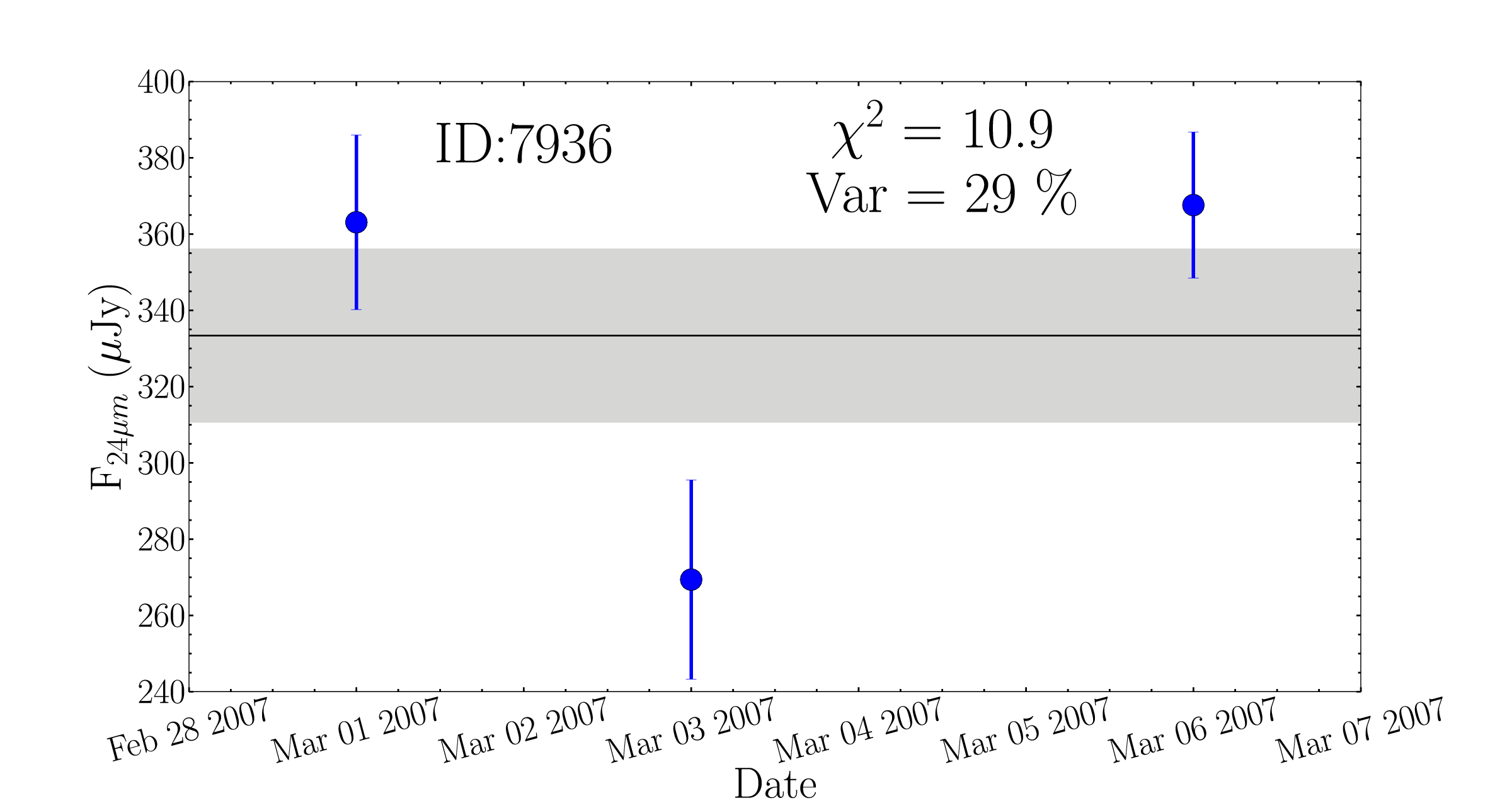}}{\hspace{0cm}}
      
      \caption{Light curves of MIPS 24 $\umu$m short-term variable candidates.} 
      \label{curvas-luz-corto}
    \end{center}
  \end{minipage}
\end{figure*}

\begin{figure*}
  \begin{minipage}{200mm}
    \begin{center}

      \subfigure {\includegraphics[width=47mm]{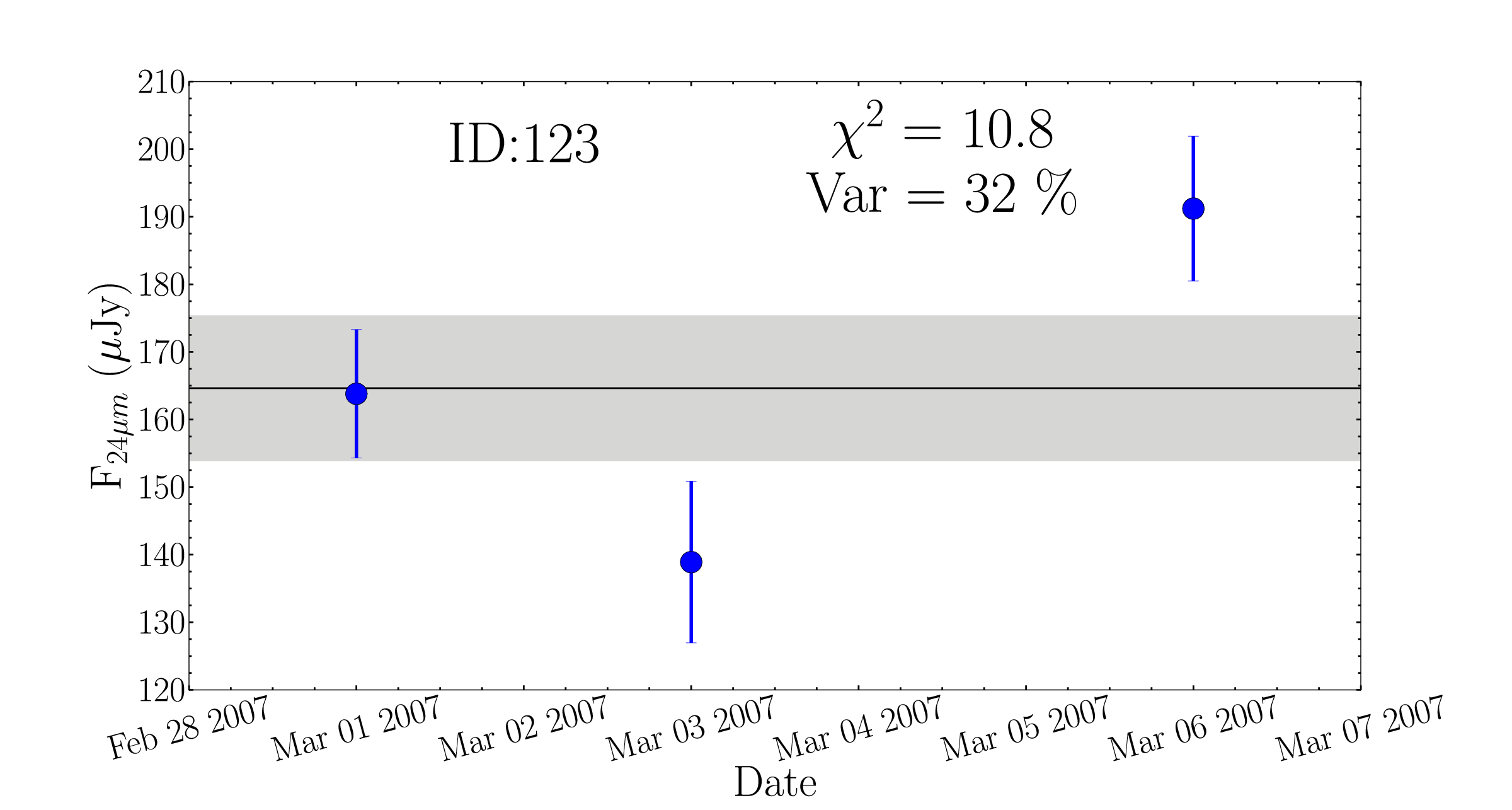}}{\hspace{0cm}}
      \subfigure {\includegraphics[width=47mm]{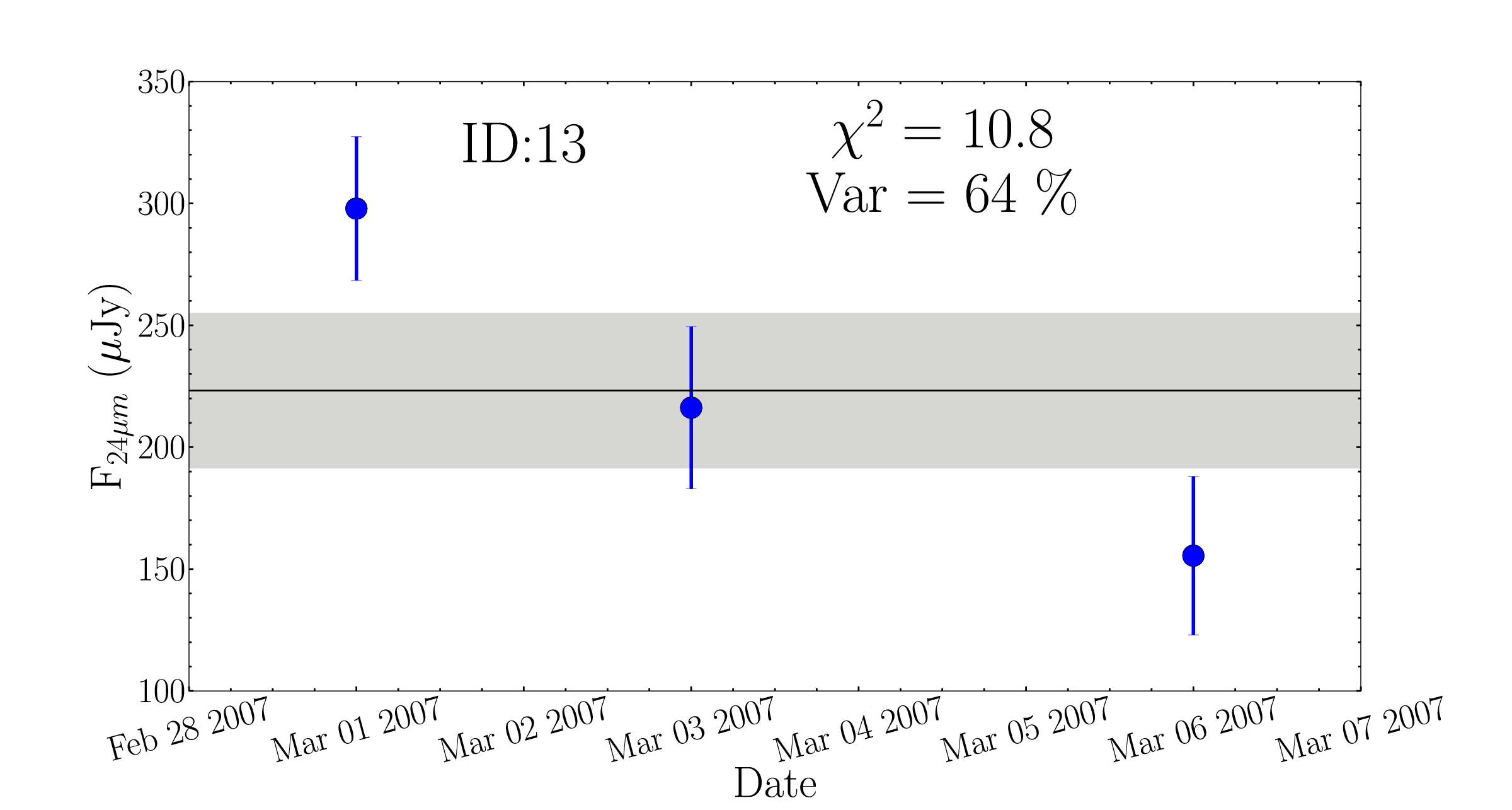}}{\hspace{0cm}}
      \subfigure {\includegraphics[width=47mm]{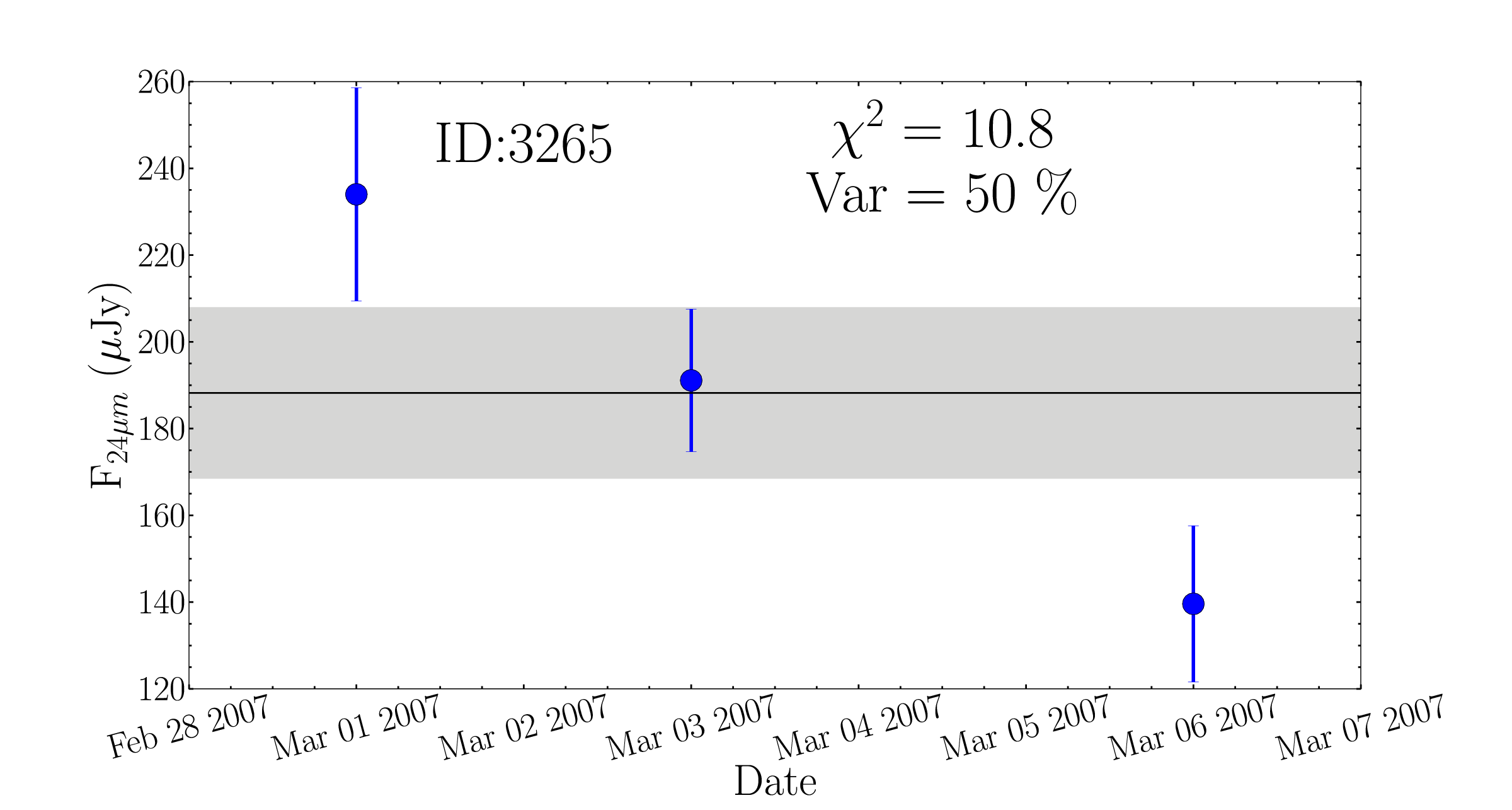}}{\hspace{0cm}}
      \subfigure {\includegraphics[width=47mm]{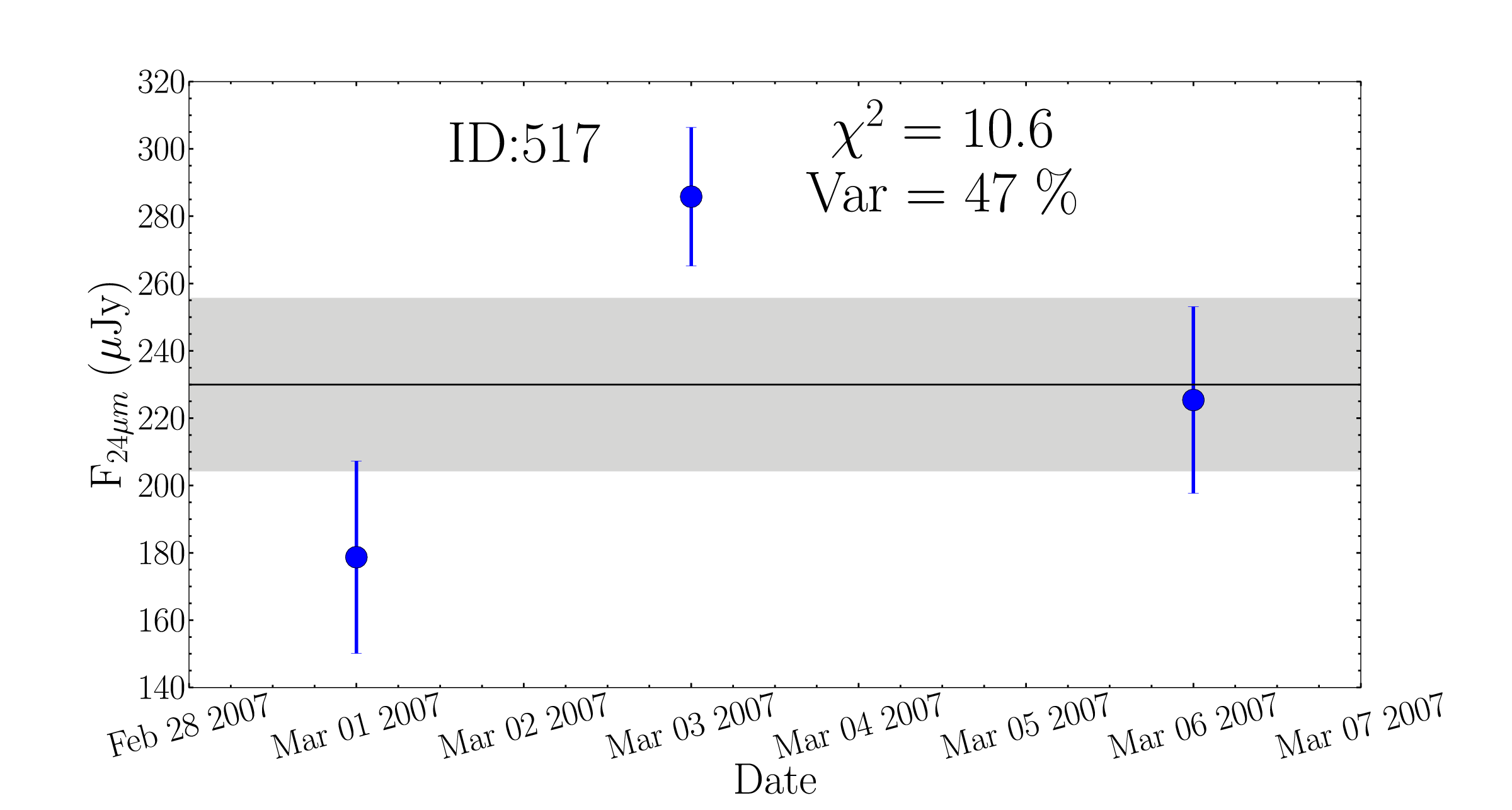}}{\hspace{0cm}}
      \subfigure {\includegraphics[width=47mm]{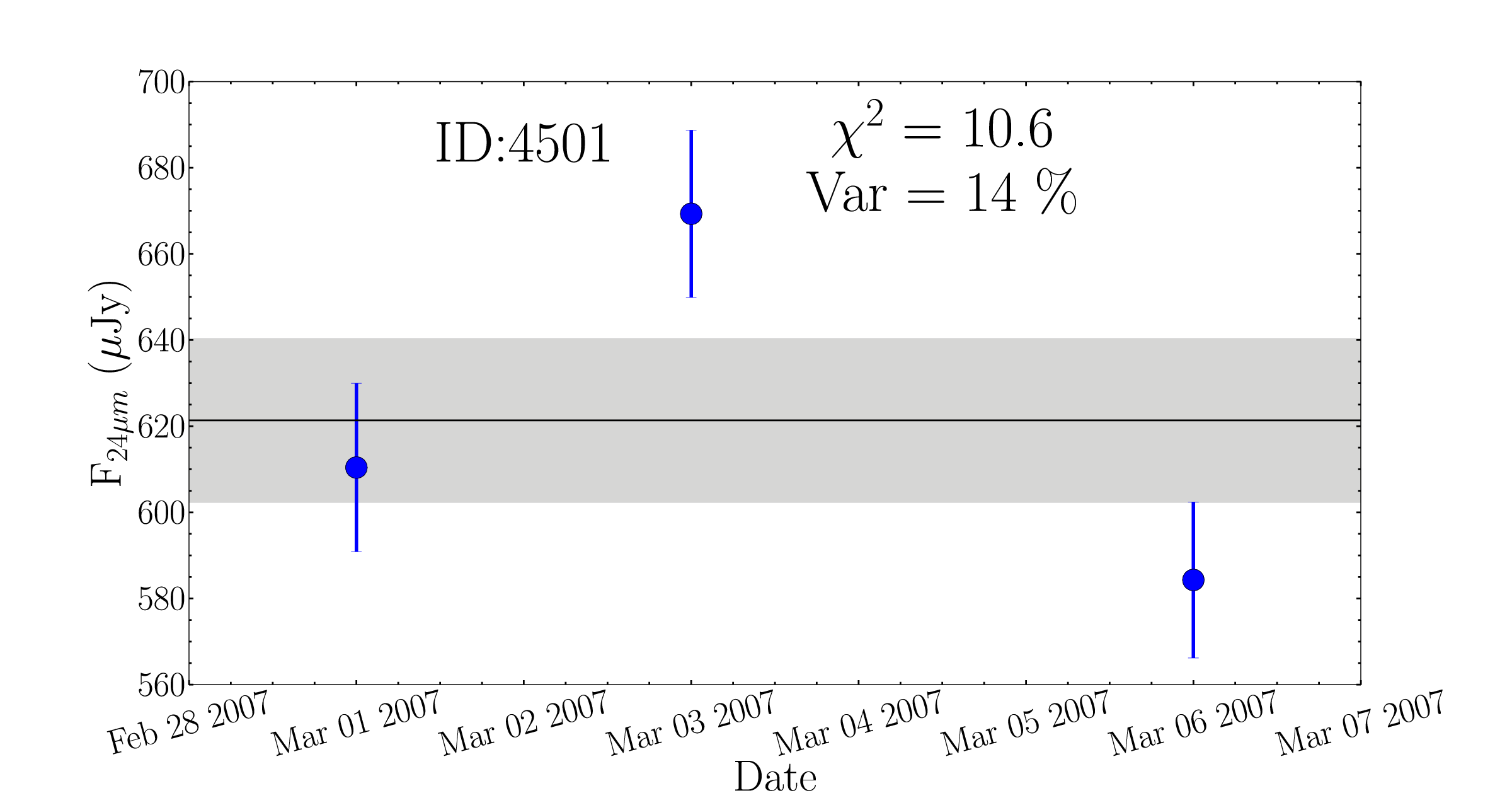}}{\hspace{0cm}}
      \subfigure {\includegraphics[width=47mm]{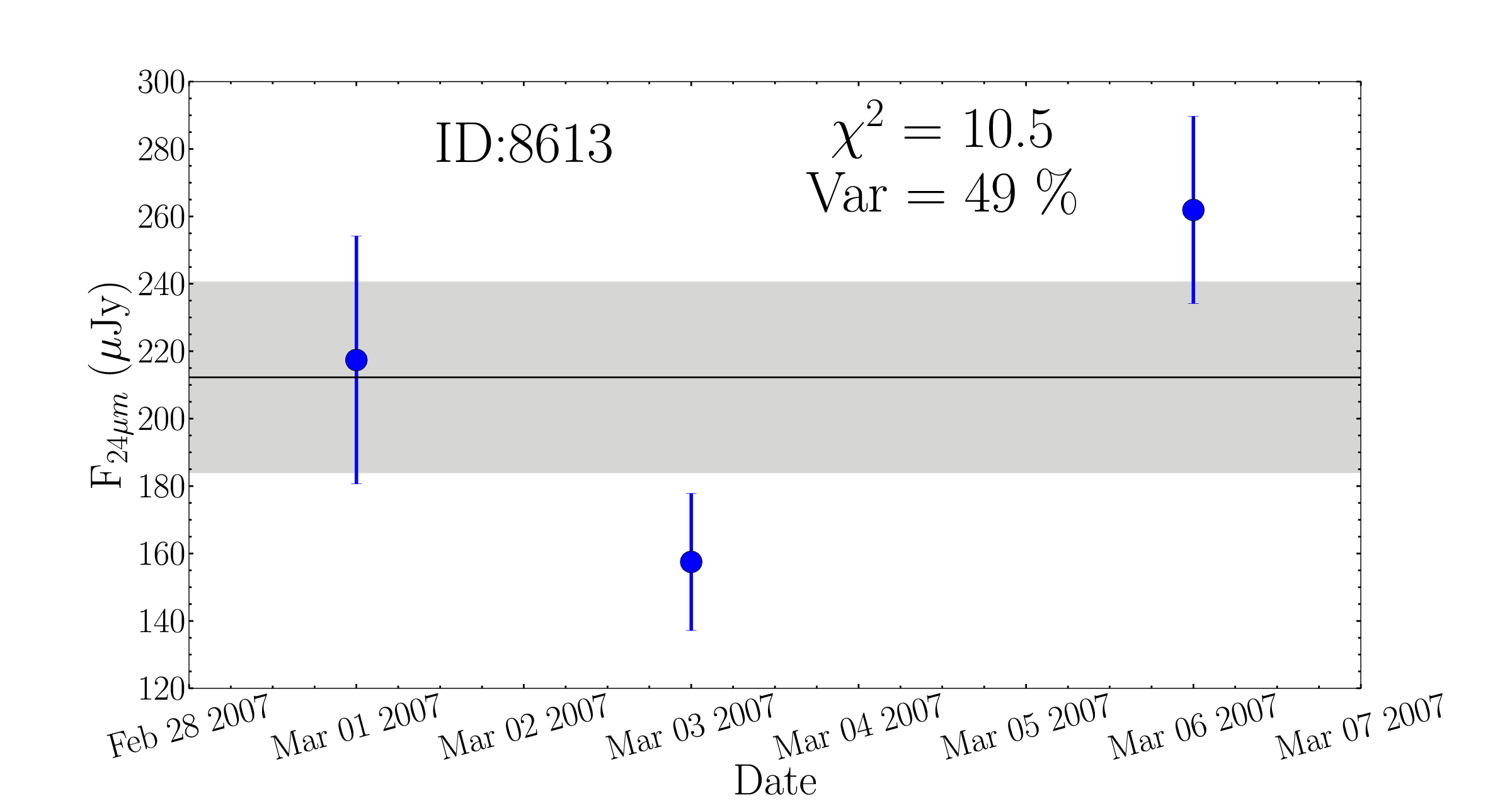}}{\hspace{0cm}}
      \subfigure {\includegraphics[width=47mm]{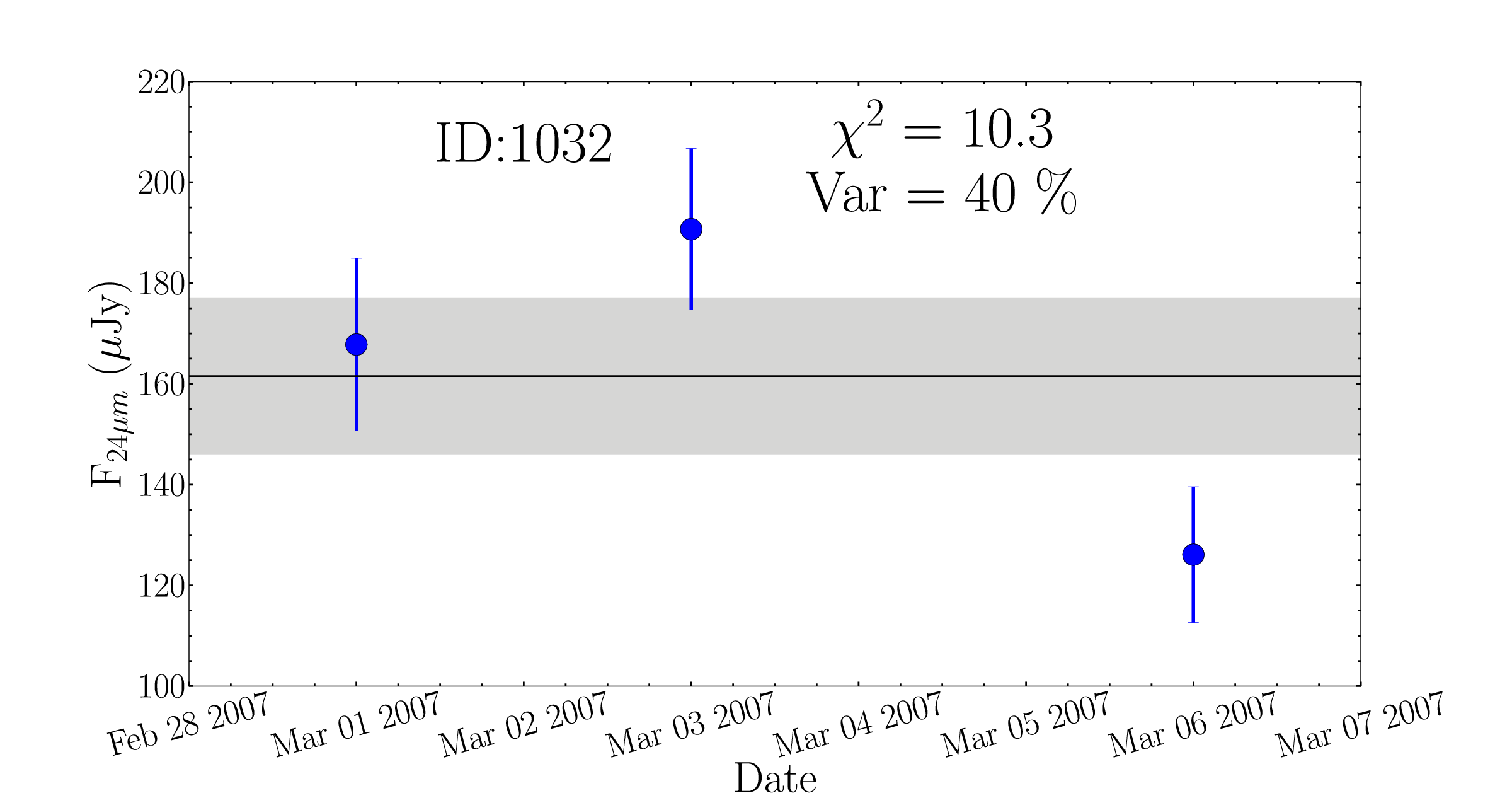}}{\hspace{0cm}}
      \subfigure {\includegraphics[width=47mm]{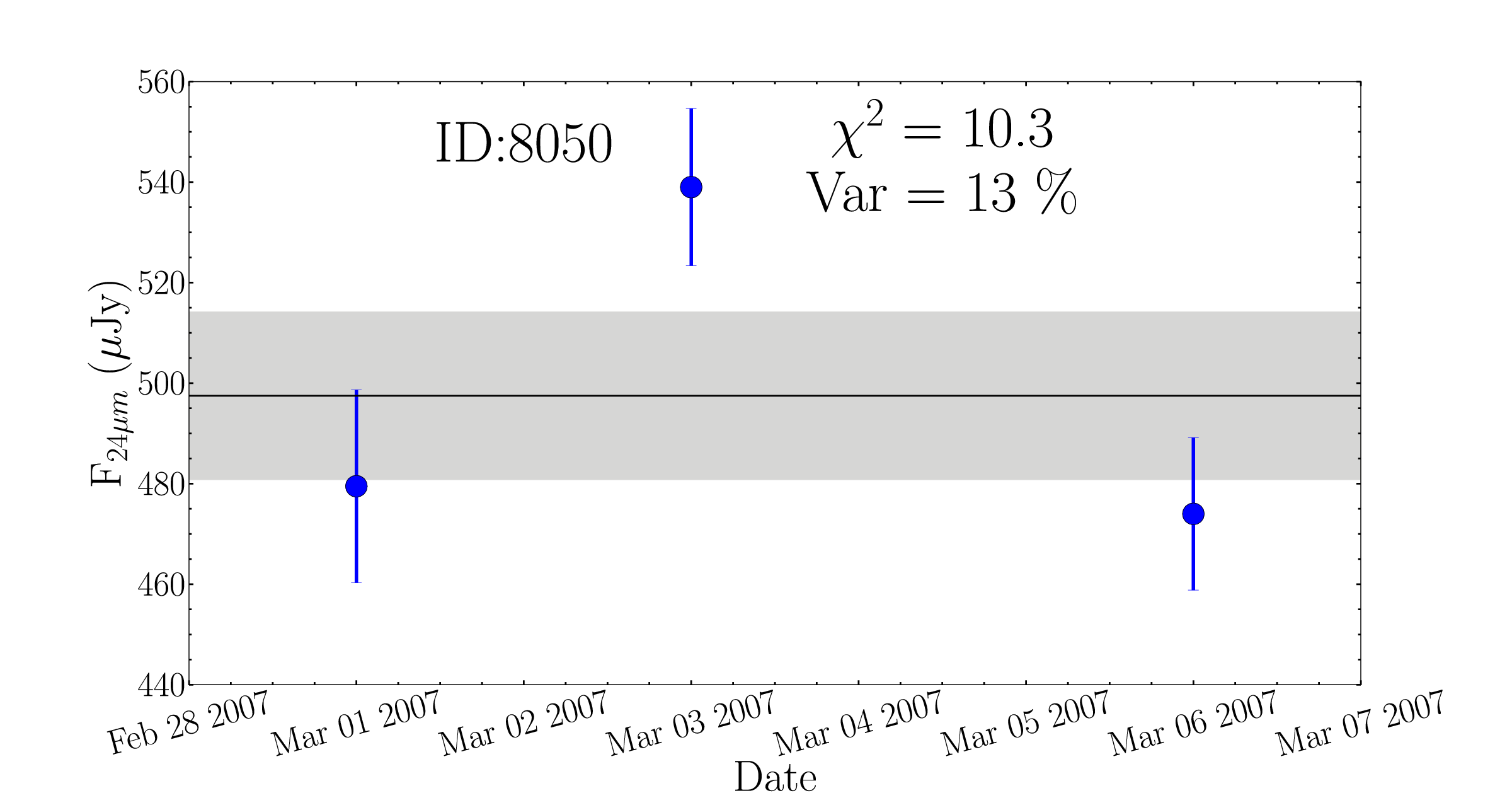}}{\hspace{0cm}}
      \subfigure {\includegraphics[width=47mm]{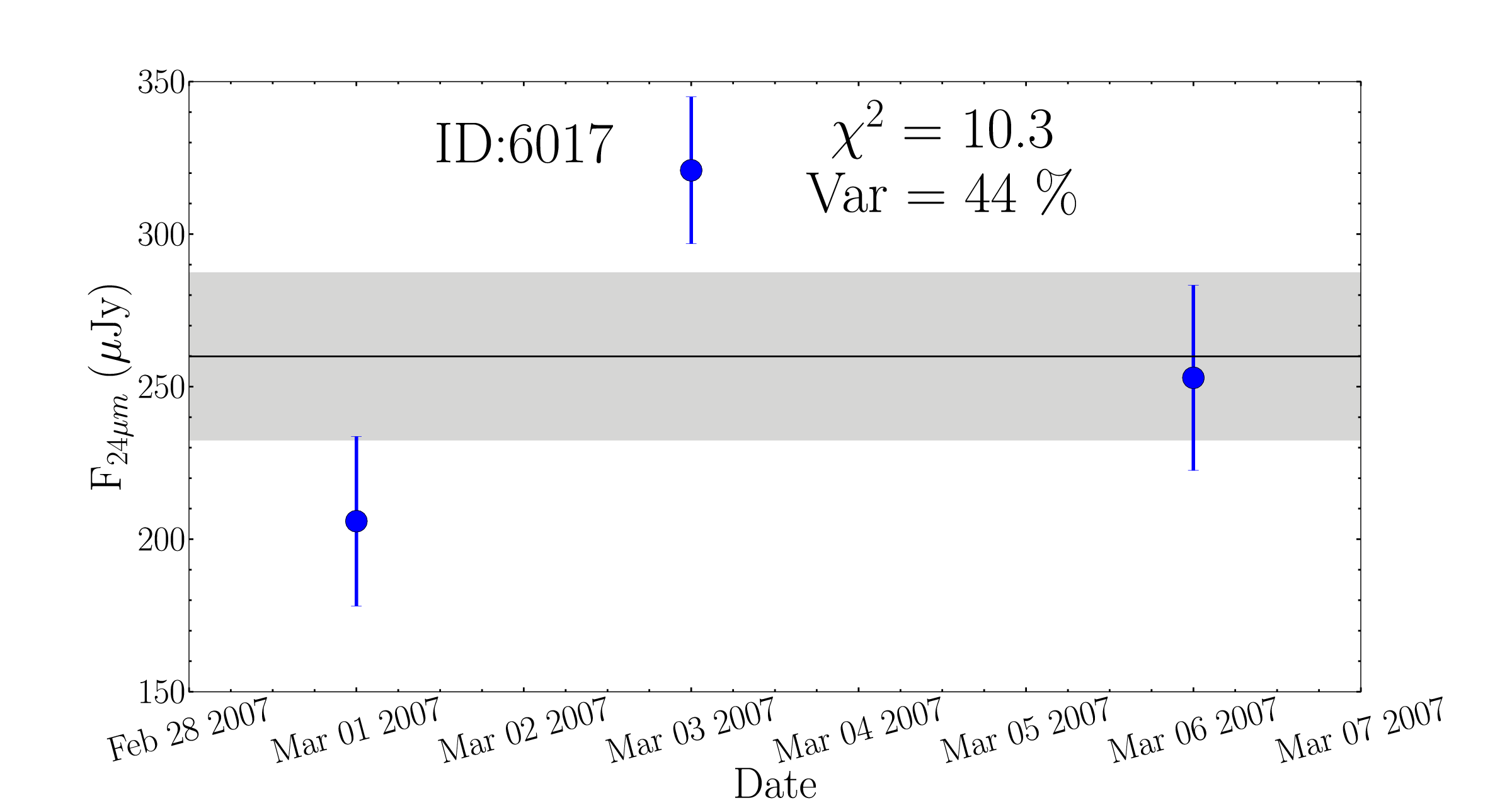}}{\hspace{0cm}}
      \subfigure {\includegraphics[width=47mm]{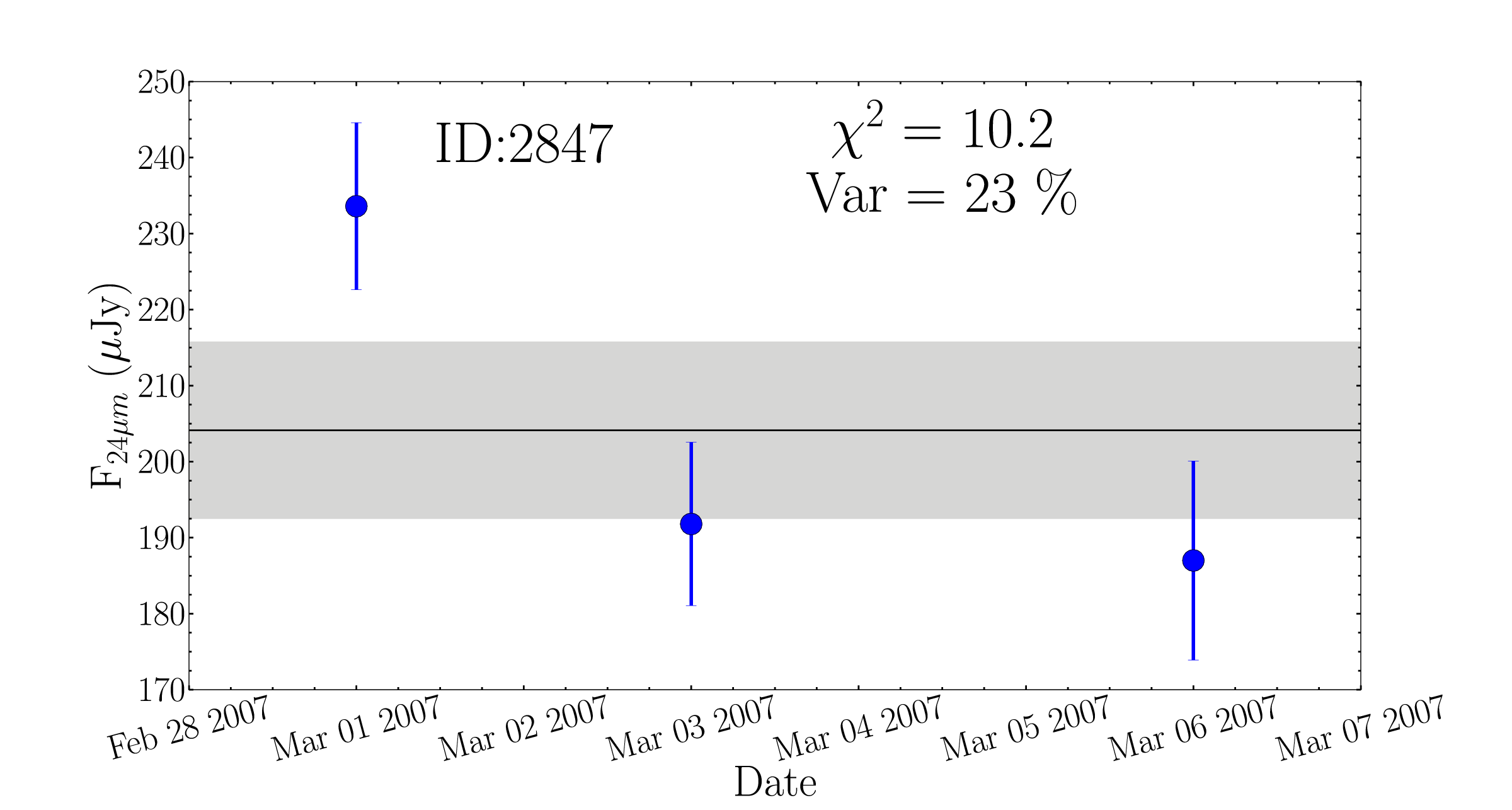}}{\hspace{0cm}}
      \subfigure {\includegraphics[width=47mm]{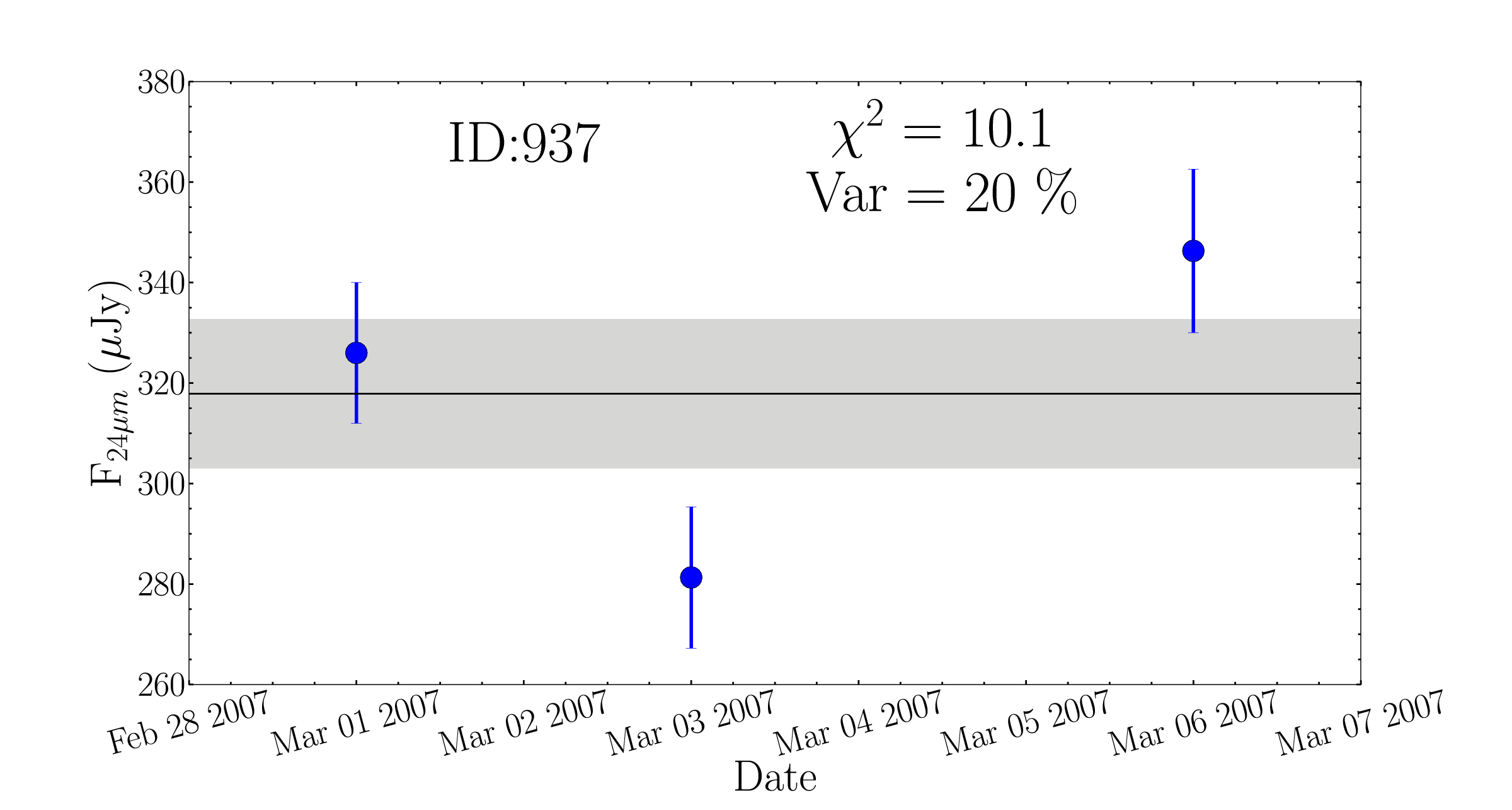}}{\hspace{0cm}}
      \subfigure {\includegraphics[width=47mm]{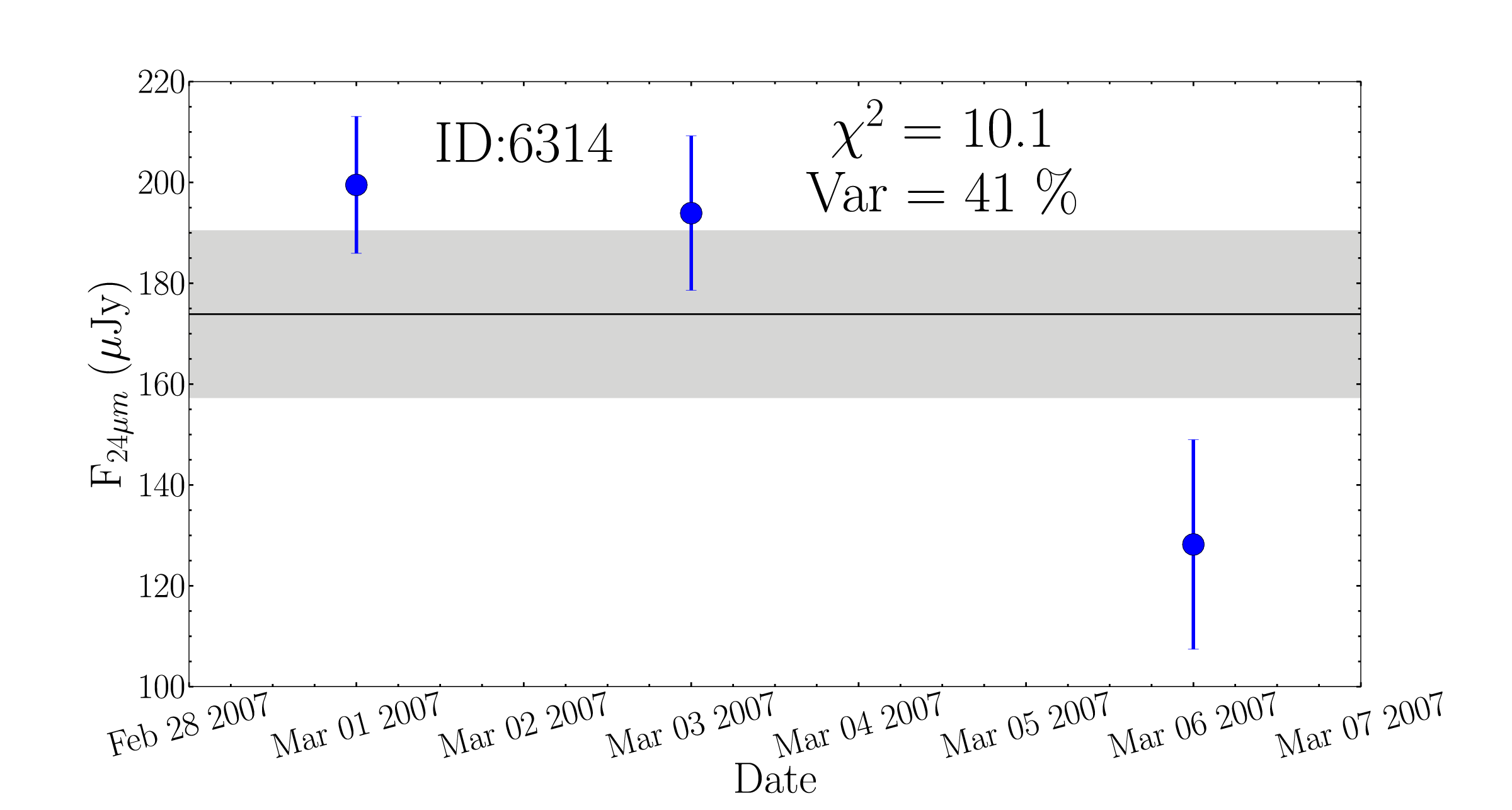}}{\hspace{0cm}}
      \subfigure {\includegraphics[width=47mm]{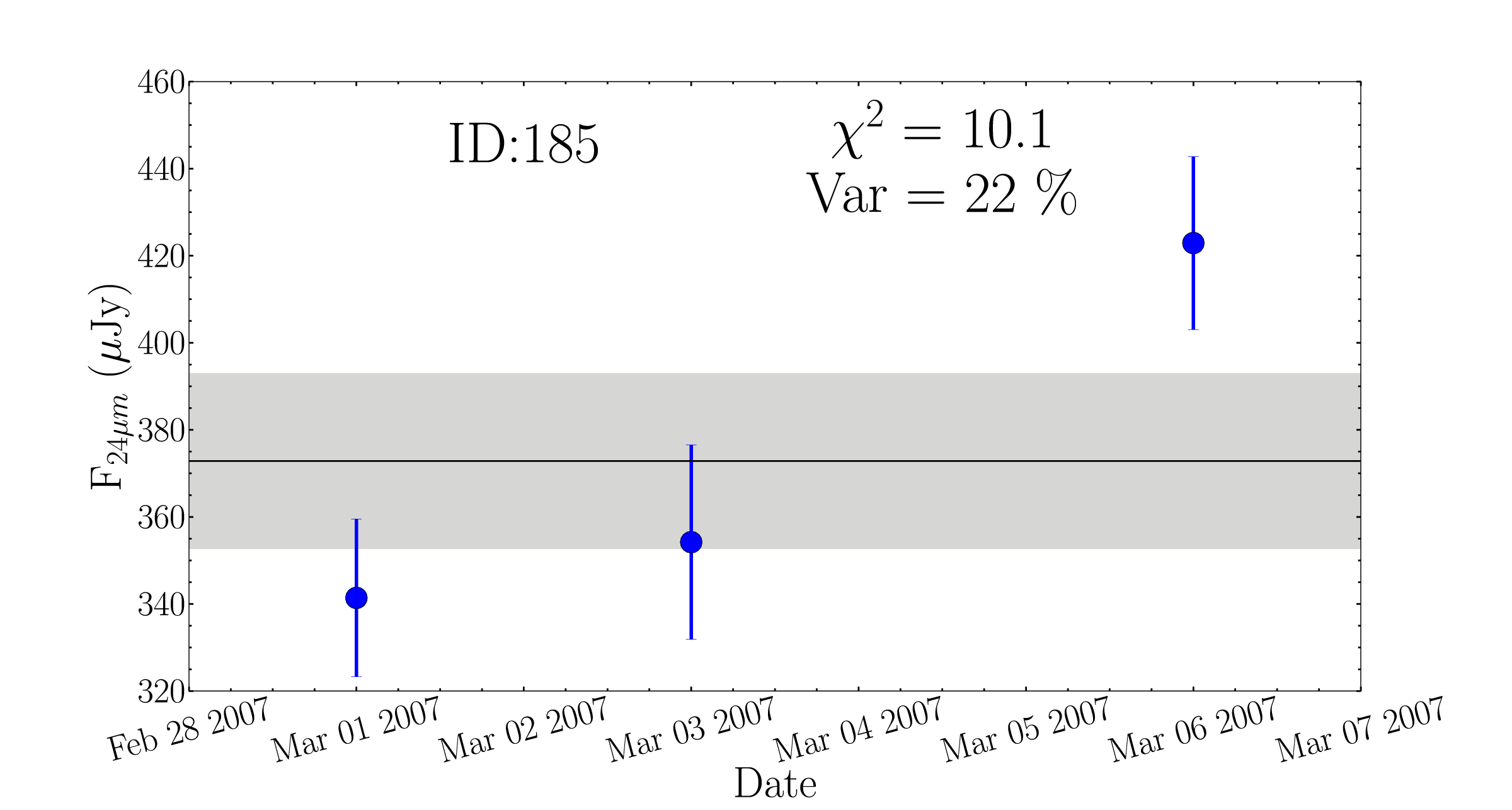}}{\hspace{0cm}}
      \subfigure {\includegraphics[width=47mm]{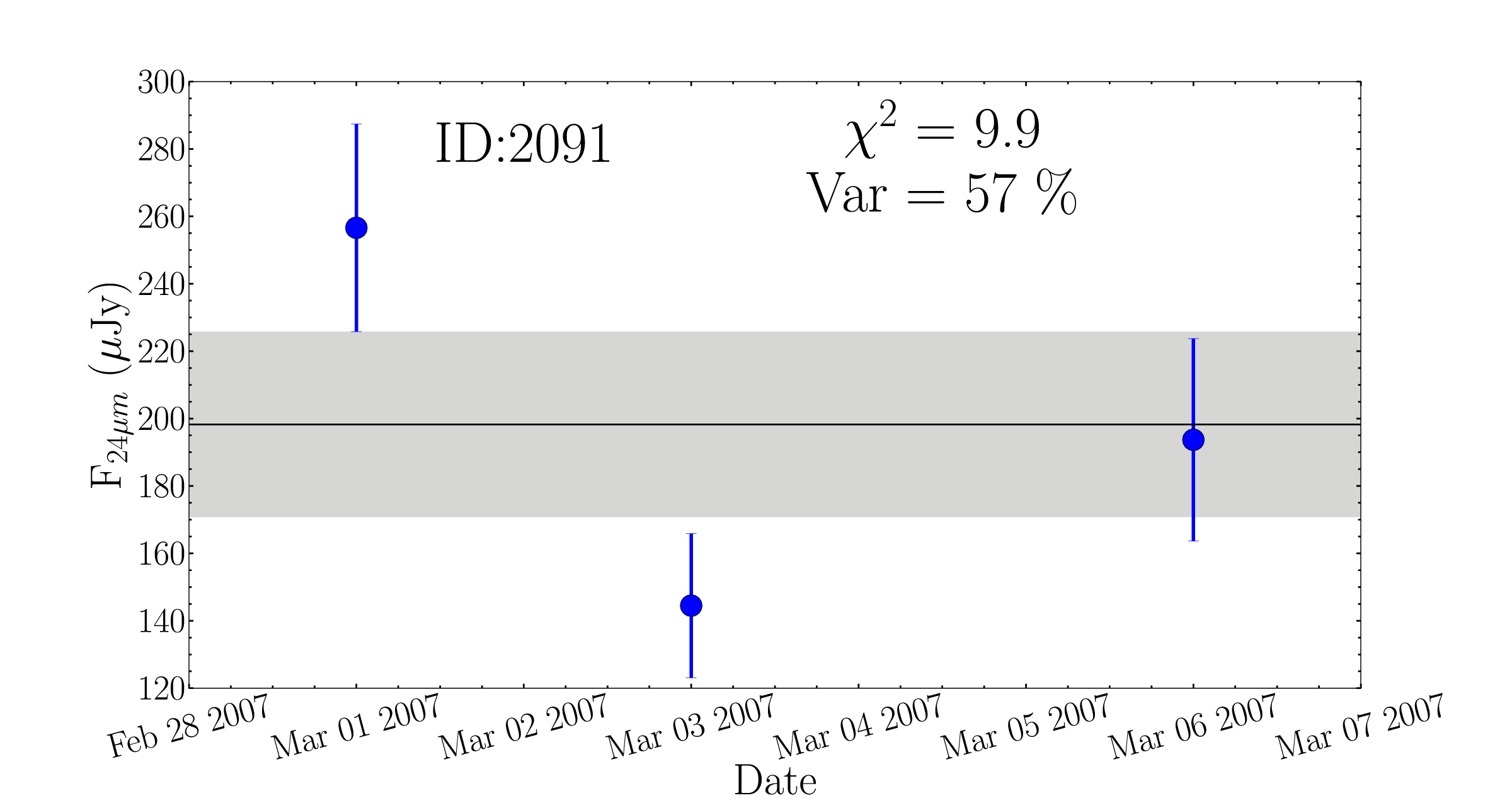}}{\hspace{0cm}}      
      \subfigure {\includegraphics[width=47mm]{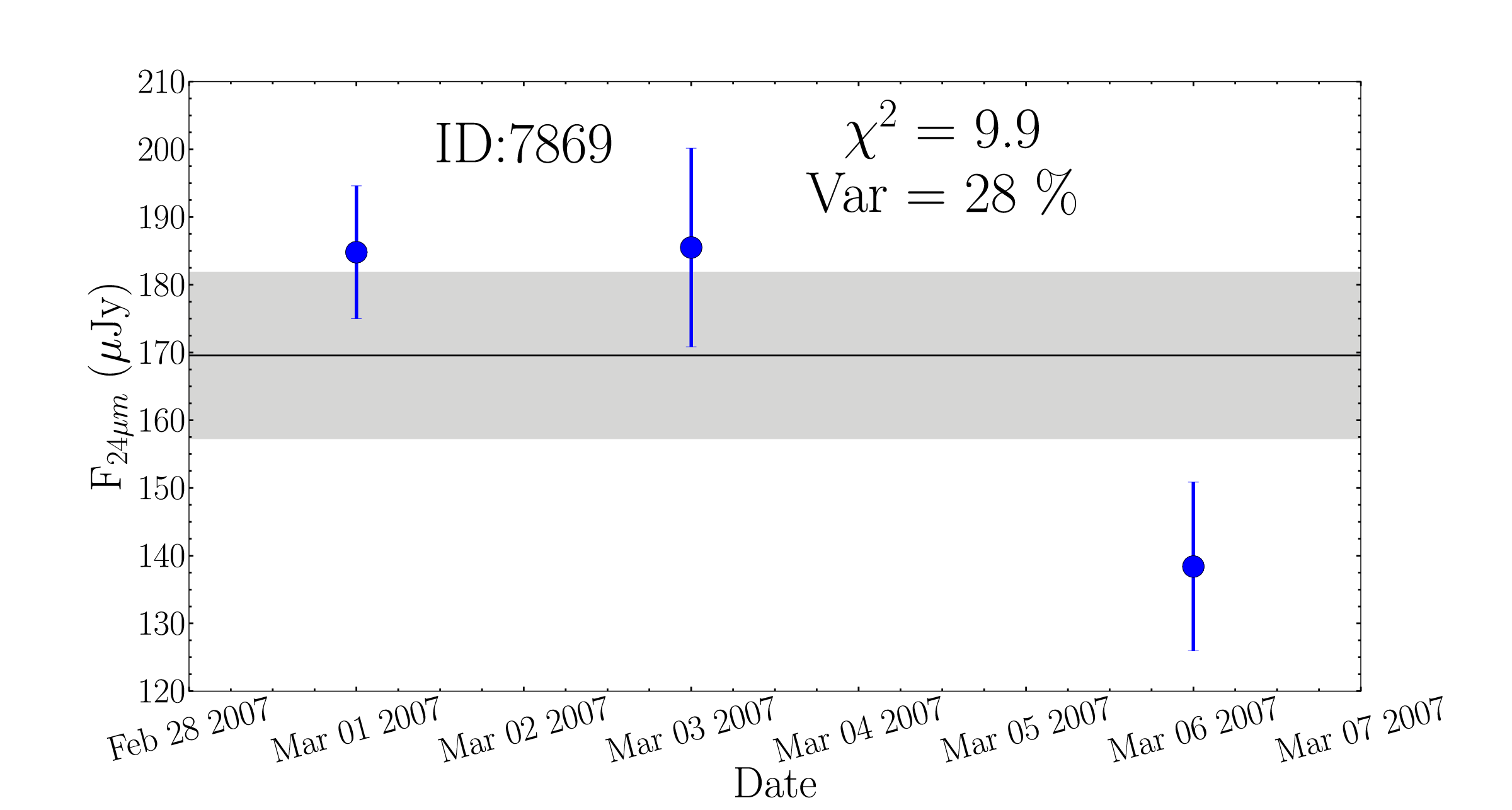}}{\hspace{0cm}}
      \subfigure {\includegraphics[width=47mm]{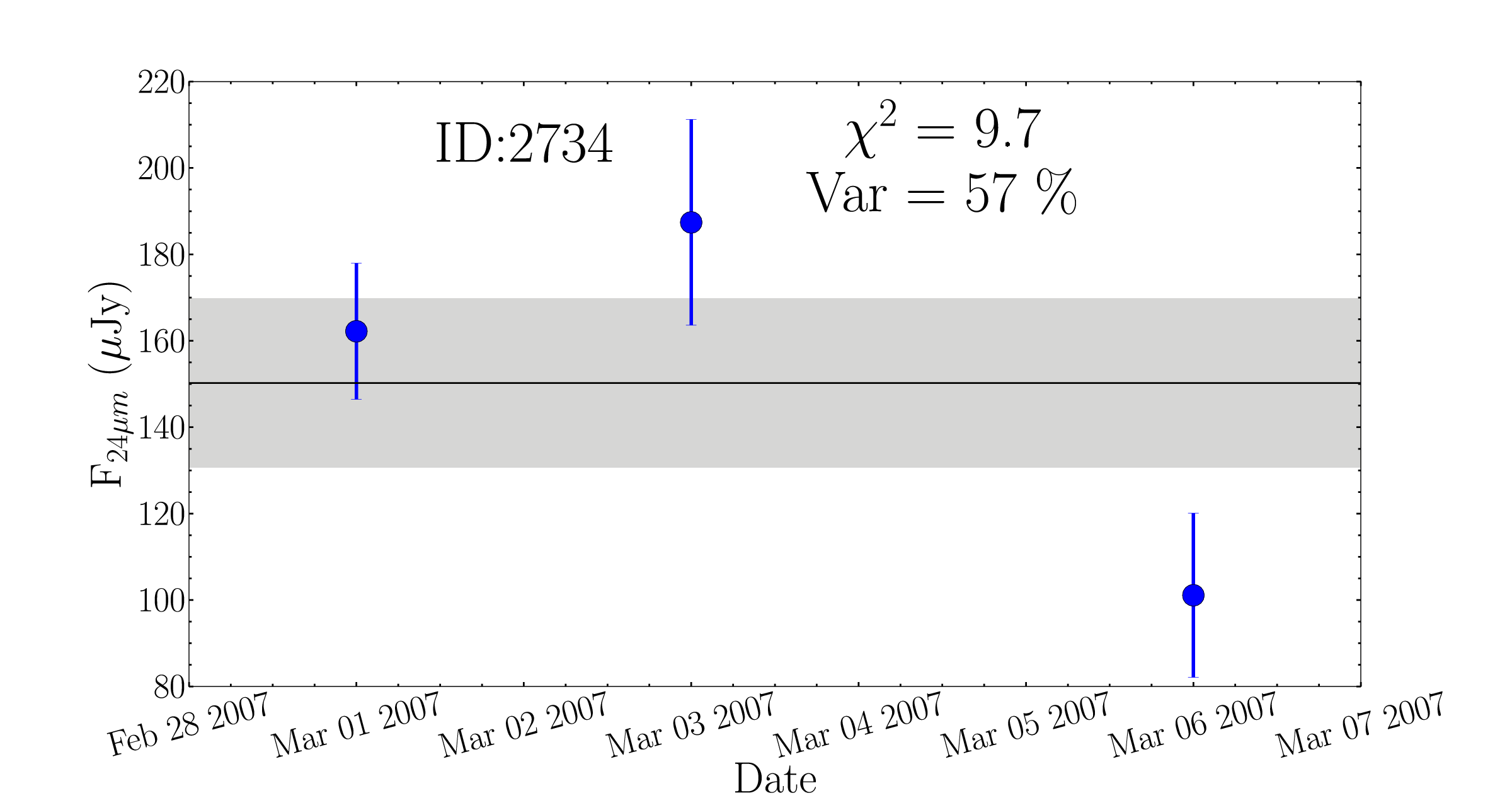}}{\hspace{0cm}}
      \subfigure {\includegraphics[width=47mm]{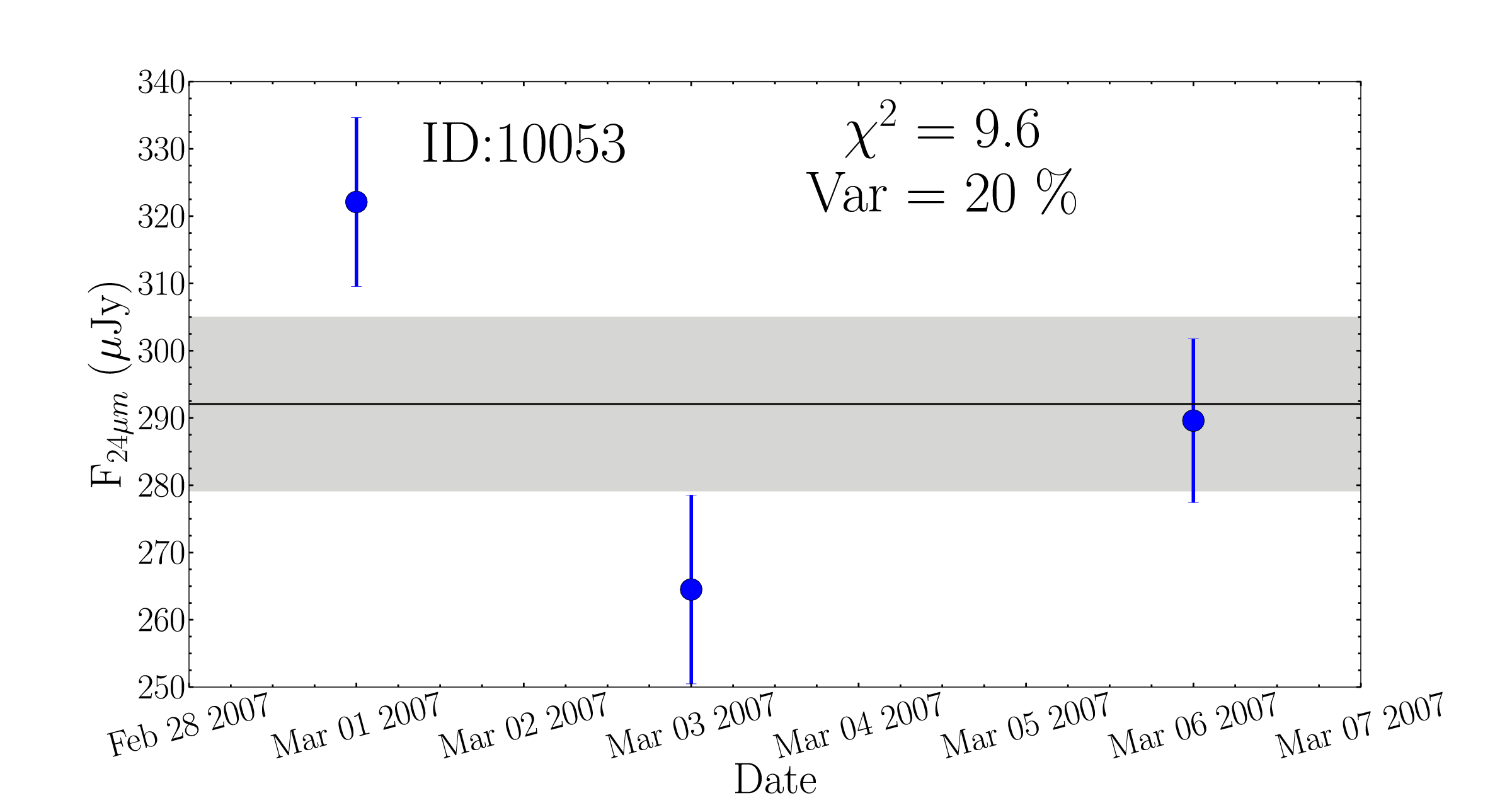}}{\hspace{0cm}}
      \subfigure {\includegraphics[width=47mm]{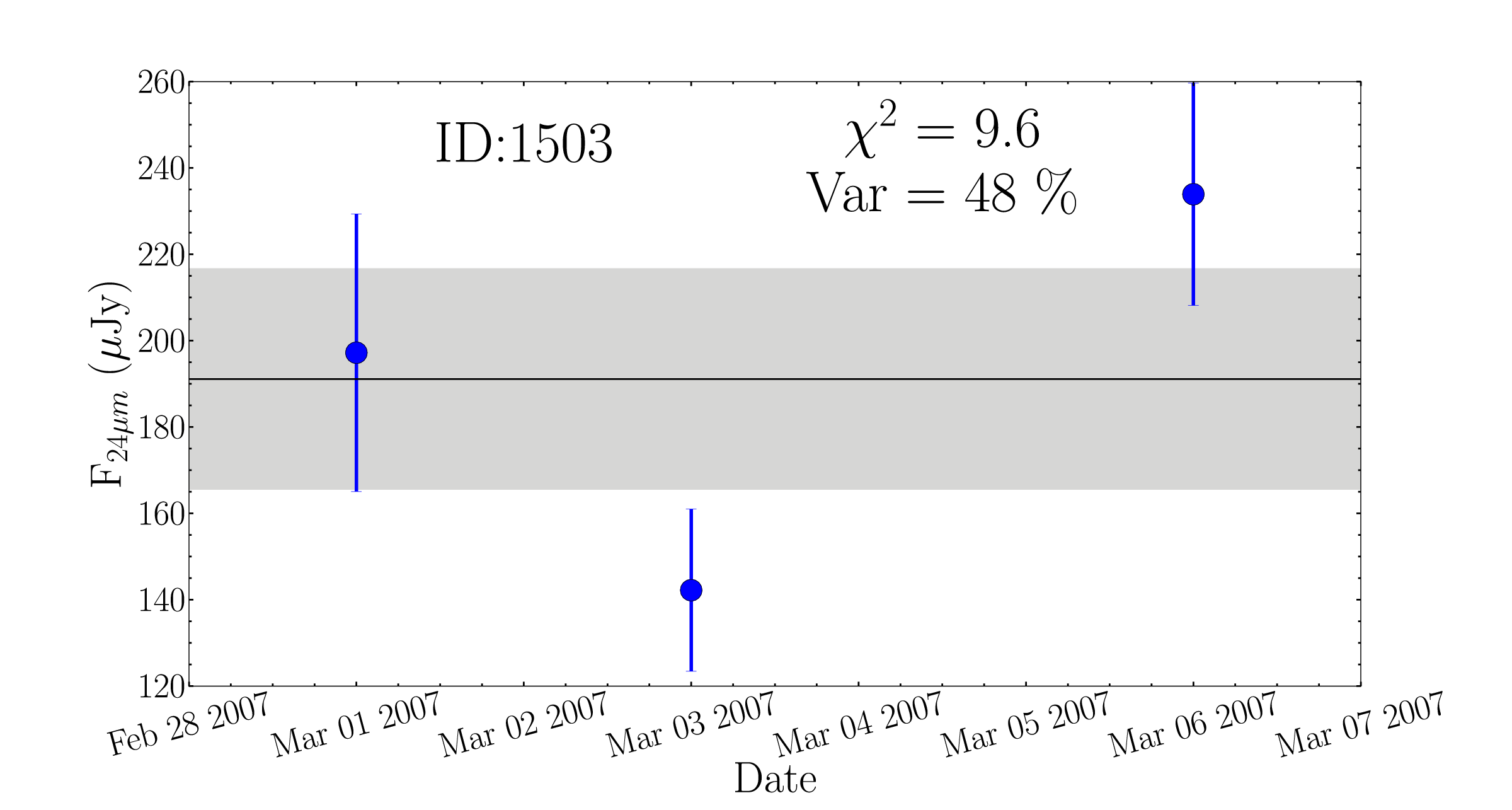}}{\hspace{0cm}}
      \subfigure {\includegraphics[width=47mm]{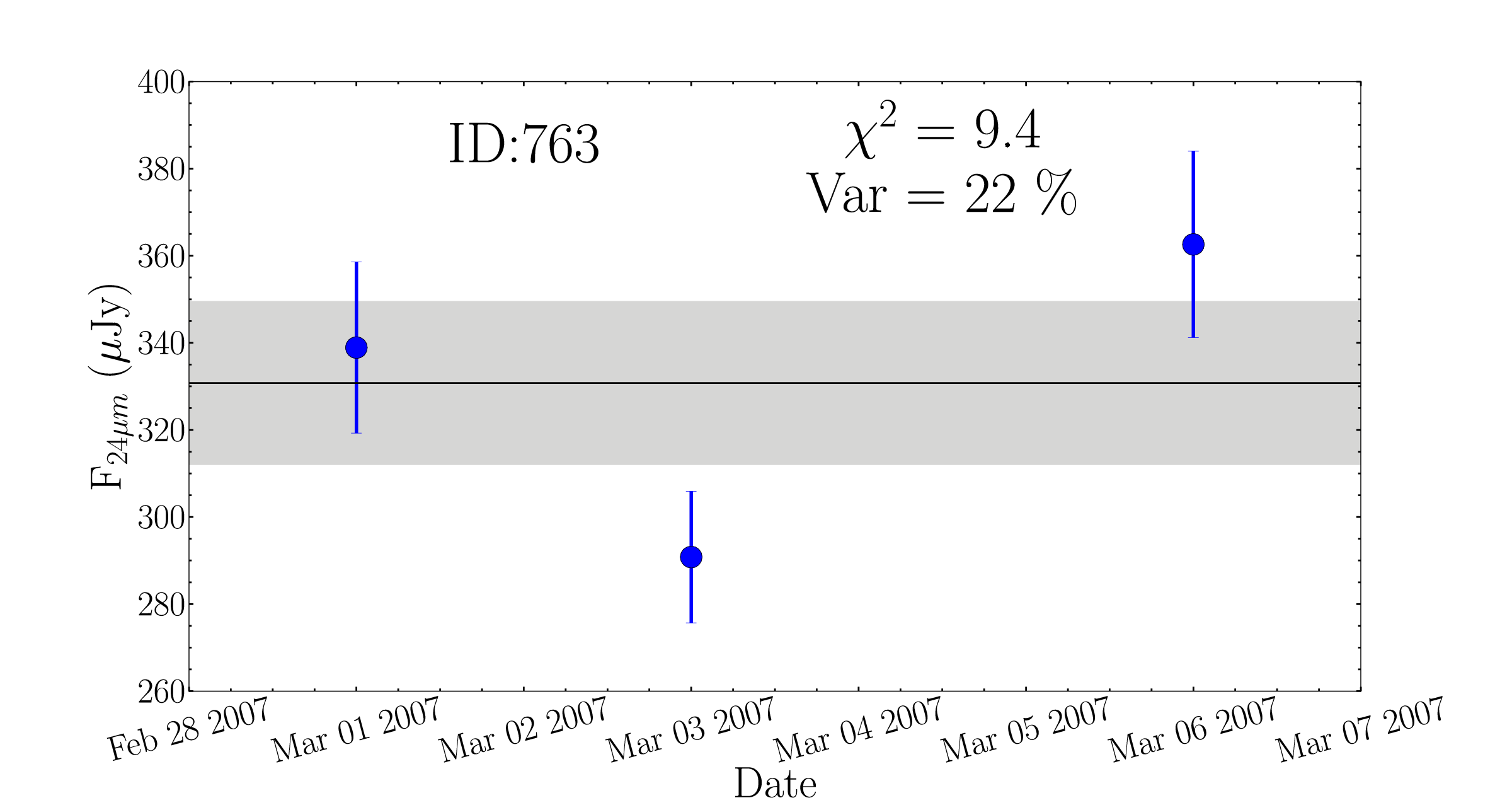}}{\hspace{0cm}}
      \subfigure {\includegraphics[width=47mm]{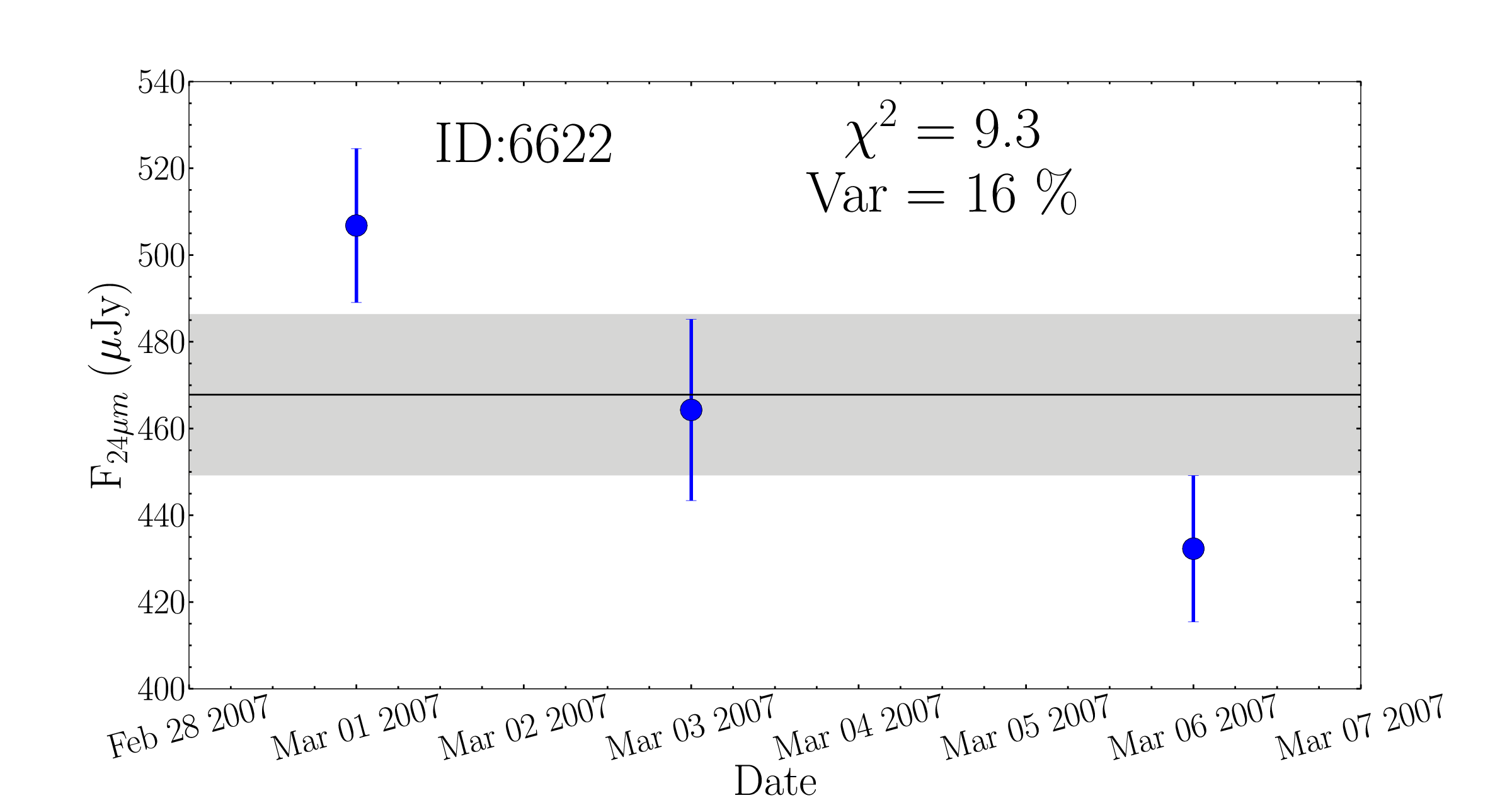}}{\hspace{0cm}}
      \subfigure {\includegraphics[width=47mm]{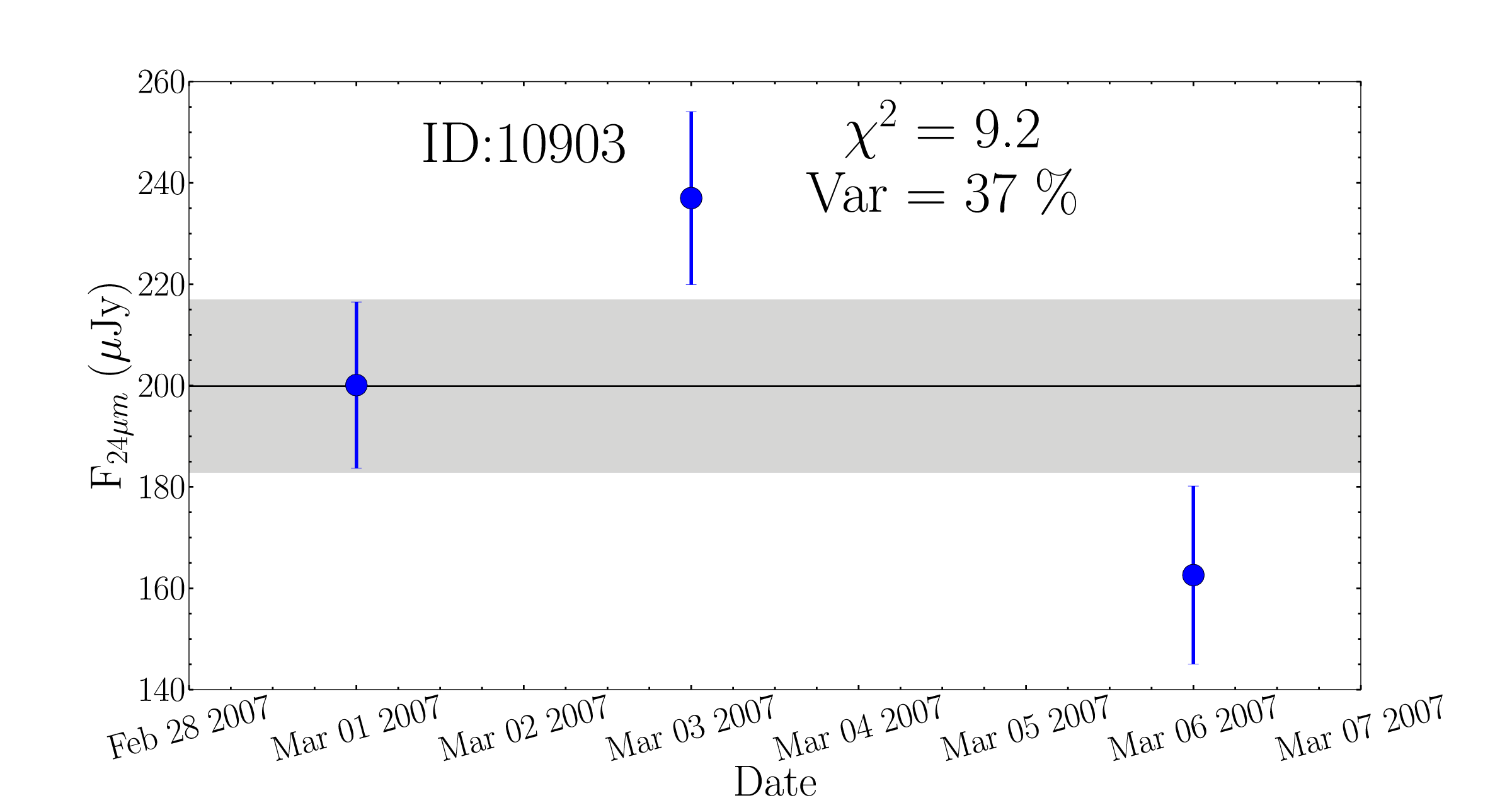}}{\hspace{0cm}}
      \subfigure {\includegraphics[width=47mm]{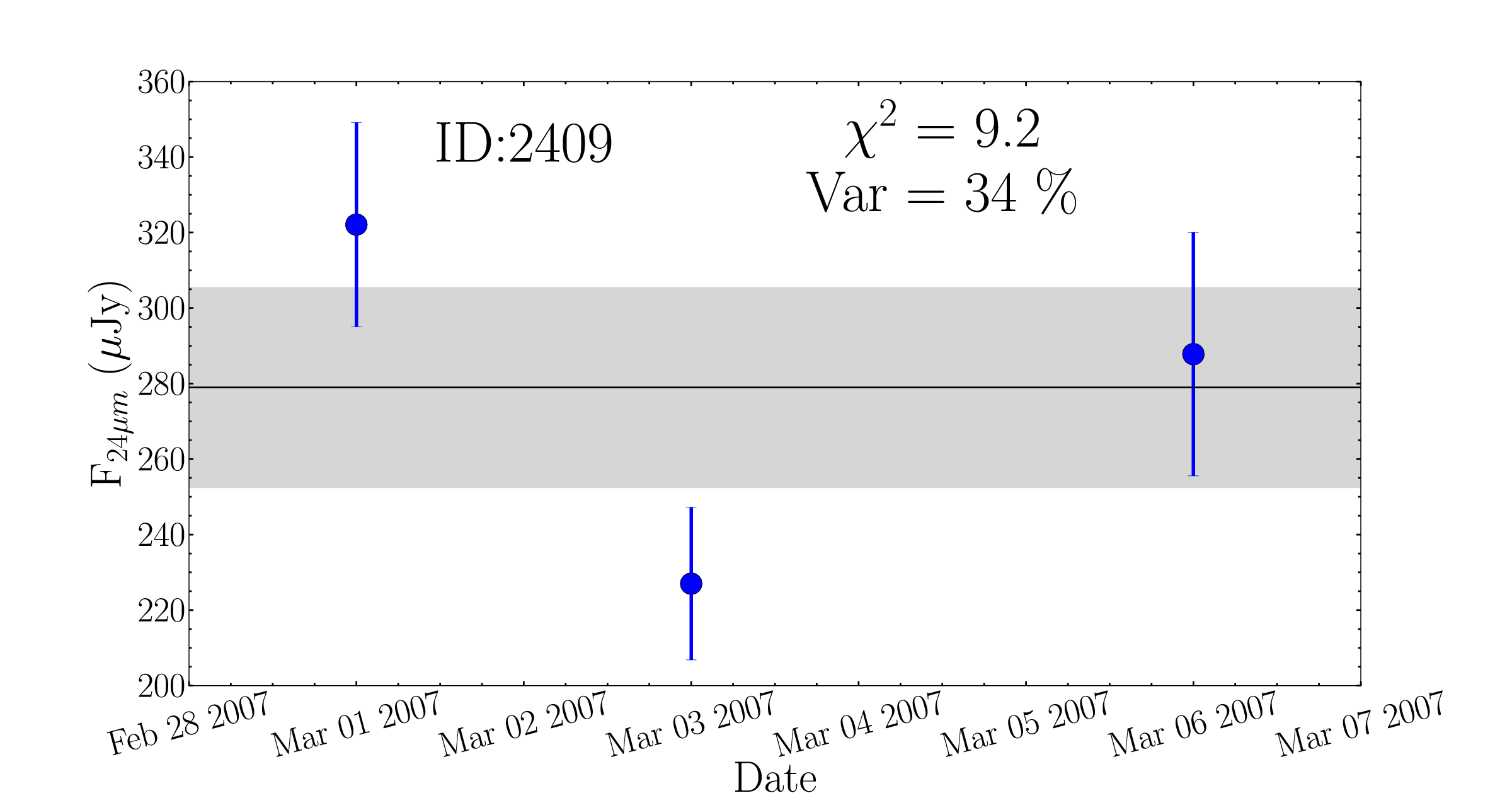}}{\hspace{0cm}}
      \subfigure {\includegraphics[width=47mm]{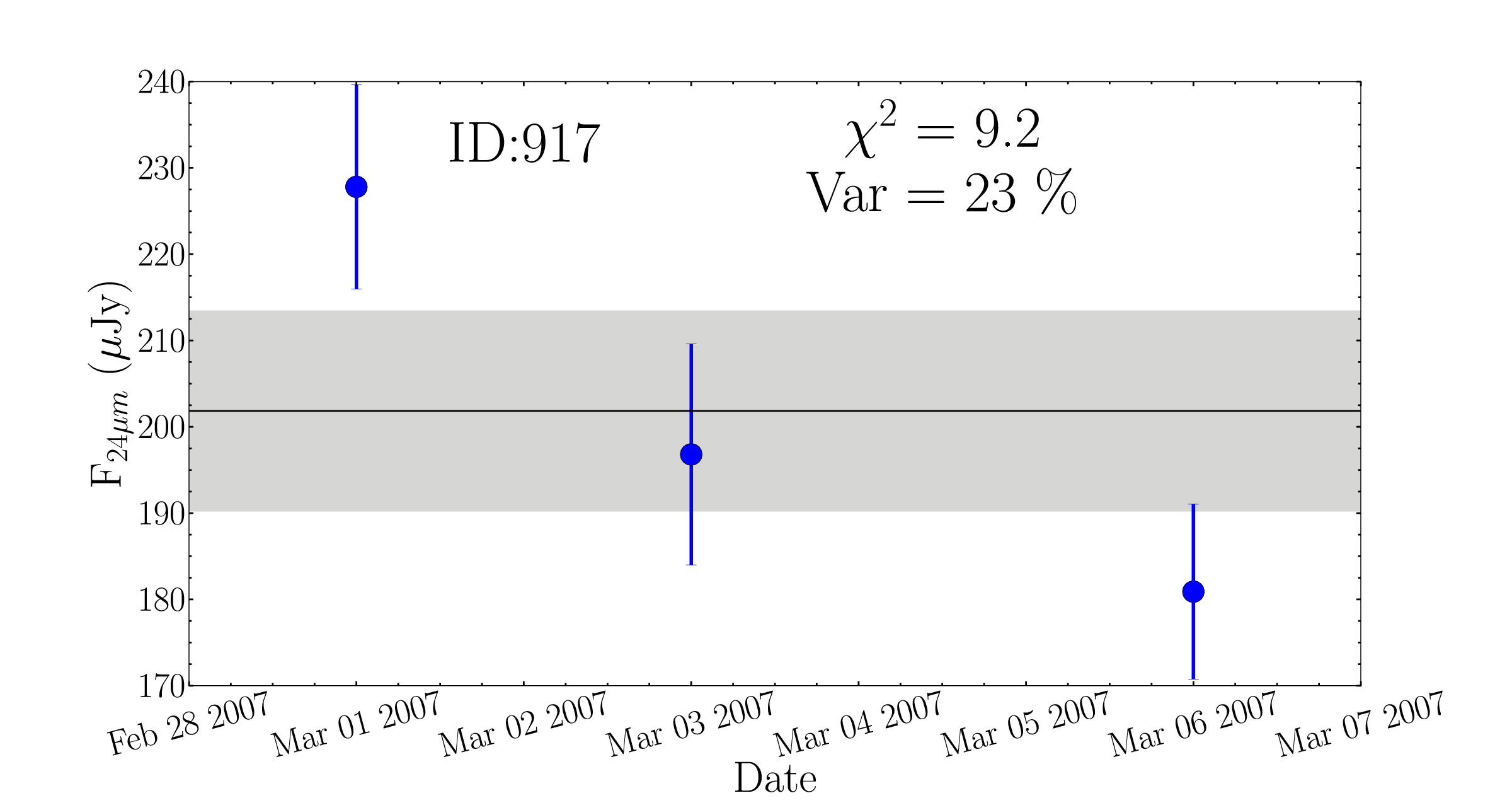}}{\hspace{0cm}}

      \contcaption{} 
      \label{curvas-luz-corto3}
    \end{center}
  \end{minipage}
\end{figure*}

\end{document}